\documentclass[12pt]{article}
\usepackage{slashed}
\usepackage{mathrsfs}
\makeatletter \@addtoreset{equation}{section} \makeatother
\renewcommand{\theequation}{\thesection.\arabic{equation}}
\newcommand{\ba}{\begin{array}}
\newcommand{\ea}{\end{array}}
\newcommand{\beq}{\begin{equation}}
\newcommand{\eeq}{\end{equation}}
\newcommand{\bea}{\begin{eqnarray}}
\newcommand{\eea}{\end{eqnarray}}

\def\bce{\begin{center}}
\def\ece{\end{center}}

\def\nonu{\nonumber}

\def\pa{\partial}
\def\al{\alpha}
\def\be{\beta}
\def\ga{\gamma}

\def\de{\delta}

\def\la{\lambda}

\def\si{\sigma}

\def\eps6{{\displaystyle \mathop{\epsilon}^{6}}{}}
\def\g6{{\displaystyle \mathop{g}^{6}}{}}

\def\nab6{{\displaystyle \mathop{\nabla}^{6}}{}}
\def\to{\rightarrow}

\def\0{{\sst{(0)}}}
\def\1{{\sst{(1)}}}
\def\2{{\sst{(2)}}}
\def\3{{\sst{(3)}}}
\def\4{{\sst{(4)}}}
\def\5{{\sst{(5)}}}
\def\6{{\sst{(6)}}}
\def\7{{\sst{(7)}}}
\def\8{{\sst{(8)}}}

\def\O #1{\overline{#1}}

\def\ba{\begin{array}}
\def\ea{\end{array}}
\def\beq{\begin{equation}}
\def\eeq{\end{equation}}
\def\be{\begin{equation}}
\def\ee{\end{equation}}

\def\to{\rightarrow}

\def\la{\lambda}
\def\eps{\epsilon}

\def\O #1{\overline{#1}}

\def\ba{\begin{array}}
\def\ea{\end{array}}
\def\beq{\begin{equation}}
\def\eeq{\end{equation}}
\def\be{\begin{equation}}
\def\ee{\end{equation}}

\def\to{\rightarrow}

\def\la{\lambda}
\def\eps{\epsilon}

\def\eps6{{\displaystyle \mathop{\epsilon}^{6}}{}}

\def\nab6{{\displaystyle \mathop{\nabla}^{6}}{}}

\newcommand{\bean}{\begin{eqnarray*}}
\newcommand{\eean}{\end{eqnarray*}}

\def\O{{\cal O}}
\usepackage{amsmath}

\usepackage{kotex}

\usepackage{amsmath}
\usepackage{graphicx}
\usepackage{amssymb}
\usepackage{amsthm}
\usepackage{grffile}
\usepackage{simplewick}
\usepackage{tikz}

\usepackage{nccmath}
\usepackage{physics}
\usepackage{wrapfig}
\usepackage{mathtools}
\usepackage{stackengine}

\usetikzlibrary{intersections,decorations.markings}

\allowdisplaybreaks

\begin{document}
\thispagestyle{empty} \addtocounter{page}{-1}
   \begin{flushright}
\end{flushright}

\vspace*{1.3cm}
  
\centerline{ \Large \bf
  Toward }
\vspace*{0.5cm}
\centerline{ \Large \bf
A Celestial Soft Symmetry Algebra 
in  the ${\cal N}=8$  Supergravity }
\vspace*{1.5cm}
\centerline{ {\bf  Changhyun Ahn$^\dagger$}
  and {\bf Man Hea Kim$^{\ddagger,\ast}$}
} 
\vspace*{1.0cm} 
\centerline{\it 
$\dagger$ Department of Physics, Kyungpook National University, Taegu
41566, Korea}
\vspace*{0.3cm}
\centerline{\it 
  $\ddagger$ Center for High Energy
  Physics, Kyungpook National University, Taegu
41566, Korea} 
\vspace*{0.3cm}
\centerline{\it 
$\ast$
Department of Physics Education,  Sunchon National University,
Sunchon 57922, Korea }
\vspace*{0.8cm} 
\centerline{\tt ahn@knu.ac.kr,
manhea.kim10000@gmail.com
} 
\vskip2cm

\centerline{\bf Abstract}
\vspace*{0.5cm}

From the classical
$SO({\cal N}=8)$ extended superconformal algebra
between the lowest ${\cal N}=8$ multiplet  in two dimensions
obtained by Ademollo et al. (1976),
we generalize it for the arbitrary ${\cal N}=8$ multiplet
with manifest $SU(8)$ symmetry containing the bosonic
$w_{1+\infty}$ algebra.
By modifying this
${\cal N}=8$ supersymmetric
$w_{1+\infty}$ algebra in two dimensions,
we propose 
a consistent
celestial soft current algebra between the graviton, the gravitinos,
the graviphotons, the graviphotinos, and
the scalars
in the ${\cal N}=8$ supergravity theory with
$SO(8)$ (or $SU(8)$) global symmetry in four dimensions
initiated by
de Wit and Freedman (at Stony Brook in 1977).
The twenty five couplings in this celestial algebra
can be written in terms of
eight arbitrary couplings via the Jacobi identity.

\vspace*{2cm}

\baselineskip=18pt
\newpage
\renewcommand{\theequation}
{\arabic{section}\mbox{.}\arabic{equation}}

\tableofcontents

\section{ Introduction
}

The extended superconformal algebra in two dimensions is a
generalization of the Virasoro algebra
by adding the additional bosonic and fermionic
currents to the bosonic spin $2$ current
which is a generator of the Virasoro algebra. 
So far, the number of supersymmetry $\cal N$ is less than or equal to
four. When the number of supersymmetry is greater than four,
the spin for the current can be negative.
This will lead to the nonunitarity of the representation
in the extended superconformal algebra.
Nevertheless, we consider the case where the number of supersymmetry
is given by eight. That is ${\cal N}=8$.
Some components of the lowest and next ${\cal N}=8$ multiplets  
can have negative spins but the remaining components
of any ${\cal N}=8$ multiplets have nonnegative spins.
For any number of supersymmetry, the extended superconformal algebra
is found in \cite{Ademolloplb}.
The main feature of this algebra is that
when we write down the (anti)commutators
in terms of the corresponding operator product expansions
(OPEs), the singular behaviors
of these OPEs between the lowest
${\cal N}=8$ multiplets
consist of the second and the first order poles
only.
Therefore, it is natural to generalize the work of
\cite{Ademolloplb} by keeping these pole structures
and increasing the spins appearing in both sides
of the OPEs at the same time in order to
have the general OPEs between any ${\cal N}=8$ multiplets. 

For example,
the OPE between
the gravitons
in the Einstein gravity \cite{PRSY,GHPS}, where the three point vertex
has the scaling dimension five,
is given by the simple pole with the
gravitational coupling constant in the holomorphic
limit, the Euler beta function and antiholomorphic descendants.
Then a family of conformally soft graviton current is defined
in order to remove all the conformally soft poles appearing in the
above OPE. The left and right weights  depend on
conformal dimension which has $1, 0, -1, \cdots$ and the previous
Euler beta function becomes the binomial coefficients.
Moreover, the infinite sum from the expansion of antiholomorphic
coordinates reduces to the finite sum where the upper bound
depends on the conformal dimension of the first graviton on the
left hand side of the OPE.
By performing four different contour integrals over
the four complex coordinates (each holomorphic and antiholomorphic
coordinates)
appearing on the left hand side of the
OPE, the corresponding commutator is obtained and the right hand side
contains the complicated mode dependent fractional expression
in addition to the linear terms in the modes.
In \cite{Strominger}, by absorbing the above fractional form
into the new soft graviton where both
the mode and the conformal dimension dependent factorial functions appear,
the commutator between the soft graviton becomes
very simple form and the right hand side contains
only linear term in the two modes. The (redefined)
right weight of soft graviton
has the positive half integral values, $\frac{3}{2}, 2, \frac{5}{2},
\cdots$.
From the constraint on the mode for the soft graviton,
the above commutator is nothing but the wedge subalgebra
of $w_{1+\infty}$ algebra \cite{Bakas}.
In other words, the modes for the soft graviton
are not arbitrary integer but constrained and the maximum number
of the mode is given by the above right weight
for the soft graviton minus one.

For the (minimal) ${\cal N}=1$
supersymmetric Einstein-Yang-Mills theory,
the OPEs between
the gravitons, the gravitinos, the gluons, and the gluinos
are given in \cite{FSTZ} explicitly.
The three point vertex
has the scaling dimension four between the gluons, the gluinos
and the gluinos while
the three point vertex has the scaling dimension five
as before for the remaining three OPEs.
On the right hand sides of the
OPEs there are the simple pole in the holomorphic
limit, the Euler beta function and antiholomorphic descendants
as before.
For example,
a family of conformally soft gravitino current can be defined
in order to remove all the conformally soft poles appearing in the
OPE between the gravitons and the gravitinos \cite{Ahn2111}.
The left and right weights  for the gravitinos depend on
conformal dimension which has $\frac{1}{2}, -\frac{1}{2},
-\frac{3}{2}, \cdots$ and the previous
Euler beta function becomes the binomial coefficients as before.
The finite sum is obtained and  the upper bound
depends on the conformal dimension of the first graviton on the
left hand side of the OPE as before.
By using four different contour integrals over
the four complex coordinates
appearing on the left hand side of the
OPE, the corresponding commutator can be obtained and
again the right hand side
contains the complicated mode dependent fractional function
besides the linear terms in the modes.
As in \cite{Strominger}, by absorbing the above fractional function
into the new soft graviton and gravitino,
the commutator between the soft graviton and gravitino becomes
simple form and the right hand side contains
only linear term in the two modes from the
graviton and the gravitino. The (redefined) right weight of soft gravitino
has the positive half integral values,
$\frac{3}{2}, 2, \frac{5}{2}, \cdots$.
It turns out that 
the two commutators between the
graviton and the gravitino are nothing but the wedge subalgebra
of ${\cal N}=1$ supersymmetric topological
$w_{\infty}$ algebra \cite{PRSS}.
In other words, the modes for the soft gravitino
are not arbitrary half integer but the maximum number
of the mode is given by the above right weight
for the soft gravitino minus one
\footnote{There exists the nontrivial OPE
between the gluinos having positive and negative helicities
in the holomorphic limit (on the right hand side, the graviton
appears). There is the nontrivial OPE
between the gravitinos having opposite helicities
in the holomorphic limit also (on the right hand side, the graviton
appears).
Moreover, the OPE between the gravitino having positive helicity
and the gluinos having negative helicity
has nontrivial singular behavior also
(on the right hand side, the gluons appear). All of these
have the scaling dimension for the three point vertex of five.
The left and right weights and the helicities
appearing on the left hand sides of the OPEs
are encoded on the right hand sides appropriately
and the role of the dimension for the three point vertex
is crucial even though this is fixed by five.
In particular, the second OPE here will be related to
the results of this paper.}.

Along the line of \cite{FSTZ}, the 
OPEs between
the gravitons, the gravitinos, the graviphotons,
the graviphotinos and the scalars
in the (maximal) ${\cal N}=8$ supergravity theory
\cite{dF} (See also \cite{CJnpb,CJplb}), where the three point vertex
has the scaling dimension five,
are given by the simple pole in the holomorphic
limit, the Euler beta function and
antiholomorphic descendants \cite{2212-2} as before.
What they did is
\begin{itemize}
\item[]  
i) to compute the various split factors
in ${\cal N}=8$ supergravity from the informations on
the split factors
from the two types of ${\cal N}=4$ super Yang-Mills theories
\cite{BSS},

ii) to compute the amplitudes to two collinear particles and

iii)to apply the Mellin transform to the standard momentum
space amplitudes by using the work of \cite{PRSY}.
\end{itemize}
The main work of \cite{2212-1} is to arrange the two types of
$SU(4)_R$ R symmetries correctly and to identify
the factorizations of the states in the
${\cal N}=8$ supergravity in terms of the states (gluons, gluinos and
scalars) in the
${\cal N}=4$ super Yang-Mills theories together with
two $SU(4)$ indices.
For example,
a family of conformally soft graviphoton current can be defined
in order to remove all the conformally soft poles appearing in the
OPE between the gravitons and the graviphotons.
The left and right weights  for the graviphotons depend on
conformal dimension which has $0, -1, -2, \cdots$ and the previous
Euler beta function becomes the binomial coefficients as before.
Similarly,
the soft graviphotinos contain all the  soft poles
$-\frac{1}{2}, -\frac{3}{2}, -\frac{5}{2}, \cdots$
appearing in the Euler beta function of the OPE between the
graviton and the graviphotinos.
Finally,
the soft scalars contain all the  soft poles
$-1, -2, -3,  \cdots$
appearing in the Euler beta function of the OPE between the
graviton and the scalars.
Therefore,  the conformal dimensions
$0, -\frac{1}{2}, -1$ from the
first elements in these infinite sequences generate
the soft symmetries corresponding to
the leading soft graviphotons, graviphotinos, and
the scalars \cite{Tropper,2212-2}.
In principle, we can do similar analysis done in
\cite{GHPS} in order to obtain the explicit celestial soft symmetry
algebra in the ${\cal N}=8$ supergravity.

In this paper,
we would like to construct the
candidate of celestial soft current algebra
in the ${\cal N}=8$ supergravity.
Our starting point is based on the earlier work of
\cite{Ademolloplb} which allows us to describe
any number of supersymmetry and reveals the linear algebra
(in other words, there are no quadratic or higher order terms
in the operators on the right hand sides).
Although this extended
superconformal algebra implies a classical algebra with
vanishing central charges, from the point of view in the
celestial holography, this is not a problem.
By noticing that the antisymmetric representations
of $SO(8)$ are given by ${\bf 1, 8_s, 28, 56_s, 35_c, 35_v, 56_s, 28,
8_s,1}$
while
those of $SU(8)$ are given by
${\bf 1, 8, 28, 56, 70, \overline{56}, \overline{28}, \overline{8}, 1}$
and the fields with $SO(8)$ indices are used to
present the chiral $SU(8)$ using the
projectors $\frac{1}{2}(1 \pm \gamma_5)$ together
with $\gamma_5$ matrix
with consistent lowest order Lagrangian and supersymmetry
transformation rules in \cite{deWit},
it is possible to obtain the extended superconformal algebra
with $SU(8)$ rather than $SO(8)$.
Then the upper indices and the lower indices
indicate that the field transforms according to
the fundamental ${\bf 8}$ and the antifundamental
$\overline{\bf 8}$ representations respectively.
According to the previous paragraphs,
the (redefined) right weights  for the graviton, the gravitinos,
the graviphotons, the graviphotinos and the scalars
having nonnegative helicities
are positive half integers.
This implies that the extended $SU(8)$ superconformal algebra
should exist for any ${\cal N}=8$ multiplets including the
lowest one.
As in the first paragraph,
we can write down 
the extended $SU(8)$ superconformal algebra
for any  ${\cal N}=8$ multiplets without destroying
the structure of the poles.
By modifying this algebra, that is,
\begin{itemize}
\item[]  
i)
in the OPEs, we should consider only the second order poles
(there are no contributions from the first order poles
and the descendants from the second order poles
are still present in  the first order poles) and

ii) there should exist an additional commutator
which vanishes when we fix the algebra for
the lowest ${\cal N}=8$ multiplet,
\end{itemize}
we will determine the final celestial
twenty five (anti)commutators which provide the
consistent and plausible candidate symmetry algebra
in the ${\cal N}=8$ $SO(8)$ supergravity theory
\footnote{In \cite{Prabhu},
the ${\cal N}=1$ superconformal algebra
is reproduced from the Lie superalgebra based on the BMS symmetries
\cite{BvM,Sachs}. This is one of the motivations of this paper.}.

In section $2$,
the $SO({\cal N})$ superconformal algebra is reviewed.
For ${\cal N}=8$, the explicit algebra is presented.
In section $3$,
the previous $SO(8)$ symmetry is generalized to
the $SU(8)$ symmetry.
The $SU(8)$ superconformal algebra is described
for general ${\cal N}=8$ multiplets.
In section $4$,
the consistent and plausible celestial soft symmetry algebra
is obtained.
In section $5$,
based on the Lagrangian of the
${\cal N}=8$ supergravity, we explain
how the previous candidate celestial algebra appears.
The Jacobi identity for the couplings
is described.
In section $6$,
we summarize what we have obtained in this paper
and the open problems related to the results of
this paper are described.
In Appendices, $A, \cdots, I$,
some details for the previous sections are described.
For the celestial holography,
see the previous works in \cite{Raclariu,Pasterski,PPR,Donnay}
together with \cite{Strominger1703}.
The Thielemans package \cite{Thielemans} is used
together with a mathematica \cite{mathematica} all
the times in this paper.

In the remaining parts of
this section, we explain the holographic interpretation of
{\it two} dimensional conformal correlator (from which
the OPE can be read off)
on the celestial two sphere
at null infinity related to the descriptions in sections $3$ and $4$
as the {\it four} dimensional S matrix (whose elements are
the scattering amplitudes) view point (and vice versa)
which is known as a celestial holography.
It is known that the celestial OPEs are encoded in the scattering
amplitudes by taking the collinear limit of
two particles. For example, see also \cite{MPR} for the
review of scattering amplitudes.

Let us first consider the bosonic case.
In \cite{GHPS}, the OPE between the two conformal primary (pure) gluons
was reproduced by i) considering four gluon MHV(Maximally Helicity
Violating) amplitude in momentum space in the presence of
the  delta function and the structure constants
of the group, ii) taking the holomorphic
collinear limit for the positive helicity gluons with the change of
variables for the two energies of collinear particles
(denoted by particle $3$ and the particle $4$), iii) using
the Mellin transform to this momentum space amplitude
and iv)
interpreting this {\it four} dimensional
{\it celestial} amplitude for the specific
helicities as the {\it two} dimensional (celestial) correlation
function of four conformal primary operators on the celestial
sphere. Moreover, the same OPE from
the behavior of more general
$(n+1)$ gluon amplitudes (when $n$ is fixed as $n=3$,
then the previous case can be reproduced)
with arbitrary general helicities
in momentum space
was obtained by using similar procedure.
See also \cite{HPS} for the expression of
$(n+1)$-point amplitudes (including fermions)
in terms of i) the two energies
of the collinear two particles, ii) the sum of these two
energies, iii) their two complex coordinates, iv) their
helicities as well as the helicity
of the third particle (in the three-point amplitude
with two collinear particles), and
v) $n$-point amplitudes
in momentum space.

The MHV $n$ graviton amplitude
without the delta function
in momentum space was given by \cite{EF} where
the $n=4$ (and $n=5$) formulas provided
the product of two four-point (and five-points) gluon amplitudes
\cite{PT}
respectively and the KLT(Kawai, Lewellen and Tye) relations \cite{KLT}
were checked.
The MHV three graviton amplitude depends on
the three energies and three complex coordinates
(and similarly the corresponding four graviton amplitude
does on four energies and four complex coordinates) \cite{Puhm}.
By taking the {\it soft limit} (the energy of the soft particle
approaches to zero)
of the fourth graviton on the
corresponding two celestial amplitudes \cite{ST} after using the
Mellin transform as before,
the conformally {\it soft} factorization of celestial amplitudes for
gravity was shown in \cite{Puhm}.
This is the celestial version of the Weinberg's soft theorem
for the gravity \cite{Weinberg}.
On the other hand,
in \cite{BDPR}, the (pure) gravity split factors (or splitting amplitudes)
were determined by the various
gluon split factors with the help of four-point and five-point
KLT relations.
Here the $(n+1)$-point graviton amplitude ($n \geq 4$) at the tree level
(and any loop order), under the holomorphic collinear limit,
factorizes into a product
of graviton split factor and lower 
$n$-point graviton amplitude. See also \cite{ADHPZ} for the checks
at the loop levels.
Moreover, in \cite{PRSY},
after the Mellin transform to the momentum space amplitude
(by restricting to the pure gravity case corresponding to pure
gluons)
was performed, the celestial amplitude as the (celestial) correlation
function of $(n+1)$ conformal primary operators provided
the OPE between the two conformal primary gravitons.
See also the split factors in the first row of Table
\ref{Split1}. See also \cite{FSTZ} for these split factors
in the context of ${\cal N}=1$ Einstein Yang-Mills theory.

Secondly, let us consider the supersymmetric case.
Because a direct computation, which is challenging problem,
using the Feynman rules for the (super)gravity
leads to a massive number (and extreme complexity) of diagrams
as we increase the number of external particles,
the above KLT relations with a double copy construction
provide the computationally  tractable way to construct
the (super)gravity amplitudes from simpler gauge theory components
with the supersymmetric Ward identities \cite{Dixon9601} which 
relate to all the different amplitudes
in modern amplitude calculations for ${\cal N}=8$ supergravity
\footnote{
\label{threefour}
For example, the four-point
scalar amplitude \cite{KLR} at the tree level
in ${\cal N}=8$ supergravity  was obtained
from the corresponding terms of the explicit Lagrangian
\cite{dF,CJnpb,CJplb} as one of the two methods.
This result as the other
of the two methods was also determined by using the method of
\cite{BEF} with the help of tensor product of the
various representation
of $SU(8)$. The {\it soft limit} analysis for the
three kinds of amplitudes described in \cite{KLR}
was determined in \cite{Liu}. Moreover, the four-point
graviphoton amplitude can be written in terms of the sum of three-point
amplitudes. For the collinear
particles three and four, the
corresponding commutator producing the scalars is
associated with the equation (\ref{eq:photon_photon}).
There exists the four-point amplitude
of graviphoton-graviphoton-scalar-scalar which is also given by
the sum of three-point amplitudes.
For the collinear particle one and particle two
which leads  to the graviton,
the commutators are given by (\ref{eq:photon_photon}) again.
When the particle two and particle three are collinear, then
the
corresponding commutator producing the graviphotons is
given by the equation (\ref{eq:photon_scalar}).}.

On the other hand, according to the
construction of \cite{HPS},
the splitting amplitude in momentum space can be determined
entirely from the interaction vertex
between the three particles.
In other words, the splitting amplitude is
nothing but  the three-point amplitude \cite{AHHH}
multiplied with some kinematic factors (i.e.,
the coupling with the square brackets). The three helicities
of three particles (the helicity of the third
particle depends on the other two helicities and
the scaling dimension of three-point vertex)
and two energies of two {\it collinear} particles (the third energy
is the sum of other two energies)
appear in the splitting amplitude \footnote{
\label{coincident}
There exists the factor
$\omega_1^{s_2-s_I-1}\, \omega_2^{s_1-s_I-1}\, (\omega_1+\omega_2)^{s_I}$
where $\omega_1 (s_1)$ and $\omega_2 (s_2)$
are the energies (the helicities) of two collinear
particles. The $s_I$ is equal to $s_3$ in later sections.
All the twenty five split factors appearing in Tables
\ref{Split1}-\ref{Split5} in Appendix $B$ contain this factor
by exhausting all the possible helicities.
Note that the first power $(s_2-s_I-1)$ and the second power
$(s_1-s_I-1)$ appear in the first and second arguments
of the Euler beta function in (\ref{OOope1})
after the Mellin transform.
The sum of these three powers $(s_1+s_2-s_I-2)$, in general, enters in
the conformal dimension of the right hand side of (\ref{OOope1})
but in this paper $(s_1+s_2-s_I-2)=0$.}. 
On the other hand, in \cite{BDPR}, the tree level gravity
splitting amplitude in ${\cal N}=8$ supergravity
can be obtained  by the gauge theory
splitting amplitudes in ${\cal N}=4$ super Yang-Mills theory
under the {\it collinear limit} on the five-point gravity amplitude
from KLT relations
as described before.
See also the footnote \ref{productofsplit} for their explicit
relations for the splitting amplitudes.
It turns out that
the two approaches for the splitting amplitude
from \cite{HPS} and \cite{BDPR} are consistent with each other
according to the footnote \ref{coincident}.

\begin{itemize}
\item[]
We are using the second approach
in section $5$ (i.e., the generalization of
\cite{PRSY} to the ${\cal N}=8$ supergravity which implies
i) {\it indirect} amplitude consideration and ii) four dimensional
amplitude-two dimensional celestial CFT
correspondence)
all the times because the
$SU(8)_R$ $R$ symmetry can be determined automatically
by the corresponding
symmetries from the two ${\cal N}=4$ super Yang-Mills theories.
In this sense, although the explicit amplitude computations
(from the Lagrangian of ${\cal N}=8$ supergravity) are
not performed explicitly in this paper,
we expect that the celestial soft symmetry algebra
dual to the four dimensional
${\cal N}=8$ supergravity
should {\it contain} the algebra found in section $4$.
\end{itemize}

On the other hand,
recently, the {\it full} celestial OPE by considering the multiparticle
exchanges was
described in \cite{BHP,GHP}
in the context of the full $n$-point tree level gluon amplitude.
See also \cite{Pate} on the role of two particle
operators for the different perspective in the gravity theory.
Therefore, our construction in this paper
will provide a useful template for testing the full celestial
soft symmetry algebra once the {\it direct} amplitude calculations
become available in the future
\footnote{We thank the referee for pointing this out.
We thank L. Dixon and A. Tropper for the discussion on these  
last seven paragraphs of this section. }.

In the first paragraph of this section, there are operators
having the negative conformal spins in the lowest and next ${\cal N}=8$
multiplets \footnote{The known example of {\it nonunitary} two dimensional
conformal field theory is given by the Yang-Lee edge singularity
\cite{YangLee,YangLee1} which contains a primary field with a negative
conformal left-weight $-\frac{1}{5}$ and right-weight $-\frac{1}{5}$
(or scaling dimension
$-\frac{2}{5}$ which is the sum of
these weights) \cite{Cardy}. See also the relevant work
\cite{Fisher} for the critical
phenomenon.}.
The former contains four operators with spins $-2,-\frac{3}{2}, -1$
and $-\frac{1}{2}$ and the latter contains
two operators with spins $-1$ and $-\frac{1}{2}$.
See also (\ref{bigPhi}).
The properties of these operators in the {\it unitary} celestial CFT
are described in the
footnote \ref{genuineornot}.
It turns out that they appear in (\ref{Negative}) and (\ref{Negative1}).
There exist the critical values for the superscripts in the
components of ${\cal N}=8$ multiplets where the corresponding
algebra between the operators with nonnegative spins is closed
for the possible superscripts in (\ref{fieldcontents})
in the sense that the above
six operators with negative spins do not appear
in this algebra.

\section{ The  $SO(8)$ superconformal algebra}

\subsection{ The  $SO({\cal N})$ superconformal algebra}

The classical $O({\cal N})$ extended superconformal algebra
is described as
\bea
&& \big[ O^{i_1 \cdots i_R}_m, O^{j_1 \cdots j_S}_n \big\}
=
\label{Ademollo}
\\
&& \mathrm{i}^{-RS}\bigg(
\big(m(2-S)-n(2-R) \big)\,O^{i_1 \cdots i_R\, j_1 \cdots j_S}_{m+n}
-\mathrm{i} \sum_{h=1}^{R} \sum_{k=1}^{S}(-1)^{h+k+S}\delta^{i_h,j_k}\,
O^{i_1 \cdots \hat{i}_h \cdots i_R\, j_1 \cdots \hat{j}_k \cdots j_S}_{m+n}
\bigg).
\nonu
\eea
The generator $O^{i_1 \cdots i_R}_m$
with the Laurent mode $m$
is a completely antisymmetric tensor of the global
$O({\cal N})$ group. The indices $i_1, \cdots, i_R$
are $O({\cal N})$ vector indices and the number of indices
$R$ in this generator is less than or equal to ${\cal N}$.
The conformal weight is $(2-\frac{R}{2})$.
For even $R$, the generator is bosonic while
for odd $R$, the generator is fermionic.
The total number of bosonic generators is given by
$\sum_{i=0}^{{\cal N}}\, \binom{\cal N}{i}=2^{{\cal N}-1}$
and similarly the total number of fermionic generators is given by
$\sum_{i=1}^{{\cal N}}\, \binom{\cal N}{i}=2^{{\cal N}-1}$.
The left hand side of (\ref{Ademollo})
becomes a commutator for even $RS$
and an anticommutator for odd $RS$.
The hatted indices on the double summation imply that
these indices without the hats are omitted.


The OPE corresponding to (\ref{Ademollo})
can be written as 
\bea
O^{i_1 \cdots i_R}(z) \, O^{j_1 \cdots j_S}(w)&=
&-\frac{\mathrm{i}^{-RS}}{(z-w)^2}(R+S-4)\,O^{i_1 \cdots i_R\, j_1 \cdots j_S}(w)
\nonu \\
&- & \frac{1}{(z-w)} \Bigg[ \mathrm{i}^{-RS}\,(R-2)\,
\partial O^{i_1 \cdots i_R\, j_1 \cdots j_S}
\label{Ademolloope}
\\
&+&   \mathrm{i}^{-RS+1}
\sum_{h=1}^{R} \sum_{k=1}^{S}(-1)^{h+k+S}\delta^{i_h,j_k}\,
O^{i_1 \cdots \hat{i}_h \cdots i_R\, j_1 \cdots \hat{j}_k \cdots j_S}
\Bigg](w)+\cdots\,.
\nonu
\eea
The above (anti)commutator (\ref{Ademollo})
can be obtained from (\ref{Ademolloope})
by using the standard procedure in the
two dimensional conformal field theory \footnote{
The Laurent mode can be described as the following
closed  contour integral
\bea
&O^{i_1 \cdots i_R}_m=\frac{1}{2\pi \mathrm{i}}\oint
d z
z^{m+1-\frac{R}{2}} O^{i_1 \cdots i_R}(z)\,.
\label{contour}
\eea
By substituting (\ref{contour}) into the
left hand side of (\ref{Ademollo}), we obtain
\bea
&& \big[ O^{i_1 \cdots i_R}_m, O^{j_1 \cdots j_S}_n \big\}
=\frac{1}{2\pi \mathrm{i}}\oint_0 w^{n+1-\frac{S}{2}}
\frac{1}{2\pi \mathrm{i}}\oint_w z^{m+1-\frac{R}{2}}
O^{i_1 \cdots i_R}(z)\, O^{j_1 \cdots j_S}(w),
\nonu
\eea
which becomes
the right hand side of (\ref{Ademollo})
after using the OPE (\ref{Ademolloope}).
%
}.


\subsection{ The  $SO(8)$ superconformal algebra}

\enlargethispage{3pt}
We focus on the ${\cal N}=8$ case
\footnote{In Appendix $A$, the lower supersymmetric cases are
described also.}.
By substituting the ${\cal N}=8$ into
(\ref{Ademollo}), we obtain the following twenty nine
(anti)commutators with antisymmetric notations 
\footnote{\label{Anti}
The antisymmetrizing $n$ indices in (\ref{SO8}) is defined by
summing over all permutations of the indices
times the sign of each permutation. Due to the $n!$
permutations, the overall factor $\frac{1}{n!}$ is also
introduced.}
\bea
\comm{O_m}{O_n}&=& 2(m-n)\,O_{m+n}
\,,
\nonu \\
\comm{O_m}{O^{A}_r}&=& (m-2r)\,O^{A}_{m+r}
\,,
\nonu \\
\comm{O_m}{O^{AB}_n}&=& -2n\,O^{AB}_{m+n}
\,,
\nonu \\
\comm{O_m}{O^{ABC}_r}&=& (-m-2r)\,O^{ABC}_{m+r}
\,,
\nonu \\
\comm{O_m}{O^{ABCD}_n}&=& 2(-m-n)\,O^{ABCD}_{m+n}
\,,
\nonu \\
\comm{O_m}{O^{ABCDE}_r}&=& (-3m-2r)\,O^{ABCDE}_{m+r}
\,,
\nonu \\
\comm{O_m}{O^{ABCDEF}_n}&=& 2(-2m-n)\,O^{ABCDEF}_{m+n}
\,,
\nonu \\
\comm{O_m}{O^{ABCDEFG}_r}&=& (-5m-2r)\,O^{ABCDEFG}_{m+r}
\,,
\nonu \\
\comm{O_m}{O^{ABCDEFGH}_n}&=& 2(-3m-n)\,O^{ABCDEFGH}_{m+n}
\,,
\nonu \\
\acomm{O^{A}_r}{O^{B}_s}&=&
\delta^{AB}\,O_{r+s}-\mathrm{i}(r-s)\,O^{AB}_{r+s}
\,,
\nonu \\
\comm{O^{A}_r}{O^{BC}_m}&=&
\mathrm{i}\,2\,\delta^{A[B}O^{C]}_{r+m}+m\,O^{ABC}_{r+m}
\,,
\nonu \\
\acomm{O^{A}_r}{O^{BCD}_s}&=&
-3\,\delta^{A[B}O_{r+s}^{CD]}-\mathrm{i}(r+s)\,O^{ABCD}_{r+s}
\,,
\nonu \\
\comm{O^{A}_r}{O^{BCDE}_m}&=&
-4\,\mathrm{i}\,\delta^{A[B}O^{CDE]}_{r+m}-(2r+m)\,O^{ABCDE}_{r+m}
\,,
\nonu \\
\acomm{O^{A}_r}{O^{BCDEF}_s}&=&
5\,\delta^{A[B}O_{r+s}^{CDEF]}+\mathrm{i}(3r+s)\,O^{ABCDEF}_{r+s}
\,,
\nonu \\
\comm{O^{A}_r}{O^{BCDEFG}_m}&=&
6\,\mathrm{i}\,\delta^{A[B}O^{CDEFG]}_{r+m}+(4r+m)\,O^{ABCDEFG}_{r+m}
\,,
\nonu \\
\acomm{O^{A}_r}{O^{BCDEFGH}_s}&=&
-7\,\delta^{A[B}O_{r+s}^{CDEFGH]}-\mathrm{i}(5r+s)\,O^{ABCDEFGH}_{r+s}
\,,
\nonu \\
\comm{O^{A}_r}{O^{BCDEFGHI}_m}&=&
-8\,\mathrm{i}\,\delta^{A[B}O^{CDEFGHI]}_{r+m}
\,,
\nonu \\
\comm{O^{AB}_m}{O^{CD}_n}&=&
-\mathrm{i}\,\delta^{AC}\,O^{BD}_{m+n}+\mathrm{i}\,\delta^{AD}\,O^{BC}_{m+n}+\mathrm{i}
\,\delta^{BC}\,O^{AD}_{m+n}-\mathrm{i}\,\delta^{BD}\,O^{AC}_{m+n}
\,,
\nonu \\
\comm{O^{AB}_m}{O^{CDE}_r}&=&
-3\,\mathrm{i}\,\Big(\delta^{A[C}O^{DE]B}_{m+r}-\delta^{B[C}O^{DE]A}_{m+r}\Big)
+m\,O^{ABCDE}_{m+r}
\,,
\nonu \\
\comm{O^{AB}_m}{O^{CDEF}_n}&=&
4\,\mathrm{i}\,\Big(\delta^{A[C}O^{DEF]B}_{m+n}
-\delta^{B[C}O^{DEF]A}_{m+n}\Big)
-2m\,O^{ABCDEF}_{m+n}
\,,
\nonu \\
\comm{O^{AB}_m}{O^{CDEFG}_r}&=&
-5\,\mathrm{i}\,\Big(\delta^{A[C}O^{DEFG]B}_{m+r}-\delta^{B[C}O^{DEFG]A}_{m+r}\Big)
+3m\,O^{ABCDEFG}_{m+r}
\,,
\nonu \\
\comm{O^{AB}_m}{O^{CDEFGH}_n}&=&
6\,\mathrm{i}\,\Big(\delta^{A[C}O^{DEFGH]B}_{m+n}-\delta^{B[C}O^{DEFGH]A}_{m+n}\Big)
-4m\,O^{ABCDEFGH}_{m+n}
\,,
\nonu \\
\comm{O^{AB}_m}{O^{CDEFGHI}_r}&=&
-7\,\mathrm{i}\,\Big(\delta^{A[C}O^{DEFGHI]B}_{m+r}-\delta^{B[C}O^{DEFGHI]A}_{m+r}\Big)
\,,
\nonu \\
\acomm{O^{ABC}_r}{O^{DEF}_s}&=&
3\,\Big(\delta^{A[D}O_{r+s}^{EF]BC}-\delta^{B[D}O_{r+s}^{EF]AC}+\delta^{C[D}O_{r+s}^{EF]AB}\Big)\nonu \\
&+ & \mathrm{i}\,(r-s)\,O_{r+s}^{ABCDEF}
\,,
\nonu \\
\comm{O^{ABC}_r}{O^{DEFG}_m}&=&
-4\,\mathrm{i}\,\Big(\delta^{A[D}O_{r+m}^{EFG]BC}-\delta^{B[D}O_{r+m}^{EFG]AC}
+\delta^{C[D}O_{r+m}^{EFG]AB}\Big)
\nonu \\
&-& (2r-m)\,O_{r+m}^{ABCDEFG}
\,,
\nonu \\
\acomm{O^{ABC}_r}{O^{DEFGH}_s}&=&
-5\,\Big(\delta^{A[D}O_{r+s}^{EFGH]BC}-\delta^{B[D}O_{r+s}^{EFGH]AC}+\delta^{C[D}O_{r+s}^{EGH]AB}\Big)
\nonu \\
&-& \mathrm{i}\,(3r-s)\,O_{r+s}^{ABCDEFGH}
\,,
\nonu \\
\comm{O^{ABC}_r}{O^{DEFGHI}_m}&=&
6\,\mathrm{i}\,\Big(\delta^{A[D}O_{r+m}^{EFGHI]BC}-\delta^{B[D}O_{r+m}^{EFGHI]AC}+\delta^{C[D}O_{r+m}^{EFGHI]AB}\Big)
\,,
\nonu \\
\comm{O^{ABCD}_m}{O^{EFGH}_n}&=&
4\,\mathrm{i}\,\Big(\delta^{A[E}O_{m+n}^{FGH]BCD}-\delta^{B[E}O_{m+n}^{FGH]ACD}+\delta^{C[E}O_{m+n}^{FGH]ABD}\nonu \\
& - & \delta^{D[E}O_{m+n}^{FGH]ABC}\Big)
- 2(m-n)\,O_{m+n}^{ABCDEFGH}
\,,
\nonu \\
\comm{O^{ABCD}_m}{O^{EFGHI}_r}&=&
5\,\mathrm{i}\,\Big(\delta^{A[E}O_{m+n}^{FGHI]BCD}-\delta^{B[E}O_{m+n}^{FGHI]ACD}+\delta^{C[E}O_{m+n}^{FGHI]ABD}
\nonu \\
&
-& \delta^{D[E}O_{m+n}^{FGHI]ABC}\Big)
\,,
\label{SO8}
\eea
where the $SO(8)$ vector indices run over
$A,B,C,D, \cdots=1,2, \cdots, 8$.
There are also trivial vanishing (anti)commutators
we do not present here.
We can also present the corresponding OPEs
from the relation in (\ref{Ademolloope}).
The conformal weights for operators having the
$SO(8)$ indices more than five are {\it negative}
from the sixth-the ninth relations in (\ref{SO8}).
The operator $\frac{1}{2}\, O_m$ plays the role of Laurent mode
for the stress energy tensor.
See also Table \ref{firsttable}.
Depending on the number of $SO(8)$ indices
of the left hand sides in (\ref{SO8}), the right hand sides
contain either the first term or the second term in (\ref{Ademollo})
\footnote{
\label{NORM}
Let us explain the normalization appearing on the
right hand sides of (\ref{SO8}).
For example, for the case of
three and four indices on the left hand side,
the mode independent term has the numerical factor $4$.
According to (\ref{Ademollo}), there exist twelve terms
for the double sum. On the right hand side of this particular
case, there are three terms. Each term should contain four terms.
Due to the antisymmetrization appearing in the footnote
\ref{Anti}, there exists an overall factor $\frac{1}{4!}$.
Moreover, for fixed index in the Kronecker delta,
there are $3!$ operators having five indices. Then
the overall factor $4$ can be multipled by this $3!$ and this
leads to the overall weight is given by $1$.
By considering other three cases for the index of Kronecker delta,
there appear three other terms with weight $1$.
Then we are left with four terms as we expected above and
each antisymmetrized Kronecker term has four terms and this
leads to the above twelve terms. Therefore, the numerical factor
$4$ should appear in the corresponding commutator of (\ref{SO8}).
We can analyze
other normalization factors similarly.}.

\section{ The $SU(8)$ superconformal algebra}

\subsection{ The  $SU(8)$ superconformal algebra
with lowest ${\cal N}=8$ multiplet}

By introducing 
\bea
2L & \equiv & O\,,
\qquad
G^{A} \equiv O^{A}\,,
\qquad
T^{AB} \equiv O^{AB}\,,
\qquad
\Gamma^{ABC} \equiv O^{ABC}\,,
\qquad
\Delta^{ABCD} \equiv O^{ABCD}\,,
\nonu \\
\Gamma_{ABC} &\equiv& \frac{1}{5!}\epsilon_{ABCDEFGH}\,O^{DEFGH}\,,
\qquad
T_{AB} \equiv \frac{1}{6!}\epsilon_{ABCDEFGH}\,O^{CDEFGH}\,,
\nonu\\
G_{A} & \equiv & \frac{1}{7!}\epsilon_{ABCDEFGH}\,O^{BCDEFGH}\,,
\qquad
D \equiv \frac{1}{8!}\epsilon_{ABCDEFGH}\,O^{ABCDEFGH}\,,
\label{Nineobject}
\eea
where the Levi Civita $\epsilon_{ABCDEFGH}$ is an $SU(8)$ invariant
tensor,
the above (anti)commutators (\ref{SO8})
can be written in terms of
\footnote{We can analyze the numerical factors
appearing on the right hand sides of (\ref{SU8}). For example,
for the example of the previous footnote \ref{NORM},
we observe that on the right hand side of (\ref{SU8}),
there is a summation over the indices $H,I$ and $J$
leading to $3!$. Then we need to multiply $\frac{1}{3!}$
into the previous coefficient $4$ which leads to $\frac{2}{3}$
as in (\ref{SU8}). We can check other numerical factors correctly.}
\bea
\comm{L_m}{L_n}&=& (m-n)\,L_{m+n}\,,
\nonu    \\
\comm{L_m}{G^{A}_r}&=&(\tfrac{1}{2}m-r)\,G^{A}_{m+r}\,,
\nonu    \\
\comm{L_m}{T^{AB}_n}&=& -n\,T^{AB}_{m+n}\,,
\nonu    \\
\comm{L_m}{\Gamma^{ABC}_r}&=&(-\tfrac{1}{2}m-r)\,\Gamma^{ABC}_{m+r}\,,
\nonu    \\   
\comm{L_m}{\Delta^{ABCD}_n}&=& (-m-n)\,\Delta^{ABCD}_{m+n}\,,
\nonu    \\
\comm{L_m}{\Gamma_{ABC,\,r}}&=& (-\tfrac{3}{2}m-r)\,\Gamma_{ABC,\,m+r}\,,
\nonu   \\
\comm{L_m}{T_{AB,\,n}}&=& (-2m-n)\,T_{AB,\,m+n}\,,
\nonu    \\
\comm{L_m}{G_{A,\,r}}&=&(-\tfrac{5}{2}m-r)\,G_{A,\,m+r}\,,
\nonu    \\
\comm{L_m}{D_n}&=&(-3m-n)\,D_{m+n}\,,
\nonu \\
\acomm{G^A_r}{G^B_s}&=&
\delta^{AB}\,2\,L_{r+s}-\mathrm{i}\,(r-s)\,T^{AB}_{r+s}\,,
\nonu \\
\comm{G^A_r}{T^{BC}_m}&=&
2\,\mathrm{i}\,\delta^{A[B}\,G^{C]}_{r+m}+m\,\Gamma^{ABC}_{r+m}\,,
\nonu \\
\acomm{G^{A}_r}{\Gamma^{BCD}_s}&=&
-3\,\delta^{A[B}\,T^{CD]}_{r+s}
-\mathrm{i}(r+s)\Delta^{ABCD}_{r+s}\,,
\nonu \\
\comm{G^A_r}{\Delta^{BCDE}_m}&=&
-4\,\mathrm{i}\,\delta^{A[B}\,\Gamma^{CDE]}_{r+m}
+(2r+m)\,\frac{1}{3!}\epsilon^{ABCDEFGH}\Gamma_{FGH,\,r+m}\,,
\nonu \\
\acomm{G^A_r}{\Gamma_{BCD,\,s}}&
=& \frac{1}{4!}\delta^{AE}\epsilon_{BCDEFGHI}\Delta^{FGHI}_{r+s}
+\mathrm{i}(3r+s)\,3\,\delta^{A}_{\,\,\,[B}T_{CD],\,r+s}\,,
\nonu \\
\comm{G^A_r}{T_{BC,\,m}}&=&
\mathrm{i}\,\delta^{AD}\,\Gamma_{DBC,\,r+m}
-(4r+m)\,2\,\delta^{A}_{\,\,\,[B}\,G_{C],\,r+m}\,,
\nonu \\
\acomm{G^A_r}{G_{B,\,s}}&=&
\delta^{AC}\,T_{CB,\,r+s}-\mathrm{i}(5r+s)\,\delta^{A}_{\,\,\,B}\,D_{r+s}\,,
\nonu \\
\comm{G^A_r}{D_m}&=&
-\mathrm{i}\,\delta^{AB}\,G_{B,\,r+m}
\,,
\nonu \\
\comm{T^{AB}_m}{T^{CD}_n}&=&
-\mathrm{i}\,\delta^{AC}\,T^{BD}_{m+n}+\mathrm{i}\,\delta^{AD}\,T^{BC}_{m+n}+\mathrm{i}\,\delta^{BC}\,T^{AD}_{m+n}-\mathrm{i}\,\delta^{BD}\,T^{AC}_{m+n}\,,
\nonu \\
\comm{T^{AB}_m}{\Gamma^{CDE}_r}&=&
-3\,\mathrm{i}\,\Big(\delta^{A[C}\Gamma^{DE]B}_{m+r}-\delta^{B[C}\Gamma^{DE]A}_{m+r}\Big)
-m\frac{1}{3!}\epsilon^{ABCDEFGH}\Gamma_{FGH,\,m+r}\,,
\nonu \\
\comm{T^{AB}_m}{\Delta^{CDEF}_n}&=&
4\,\mathrm{i}\,\Big(\delta^{A[C}\Delta^{DEF]B}_{m+n}-\delta^{B[C}\Delta^{DEF]A}_{m+n}\Big)
-m\,\epsilon^{ABCDEFGH}\,T_{GH,\,m+n}\,,
\nonu \\
\comm{T^{AB}_m}{\Gamma_{CDE,\,r}}&=&
-3\,\mathrm{i}\,\Big(\delta^{A}_{\,\,\,[C}\Gamma_{DE]F,m+r}\,
\delta^{FB}-\delta^{B}_{\,\,\,[C}\Gamma_{DE]F,\,m+r}\,\delta^{FA}\Big)
-18m\,\delta^{A}_{\,\,\,[C}G_{D,\,m+r}\delta_{E]}^{\,\,\,B}\,,
\nonu \\
\comm{T^{AB}_m}{T_{CD,\,n}}&=&
-\mathrm{i}\,\delta^{A}_{\,\,\,C}\,\delta^{BE}\,T_{ED,\,m+n}+\mathrm{i}\,\delta^{A}_{\,\,\,D}\,\delta^{BE}\,T_{EC,\,m+n}+\mathrm{i}\,\delta^{B}_{\,\,\,C}\,\delta^{AE}\,T_{ED,\,m+n}
\nonu \\
&-& \mathrm{i}\,\delta^{B}_{\,\,\,D}\,\delta^{AE}\,T_{EC,\,m+n}
-4m
\delta^{A B}_{C D}
\,D_{m+n}\,,
\nonu \\
\comm{T^{AB}_m}{G_{C,\,r}}&=&
-2\,\mathrm{i}\,\delta^{AD}\,\delta_{C[D}G_{E],\,m+r}\,\delta^{EB}\,,
\nonu \\
\acomm{\Gamma^{ABC}_r}{\Gamma^{DEF}_s}&=&
3\,\Big(\delta^{A[D}\Delta_{r+s}^{EF]BC}-\delta^{B[D}\Delta_{r+s}^{EF]AC}+\delta^{C[D}\Delta_{r+s}^{EF]AB}\Big)
\nonu \\
&+&
\mathrm{i}\,(r-s)\,\frac{1}{2}\,\epsilon^{ABCDEFGH}\,T_{GH,\,r+s}\,,
\nonu
\\
\comm{\Gamma^{ABC}_r}{\Delta^{DEFG}_m}&=&
\frac{2}{3}\,\mathrm{i}\,\Big(\delta^{A[D}\epsilon^{EFG]BCHIJ} 
-\delta^{B[D}\epsilon^{EFG]ACHIJ} 
+\delta^{C[D}\epsilon^{EFG]ABHIJ} \Big)\Gamma_{HIJ,\,r+m}
\nonu \\
&+& (2r-m)\,\epsilon^{ABCDEFGH}\,G_{H,\,r+m}\,,
\nonu \\
\acomm{\Gamma^{ABC}_r}{\Gamma_{DEF,\,s}}
&=&
-\frac{\mathrm{i}}{5!}(3r-s)\epsilon^{ABCGHIJK}\epsilon_{DEFGHIJK}\,D_{r+s}
-9\,\delta^{G[A}\delta^{BC]}_{[EF}T_{D]G,\,r+s}\, ,
\nonu \\
\comm{\Gamma^{ABC}_r}{T_{DE,\,m}}&=&
3\,\mathrm{i}\, \delta^{[A B}_{D E}\delta^{C] F}
\,G_{F,\,r+m}\,,
\nonu \\
\comm{\Delta^{ABCD}_m}{\Delta^{EFGH}_n}&=&
2\,\mathrm{i}\,\Big(
\delta^{A[E}\epsilon^{FGH]BCDIJ}
-\delta^{B[E}\epsilon^{FGH]ACDIJ}   
\nonu \\
& 
+& \delta^{C[E}\epsilon^{FGH]ABDIJ}   
-\delta^{D[E}\epsilon^{FGH]ABCIJ}   \Big) \, T_{IJ,\,m+n}
\nonu \\
&
-& 2(m-n)\epsilon^{ABCDEFGH}\,D_{m+n}\,,
\nonu \\
\comm{\Delta^{ABCD}_m}{\Gamma_{EFG,\,r}}
&=&
\frac{\mathrm{i}}{4!}\,\epsilon^{ABCD H I J K}\epsilon_{EFG H I J K L}\,\delta^{LM}\,G_{M,\,m+r}\, .
\label{SU8}
\eea
We present
the conformal weights for the nine operators which
are the elements of the lowest ${\cal N}=8$ multiplet in Table
\ref{firsttable}.

\begin{table}[tbp]
\centering
\renewcommand{\arraystretch}{1.7}
\begin{tabular}{ |c|| c| c| c|c| c| c| c|c|c|c|  }
\hline
$\text{ Op.}$ & 
$L$  & 
$G^A$  & 
$T^{AB}$  & 
$\Gamma^{ABC}$  & 
$\Delta^{ABCD}$ & 
$\Gamma_{ABC}$ & 
$T_{AB}$ & 
$G_{A}$ & 
$D$\\
\hline
$\text{ Other not.}$ & 
${\scriptstyle \Phi_{+2}^{(0)}} $  & 
${\scriptstyle \Phi_{\frac{3}{2}}^{(0),A}} $  & 
${\scriptstyle \Phi_{+1}^{(0),AB} }$  & 
${\scriptstyle \Phi_{+\frac{1}{2}}^{(0),ABC} }$  & 
${\scriptstyle \Phi_{0}^{(0),ABCD} }$ & 
${\scriptstyle \Phi^{(0)}_{ABC,-\frac{1}{2}}}$ & 
${\scriptstyle \Phi^{(0)}_{AB,-1} }$ & 
${\scriptstyle \Phi^{(0)}_{A,-\frac{3}{2}}}$ & 
${\scriptstyle \Phi_{-2}^{(0)} }$\\
\hline
$\text{Conf. spins}$ & 
$2$ & 
$\frac{3}{2}$ &
$1$ &
$\frac{1}{2}$ &
$0$ &
$-\frac{1}{2}$ &
$-1$ &
$-\frac{3}{2}$ &
$-2$ 
\\
\hline
\end{tabular}
\caption{
We list the conformal (Conf.) spins for each operator (Op.)
(from the first nine equations of (\ref{SU8})
under the stress energy tensor $L$)  
and the upper indices and the lower indices
are $SU(8)$ fundamental and antifundamental representations.
In the second row, we present other notations (Other not.) for the same
operators in order to generalize for any ${\cal N}=8$ multiplet.
The operators having more than two indices
are totally antisymmetric under the exchange of any two indices
(We can also apply this property for the operators having
zero or one index in the view point of $SU(8)$ Young tableaux
because the former corresponds to the complex representation of
the operator with eight indices while the latter does to
the complex representation of the operator with seven indices.).}
\label{firsttable}
\end{table}

\subsection{The extension of
$SU(8)$ superconformal algebra
}

In order to understand the celestial holography from the introduction
on the possibility of infinite number of positive half integers,
we generalize the previous (anti)commutators
which hold for the lowest ${\cal N}=8$ multiplet
(\ref{Nineobject}) to those for any ${\cal N}=8$ multiplet.
We simply preserve the pole structures of OPE in
(\ref{Ademolloope}) by increasing the conformal weights
both sides.
From the experience of \cite{AK2501}, we can write down
the corresponding (anti)commutators as follows
\footnote{
We expand ${\cal N}=8$ multiplet in terms of
the Grassmann variables
\bea
\mathbf{\Phi}^{(h-2)}(Z)
& = & \Phi^{(h)}_{-2}(z)
+\theta^A\,\Phi^{(h)}_{A,-\frac{3}{2}}(z)
+ \theta^{AB}\,\Phi^{(h)}_{AB,-1}(z)
+\theta^{ABC}\,\Phi^{(h)}_{ABC,-\frac{1}{2}}(z)
\nonu \\
&+ & \frac{1}{4!} \, \theta^{ABCD}\,\epsilon_{ABCDEFGH}\, \Phi^{(h),EFGH}_{0}(z)
\nonu \\
& + &  \frac{1}{3!} \,
\theta^{ABCDE}\,\epsilon_{ABCDEFGH}\, \Phi^{(h),FGH}_{+\frac{1}{2}}(z)
+ \frac{1}{2!} \,
\theta^{ABCDEF}\,\epsilon_{ABCDEFGH} \, \Phi^{(h),GH}_{+1}(z)
\nonu \\
&+ & \theta^{ABCDEFG}\,\epsilon_{ABCDEFGH}\, \Phi^{(h),H}_{+\frac{3}{2}}(z)
+\theta^{ABCDEFGH}\,\epsilon_{ABCDEFGH}\, \Phi^{(h)}_{+2}(z)\, .
\label{bigPhi}
\eea
The Grassmann coordinate $\theta^A$
with $SU(8)$ fundamental index $A$ has a conformal
spin $-\frac{1}{2}$. Each term in (\ref{bigPhi}) has the conformal
spin $(h-2)$. This ${\cal N}=8$ multiplet
is a chiral superfield because it depends on the complex
coordinate $z$ only. From the definition of right-weight denoted by $\bar{h}$,
the vanishing of right-weight implies that the conformal dimension $\Delta$
is equal to the conformal spin $s$. If we substitute this condition
into the definition of left-weight denoted by $h$ (which is not the same as the
one in (\ref{bigPhi}) but is related to each other and see also Table
\ref{BigPhiassign}), then we are left with
the fact that
the left-weight $h$, the conformal dimension $\Delta$ and
the conformal spin $s$
are equal to each other. In this sense, we are using
the conformal dimension or equivalently the conformal spin above.}:
\bea
\comm{(\Phi^{(h_1)}_{+2})_m}{(\Phi^{(h_2)}_{+2})_n}&=&
\Big((h_2+1)m-(h_1+1)n\Big)\,(\Phi^{(h_1+h_2)}_{+2})_{m+n}\,,
\nonu    \\
\comm{(\Phi^{(h_1)}_{+2})_m}{(\Phi^{(h_2),A}_{+\frac{3}{2}})_r}&=&
\Big((h_2+\tfrac{1}{2})m-(h_1+1)r\Big)\,(\Phi^{(h_1+h_2),A}_{+\frac{3}{2}})_{m+r}\,,
\nonu    \\
\comm{(\Phi^{(h_1)}_{+2})_m}{(\Phi^{(h_2),AB}_{+1})_n}&=&
\Big(h_2\,m-(h_1+1)n \Big)\,(\Phi^{(h_1+h_2),AB}_{+1})_{m+n}\,,
\nonu    \\
\comm{(\Phi^{(h_1)}_{+2})_m}{(\Phi^{(h_2),ABC}_{+\frac{1}{2}})_r}&=&
\Big((h_2-\tfrac{1}{2})m-(h_1+1)r\Big)\,(\Phi^{(h_1+h_2),ABC}_{+\frac{1}{2}})_{m+r}\,,
\nonu    \\   
\comm{(\Phi^{(h_1)}_{+2})_m}{(\Phi^{(h_2),ABCD}_{0})_n}&=&
\Big((h_2-1)m-(h_1+1)n\Big)\,(\Phi^{(h_1+h_2),ABCD}_0)_{m+n}\,,
\nonu    \\
\comm{(\Phi^{(h_1)}_{+2})_m}{(\Phi^{(h_2)}_{ABC,-\frac{1}{2}})_r}&=&
\Big((h_2-\tfrac{3}{2})m-(h_1+1)r\Big)\,(\Phi^{(h_1+h_2)}_{ABC,-\frac{1}{2}})_{m+r}\,,
\nonu   \\
\comm{(\Phi^{(h_1)}_{+2})_m}{(\Phi^{(h_2)}_{AB,-1})_n}&=&
\Big((h_2-2)m-(h_1+1)n\Big)\,(\Phi^{(h_1+h_2)}_{AB,-1})_{m+n}\,,
\nonu    \\
\comm{(\Phi^{(h_1)}_{+2})_m}{(\Phi^{(h_2)}_{A,-\frac{3}{2}})_r}&=&
\Big((h_2-\tfrac{5}{2})m-(h_1+1)r\Big)\,(\Phi^{(h_1+h_2)}_{A,-\frac{3}{2}})_{m+r}\,,
\nonu    \\
\comm{(\Phi^{(h_1)}_{+2})_m}{(\Phi^{(h_2)}_{-2})_n}&=&
\Big((h_2-3)m-(h_1+1)n\Big)\,(\Phi^{(h_1+h_2)}_{-2})_{m+n}\,,
\nonu \\
\acomm{(\Phi^{(h_1),A}_{+\frac{3}{2}})_r}{(\Phi^{(h_2),B}_{+\frac{3}{2}})_s}&=&
\delta^{AB}\,2\,(\Phi^{(h_1+h_2)}_{+2})_{r+s}\nonu \\
&- & 2\,\mathrm{i}\,
\Big((h_2+\tfrac{1}{2})r-(h_1+\tfrac{1}{2})s\Big)\,(\Phi^{(h_1+h_2),AB}_{+1})_{r+s}\,,
\nonu \\
\comm{(\Phi^{(h_1),A}_{+\frac{3}{2}})_r}{(\Phi^{(h_2),BC}_{+1})_m}
&=&
2\,\mathrm{i}\,\delta^{A[B}\,(\Phi^{(h_1+h_2),C]}_{+\frac{3}{2}})_{r+m}
\nonu \\
&- & 2\Big(h_2\,r-(h_1+\tfrac{1}{2})m\Big)\,(\Phi^{(h_1+h_2),ABC}_{+\frac{1}{2}})_{r+m}\,,
\nonu \\
\acomm{(\Phi^{(h_1),A}_{+\frac{3}{2}})_r}
{(\Phi^{(h_2),BCD}_{+\frac{1}{2}})_s}&=&
-3\,\delta^{A[B}\,(\Phi^{(h_1+h_2),CD]}_{+1})_{r+s}
\nonu \\
&+&
2\,\mathrm{i}\Big((h_2-\tfrac{1}{2})r-(h_1+\tfrac{1}{2})s\Big)(\Phi^{(h_1+h_2),ABCD}_0)_{r+s}\,,
\nonu \\
\comm{(\Phi^{(h_1),A}_{+\frac{3}{2}})_r}{(\Phi^{(h_2),BCDE}_0)_m}&=&
-4\,\mathrm{i}\,\delta^{A[B}\,(\Phi^{(h_1+h_2),CDE]}_{+\frac{1}{2}})_{r+m}
\nonu \\
&-&
2\Big((h_2-1)r-(h_1+\tfrac{1}{2})m\Big)\,\frac{1}{3!}\epsilon^{ABCDEFGH}(\Phi^{(h_1+h_2)}_{FGH,-\frac{1}{2}})_{r+m}\,,
\nonu \\
\acomm{(\Phi^{(h_1),A}_{+\frac{3}{2}})_r}{(\Phi^{(h_2)}_{BCD,-\frac{1}{2}})_s}&
=&
\frac{1}{4!}\delta^{AE}\epsilon_{BCDEFGHI}\,(\Phi^{(h_1+h_2),FGHI}_{0})_{r+s}
\nonu \\
&-&
2\,\mathrm{i}\Big((h_2-\tfrac{3}{2})r-(h_1+\tfrac{1}{2})s\Big)\,3\,\delta^{A}_{\,\,\,[B}(\Phi^{(h_1+h_2)}_{CD],-1})_{r+s}\,,
\nonu \\
\comm{(\Phi^{(h_1),A}_{+\frac{3}{2}})_r}{(\Phi^{(h_2)}_{BC,-1})_m}&=&
\mathrm{i}\,\delta^{AD}\,(\Phi^{(h_1+h_2)}_{DBC,-\frac{1}{2}})_{r+m}
\nonu \\
&+& 2\Big((h_2-2)r-(h_1+\tfrac{1}{2})m\Big)\,2\,\delta^{A}_{\,\,\,[B}\,
(\Phi^{(h_1+h_2)}_{C],-\frac{3}{2}})_{r+m}\,,
\nonu \\
\acomm{(\Phi^{(h_1),A}_{+\frac{3}{2}})_r}{(\Phi^{(h_2)}_{B,-\frac{3}{2}})_s}&=&
\delta^{AC}\,(\Phi^{(h_1+h_2)}_{CB,-1})_{r+s}\nonu \\
&+ &
2\,\mathrm{i}\Big((h_2-\tfrac{5}{2})r-(h_1+\tfrac{1}{2})s\Big)\,\delta^{A}_{\,\,\,B}\,(\Phi^{(h_1+h_2)}_{-2})_{r+s}\,,
\nonu \\
\comm{(\Phi^{(h_1),A}_{+\frac{3}{2}})_r}{(\Phi^{(h_2)}_{-2})_m}&=&
-\mathrm{i}\,\delta^{AB}\,(\Phi^{(h_1+h_2)}_{B,-\frac{3}{2}})_{r+m}
\,,
\nonu \\
\comm{(\Phi^{(h_1),AB}_{+1})_m}{(\Phi^{(h_2),CD}_{+1})_n}&=&
-\mathrm{i}\,\delta^{AC}\,(\Phi^{(h_1+h_2),BD}_{+1})_{m+n}
+\mathrm{i}\,\delta^{AD}\,(\Phi^{(h_1+h_2),BC}_{+1})_{m+n}
\nonu \\
&
+& \mathrm{i}\,\delta^{BC}\,(\Phi^{(h_1+h_2),AD}_{+1})_{m+n}
-\mathrm{i}\,\delta^{BD}\,(\Phi^{(h_1+h_2),AC}_{+1})_{m+n}\,,
\nonu \\
\comm{(\Phi^{(h_1),AB}_{+1})_m}{(\Phi^{(h_2),CDE}_{+\frac{1}{2}})_r}&=&
-3\,\mathrm{i}\,\Big(\delta^{A[C}(\Phi^{(h_1+h_2),DE]B}_{+\frac{1}{2}})_{m+r}-\delta^{B[C}(\Phi^{(h_1+h_2),DE]A}_{+\frac{1}{2}})_{m+r}\Big)
\nonu \\
&+&
2\Big((h_2-\tfrac{1}{2})m-h_1\,r\Big)\frac{1}{3!}\epsilon^{ABCDEFGH}(\Phi^{(h_1+h_2)}_{FGH,-\frac{1}{2}})_{m+r}\,,
\nonu \\
\comm{(\Phi^{(h_1),AB}_{+1})_m}{(\Phi^{(h_2),CDEF}_0)_n}&=&
4\,\mathrm{i}\,\Big(\delta^{A[C}(\Phi^{(h_1+h_2),DEF]B}_{0})_{m+n}-\delta^{B[C}(\Phi^{(h_1+h_2),DEF]A}_0)_{m+n}\Big)
\nonu \\
&+&
\Big((h_2-1)m-h_1\,n\Big)\,\epsilon^{ABCDEFGH}\,(\Phi^{(h_1+h_2)}_{GH,-1})_{m+n}\,,
\nonu \\
\comm{(\Phi^{(h_1),AB}_{+1})_m}{(\Phi^{(h_2)}_{CDE,-\frac{1}{2}})_r}&=&
-3\,\mathrm{i}\,\Big(\delta^{A}_{\,\,\,[C}(\Phi^{(h_1+h_2)}_{DE]F,-\frac{1}{2}})_{m+r}\,\delta^{FB}-\delta^{B}_{\,\,\,[C}(\Phi^{(h_1+h_2)}_{DE]F,-\frac{1}{2}})_{m+r}\,\delta^{FA}\Big)
\nonu \\
&+& 12\Big((h_2-\tfrac{3}{2})m-h_1\,r\Big)\,\delta^{A}_{\,\,\,[C}(\Phi^{(h_1+h_2)}_{D,-\frac{3}{2}})_{m+r}\delta_{E]}^{\,\,\,B}\,,
\nonu \\
\comm{(\Phi^{(h_1),AB}_{+1})_m}{(\Phi^{(h_2)}_{CD,-1})_n}&=&
-\mathrm{i}\,\delta^{A}_{\,\,\,C}\,\delta^{BE}\,(\Phi^{(h_1+h_2)}_{ED,-1})_{m+n}
+\mathrm{i}\,\delta^{A}_{\,\,\,D}\,\delta^{BE}\,(\Phi^{(h_1+h_2)}_{EC,-1})_{m+n}
\nonu \\
&+&
\mathrm{i}\,\delta^{B}_{\,\,\,C}\,\delta^{AE}\,(\Phi^{(h_1+h_2)}_{ED,-1})_{m+n}
-\mathrm{i}\,\delta^{B}_{\,\,\,D}\,\delta^{AE}\,(\Phi^{(h_1+h_2)}_{EC,-1})_{m+n}
\nonu \\
&
+& 2\Big((h_2-2)m-h_1\,n\Big)
\delta^{A B}_{C D}
\,(\Phi^{(h_1+h_2)}_{-2})_{m+n}\,,
\nonu \\
\comm{(\Phi^{(h_1),AB}_{+1})_m}{(\Phi^{(h_2)}_{C,-\frac{3}{2}})_r}&=&
-2\,\mathrm{i}\,\delta^{AD}\,\delta_{C[D}(\Phi^{(h_1+h_2)}_{E],-\frac{3}{2}})_{m+r}\,\delta^{EB}\,,
\nonu \\
\acomm{(\Phi^{(h_1),ABC}_{+\frac{1}{2}})_r}{(\Phi^{(h_2),DEF}_{+\frac{1}{2}})_s}&=&
3\,\Big(\delta^{A[D}(\Phi^{(h_1+h_2),EF]BC}_0)_{r+s}-\delta^{B[D}(\Phi^{(h_1+h_2),EF]AC}_0)_{r+s}
\nonu \\
&+& \delta^{C[D}(\Phi^{(h_1+h_2),EF]AB}_0)_{r+s}\Big)
\nonu \\
&-&
2\,\mathrm{i}\,\Big((h_2-\tfrac{1}{2})r-(h_1-\tfrac{1}{2})s\Big)\,\frac{1}{2}\,\epsilon^{ABCDEFGH}\,(\Phi^{(h_1+h_2)}_{GH,-1})_{r+s}\,,
\nonu \\
\comm{(\Phi^{(h_1),ABC}_{+\frac{1}{2}})_r}{(\Phi^{(h_2),DEFG}_0)_m}&=&
\frac{2}{3}\,\mathrm{i}\,\Big(\delta^{A[D}\epsilon^{EFG]BCHIJ} 
-\delta^{B[D}\epsilon^{EFG]ACHIJ} 
\nonu \\
& 
+& \delta^{C[D}\epsilon^{EFG]ABHIJ}\Big)(\Phi^{(h_1+h_2)}_{HIJ,-\frac{1}{2}})_{r+m}
\nonu \\
&-&
2\Big((h_2-1)r-(h_1-\tfrac{1}{2})m\Big)\,\epsilon^{ABCDEFGH}\,(\Phi^{(h_1+h_2)}_{H,-\frac{3}{2}})_{r+m}\,,
\nonu \\
\acomm{(\Phi^{(h_1),ABC}_{+\frac{1}{2}})_r}{(\Phi^{(h_2)}_{DEF,-\frac{1}{2}})_s}
&=&
\frac{\mathrm{2\,i}}{5!}\,\Big((h_2-\tfrac{3}{2})r-(h_1-\tfrac{1}{2})s\Big)\epsilon^{ABCGHIJK}\epsilon_{DEFGHIJK}
\nonu \\
& \times & (\Phi^{(h_1+h_2)}_{-2})_{r+s} -
9\,\delta^{G[A}\delta^{BC]}_{[EF}(\Phi^{(h_1+h_2)}_{D]G,\,-1})_{r+s} \, ,
\nonu \\
\comm{(\Phi^{(h_1),ABC}_{+\frac{1}{2}})_r}{(\Phi^{(h_2)}_{DE,-1})_m}&=&
3\,\mathrm{i}\, \delta^{[A B}_{D E} \delta^{C]F}
\, (\Phi^{(h_1+h_2)}_{F,-\frac{3}{2}})_{r+m}\,,
\nonu \\
\comm{(\Phi^{(h_1),ABCD}_0)_m}{(\Phi^{(h_2),EFGH}_0)_n}&=&
2\,\mathrm{i}\,\Big(
\delta^{A[E}\epsilon^{FGH]BCDIJ}
-\delta^{B[E}\epsilon^{FGH]ACDIJ}  
\nonu \\
& 
+& \delta^{C[E}\epsilon^{FGH]ABDIJ}   
-\delta^{D[E}\epsilon^{FGH]ABCIJ}   \Big) \, (\Phi^{(h_1+h_2)}_{IJ,-1})_{m+n}
\nonu \\
&
+&
2\Big((h_2-1)m-(h_1-1)n\Big)\epsilon^{ABCDEFGH}\,(\Phi^{(h_1+h_2)}_{-2})_{m+n}\,,
\nonu \\
\comm{(\Phi^{(h_1),ABCD}_0)_m}{(\Phi^{(h_2)}_{EFG,-\frac{1}{2}})_r}
&=&
\frac{\mathrm{i}}{4!}\,\epsilon^{ABCD H I J K}\epsilon_{EFG H I J K L}\,\delta^{LM}\,
(\Phi^{(h_1+h_2)}_{M,-\frac{3}{2}})_{m+r}\, .
%
\label{eq:calN8_et}
\eea
There are twenty terms for the mode independent Kronecker delta
on the right hand sides of (\ref{eq:calN8_et}).
The corresponding (anti)commutators for the five of them do not
have mode dependent terms.
Of course, for $h_1=0=h_2$, the above equations (\ref{eq:calN8_et})
lead to the previous expressions in (\ref{SU8}). 
It is not obvious to observe how
each mode dependent term having both $h_1$ and $h_2$
respectively occurs because in the lowest ${\cal N}=8$
multiplet (that is $h_1=0=h_2$) this is identically zero.
We can see that the  operator having the subscript
$+1$ on the left hand sides of (\ref{eq:calN8_et})
contains the right hand side where the mode dependent term
has the exact coefficient $h_1$ or $h_2$.
Therefore, this kind of operator
with $+1$ subscript will play an important role
in next section
\footnote{
\label{sanitycheck}
So far, we did not discuss about any OPEs corresponding to
the well known collinear limits in the ${\cal N}=8$ supergravity.
Our claim for the identification of the (anti)commutators described
in this section as the celestial OPEs explained in next sections
can be strengthened by at least a minimal sanity check.
In other words, we could choose some OPEs
corresponding to
the collinear limits in the ${\cal N}=8$ supergravity
and show that our proposed algebra reproduces the expected
structure constants at leading order
(even up to an overall normalization).
This would provide the additional reassurance that
our algebra captures the correct physical contents. See also the footnotes
\ref{OPEanticommutator} and \ref{examples}.
We thank the referee for pointing this out.}.

\section{The celestial superconformal algebra with $SU(8)$ symmetry}

We expect that the above 
twenty nine (anti)commutators (\ref{eq:calN8_et})
should provide the celestial soft
symmetry algebra in the ${\cal N}=8$ supergravity.
Then how do we assign the helicities? 
We consider {\it the subscripts} appearing in the components of
the ${\cal N}=8$
multiplet $\Phi^{(h-2)}(\bar{Z})$ in terms of
antichiral coordinates (see also (\ref{bigPhi})) as {\it its helicities}
in four dimensions.
In Table \ref{BigPhiassign}, we present
the right-weights, the helicities (corresponding to
two dimensional conformal spins), the left-weights and
conformal dimensions.
By realizing the sum of the upper and lower indices
as the conformal spins in the antichiral sector
(done in (\ref{bigPhi})) denoted by
$q$ (or $p$), the right-weights $\bar{h}$ can be
obtained by the relation $\bar{h}=1-q$ \cite{Strominger}.
For the helicities appearing as the subscripts,
we can determine the left-weights by adding
the right-weights to the helicities.
Moreover, by adding the left-weights to the right-weights,
the conformal dimensions $\Delta$ corresponding to
the energy of the particle in four dimensions can be obtained \footnote{
Later, we expand the above ${\cal N}=8$ multiplet
in terms of the Grassmann variables $\eta_{A}$ in
(\ref{etaexpansion}).
After choosing the left-weight and the right-weight
for  $\eta_{A}$ as $(0,-\frac{1}{2})$ (in other words, the helicity
$s=+\frac{1}{2}$ and  $\Delta=-\frac{1}{2}$),
each term including the Grassmann variables 
has $\Delta=-2h$ and the helicities $s=+2$ (or
the left-weight and the right-weight are given by $(1-h)$
and $(-1-h)$ respectively). See also \cite{Jiang2105,Tropper1}.}.

\begin{table}[tbp]
\centering
\renewcommand{\arraystretch}{1.7}
\begin{tabular}{ |c|| c| c| c|c| c| c| c|c|c|c|  }
\hline
$\text{ Op.}$ & 
${\scriptstyle \Phi_{+2}^{(h)}}$  & 
${\scriptstyle \Phi_{+\frac{3}{2}}^{(h),A}}$  & 
${\scriptstyle \Phi_{+1}^{(h),AB}}$  & 
${\scriptstyle \Phi_{+\frac{1}{2}}^{(h),ABC}}$  & 
${ \scriptstyle \Phi_{0}^{(h),ABCD}}$ & 
${\scriptstyle \Phi^{(h)}_{ABC, -\frac{1}{2}}}$ & 
${\scriptstyle \Phi^{(h)}_{AB,-1}}$ & 
${\scriptstyle \Phi^{(h)}_{A,-\frac{3}{2}}}$ & 
${\scriptstyle \Phi_{-2}^{(h)}}$\\
\hline
$\text{R.-wt.}$ & 
$-1-h$ & 
$-\frac{1}{2}-h$ &
$-h$ &
$\frac{1}{2}-h$ &
$1-h$ &
$\frac{3}{2}-h$ &
$2-h$ &
$\frac{5}{2}-h$ &
$3-h$ 
\\
\hline
$\text{Hel.}$ & $+2$ & $+\frac{3}{2}$ &$+1$ &$+\frac{1}{2}$ &
$0$ &$-\frac{1}{2}$ &$-1$ & $-\frac{3}{2}$ & $-2$
\\
\hline
\hline
$\text{L.-wt.}$ & 
$1-h$ & 
$1-h$ &
$1-h$ &
$1-h$ &
$1-h$ &
$1-h$ &
$1-h$ &
$1-h$ &
$1-h$
\\
\hline
$\text{Dim.}$ & 
${\scriptstyle -2h}$ & 
${\scriptstyle \frac{1}{2}-2h}$ &
${\scriptstyle 1-2h}$ &
${\scriptstyle \frac{3}{2}-2h}$ &
${\scriptstyle 2-2h}$ &
${\scriptstyle \frac{5}{2}-2h}$ &
${\scriptstyle 3-2h}$ &
${\scriptstyle \frac{7}{2}-2h}$ &
${\scriptstyle 4-2h}$
\\
\hline
\end{tabular}
\caption{
The upper indices $A,B,C,D, \cdots $
and the lower indices $A,B,C,D, \cdots $ 
are $SU(8)$ fundamental and antifundamental representations
respectively.
For each operator (Op.), we list the right-weight (R.-wt.) or
antiholomorphic weight, the helicity (Hel.),
the left-weight (L.-wt.) or
holomorphic weight, and the conformal dimension (Dim.) at each row.
The right-weight is given by one minus the index $q$ \cite{Strominger}
which plays the role of the spins of operators (sum of both the superscript
and the subscript), $(h+2)$,
$(h+\frac{3}{2})$, $(h+1)$, $(h+0)$, $(h-\frac{1}{2})$, $(h-1)$,
$(h-\frac{3}{2})$ and $(h-2)$ respectively. Then
the left-weight is given by the sum of the right-weight and the
helicity $s$.
Finally the dimension $\Delta$
is the sum of the left-weight and the right-weight.
For one parameter $h=-1, 0, 1, 2, \cdots$, all the relevant quantities
are fixed in this Table. For $h=2$, the values
of the conformal dimension $\Delta$ in the first five are given by
$-4$, $-\frac{7}{2}$, $-3$, $-\frac{5}{2}$ and $-2$ and appear in the
infinite sequences for the graviton, the gravitinos,
the graviphotons, the graviphotinos and the scalars
respectively in the introduction.}
\label{BigPhiassign}
\end{table}

By
\begin{itemize}
\item[]
i) ignoring the
Kronecker delta terms appearing in (\ref{eq:calN8_et}) (the sum of
helicities is not equal to $2$),

ii) introducing the additional commutator
for the eighteenth of (\ref{eq:calN8_et}) and

iii) putting the
structure constants $\kappa$
(under the assumption of celestial holography),
\end{itemize} 
we obtain the following twenty five (anti)commutators
\bea
\comm{(\Phi^{(h_1)}_{+2})_m}{(\Phi^{(h_2)}_{+2})_n}&=&
\kappa_{+2,+2,-2}\,\Big((h_2+1)m-(h_1+1)n\Big)\,(\Phi^{(h_1+h_2)}_{+2})_
{m+n}
\, :\text{eq.1},
\nonu    \\
\comm{(\Phi^{(h_1)}_{+2})_m}{(\Phi^{(h_2),A}_{+\frac{3}{2}})_r}&=&
\kappa_{+2,+\frac{3}{2},-\frac{3}{2}}\,\Big((h_2+\tfrac{1}{2})m-(h_1+1)r\Big)\,(\Phi^{(h_1+h_2),A}_{+\frac{3}{2}})_{m+r}
\, :\text{eq.2},
\nonu    \\
\comm{(\Phi^{(h_1)}_{+2})_m}{(\Phi^{(h_2),AB}_{+1})_n}&=&
\kappa_{+2,+1,-1}\,\Big(h_2\,m-(h_1+1)n \Big)\,(\Phi^{(h_1+h_2),AB}_{+1})_{m+n}
\, :\text{eq.3},
\nonu    \\
\comm{(\Phi^{(h_1)}_{+2})_m}{(\Phi^{(h_2),ABC}_{+\frac{1}{2}})_r}&=&
\kappa_{+2,+\frac{1}{2},-\frac{1}{2}}\,\Big((h_2-\tfrac{1}{2})m-(h_1+1)r\Big)\,
(\Phi^{(h_1+h_2),ABC}_{+\frac{1}{2}})_{m+r}
\, \nonu \\
& : & \text{eq.4},
\nonu    \\   
\comm{(\Phi^{(h_1)}_{+2})_m}{(\Phi^{(h_2),ABCD}_{0})_n}
&=&
\kappa_{+2,0,0}\,\Big((h_2-1)m-(h_1+1)n\Big)\,(\Phi^{(h_1+h_2),ABCD}_{0})_{m+n}
\, :\text{eq.5},
\nonu    \\
\comm{(\Phi^{(h_1)}_{+2})_m}{(\Phi^{(h_2)}_{ABC,-\frac{1}{2}})_r}
&=&\kappa_{+2,-\frac{1}{2},+\frac{1}{2}}\,\,\Big((h_2-\tfrac{3}{2})m-(h_1+1)r\Big)\,(\Phi^{(h_1+h_2)}_{ABC,-\frac{1}{2}})_{m+r}
\, :\text{eq.6},
\nonu   \\
\comm{(\Phi^{(h_1)}_{+2})_m}{(\Phi^{(h_2)}_{AB,-1})_n}
&=&\kappa_{+2,-1,+1}\,\Big((h_2-2)m-(h_1+1)n\Big)\,(\Phi^{(h_1+h_2)}_{AB,-1})_{m+n}
\, :\text{eq.7},
\nonu    \\
\comm{(\Phi^{(h_1)}_{+2})_m}{(\Phi^{(h_2)}_{A,-\frac{3}{2}})_r}
&=&
\kappa_{+2,-\frac{3}{2},+\frac{3}{2}}\,\Big((h_2-\tfrac{5}{2})m-(h_1+1)r\Big)\,(\Phi^{(h_1+h_2)}_{A,-\frac{3}{2}})_{m+r}
\, :\text{eq.8},
\nonu    \\
\comm{(\Phi^{(h_1)}_{+2})_m}{(\Phi^{(h_2)}_{-2})_n}&=&
\kappa_{+2,-2,+2}\,\Big((h_2-3)m-(h_1+1)n\Big)\,(\Phi^{(h_1+h_2)}_{-2})_{m+n}
\, :\text{eq.9},
\nonu \\
\acomm{(\Phi^{(h_1),A}_{+\frac{3}{2}})_r}{(\Phi^{(h_2),B}_{+\frac{3}{2}})_s}
&=& \kappa_{+\frac{3}{2},+\frac{3}{2},-1}\,
\Big((h_2+\tfrac{1}{2})r-(h_1+\tfrac{1}{2})s\Big)\,(\Phi^{(h_1+h_2),AB}_{+1})_{r+s}\,,
\nonu \\
&:& \text{eq.10},
\nonu \\
\comm{(\Phi^{(h_1),A}_{+\frac{3}{2}})_r}{(\Phi^{(h_2),BC}_{+1})_m}
&=&
\kappa_{+\frac{3}{2},+1,-\frac{1}{2}}\,\Big(h_2\,r-(h_1+\tfrac{1}{2})m\Big)\,(\Phi^{(h_1+h_2),ABC}_{+\frac{1}{2}})_{r+m}
\, :\text{eq.11},
\nonu \\
\acomm{(\Phi^{(h_1),A}_{+\frac{3}{2}})_r}{(\Phi^{(h_2),BCD}_{+\frac{1}{2}})_s}
&=&
\kappa_{+\frac{3}{2},+\frac{1}{2},0}\,\Big((h_2-\tfrac{1}{2})r-(h_1+\tfrac{1}{2})s\Big)(\Phi^{(h_1+h_2),ABCD}_0)_{r+s}
\nonu \\
&:& \text{eq.12},
\nonu \\
\comm{(\Phi^{(h_1),A}_{+\frac{3}{2}})_r}{(\Phi^{(h_2),BCDE}_0)_m}&=&
\kappa_{+\frac{3}{2},0,+\frac{1}{2}}\,\Big((h_2-1)r-(h_1+\tfrac{1}{2})m\Big)\,\frac{1}{3!}\epsilon^{ABCDEFGH}\nonu \\
& \times & (\Phi^{(h_1+h_2)}_{FGH,-\frac{1}{2}})_{r+m}
\, :  \text{eq.13},
\nonu \\
\acomm{(\Phi^{(h_1),A}_{+\frac{3}{2}})_r}{(\Phi^{(h_2)}_{BCD,-\frac{1}{2}})_s}
&
=& \kappa_{+\frac{3}{2},-\frac{1}{2},+1}\,
\Big((h_2-\tfrac{3}{2})r-(h_1+\tfrac{1}{2})s\Big)\,3
\delta^{A}_{\,\,\,[B}(\Phi^{(h_1+h_2)}_{CD],-1})_{r+s}\,
\nonu \\
&:& \text{eq.14},
\nonu \\
\comm{(\Phi^{(h_1),A}_{+\frac{3}{2}})_r}{(\Phi^{(h_2)}_{BC,-1})_m}
&=&
\kappa_{+\frac{3}{2},-1,+\frac{3}{2}}\,\Big((h_2-2)r-(h_1+\tfrac{1}{2})m\Big)\,
\nonu \\
& \times & 2!\, \delta^{A}_{\,\,\,[B}\,
(\Phi^{(h_1+h_2)}_{C],-\frac{3}{2}})_{r+m}\, :\text{eq.15},
\nonu \\
\acomm{(\Phi^{(h_1),A}_{+\frac{3}{2}})_r}{(\Phi^{(h_2)}_{B,-\frac{3}{2}})_s}
&=&
\kappa_{+\frac{3}{2},-\frac{3}{2},+2}\,\Big((h_2-\tfrac{5}{2})r-(h_1+\tfrac{1}{2})s\Big)\,\delta^{A}_{\,\,\,B}\,(\Phi^{(h_1+h_2)}_{-2})_{r+s}
\nonu \\
&:& \text{eq.16},
\nonu \\
\comm{(\Phi^{(h_1),AB}_{+1})_m}{(\Phi^{(h_2),CD}_{+1})_n}&=&
\kappa_{+1,+1,0}\,\Big(h_2\,m-h_1\,n\Big)\,(\Phi^{(h_1+h_2),ABCD}_{0})_{m+n}
\, :\text{eq.17},
\nonu \\
\comm{(\Phi^{(h_1),AB}_{+1})_m}{(\Phi^{(h_2),CDE}_{+\frac{1}{2}})_r}&=&
\kappa_{+1,+\frac{1}{2},+\frac{1}{2}}\,\Big((h_2-\tfrac{1}{2})m-h_1\,r\Big)\frac{1}{3!}\,\epsilon^{ABCDEFGH}(\Phi^{(h_1+h_2)}_{FGH,-\frac{1}{2}})_{m+r}
\nonu \\
&: & \text{eq.18},
\nonu \\
\comm{(\Phi^{(h_1),AB}_{+1})_m}{(\Phi^{(h_2),CDEF}_0)_n}&=&
\kappa_{+1,0,+1}\,\Big((h_2-1)m-h_1\,n\Big)\,\epsilon^{ABCDEFGH}\,\frac{1}{2!}\,(\Phi^{(h_1+h_2)}_{GH,-1})_{m+n}
\nonu \\
&: & \text{eq.19},
\nonu \\
\comm{(\Phi^{(h_1),AB}_{+1})_m}{(\Phi^{(h_2)}_{CDE,-\frac{1}{2}})_r}&=&
\kappa_{+1,-\frac{1}{2},+\frac{3}{2}}\,\Big((h_2-\tfrac{3}{2})m-h_1\,r\Big)\,
3! \delta^{A}_{\,\,\,[C}(\Phi^{(h_1+h_2)}_{D,-\frac{3}{2}})_{m+r}\delta_{E]}^{\,\,\,B}
\, \nonu \\
&: & \text{eq.20},
\nonu \\
\comm{(\Phi^{(h_1),AB}_{+1})_m}{(\Phi^{(h_2)}_{CD,-1})_n}&=&
\kappa_{+1,-1,+2}\,\Big((h_2-2)m-h_1\,n\Big)\,
\delta^{AB}_{CD} \, (\Phi^{(h_1+h_2)}_{-2})_{m+n}
\, : \text{eq.21},
\nonu \\
\acomm{(\Phi^{(h_1),ABC}_{+\frac{1}{2}})_r}{(\Phi^{(h_2),DEF}_{+\frac{1}{2}})_s}&=&
\kappa_{+\frac{1}{2},+\frac{1}{2},+1}\,\Big((h_2-\tfrac{1}{2})r-(h_1-\tfrac{1}{2})s\Big)\,\frac{1}{2!}\,\epsilon^{ABCDEFGH}
\nonu \\
& \times & (\Phi^{(h_1+h_2)}_{GH,-1})_{r+s}
\, :  \text{eq.22},
\nonu \\
\comm{(\Phi^{(h_1),ABC}_{+\frac{1}{2}})_r}{(\Phi^{(h_2),DEFG}_0)_m}&=&
\kappa_{+\frac{1}{2},0,+\frac{3}{2}}\,\Big((h_2-1)r-(h_1-\tfrac{1}{2})m\Big)\,\epsilon^{ABCDEFGH}
\nonu \\
& \times & (\Phi^{(h_1+h_2)}_{H,-\frac{3}{2}})_{r+m}
\, :  \text{eq.23},
\nonu \\
\acomm{(\Phi^{(h_1),ABC}_{+\frac{1}{2}})_r}{(\Phi^{(h_2)}_{DEF,-\frac{1}{2}})_s}
&=&\kappa_{+\frac{1}{2},-\frac{1}{2},+2}\,\Big((h_2-\tfrac{3}{2})r-(h_1-\tfrac{1}{2})s\Big)\,
\delta^{ABC}_{DEF}\,(\Phi^{(h_1+h_2)}_{-2})_{r+s}
\nonu \\
&: & \text{eq.24},
\nonu \\
\comm{(\Phi^{(h_1),ABCD}_0)_m}{(\Phi^{(h_2),EFGH}_0)_n}&=&
\kappa_{0,0,+2}\,\Big((h_2-1)m-(h_1-1)n\Big)\epsilon^{ABCDEFGH}\,(\Phi^{(h_1+h_2)}_{-2})_{m+n}
\nonu \\
& : & \text{eq.25}\,.
%
\label{Eqs}
\eea
Note that in (\ref{Eqs}),
the numerical factors appearing on the right hand sides
of (\ref{Eqs}) are introduced  in order to have the modes
with the coefficients one (up to signs) when we fix
the free $SU(8)$ indices appearing in both sides
\footnote{From the tensor product of $SU(8)$ \cite{FKS}, it is known that
\bea
{\bf 1} \otimes {\bf 1} &=& {\bf 1} ,
\qquad
{\bf 1} \otimes {\bf 8} = {\bf 8} ,
\qquad
{\bf 1} \otimes {\bf 28} = {\bf 28} ,
\qquad
{\bf 1} \otimes {\bf 56} = {\bf 56} ,
\nonu \\
{\bf 1} \otimes {\bf 70} &=& {\bf 70} ,
\qquad
{\bf 1} \otimes \overline{\bf 56} = \overline{\bf 56} ,
\qquad
{\bf 1} \otimes \overline{\bf 28} = \overline{\bf 28} ,
\qquad
{\bf 1} \otimes \overline{\bf 8} = \overline{\bf 8} ,
\qquad
{\bf 1} \otimes \overline{\bf 1} = \overline{\bf 1},
\nonu \\
{\bf 8} \otimes {\bf 8} &=& {\bf 28} \oplus  \cdots,
\qquad
{\bf 8} \otimes {\bf 28} = {\bf 56} \oplus  \cdots,
\qquad
{\bf 8} \otimes {\bf 56} = {\bf 70} \oplus  \cdots,
\nonu \\
{\bf 8} \otimes {\bf 70} & = & \overline{\bf 56} \oplus  \cdots,
\qquad
{\bf 8} \otimes \overline{\bf 56} = \overline{\bf 28} \oplus 
\cdots,
\qquad
{\bf 8} \otimes \overline{\bf 28} = \overline{\bf 8} \oplus  \cdots,
\qquad
{\bf 8} \otimes \overline{\bf 8} = {\bf 1} \oplus  \cdots,
\nonu \\
{\bf 28} \otimes {\bf 28} &=& {\bf 70} \oplus  \cdots,
\qquad
{\bf 28} \otimes {\bf 56} = \overline{\bf 56} \oplus  \cdots,
\qquad
{\bf 28} \otimes {\bf 70} = \overline{\bf 28} \oplus  \cdots,
\qquad
{\bf 28} \otimes \overline{\bf 56} = \overline{\bf 8} \oplus  \cdots,
\nonu \\
{\bf 28} \otimes \overline{\bf 28} &=& {\bf 1} \oplus  \cdots,
\qquad
{\bf 56} \otimes {\bf 56} = \overline{\bf 28} \oplus  \cdots,
\qquad
{\bf 56} \otimes {\bf 70} = \overline{\bf 8} \oplus  \cdots,
\qquad
{\bf 56} \otimes \overline{\bf 56} = {\bf 1} \oplus  \cdots,
\nonu \\
{\bf 70} \otimes {\bf 70} &=& {\bf 1} \oplus  \cdots\, ,
\label{branching}
\eea
where the higher dimensional representations
appearing on the right hand sides are ignored.
Therefore, the above (anti)commutators correspond to
this $SU(8)$ tensor product (\ref{branching}) if we focus on the
$SU(8)$ indices only.
For example, the eq. $13$ has the Levi Civita $\epsilon^{ABCDEFGH}$
contracted with the lower indices $FGH$ and this implies
that the right hand side plays the role of
the upper antisymmetric five indices which is equivalent to
the complex representation of the lower antisymmetric three indices.
The relevant one is given by the thirteenth equation of
(\ref{branching}) where the right hand side contains $\overline{\bf 56}$.
We can analyze other cases similarly.}.
The previous numerical coefficients
appearing in (\ref{eq:calN8_et}) are absorbed in the
coupling $\kappa$
\footnote{By interchanging 
the two operators with different helicities
appearing on the left hand sides,
the corresponding twenty couplings
are related to the ones in (\ref{Eqs})
explicitly. That is,
there are
$\kappa_{+\frac{3}{2},-\frac{1}{2},+1}=-\kappa_{-\frac{1}{2},+\frac{3}{2},+1}$,
$\kappa_{+\frac{3}{2},-1,+\frac{3}{2}}=-\kappa_{-1,+\frac{3}{2},+\frac{3}{2}}$,
$\kappa_{+\frac{3}{2},-\frac{3}{2},+2}=-\kappa_{-\frac{3}{2},+\frac{3}{2},+2}$
and $\kappa_{+\frac{1}{2},-\frac{1}{2},+2}=-\kappa_{-\frac{1}{2},+\frac{1}{2},+2}$\
and the remaining sixteen couplings satisfy
$\kappa_{s_1,s_2,-s_3}=\kappa_{s_2,s_1,-s_3}$.
Note that there are forty five cases \cite{AK2407} satisfying the
condition $s_1+s_2-s_3=2$.
}. Note that there exists
a new commutator characterized by eq. $17$
in (\ref{Eqs}) which does not appear in (\ref{eq:calN8_et}).
When we put $h_1=0=h_2$ into this equation, the right hand side
of this commutator vanishes. This commutator is necessary
in order to explain the celestial soft symmetry algebra later.
Moreover, compared to the ${\cal N}=4$
$SO(4)$ supergravity \cite{AK2501,Das,CS1,CS2,CSF}, the above eq. $18$ and eq. $22$
are new.

Except the  five commutators
having the scalars on the left hand sides (denoted by
the eq. $5$, the eq. $13$, the eq. $19$, the eq. $23$ and
eq. $25$), it is obvious to observe that
the remaining $18$ (anti)commutators
(that is $18=25-5-2$ where $2$ come from the eq. $18$ and the eq. $22$
we mentioned before)
can be seen from the analysis given in \cite{AK2501} for the
${\cal N}=4$ supergravity.

Then what happens for the remaining six
commutators in the construction of \cite{AK2501}?
The reason is as follows.
For the commutators
between the helicities $(+\frac{3}{2}, +1, +\frac{1}{2},0)$
with the scalars corresponding to
the eq. $13$, the
eq. $19$, the eq. $23$ and the eq. $25$, the right hand sides contain
the Levi Civita $\epsilon^{ABCDEFGH}$.
When we consider the $1,2,3,4$ indices (from the first
$SU(4)$ fundamental indices in the branching of $SU(8) \rightarrow
SU(4) \times SU(4) \times U(1)$) for the scalars
appearing in the second element on the left hand sides, then
the gravitinos, the graviphotons, the
graviphotinos and the scalars for the first
element on the left hand sides should have the indices from the
remaining ones $5,6,7,8$ indices (can be interpreted as
the second $SU(4)$ fundamental indices) and vice versa
in order to have nonzero
commutators.
However, the graviton with scalars
corresponding to the eq. $5$ can provide two commutators
when we fix the above two types of indices for the scalars.
Therefore, because the above five commutators
play the role of six commutators, the previous
twenty four (anti)commutators in \cite{AK2501}
can be seen from this analysis by including the eighteen
ones in previous paragraph.
See also the last Appendix $I$
for the complete algebra \footnote{Precisely speaking,
when we take the indices for $ABCD=1234$ and $EFGH=5678$ at
the eq. $25$ of (\ref{Eqs}), then
in the view point of the first $SU(4)$ fundamental indices
corresponding to ${\cal N}=4$ supergravity,
the scalar having $5678$ indices is really a scalar and
the left hand sides of eq. $13$, the
eq. $19$, the eq. $23$ and the eq. $25$ contain
this $5678$ indices and the eq. $5$
has both $1234$ indices and $5678$ indices.}.

Of course the above eq. $18$ and eq. $22$
contain the  $\epsilon^{ABCDEFGH}$ and
the indices on the left hand sides should be different
in order to have nonzero (anti)commutators.
In next section, we would like to observe the
previous abstract algebra (\ref{Eqs}) obtained in two dimensions
from the bulk point of view with the 
${\cal N}=8$ supergravity in four dimensions
\footnote{We are using the following 
normalizations in $\delta^{AB}_{CD}=2! \delta^{[A}_{C} \delta^{B]}_{D}$
and $\delta^{ABC}_{DEF}=3!  \delta^{[A}_{D} \delta^{B}_{E} \delta^{C]}_F$.

From the tensor product of $SO(8)$, it is known in \cite{FKS} that
\bea
{\bf 1} \otimes {\bf 1} &=& {\bf 1},
\qquad
{\bf 1} \otimes {\bf 8_s} = {\bf 8_s},
\qquad
{\bf 1} \otimes {\bf 28} = {\bf 28},
\qquad
{\bf 1} \otimes {\bf 56_s} = {\bf 56_s},
\qquad
{\bf 1} \otimes {\bf 35_c} = {\bf 35_c},
{\bf 1} \otimes {\bf 35_v} = {\bf 35_v},
\nonu \\
{\bf 8_s} \otimes {\bf 8_s} & = & {\bf 1} \oplus {\bf 28} \oplus  \cdots,
\qquad
{\bf 8_s} \otimes {\bf 28} = {\bf 8_s} \oplus {\bf 56_s}
\oplus  \cdots,
\qquad
{\bf 8_s} \otimes {\bf 56_s} = {\bf 28}\oplus {\bf 35_c} \oplus
{\bf 35_v}\cdots,
\nonu \\
{\bf 8_s} \otimes {\bf 35_c}  &=&  {\bf 56_s} \oplus  \cdots,
\qquad
{\bf 8_s} \otimes {\bf 35_v}  =  {\bf 56_s} \oplus  \cdots,
\qquad
{\bf 28} \otimes {\bf 28} ={\bf 1} \oplus {\bf 35_c}\oplus
{\bf 35_v} \oplus  \cdots,
\nonu \\
{\bf 28} \otimes {\bf 56_s} & = & {\bf 8_s} \oplus
{\bf 56_s} \oplus {\bf 56_s}  \cdots,
\qquad
{\bf 28} \otimes {\bf 35_c} = {\bf 28} \oplus  \cdots,
\qquad
{\bf 28} \otimes {\bf 35_v} = {\bf 28} \oplus  \cdots,
\nonu \\
{\bf 56_s} \otimes {\bf 56_s} &=& {\bf 1}\oplus {\bf 28}\oplus
{\bf 28} \oplus  \cdots,
\qquad
{\bf 56_s} \otimes {\bf 35_c} = {\bf 8_s} \oplus  \cdots,
\qquad
{\bf 56_s} \otimes {\bf 35_v} = {\bf 8_s} \oplus  \cdots,
\nonu \\
{\bf 35_c} \otimes {\bf 35_c} &=& {\bf 1} \oplus  \cdots,
\qquad
{\bf 35_v} \otimes {\bf 35_v} = {\bf 1} \oplus  \cdots\,      ,
\label{branching1}
\eea
where the higher dimensional representations
appearing on the right hand sides are ignored.
Note that there are more than two kinds of representations
on the right hand sides between the
representations ${\bf 8_s}$, ${\bf 28}$ and ${\bf 56_s}$.
The six tensor products of these three representations
correspond to those for twelve $SU(8)$ tensor products
in (\ref{branching})
between ${\bf 8}$, $\bar{\bf 8}$, ${\bf 28}$, $\overline{\bf 28}$,
${\bf 56}$, and $\overline{\bf 56}$.
Also we can associate the last eight tensor products having the
scalars ${\bf 35_c}$ and ${\bf 35_v}$ in (\ref{branching1})
(tenth, eleventh, fourteenth, fifteenth, seventeenth-twentieth) with
those four tensor product having the scalars ${\bf 70}$ in
(\ref{branching}).
}.

\section{The celestial holography in the ${\cal N}=8$ supergravity}

From the particle contents in the ${\cal N}=8$ supergravity,
we associate the scalars
with helicity $0$, the graviphotinos with helicities $\pm \frac{1}{2}$,
the
graviphotons with helicities $\pm 1$, the gravitinos
with helicities $\pm \frac{3}{2}$ and the graviton with helicities $\pm 2$
with the previous celestial operators
as follows:
\bea
(A + i B)^{ABCD} & \longleftrightarrow  &  \Phi^{(h),ABCD}_{0} \, ,
\nonu \\
\chi^{ABC} & \longleftrightarrow &
\Phi^{(h),ABC}_{+\frac{1}{2}} \, ,
\nonu \\
\chi_{ABC} & \longleftrightarrow &
\Phi^{(h)}_{ABC,-\frac{1}{2}} \, ,
\nonu \\
A^{AB}_{\mu} & \longleftrightarrow &  \Phi^{(h),AB}_{+1} \, ,
\nonu \\
A_{\mu AB} & \longleftrightarrow &  \Phi^{(h)}_{AB,-1} \, ,
\nonu \\
\psi^{A}_{\mu} & \longleftrightarrow &
\Phi^{(h),A}_{+\frac{3}{2}} \, ,
\nonu  \\
\psi_{\mu A} & \longleftrightarrow &
\Phi^{(h)}_{A,-\frac{3}{2}} \, ,
\nonu  \\ 
e_{\mu}^a & \longleftrightarrow  &  \Phi^{(h)}_{\pm 2} \, .
\label{CORR}
\eea
On the left hand side of (\ref{CORR}),
the helicities can be encoded by the locations of
$SU(8)$ indices while on the right hand side of
(\ref{CORR}) those $SU(8)$ indices appear exactly in the
same positions \footnote{The $SU(8)$ self duality constraint
for the seventy scalars can be described
as \cite{deWit}
\bea
(A + i B)^{ABCD} = \frac{1}{4!}\, \eta\, \epsilon^{ABCDEFGH}\,
(A - i B)_{EFGH} \,, \qquad
(A - i B)_{ABCD} = \frac{1}{4!}\, \eta\, \epsilon_{ABCDEFGH}\,
(A + i B)^{EFGH} \, , 
\nonu
\eea
where $\eta$ takes the values
$\eta =\pm 1$.
Here the scalars $A^{ABCD}$ are thirty five
scalars while the pseudoscalars
$B^{ABCD}$ are thirty five pseudoscalars
with $A^{ABCD} = \frac{1}{4!}\, \eta\, \epsilon^{ABCDEFGH}
\, A_{EFGH}$ and
$B^{ABCD} = -\frac{1}{4!}\, \eta\, \epsilon^{ABCDEFGH}
\, B_{EFGH}$ in the $SO(8)$. Note that
the ${\bf 70}$ dimensional representations of
$SO(8)$ is reducible to
the sum of two irreducible
thirty five dimensional representations:
${\bf 35_c} \oplus {\bf 35_v}$ \cite{dF}.
See also the equation (\ref{branching1}).
}. For the upper indices, the helicity
is positive (particles with the helicity $(2-\frac{n}{2})$
transform as $n$-th rank antisymmetric tensor representations of the
$SO(8)$ (or $SU(8)$) \cite{deWit}) where $n \leq 4$
while for the lower indices (in this case $n \geq 5$
and the corresponding Young tableaux implies that
we can use
the $(8-n)$ number of boxes in the first
column of Young tableaux
with
the complex
representations
$\overline{\bf 56}$ ($n=5$),
$\overline{\bf 28}$ ($n=6$),
$\overline{\bf 8}$ ($n=7$),
and ${\bf 1}$ ($n=8$) for the $n$ number of boxes in the first
column of Young tableaux)
the helicity is negative.
Note that on the right hand sides of (\ref{CORR}), 
the subscripts are present already and they
appear in the decomposition of Grassmann variable of (\ref{bigPhi})
such that the sum of the number of these fermionic variables
multiplied by $-\frac{1}{2}$
and the numerical value of the subscript is equal to $-2$.
These subscripts on the right hand sides of (\ref{CORR})
are exactly the same as the helicities on the left hand sides.
Therefore, we do not have to introduce  the additional
subscripts for the helicities.
Instead of writing down $0, +\frac{1}{2}, -\frac{1}{2}, +1, -1,
+\frac{3}{2}, -\frac{3}{2}, \pm 2$ in the subscripts additionally on the
right hand sides respectively, we
denote the helicities as the numerical subscripts.
In other words, those subscripts appearing in the
Grassmann expansion in two dimensional conformal field theory
play the role of the helicities in the celestial conformal
field theory.

The 
tree level ${\cal N}=8$ supergravity splitting amplitude
is given by the product of two tree level ${\cal N}=4$ super
Yang-Mills theory 
splitting amplitudes multiplied by
the angle bracket, the square bracket
between the two collinear particles together with minus sign
\cite{BDPR,BBJ}.
The helicities of two particles
in the ${\cal N}=8$
supergravity theory splitting amplitude
are given by the sum of the helicities of each particle
in the ${\cal N}=4$ super
Yang-Mills theory 
splitting amplitudes
respectively.
The third helicity
of the ${\cal N}=8$
supergravity theory splitting amplitude
is given by the sum of the third helicity of each particle
in the ${\cal N}=4$ super
Yang-Mills theory 
splitting amplitudes.
The helicities for the gluons, the gluinos and the scalars
are given by $(\pm 1, \pm \frac{1}{2}, 0)$
in the ${\cal N}=4$ super Yang-Mills theory.
By linear combinations of these two sets of
helicities, we are left with the helicities $(\pm 2,
\pm \frac{3}{2}, \pm 1, \pm \frac{1}{2}, 0)$
corresponding to the helicities of the graviton,
the gravitinos, the graviphotons, the graviphotinos and
scalars in the ${\cal N}=8$ supergravity theory.

After we parametrize each momentum of the two collinear
particles in terms of their energies with their null vectors, 
the sum of these momenta can be written in terms of
the sum of these energies multiplied by the above first (or second)
null vector.
Then one of the complex coordinates in the celestial two sphere
can be written in terms of the ratio of the energy of first particle
and the sum the energies of the two particles.
Then the above angle and square brackets
can be written as the two energies and
the two complex coordinates in the the celestial sphere
by using  the spinor helicity formalism.
Therefore, the split factor in the ${\cal N}=8$
supergravity theory is given by 
the two energies, the sum of these two energies and
the two complex  coordinates for each
collinear particle appearing in the
two null vectors. See also the second and the third
columns of Tables \ref{Split1}, \ref{Split2},
\ref{Split3}, \ref{Split4}
and \ref{Split5}.

The power of the sum of two energies provides
the helicity of the collinear channel momentum inside of
$(n-1)$ point amplitude (Here $n$
is nothing to do with the above
number of antisymmetric representations of
$SU(8)$) and gives the opposite of the
subscript of the split factor
of the ${\cal N}=8$ supergravity.
The celestial amplitude can be obtained
from the conventional momentum space amplitude by 
Mellin transform.
The $n$ point celestial correlator
can be written in terms of
\begin{itemize}
\item[]
i) the complex coordinate
dependence,

ii) the Euler beta function, and

iii)
the $(n-1)$ point correlator \cite{PRSY}.
\end{itemize}
Then the OPE corresponding to
the two operators having the specific helicities
can be read off and the power of the first particle energy
(or conformal dimension)
and the power of the second particle energy
(or conformal dimension) appear
in the first argument and second argument of the Euler beta
function respectively.
See also the last
columns of Tables \ref{Split1}, \ref{Split2},
\ref{Split3}, \ref{Split4}
and \ref{Split5}.

The celestial OPE with the constraint
$s_1+s_2-s_3=2$ (for the time being
the $SU(8)$ indices are abbreviated)
from  a three-point interaction with the bulk dimension
of three point vertex $d_V=5$ takes the form \cite{HPS,MRSV,Jiang2108}
\bea
&& {\cal O}_{\Delta_1, s_1}(z_1, \bar{z}_1)\,
{\cal O}_{\Delta_2, s_2}(z_2,\bar{z}_2)=
\nonu \\
&&
\frac{\kappa_{s_1,s_2,-s_3}}{z_{12}}\, \sum_{\alpha=0}^{\infty}\,
B(\Delta_1+1-s_1+\alpha, \Delta_2+1-s_2)\,
\frac{\bar{z}_{12}^{\alpha+1}}{\alpha!}\, \bar{\pa}^{\alpha} \, 
{\cal O}_{\Delta_1+\Delta_2, s_3}(z_2,\bar{z}_2) + \cdots \, ,
\label{OOope1}
\eea
where $z_{12} \equiv z_1-z_2$, $\bar{z}_{12} \equiv
\bar{z}_1-\bar{z}_2$
and $\kappa_{s_1,s_2,-s_3}$ is the coupling
constant appearing in the bulk
three point scattering amplitude of
massless particles with helicities
$s_1, s_2$ and $s_3=s_1+s_2-2$.
Note that the dummy variable $\al$ also appears
inside of
the Euler beta function in the expansion of $\bar{z}_{12}$.
The connection in the celestial holography
comes from the presence of this bulk three point 
coupling constant $\kappa_{s_1,s_2,-s_3}$ in the celestial OPE.
In general, the conformal dimension $\Delta_3$ for the operator
appearing on the right hand side
is constrained and summed over
\begin{itemize}
\item[]  
i) the dimension $\Delta_1$,
  
ii) the dimension $\Delta_2$ and

iii) the $(d_V-5)$.
\end{itemize}
Moreover, the helicity $s_3$ is, in general, given by
$s_3=-(d_V-3)+s_1+s_2$ and the general expression for the
power of $\bar{z}_{12}$,
$(d_V-4)$, in (\ref{OOope1})  is replaced by $1$.
We can check that both sides of (\ref{OOope1})
are preserved under the left and right conformal weights.
From the argument in the Euler beta function,
the OPE coefficient has the infinite number of poles
and the conformally soft limits occur at the negative
integers or zero for the two arguments of
Euler beta function.  

The $SU(8)$ fundamental indices are denote by
$A,B,C,D, \cdots = 1,2, \cdots, 8$
while the first $SU(4)$ fundamental indices
are given by $a,b,c,d, \cdots =1,2,3,4$
and the second $SU(4)$ fundamental indices
are given by $r,s,t,u, \cdots =5,6,7,8$ \cite{BEF,2212-1,2212-2}. 
The factorization of both $SU(8)_R$ $R$ symmetry indices
and the helicities in the ${\cal N}=8$ supergravity
is given by (\ref{abcdrstu}).
See for the complete OPEs in Appendix $C$ where the $SU(8)$
indices are included
\footnote{
\label{su4case}
The
branching rule for
the  $SU(8) \rightarrow SU(4) \times SU(4) \times U(1)$
\cite{FKS}
is given by
\bea
\Phi_{+2}^{(h)}:{\bf 1} &
\rightarrow & ({\bf 1}, {\bf 1})_0 \, ,
\nonu \\
\Phi_{+\frac{3}{2}}^{(h),A}:{\bf 8} &
\rightarrow & [({\bf 4, 1})_1 \oplus ({\bf 1, 4})_{-1}]
\leftrightarrow  \Phi_{+\frac{3}{2}}^{(h),a} \oplus
\Phi_{+\frac{3}{2}}^{(h),r}\, ,
\nonu \\
\Phi_{+1}^{(h),AB}:{\bf 28} &
\rightarrow & [({\bf 6, 1})_2 \oplus ({\bf 1, 6})_{-2}]
\oplus ({\bf 4, 4})_{0}
\leftrightarrow  \Phi_{+1}^{(h),ab} \oplus
\Phi_{+1}^{(h),rs} \oplus
\Phi_{+1}^{(h),ar}
\, ,
\nonu \\
\Phi_{+\frac{1}{2}}^{(h),ABC}:{\bf 56} &
\rightarrow & [(\overline{\bf 4}, {\bf 1})_3
\oplus ({\bf 1},\overline{\bf  4})_{-3}]
\oplus ({\bf 4, 6})_{-1} \oplus
({\bf 6, 4})_{1}
\leftrightarrow  \Phi_{+\frac{1}{2}}^{(h),abc} \oplus
\Phi_{+\frac{1}{2}}^{(h),rst} \oplus
\Phi_{+\frac{1}{2}}^{(h),ars} \oplus
\Phi_{+\frac{1}{2}}^{(h),abr}
\, ,
\nonu \\
\Phi_{0}^{(h),ABCD}:{\bf 70} & \rightarrow & [({\bf 1}, {\bf 1})_4
\oplus ({\bf 1},{\bf  1})_{-4}]
\oplus ({\bf 4}, \overline{\bf 4})_{-2} \oplus
(\overline{\bf 4}, {\bf 4})_{2} \oplus
({\bf 6}, {\bf 6})_{0}
\nonu \\
& \leftrightarrow &  \Phi_{0}^{(h),abcd} \oplus
\Phi_{0}^{(h),rstu} \oplus
\Phi_{0}^{(h),arst} \oplus
\Phi_{0}^{(h),abcr} \oplus
\Phi_{0}^{(h),abrs} 
\, ,
\nonu \\
\Phi_{-2}^{(h)}:{\bf 1} &
\rightarrow & ({\bf 1}, {\bf 1})_0 \, ,
\nonu \\
\Phi_{A,-\frac{3}{2}}^{(h)}:
\overline{\bf 8} & \rightarrow & [(\overline{\bf 4}, {\bf 1})_{-1}
\oplus ({\bf 1}, \overline{\bf 4})_{1}]
\leftrightarrow  \Phi_{a,-\frac{3}{2}}^{(h)} \oplus
\Phi_{r,-\frac{3}{2}}^{(h)}
\, ,
\nonu \\
\Phi_{AB,-1}^{(h)}:
\overline{\bf 28} & \rightarrow & [({\bf 6},{\bf  1})_{-2}
\oplus ({\bf 1},{\bf  6})_{2}]
\oplus (\overline{\bf 4},\overline{\bf 4})_{0}
\leftrightarrow  \Phi_{ab,-1}^{(h)} \oplus
\Phi_{rs,-1}^{(h)} \oplus
\Phi_{ar,-1}^{(h)}
\, ,
\label{su4branching}
\\
\Phi_{ABC,-\frac{1}{2}}^{(h)}:
\overline{\bf 56} & \rightarrow & [({\bf 4}, {\bf 1})_{-3}
\oplus ({\bf 1},{\bf  4})_{3}]
\oplus (\overline{\bf 4}, {\bf 6})_{1} \oplus
({\bf 6}, \overline{\bf  4})_{-1}
\leftrightarrow  \Phi_{abc,-\frac{1}{2}}^{(h)} \oplus
\Phi_{rst,-\frac{1}{2}}^{(h)} \oplus
\Phi_{ars,-\frac{1}{2}}^{(h)} \oplus
\Phi_{abr,-\frac{1}{2}}^{(h)}
\, .
\nonu
\eea
We collect the singlets in the second
$SU(4)$ and the singlets in the first $SU(4)$
inside the brackets.
Then by considering two additional singlets
denoted by $({\bf 1},{\bf 1})_0$
under the $SU(4) \times SU(4)$
having  a vanishing $U(1)$ charge,
we can associate
$({\bf 1},{\bf 1})_0$,
$({\bf 4},{\bf 1})_1$,
$({\bf 6},{\bf 1})_2$,
$(\overline{\bf 4},{\bf 1})_3$,
and ${({\bf 1},{\bf 1})_4}$
with the graviton with $+2$ helicity,
the gravitinos with $+\frac{3}{2}$ helicity,
the graviphotons with $+1$ helicity,
the graviphotinos with $+\frac{1}{2}$ helicity,
and the complex scalar
with $+0$ helicity of the ${\cal N}=4$ supergravity.
Similarly,
$({\bf 1},{\bf 1})_0$,
$(\overline{\bf 4},{\bf 1})_{-1}$,
$({\bf 6},{\bf 1})_{-2}$,
$({\bf 4},{\bf 1})_{-3}$,
and ${({\bf 1},{\bf 1})_{-4}}$
correspond to
the graviton with $-2$ helicity,
the gravitinos with $-\frac{3}{2}$ helicity,
the graviphotons with $-1$ helicity,
the graviphotinos with $-\frac{1}{2}$ helicity,
and the conjugated complex scalar with $-0$ helicity.
In $SU(4)$, we have $\overline{\bf 6}={\bf 6}$
which are used all the time.
In $SU(8)$, there exists $\overline{\bf 70}={\bf 70}$.
The $U(1)$ charge here is the twice of
the $U(1)$ charge in \cite{AK2501}.
See also the last Appendix $I$.
}.

Generally speaking, there exist holomorphic
and antiholomorphic mode expansions for the soft current.
We consider the particular holomorphic mode which leads to
the simple pole on the holomorphic coordinate, along the lines of
\cite{HPS,Ahn2111}. This implies that there is no loop counting parameter
appeared in \cite{AMS}, in our soft symmetry current algebra,
because the holomorphic mode is not arbitrary but fixed by
one minus the left conformal weight.

Then it is straightforward to compute the
(anti)commutators from (\ref{OOope1})
by performing the various contour integrals described in \cite{GHPS}
\footnote{
\label{OPEanticommutator}
We can calculate
  the following (anti)commutator (ignoring $SU(8)$ indices)
  between the soft currents
  from the celestial OPE
  \cite{HPS,MRSV} with $s_1+s_2-s_3=2$ by absorbing the infinities
appearing in the Euler beta function (we are using the hatted operators)  
\bea
{\cal \hat{O}}_{\Delta_1,s_1}(z_1,\bar{z}_1)
{\cal \hat{O}}_{\Delta_2,s_2}(z_2,\bar{z}_2)
& = &
 \frac{1}{z_{12}}
\sum_{\al=0}^{\infty}
\binom{-\Delta_1-\Delta_2+s_1+s_2-\al-2}{-\Delta_2+s_2-1}\,
\frac{\bar{z}_{12}^{1+\al}}{\al!}
\, \partial^{\al}_{\bar{z}_2}\, {\cal
\hat{O}}_{\Delta_3,s_3}(z_2,\bar{z}_2) + \cdots
\nonumber \\
&=&
 \frac{\bar{z}_{12}}{z_{12}}
\binom{-\Delta_1-\Delta_2+s_1+s_2-2}{-\Delta_2+s_2-1}\,
\, {\cal \hat{O}}_{\Delta_3,s_3}(z_2,\bar{z}_2) + \cdots,
\nonu
\eea
by using the procedure in \cite{Strominger,GHPS} (the dummy variable
$\al=0$)
and it turns out (See also \cite{Tropper1,AK2501})
that the corresponding (anti)commutator leads to
\bea
&& \bigg[ ({\cal \hat{O}}_{\Delta_1,s_1})_m, ({\cal \hat{O}}_{\Delta_2,s_2})_n \bigg\}  
=
-2\,\bigg(\bigg[-\tfrac{\Delta_2-s_2}{2}\bigg]\,m-
\bigg[-\tfrac{\Delta_1-s_1}{2}\bigg]\,n\bigg)
\nonu \\
&& \times \Bigg[
  \frac{(-\tfrac{\Delta_1-s_1}{2}-\tfrac{\Delta_2-s_2}{2}-m-n-1)!(
    -\tfrac{\Delta_1-s_1}{2}-\tfrac{\Delta_2-s_2}{2}+m+n-1)!}{
    (-\tfrac{\Delta_1-s_1}{2}-m)!(-\tfrac{\Delta_2-s_2}{2}-n)!(-
    \tfrac{\Delta_1-s_1}{2}+m)!(-\tfrac{\Delta_2-s_2}{2}+n)!}
\Bigg]
({\cal \hat{O}}_{\Delta_1+\Delta_2,s_1+s_2-2})_{m+n}.
\label{anticommfootnote}
\eea
Finally, by multiplying the denominator of the right hand in this equation
and redefining each two factorials multiplied each celestial operator
as the new celestial operator,
we obtain the final (anti)commutators in (\ref{Eqs})
by adding the couplings.
The overall factor $-2$ is absorbed inside the coupling.

Then we obtain, by using the different notations for the operators,
\bea
\bigg[ (\Phi^{(h_1)}_{s_1})_m, (\Phi^{(h_2)}_{s_2})_n \bigg\}&=&
\kappa_{s_1,s_2,-s_3}\,\Big(\bigg[(h_2+s_2)-1\bigg]m-
\bigg[(h_1+s_1)-1\bigg]n\Big)\,
(\Phi^{(h_1+h_2)}_{-s_1-s_2+2})_{m+n} \, .
\label{Anticomm}
\eea
Note that
the coefficients (i.e., the minus
right conformal weights) in the modes $m$ and $n$
for the first factor on the right hand side of (\ref{anticommfootnote})
are replaced by $(q_2-1)$ and
$(q_1-1)$
respectively in (\ref{Eqs}) with proper $SU(8)$ indices.
The $q_1$ and $q_2$ are defined in Table \ref{BigPhiassign}.
We thank the referee for pointing this out.}.
In the following subsections, we apply (\ref{Eqs})
to the ${\cal N}=8$ supergravity and the claim given in the abstract
can be shown
\footnote{\label{examples}
We can analyze further.
i) For example, from the footnote \ref{threefour},
the $SU(8)$ indices in the interaction between
the two graviphotons leading to the scalars appear on the scalars.
After substituting $s_1=+1$ with $AB$ indices
and $s_2=+1$ with $CD$ indices on the left hand side  
and putting $ABCD$ indices on the right hand side of (\ref{Anticomm}),
we obtain the eq. $17$ of (\ref{Eqs}).
ii) Similarly, in the interaction between
the two graviphotons leading to the gravitons
the $SU(8)$ indices appear on the mixed Kronecker delta.
By substituting $s_1=+1$ with upper $AB$ indices
and $s_2=-1$ with lower $CD$ indices on the left hand side  
and putting the Kronecker delta
on the right hand side of (\ref{Anticomm}), then
the eq. $21$ of (\ref{Eqs}) can be obtained.
iii) Finally,  in the interaction between
the graviphotons and the scalars leading to the graviphotons
the $SU(8)$ indices appear on the Levi Civita contracted with
the two indices of the graviphotons.
By substituting $s_1=+1$ with upper $AB$ indices
and $s_2=0$ with upper $CDEF$ indices on the left hand side  
and putting the Levi Civita with
upper $ABCDEFGH$ indices together with lower $GH$ indices
in the grviphotons
on the right hand side of (\ref{Anticomm}), then
the eq. $19$ of (\ref{Eqs}) can be determined.
Note that for the last two cases ii) and iii),
the amplitude factors depend on the energies of two collinear particles
associated with the footnote \ref{coincident}
and after the Mellin transform
these dependences occur in the arguments of the Euler beta function.
As in the footnote \ref{sanitycheck},
the physical identification of
the resulting (anti)commutators as the celestial OPEs (and vice versa)
is strengthened by these three sanity checks.
We thank the referee for pointing this out.
}.

\subsection{The graviton-the graviton}

From the eq. $1$ and eq. $9$ of (\ref{Eqs}),
we obtain the following celestial commutators
\bea
\comm{(\Phi^{(h_1)}_{+2})_m}{(\Phi^{(h_2)}_{+2})_n}&=&
\kappa_{+2,+2,-2}\,\Big((h_2+1)m-(h_1+1)n\Big)\,
(\Phi^{(h_1+h_2)}_{+2})_{m+n}\, ,
\nonu \\
\comm{(\Phi^{(h_1)}_{+2})_m}{(\Phi^{(h_2)}_{-2})_n}&=&
\kappa_{+2,-2,+2}\,\Big((h_2-3)m-(h_1+1)n\Big)\,
(\Phi^{(h_1+h_2)}_{-2})_{m+n} \, .
\label{eq:graviton_graviton}
\eea

$\bullet$ The first $\frac{1}{\kappa^2} \, e \, R$ term

The $e$ is the determinant
of vierbein ($e \equiv \mbox{det} \, e_{\mu}^a$),
the $R$ is a scalar curvature and $\kappa$ is a gravitational coupling
constant.
The metric (inverse metric, its determinant, affine connection
and a scalar curvature) and 
the vierbein (and its determinant) 
can be expanded around the flat Minkowski spacetimes
\cite{DeWitt,BG,Woodard,CSS}.
Note that the cubic gravitons with two derivatives
appear in the linear $\kappa$ term of the expansion of
this term of the Lagrangian.
Then the scaling dimension of
three point vertex $d_V$ is given by
the sum of three from the three gravitons and
two from two derivatives \cite{PRSY}.
Moreover, the sum of helicities for
these gravitons is given by $(d_V-3)$ \cite{HPS}.
This leads to the fact that
the sum of helicities of those gravitons becomes two
\footnote{We focus on the $d_V=5$ case mainly.}.  
Therefore, 
the helicities $(+2,\pm 2, \mp 2)$ for three gravitons
with two derivatives 
should appear in the coupling of this three point graviton
amplitude and 
the corresponding two celestial commutators are given by
the above equations in (\ref{eq:graviton_graviton}) \footnote{
Note that the helicities on the
right hand sides (in our convention) appear negatively \cite{PRSY}.
The singular behavior for the corresponding OPEs in the
antiholomorphic sector contains the second and the first order
poles. Note that due to the difference between the $+2$
helicity and the $-2$ helicity, the $m$ dependence
(the coefficient is given by
the sum of the superscript and the subscript minus one)
on the right hand sides of (\ref{eq:graviton_graviton})
is different from each other. See also Table \ref{BigPhiassign}.
}.

\subsection{The graviton-the gravitinos
\label{23half}}

From the eq. $2$ and eq. $8$ of (\ref{Eqs}),
we write down the following celestial commutators
\bea
\comm{(\Phi^{(h_1)}_{+2})_m}{(\Phi^{(h_2),a}_{+\frac{3}{2}})_n}&=&
\kappa_{+2,+\frac{3}{2},-\frac{3}{2}}\,\Big((h_2+\tfrac{1}{2})m-(h_1+1)n\Big)\,(\Phi^{(h_1+h_2),a}_{+\frac{3}{2}})_{m+n} \, ,
\nonu
\\
\comm{(\Phi^{(h_1)}_{+2})_m}{(\Phi^{(h_2),r}_{+\frac{3}{2}})_n}&=&
\kappa_{+2,+\frac{3}{2},-\frac{3}{2}}\,\Big((h_2+\tfrac{1}{2})m-(h_1+1)n\Big)\,(\Phi^{(h_1+h_2),r}_{+\frac{3}{2}})_{m+n} \, ,
\nonu
\\
\comm{(\Phi^{(h_1)}_{+2})_m}{(\Phi^{(h_2)}_{a,\,-\frac{3}{2}})_n}&=&
\kappa_{+2,-\frac{3}{2},+\frac{3}{2}}\,\Big((h_2-\tfrac{5}{2})m-(h_1+1)n\Big)\,(\Phi^{(h_1+h_2)}_{a,\,-\frac{3}{2}})_{m+n} \, ,
\nonu
\\
\comm{(\Phi^{(h_1)}_{+2})_m}{(\Phi^{(h_2)}_{r,\,-\frac{3}{2}})_n}&=&
\kappa_{+2,-\frac{3}{2},+\frac{3}{2}}\,\Big((h_2-\tfrac{5}{2})m-(h_1+1)n\Big)\,(\Phi^{(h_1+h_2)}_{r,\,-\frac{3}{2}})_{m+n}\, .
\label{eq:graviton_gravitino}
\eea
The first and
the third where $a=1,2,3,4$
are related to the first $SU(4)$ global symmetry
while 
the second and
the fourth where $r=5,6,7,8$ are related to the second $SU(4)$
global symmetry from the footnote \ref{su4case}.

$\bullet$ The second 
$\epsilon^{\mu \nu \sigma \rho} \, \bar{\psi}^A_{\mu} \, \ga_5\, \ga_{\nu}\,
D_{\rho}\, \psi_{\si}^A$  term (or
$SU(8)$ invariant
$\epsilon^{\mu \nu \sigma \rho} \, \bar{\psi}^A_{\mu} \, \ga_{\nu}\,
D_{\rho}\, \psi_{\si A}$  term)

The Majorana conjugate of the gravitinos
is given by its transpose and the charge conjugation matrix.
Note that the $\ga_5$ matrix is a constant.
The $\ga_{\nu}$ matrix's
expansion around the flat Minkowski spacetimes \cite{Woodard} 
leads to the linear $\kappa$ term with a graviton.
The scaling dimension of the three point vertex between
the graviton and two gravitinos with a
single derivative is given by
$d_V = 1+ 2 \times \frac{3}{2}+1=5$. Note that
the scaling dimension of  a gravitino is equal to $\frac{3}{2}$.
The helicities $(+2,\pm \frac{3}{2}, \mp \frac{3}{2})$
for the graviton and two gravitinos
with a single derivative 
appear in the coupling of this three point amplitude and 
the corresponding celestial commutators
are given by (\ref{eq:graviton_gravitino}).
Recall that the coefficients of the mode $m$
are given by the sum of the superscript and the
subscript of the second operators
appearing on the left hand sides (as before) minus one.

\subsection{The graviton-the graviphotons
\label{2one}}

Similarly from the eq. $3$ and eq. $7$ of (\ref{Eqs}),
the following celestial commutators
are obtained
\bea
\comm{(\Phi^{(h_1)}_{+2})_m}{(\Phi^{(h_2,ab)}_{+1})_n}&=&
\kappa_{+2,+1,-1}\,\Big(h_2\,m-(h_1+1)n\Big)\,(\Phi^{(h_1+h_2),ab}_{+1})_{m+n} \, ,
\nonu
\\
\comm{(\Phi^{(h_1)}_{+2})_m}{(\Phi^{(h_2,ar)}_{ar,\,+1})_n}&=&
\kappa_{+2,+1,-1}\,\Big(h_2\,m-(h_1+1)n\Big)\,(\Phi^{(h_1+h_2),ar}_{+1})_{m+n} \, ,
\nonu
\\
\comm{(\Phi^{(h_1)}_{+2})_m}{(\Phi^{(h_2),rs}_{\,+1})_n}&=&
\kappa_{+2,+1,-1}\,\Big(h_2\,m-(h_1+1)n\Big)\,(\Phi^{(h_1+h_2),rs}_{+1})_{m+n} \, ,
\nonu
\\
\comm{(\Phi^{(h_1)}_{+2})_m}{(\Phi^{(h_2)}_{ab,\,-1})_n}&=&
\kappa_{+2,-1,+1}\,\Big((h_2-2)m-(h_1+1)n\Big)\,(\Phi^{(h_1+h_2)}_{ab,\,-1})_{m+n} \, ,
\nonu
\\
\comm{(\Phi^{(h_1)}_{+2})_m}{(\Phi^{(h_2)}_{ar,\,-1})_n}&=&
\kappa_{+2,-1,+1}\,\Big((h_2-2)m-(h_1+1)n\Big)\,(\Phi^{(h_1+h_2)}_{ar,\,-1})_{m+n} \, ,
\nonu
\\
\comm{(\Phi^{(h_1)}_{+2})_m}{(\Phi^{(h_2)}_{rs,\,-1})_n}&=&
\kappa_{+2,-1,+1}\,\Big((h_2-2)m-(h_1+1)n\Big)\,(\Phi^{(h_1+h_2)}_{rs,\,-1})_{m+n} \, .
\label{eq:graviton_photon}
\eea
The first and
the fourth are related to the first $SU(4)$ global symmetry
while 
the third and
the last are related to the second $SU(4)$
global symmetry.
The remaining ones have mixed two
$SU(4)$ indices.
See also the third relation of  (\ref{su4branching})
where the ${\bf 28}$ (and $\overline{\bf 28}$) representation of $SU(8)$
is breaking into each representation of
$SU(4) \times SU(4) \times U(1)$.

$\bullet$ The third
$e \, g^{\mu \rho} \, g^{\nu \si}\, F_{\mu \nu}^{AB} \, F_{\rho \si}^{ AB}$
term (or $SU(8)$ invariant $e \, g^{\mu \rho} \, g^{\nu \si}\,
F_{\mu \nu}^{AB} \, F_{\rho \si AB}$
term)

The field strength $F_{\mu \nu}^{AB}$ is given by
$F_{\mu \nu}^{AB} \equiv \pa_{\mu} \, A_{\nu}^{AB} - \pa_{\nu}\,
A_{\mu}^{AB}$.
As for three gravitons case, the determinant of vierbein $e$
contains the linear term in the $\kappa$.
The scaling dimension of three point vertex
between the graviton and two graviphotons together with two
derivatives is given by
$d_V=5$ (the scaling dimension of 
graviphotons is one). The sum of helicities
is given by two again and
the helicities $(2, \pm 1, \mp 1)$ appear in
the celestial commutators (\ref{eq:graviton_photon}).

\subsection{The graviton-the graviphotinos
\label{22half}}

From the eq. $4$ and eq. $6$ of (\ref{Eqs}),
the following celestial commutators
can be determined 
\bea
\comm{(\Phi^{(h_1)}_{+2})_m}{(\Phi^{(h_2),abc}_{+\frac{1}{2}})_n}&=&
\kappa_{+2,+\frac{1}{2},-\frac{1}{2}}\,\Big((h_2-\tfrac{1}{2})m-(h_1+1)n\Big)\,(\Phi^{(h_1+h_2),abc}_{+\frac{1}{2}})_{m+n} \, ,
\nonu
\\
\comm{(\Phi^{(h_1)}_{+2})_m}{(\Phi^{(h_2),abr}_{+\frac{1}{2}})_n}&=&
\kappa_{+2,+\frac{1}{2},-\frac{1}{2}}\,\Big((h_2-\tfrac{1}{2})m-(h_1+1)n\Big)\,(\Phi^{(h_1+h_2),abr}_{+\frac{1}{2}})_{m+n} \, ,
\nonu
\\
\comm{(\Phi^{(h_1)}_{+2})_m}{(\Phi^{(h_2),ars}_{+\frac{1}{2}})_n}&=&
\kappa_{+2,+\frac{1}{2},-\frac{1}{2}}\,\Big((h_2-\tfrac{1}{2})m-(h_1+1)n\Big)\,(\Phi^{(h_1+h_2),ars}_{+\frac{1}{2}})_{m+n} \, ,
\nonu
\\
\comm{(\Phi^{(h_1)}_{+2})_m}{(\Phi^{(h_2),rst}_{+\frac{1}{2}})_n}&=&
\kappa_{+2,+\frac{1}{2},-\frac{1}{2}}\,\Big((h_2-\tfrac{1}{2})m-(h_1+1)n\Big)\,(\Phi^{(h_1+h_2),rst}_{+\frac{1}{2}})_{m+n} \, ,
\nonu
\\
\comm{(\Phi^{(h_1)}_{+2})_m}{(\Phi^{(h_2)}_{abc,\,-\frac{1}{2}})_n}&=&
\kappa_{+2,-\frac{1}{2},+\frac{1}{2}}\,\Big((h_2-\tfrac{3}{2})m-(h_1+1)n\Big)\,(\Phi^{(h_1+h_2)}_{abc,\,-\frac{1}{2}})_{m+n} \, ,
\nonu
\\
\comm{(\Phi^{(h_1)}_{+2})_m}{(\Phi^{(h_2)}_{abr,\,-\frac{1}{2}})_n}&=&
\kappa_{+2,-\frac{1}{2},+\frac{1}{2}}\,\Big((h_2-\tfrac{3}{2})m-(h_1+1)n\Big)\,(\Phi^{(h_1+h_2)}_{abr,\,-\frac{1}{2}})_{m+n} \, ,
\nonu
\\
\comm{(\Phi^{(h_1)}_{+2})_m}{(\Phi^{(h_2)}_{ars,\,-\frac{1}{2}})_n}&=&
\kappa_{+2,-\frac{1}{2},+\frac{1}{2}}\,\Big((h_2-\tfrac{3}{2})m-(h_1+1)n\Big)\,(\Phi^{(h_1+h_2)}_{ars,\,-\frac{1}{2}})_{m+n} \, ,
\nonu
\\
\comm{(\Phi^{(h_1)}_{+2})_m}{(\Phi^{(h_2)}_{rst,\,-\frac{1}{2}})_n}&=&
\kappa_{+2,-\frac{1}{2},+\frac{1}{2}}\,\Big((h_2-\tfrac{3}{2})m-(h_1+1)n\Big)\,(\Phi^{(h_1+h_2)}_{rst,\,-\frac{1}{2}})_{m+n} \, .
\label{eq:graviton_photino}
\eea
The first and
the fifth are related to the first $SU(4)$ global symmetry
while 
the fourth and
the last are related to the second $SU(4)$
global symmetry.
The remaining four have mixed two
$SU(4)$ indices.
See also the fourth relation of  (\ref{su4branching})
where the ${\bf 56}$ (and $\overline{\bf 56}$)
representation of $SU(8)$
is breaking into each representation of
$SU(4) \times SU(4) \times U(1)$.

$\bullet$ The fourth
$e \, \bar{\chi}^{ABC} \, \ga^{\mu} \, D_{\mu}\, \chi^{ABC}$
term (or $SU(8)$ invariant
$e \, \bar{\chi}^{ABC} \, \ga^{\mu} \, D_{\mu}\, \chi_{ABC}$
term)

There exists a linear term in the
$\kappa$ from the expansion of the determinant
of vierbein around the flat Minkowski spacetimes.
The scaling dimension of the three point vertex between
the graviton and two graviphotinos together with a
single derivative is given by
$d_V = 1+ 2 \times \frac{3}{2}+1=5$ where the number $\frac{3}{2}$
comes from the scaling dimension from a graviphotino.
Again, the sum of helicities should be equal to
two.
The helicities $(+2,\pm \frac{1}{2}, \mp \frac{1}{2})$
for  the graviton and two graviphotinos
with a single derivative  should  appear in the celestial
commutators given by (\ref{eq:graviton_photino}).

\subsection{The graviton-the scalars
\label{twozero}}

From the eq. $5$ of (\ref{Eqs}),
we obtain the following celestial commutators
\bea
\comm{(\Phi^{(h_1)}_{+2})_m}{(\Phi^{(h_2),abcd}_{0})_n}&=&
\kappa_{+2,0,0}\,\Big((h_2-1)m-(h_1+1)n\Big)\,(\Phi^{(h_1+h_2),abcd}_{0})_{m+n} \, ,
\nonu
\\
\comm{(\Phi^{(h_1)}_{+2})_m}{(\Phi^{(h_2),abcr}_{0})_n}&=&
\kappa_{+2,0,0}\,\Big((h_2-1)m-(h_1+1)n\Big)\,(\Phi^{(h_1+h_2),abcr}_{0})_{m+n} \, ,
\nonu\\
\comm{(\Phi^{(h_1)}_{+2})_m}{(\Phi^{(h_2),abrs}_{0})_n}&=&
\kappa_{+2,0,0}\,\Big((h_2-1)m-(h_1+1)n\Big)\,(\Phi^{(h_1+h_2),abrs}_{0})_{m+n} \, ,
\nonu\\
\comm{(\Phi^{(h_1)}_{+2})_m}{(\Phi^{(h_2),arst}_{0})_n}&=&
\kappa_{+2,0,0}\,\Big((h_2-1)m-(h_1+1)n\Big)\,(\Phi^{(h_1+h_2),arst}_{0})_{m+n} \, ,
\nonu\\
\comm{(\Phi^{(h_1)}_{+2})_m}{(\Phi^{(h_2),rstu}_{0})_n}&=&
\kappa_{+2,0,0}\,\Big((h_2-1)m-(h_1+1)n\Big)\,(\Phi^{(h_1+h_2),rstu}_{0})_{m+n} \, .
\label{eq:graviton_scalar}
\eea
The first is related to the first $SU(4)$ global symmetry
while 
the last  is related to the second $SU(4)$
global symmetry.
The remaining three have mixed two
$SU(4)$ indices.
See also the fifth relation of  (\ref{su4branching})
where the ${\bf 70}$
representation of $SU(8)$
is breaking into each representation of
$SU(4) \times SU(4) \times U(1)$.

$\bullet$ The fifth
$e \, g^{\mu \nu}\, (\pa_{\mu}\, A^{ABCD}) \, (\pa_{\nu}\, A^{ABCD})$
and sixth 
$e \, g^{\mu \nu}\, (\pa_{\mu}\, B^{ABCD}) \, (\pa_{\nu}\, B^{ABCD})$
terms (or $SU(8)$ invariant $e \, g^{\mu \nu}\, \epsilon_{ABCDEFGH}
(\pa_{\mu}\, \phi^{\ast ABCD}) \,
(\pa_{\nu}\, \phi^{\ast EFGH})$ term)

After identifying
the lowest order term
in the expansion of the determinant of vierbein
around the flat Minkowski spacetime
with the kinetic term,
the next linear $\kappa$ term provides the interaction
between the graviton and two scalars 
with two derivatives.
The scaling dimension of three point vertex
between the graviton and two scalars together with two
derivatives is given by
$d_V=5$.
The sum of helicities should be
equal to two as before. 
The celestial commutators (\ref{eq:graviton_scalar}) shows 
the helicities $(+2, 0, 0)$
for the graviton and two scalars.

\subsection{The gravitinos-the gravitinos
\label{3half3half}}

From the eq. $10$ and eq. $16$ of (\ref{Eqs}),
the following celestial anticommutators
can be obtained
\bea
\acomm{(\Phi^{(h_1),a}_{+\frac{3}{2}})_m}{(\Phi^{(h_2),b}_{+\frac{3}{2}})_n}
&= & \kappa_{+\frac{3}{2},+\frac{3}{2},-1}\,\Big((h_2+\tfrac{1}{2})m-(h_1+\tfrac{1}{2})n\Big)\,(\Phi^{(h_1+h_2),ab}_{+1})_{m+n} \, ,
\nonu\\
\acomm{(\Phi^{(h_1),r}_{+\frac{3}{2}})_m}{(\Phi^{(h_2),s}_{+\frac{3}{2}})_n}
&= & \kappa_{+\frac{3}{2},+\frac{3}{2},-1}\,\Big((h_2+\tfrac{1}{2})m-(h_1+\tfrac{1}{2})n\Big)\,(\Phi^{(h_1+h_2),rs}_{+1})_{m+n} \, ,
\nonu\\
\acomm{(\Phi^{(h_1),a}_{+\frac{3}{2}})_m}{(\Phi^{(h_2),r}_{+\frac{3}{2}})_n}
& = & \kappa_{+\frac{3}{2},+\frac{3}{2},-1}\,\Big((h_2+\tfrac{1}{2})m-(h_1+\tfrac{1}{2})n\Big)\,(\Phi^{(h_1+h_2),ar}_{+1})_{m+n} \, ,
\nonu\\
\acomm{(\Phi^{(h_1),a}_{+\frac{3}{2}})_m}{(\Phi^{(h_2)}_{b,\,-\frac{3}{2}})_n}
&=&\kappa_{+\frac{3}{2},-\frac{3}{2},+2}\,\Big((h_2-\tfrac{5}{2})m-(h_1+\tfrac{1}{2})n\Big)\delta^{a}_{b}\,(\Phi^{(h_1+h_2)}_{-2})_{m+n} \, ,
\nonu\\
\acomm{(\Phi^{(h_1),r}_{+\frac{3}{2}})_m}{(\Phi^{(h_2)}_{s,\,-\frac{3}{2}})_n}
& = & \kappa_{+\frac{3}{2},-\frac{3}{2},+2}\,\Big((h_2-\tfrac{5}{2})m-(h_1+\tfrac{1}{2})n\Big)\delta^{r}_{s}\,(\Phi^{(h_1+h_2)}_{-2})_{m+n} \, .
\label{eq:gravitino_gravitino}
\eea
The first and
the fourth are related to the first $SU(4)$ global symmetry
while 
the second and
the last are related to the second $SU(4)$
global symmetry.
For the first three cases,
when we interchange the $h_1 \leftrightarrow h_2$,
$a(r)[a] \leftrightarrow b(s)[r]$, $m \leftrightarrow n$,
we have the invariance of the right hand sides
because the graviphotons are antisymmetric
under the $SU(8)$ indices and this will act on the
mode dependent terms also. See also the footnote $1$.
Among seven
relations in the tensor product ${\bf 8} \otimes {\bf 8}$
and in the tensor product ${\bf 8} \otimes \overline{\bf 8}$,
we are left with the five nontrivial anticommutators
\footnote{
\label{17}
In the tensor product
${\bf 8} \otimes {\bf 8}$, there are
three tensor products between
the $SU(4) \times SU(4) \times U(1)$, 
$({\bf 4},{\bf 1})_{1} \otimes ({\bf 4},{\bf 1})_{1}$
from which there is an antisymmetric representation
$({\bf 6},{\bf 1})_2$,
$({\bf 1},{\bf 4})_{-1} \otimes ({\bf 1},{\bf 4})_{-1}$
from which there exists an antisymmetric
representation $({\bf 1},{\bf 6})_{-2}$,
and
$({\bf 4},{\bf 1})_{1} \otimes ({\bf 1},{\bf 4})_{-1}$
where $({\bf 4},{\bf 4})_0$ appears.
Similarly,
in
${\bf 8} \otimes \overline{\bf 8}$, there are
two tensor products under the $SU(4) \times SU(4) \times U(1)$, 
$({\bf 4},{\bf 1})_{1} \otimes (\overline{\bf 4},{\bf 1})_{-1}$
from which there is
$({\bf 1},{\bf 1})_0$
and
$({\bf 1},{\bf 4})_{-1} \otimes ({\bf 1},\overline{\bf 4})_{1}$
from which again there exists
$({\bf 1},{\bf 1})_0$.
For the remaining
two tensor products having
nonzero $U(1)$ charges, there are trivial vanishing
anticommutators  
$\acomm{(\Phi^{(h_1),a}_{+\frac{3}{2}})_m}{(\Phi^{(h_2)}_{r,\,-\frac{3}{2}})_n}_2=0=
\acomm{(\Phi^{(h_1),r}_{+\frac{3}{2}})_m}{(\Phi^{(h_2)}_{a,\,-\frac{3}{2}})_n}_{-2}$
from the eq. $16$ of (\ref{Eqs}) with $\de^a_r=0=\de^r_a$.
Here the subscript in the anticommutators
stands for the $U(1)$ charge.}.

$\bullet$ The seventh
$\kappa \,  e\, g^{\mu \rho} \, g^{\nu \si}\,
\bar{\psi}^A_{\mu} \,  F_{\rho \si}^{ AB}\, \psi_{\nu}^B$
and eighth
$\kappa \, e\,  \epsilon^{\mu \nu \rho \si} \,
\bar{\psi}^A_{\mu} \,  \ga_5\,
F_{\rho \si}^{ AB}\, \psi_{\nu}^B$
terms (or $SU(8)$ invariant
$\kappa \,  e\, g^{\mu \rho} \, g^{\nu \si}\,
\bar{\psi}^A_{\mu} \,  F_{\rho \si AB}\, \psi_{\nu}^B$ term)

We consider the interactions between two gravitinos and the graviphotons.
Note that the $SU(8)$ indices for the two gravitinos
are contracted with the ones for the graviphotons.
The sum of helicities should be
equal to two as before and 
the first three celestial anticommutators of
(\ref{eq:gravitino_gravitino}) gives 
the helicities $(+\frac{3}{2}, +\frac{3}{2}, -1)$
for  two gravitinos and the graviphotons
while
the helicities $(+ \frac{3}{2}, - \frac{3}{2},+2)$
for  two gravitinos and the graviton appear
in the last two anticommutators given by (\ref{eq:gravitino_gravitino})
where the corresponding term of the Lagrangian appears in the subsection
\ref{23half}.

\subsection{The gravitinos-the graviphotons
\label{3halfone}}

From the eq. $11$ and eq. $15$ of (\ref{Eqs}),
we describe the following celestial commutators
\bea
\comm{(\Phi^{(h_1),a}_{+\frac{3}{2}})_m}{(\Phi^{(h_2),bc}_{+1})_n}&=&\kappa_{+\frac{3}{2},+1,-\frac{1}{2}}\,\Big(h_2\,m-(h_1+\tfrac{1}{2})n\Big)\,(\Phi^{(h_1+h_2),abc}_{+\frac{1}{2}})_{m+n} \, ,
\nonu\\
\comm{(\Phi^{(h_1),a}_{+\frac{3}{2}})_m}{(\Phi^{(h_2),br}_{+1})_n}&=&
\kappa_{+\frac{3}{2},+1,-\frac{1}{2}}\,\Big(h_2\,m-(h_1+\tfrac{1}{2})n\Big)\,(\Phi^{(h_1+h_2),abr}_{+\frac{1}{2}})_{m+n} \, ,
\nonu\\
\comm{(\Phi^{(h_1),a}_{+\frac{3}{2}})_m}{(\Phi^{(h_2),rs}_{+1})_n}
&=&\kappa_{+\frac{3}{2},+1,-\frac{1}{2}}\,\Big(h_2\,m-(h_1+\tfrac{1}{2})n\Big)\,(\Phi^{(h_1+h_2),ars}_{+\frac{1}{2}})_{m+n} \, ,
\nonu\\
\comm{(\Phi^{(h_1),r}_{+\frac{3}{2}})_m}{(\Phi^{(h_2),ab}_{+1})_n}&=&
\kappa_{+\frac{3}{2},+1,-\frac{1}{2}}\,\Big(h_2\,m-(h_1+\tfrac{1}{2})n\Big)\,(\Phi^{(h_1+h_2),rab}_{+\frac{1}{2}})_{m+n} \, ,
\nonu\\
\comm{(\Phi^{(h_1),r}_{+\frac{3}{2}})_m}{(\Phi^{(h_2),as}_{+1})_n}&=&
\kappa_{+\frac{3}{2},+1,-\frac{1}{2}}\,\Big(h_2\,m-(h_1+\tfrac{1}{2})n\Big)\,(\Phi^{(h_1+h_2),ras}_{+\frac{1}{2}})_{m+n} \, ,
\nonu\\
\comm{(\Phi^{(h_1),r}_{+\frac{3}{2}})_m}{(\Phi^{(h_2),st}_{+1})_n}&=&
\kappa_{+\frac{3}{2},+1,-\frac{1}{2}}\,\Big(h_2\,m-(h_1+\tfrac{1}{2})n\Big)\,(\Phi^{(h_1+h_2),rst}_{+\frac{1}{2}})_{m+n} \, ,
\nonu\\
\comm{(\Phi^{(h_1),a}_{+\frac{3}{2}})_m}{(\Phi^{(h_2)}_{bc,\,-1})_n}&=&
\kappa_{+\frac{3}{2},-1,+\frac{3}{2}}\,\Big((h_2-2)m-(h_1+\tfrac{1}{2})n\Big)\,
2!\, \delta^a_{[b}(\Phi^{(h_1+h_2)}_{c],\,-\frac{3}{2}})_{m+n} \, ,
\nonu\\
\comm{(\Phi^{(h_1),a}_{+\frac{3}{2}})_m}{(\Phi^{(h_2)}_{br,\,-1})_n}&=&
\kappa_{+\frac{3}{2},-1,+\frac{3}{2}}\,\Big((h_2-2)m-(h_1+\tfrac{1}{2})n\Big)\,
\delta^a_b(\Phi^{(h_1+h_2)}_{r,\,-\frac{3}{2}})_{m+n} \, ,
\nonu\\
\comm{(\Phi^{(h_1),r}_{+\frac{3}{2}})_m}{(\Phi^{(h_2)}_{as,\,-1})_n}&=&
-\kappa_{+\frac{3}{2},-1,+\frac{3}{2}}\,\Big((h_2-2)m-(h_1+\tfrac{1}{2})n\Big)
\,\delta^r_s(\Phi^{(h_1+h_2)}_{a,\,-\frac{3}{2}})_{m+n} \, ,
\nonu\\
\comm{(\Phi^{(h_1),r}_{+\frac{3}{2}})_m}{(\Phi^{(h_2)}_{st,\,-1})_n}&=&
\kappa_{+\frac{3}{2},-1,+\frac{3}{2}}\,\Big((h_2-2)m-(h_1+\tfrac{1}{2})n\Big)\,
2!\, \delta^r_{[s}(\Phi^{(h_1+h_2)}_{t],\,-\frac{3}{2}})_{m+n} \, .
\label{eq:gravitino_photon}
\eea
The first and
the seventh are related to the first $SU(4)$ global symmetry
while 
the sixth and
the last are related to the second $SU(4)$
global symmetry.
Note that the minus sign
appearing in the second from the below
can be understood from the Kronecker delta in the eq. $15$
of (\ref{Eqs}) \footnote{
\label{18}
In
the tensor product ${\bf 8} \otimes {\bf 28}$, there are
six tensor products
between the $SU(4) \times SU(4) \times U(1)$,
$({\bf 4},{\bf 1})_{1} \otimes ({\bf 6},{\bf 1})_{2}$
from which there exists an antisymmetric
representation $(\overline{\bf 4},{\bf 1 })_3$,
$({\bf 1},{\bf 4})_{-1} \otimes ({\bf 6},{\bf 1})_{2}$
where $({\bf 6},{\bf 4})_1$ appears,
$({\bf 4},{\bf 1})_{1} \otimes ({\bf 1},{\bf 6})_{-2}$
where there exists $({\bf 4},{\bf 6})_{-1}$,
$({\bf 4},{\bf 1})_{1} \otimes ({\bf 4},{\bf 4})_{0}$
from which there is a representation
$({\bf 6},{\bf 4} )_{1}$,
$({\bf 1},{\bf 4})_{-1} \otimes ({\bf 4},{\bf 4})_{0}$
from which there is a  representation
$({\bf 4},{\bf 6} )_{-1}$,
and $({\bf 1},{\bf 4})_{-1} \otimes ({\bf 1},{\bf 6})_{-2}$
from which there is an antisymmetric
representation $({\bf 1},\overline{\bf 4})_{-3}$.
Similarly,
in the tensor product
${\bf 8} \otimes \overline{\bf 28}$,
there are
four tensor products, 
$({\bf 4},{\bf 1})_{1} \otimes ({\bf 6},{\bf 1})_{-2}$
from which there exists an antisymmetric
representation $(\overline{\bf 4},{\bf 1 })_{-1}$,
$({\bf 4},{\bf 1})_{1} \otimes (\overline{\bf 4},\overline
{\bf 4})_{0}$
from which there is an antisymmetric representation
$({\bf 1},\overline{\bf 4} )_{1}$,
$({\bf 1},{\bf 4})_{-1} \otimes (\overline{\bf 4},
\overline{\bf 4})_{0}$
from which there is an antisymmetric representation
$(\overline{\bf 4},{\bf 1} )_{-1}$,
and $({\bf 1},{\bf 4})_{-1} \otimes ({\bf 1},{\bf 6})_{2}$
from which there is an antisymmetric
representation $({\bf 1},\overline{\bf 4})_{1}$.
For the remaining two tensor products
having $\pm 3$ $U(1)$ charges,
there exist
vanishing commutators
$\comm{(\Phi^{(h_1),a}_{+\frac{3}{2}})_m}{(\Phi^{(h_2)}_{rs,\,-1})_n}_3=0=
\comm{(\Phi^{(h_1),r}_{+\frac{3}{2}})_m}{(\Phi^{(h_2)}_{ab,\,-1})_n}_{-3}$
from the eq. $15$ of (\ref{Eqs}).}.

$\bullet$ The ninth
$\kappa \, e  \, \bar{\psi}^A_{\la} \,
\sigma^{\mu \nu} \, F^{BC}_{\mu \nu}\, \gamma^{\la}\, \chi^{ABC}$
term (or $SU(8)$ invariant $\kappa \, e  \, \bar{\psi}^A_{\la} \,
\sigma^{\mu \nu} \, F^{BC}_{\mu \nu}\, \gamma^{\la}\, \chi_{ABC}$
term)

There exists a single derivative from the graviphotons.
The $\sigma^{\mu \nu}$ is the antisymmetric
gamma matrices and is contracted with the graviphotons.
The single gamma matrix is contracted with the gravitino.
The first six celestial commutators of (\ref{eq:gravitino_photon})
provide the helicities $(+\frac{3}{2}, +1, -\frac{1}{2})$
between the gravitino, graviphoton and the graviphotino.
Similarly, the last four celestial
commutators of (\ref{eq:gravitino_photon}) give
the helicities $(+\frac{3}{2},-1, +\frac{3}{2})$
for gravitino, graviphoton and gravitino
with the term of Lagrangian appeared in the subsection
\ref{3half3half}.

\subsection{The gravitinos-the graviphotinos
\label{3half2half}}

From the eq. $12$ and eq. $14$ of (\ref{Eqs}),
we obtain the following celestial anticommutators
\bea
\acomm{(\Phi^{(h_1),a}_{+\frac{3}{2}})_m}{(\Phi^{(h_2),bcd}_{+\frac{1}{2}})_n}
& = & \kappa_{+\frac{3}{2},+\frac{1}{2},0}\,\Big((h_2-\tfrac{1}{2})m-(h_1+\tfrac{1}{2})n\Big)\,(\Phi^{(h_1+h_2),abcd}_{0})_{m+n} \, ,
\nonu\\
\acomm{(\Phi^{(h_1),a}_{+\frac{3}{2}})_m}{(\Phi^{(h_2),bcr}_{+\frac{1}{2}})_n}
&= & \kappa_{+\frac{3}{2},+\frac{1}{2},0}\,\Big((h_2-\tfrac{1}{2})m-(h_1+\tfrac{1}{2})n\Big)\,(\Phi^{(h_1+h_2),abcr}_{0})_{m+n} \, ,
\nonu\\
\acomm{(\Phi^{(h_1),a}_{+\frac{3}{2}})_m}{(\Phi^{(h_2),brs}_{+\frac{1}{2}})_n}
& = & \kappa_{+\frac{3}{2},+\frac{1}{2},0}\,\Big((h_2-\tfrac{1}{2})m-(h_1+\tfrac{1}{2})n\Big)\,(\Phi^{(h_1+h_2),abrs}_{0})_{m+n} \, ,
\nonu\\
\acomm{(\Phi^{(h_1),a}_{+\frac{3}{2}})_m}{(\Phi^{(h_2),rst}_{+\frac{1}{2}})_n}
& = & \kappa_{+\frac{3}{2},+\frac{1}{2},0}\,\Big((h_2-\tfrac{1}{2})m-(h_1+\tfrac{1}{2})n\Big)\,(\Phi^{(h_1+h_2),arst}_{0})_{m+n} \, ,
\nonu\\
\acomm{(\Phi^{(h_1),r}_{+\frac{3}{2}})_m}{(\Phi^{(h_2),abc}_{+\frac{1}{2}})_n}
&= & \kappa_{+\frac{3}{2},+\frac{1}{2},0}\,\Big((h_2-\tfrac{1}{2})m-(h_1+\tfrac{1}{2})n\Big)\,(\Phi^{(h_1+h_2),rabc}_{0})_{m+n} \, ,
\nonu\\
\acomm{(\Phi^{(h_1),r}_{+\frac{3}{2}})_m}{(\Phi^{(h_2),abs}_{+\frac{1}{2}})_n}
&= & \kappa_{+\frac{3}{2},+\frac{1}{2},0}\,\Big((h_2-\tfrac{1}{2})m-(h_1+\tfrac{1}{2})n\Big)\,(\Phi^{(h_1+h_2),rabs}_{0})_{m+n} \, ,
\nonu\\
\acomm{(\Phi^{(h_1),r}_{+\frac{3}{2}})_m}{(\Phi^{(h_2),ast}_{+\frac{1}{2}})_n}
&= & \kappa_{+\frac{3}{2},+\frac{1}{2},0}\,\Big((h_2-\tfrac{1}{2})m-(h_1+\tfrac{1}{2})n\Big)\,(\Phi^{(h_1+h_2),rast}_{0})_{m+n} \, ,
\nonu\\
\acomm{(\Phi^{(h_1),r}_{+\frac{3}{2}})_m}{(\Phi^{(h_2),stu}_{+\frac{1}{2}})_n}
& = & \kappa_{+\frac{3}{2},+\frac{1}{2},0}\,\Big((h_2-\tfrac{1}{2})m-(h_1+\tfrac{1}{2})n\Big)\,(\Phi^{(h_1+h_2),rstu}_{0})_{m+n} \, ,
\nonu\\
\acomm{(\Phi^{(h_1),a}_{+\frac{3}{2}})_m}{(\Phi^{(h_2)}_{bcd,\,-\frac{1}{2}})_n}
&= & \kappa_{+\frac{3}{2},-\frac{1}{2},+1}\,\Big((h_2-\tfrac{3}{2})m-(h_1+\tfrac{1}{2})n\Big)\,3\delta^{a}_{[b}(\Phi^{(h_1+h_2)}_{cd],\,-1})_{m+n} \, ,
\nonu\\
\acomm{(\Phi^{(h_1),a}_{+\frac{3}{2}})_m}{(\Phi^{(h_2)}_{bcr,\,-\frac{1}{2}})_n}
&= &
\kappa_{+\frac{3}{2},-\frac{1}{2},+1}\,\Big((h_2-\tfrac{3}{2})m-(h_1+\tfrac{1}{2})n\Big)\,2!\, \delta^{a}_{[b}(\Phi^{(h_1+h_2)}_{c]r,\,-1})_{m+n} \, ,
\nonu\\
\acomm{(\Phi^{(h_1),a}_{+\frac{3}{2}})_m}{(\Phi^{(h_2)}_{brs,\,-\frac{1}{2}})_n}
&= &
\kappa_{+\frac{3}{2},-\frac{1}{2},+1}\,\Big((h_2-\tfrac{3}{2})m-(h_1+\tfrac{1}{2})n\Big)\,\delta^{a}_{b}
(\Phi^{(h_1+h_2)}_{rs,\,-1})_{m+n} \, ,
\nonu\\
\acomm{(\Phi^{(h_1),r}_{+\frac{3}{2}})_m}{(\Phi^{(h_2)}_{abs,\,-\frac{1}{2}})_n}
&= &
\kappa_{+\frac{3}{2},-\frac{1}{2},+1}\,\Big((h_2-\tfrac{3}{2})m-(h_1+\tfrac{1}{2})n\Big)\,\delta^{r}_{s}
(\Phi^{(h_1+h_2)}_{ab,\,-1})_{m+n} \, ,
\nonu\\
\acomm{(\Phi^{(h_1),r}_{+\frac{3}{2}})_m}{(\Phi^{(h_2)}_{ast,\,-\frac{1}{2}})_n}
& = & \kappa_{+\frac{3}{2},-\frac{1}{2},+1}\,\Big((h_2-\tfrac{3}{2})m-(h_1+\tfrac{1}{2})n\Big)\,2!\, \delta^{r}_{[s}(\Phi^{(h_1+h_2)}_{t]a,\,-1})_{m+n} \, ,
\nonu\\
\acomm{(\Phi^{(h_1),r}_{+\frac{3}{2}})_m}{(\Phi^{(h_2)}_{stu,\,-\frac{1}{2}})_n}
& = & \kappa_{+\frac{3}{2},-\frac{1}{2},+1}\,\Big((h_2-\tfrac{3}{2})m-(h_1+\tfrac{1}{2})n\Big)\,3\delta^{r}_{[s}(\Phi^{(h_1+h_2)}_{tu],\,-1})_{m+n} \, .
\label{eq:gravitino_photino}
\end{eqnarray}
The first and
the ninth are related to the first $SU(4)$ global symmetry
while 
the eighth and
the last are related to the second $SU(4)$
global symmetry
\footnote{
\label{19}  
In the tensor product
${\bf 8} \otimes {\bf 56}$, there are
eight tensor products, 
$({\bf 4},{\bf 1})_{1} \otimes (\overline{\bf 4},{\bf 1})_{3}$
from which there is $({\bf 1},{\bf 1})_{4}$,
$({\bf 4},{\bf 1})_{1} \otimes ({\bf 6},{\bf 4})_{1}$
from which there exists $(\overline{\bf 4},{\bf 4})_{2}$,
$({\bf 4},{\bf 1})_{1} \otimes ({\bf 4},{\bf 6})_{-1}$
from which there is $({\bf 6},{\bf 6})_{0}$,
$({\bf 4},{\bf 1})_{1} \otimes ({\bf 1},\overline{\bf 4})_{-3}$
from which there appears $({\bf 4},\overline{\bf 4})_{-2}$,
$({\bf 1},{\bf 4})_{-1} \otimes (\overline{\bf 4},{\bf 1})_{3}$
where there exists  $(\overline{\bf 4},{\bf 4})_{2}$,
$({\bf 1},{\bf 4})_{-1} \otimes ({\bf 6},{\bf 4})_{1}$
where there is a representation $({\bf 6},{\bf 6})_{0}$,
$({\bf 1},{\bf 4})_{-1} \otimes ({\bf 4},{\bf 6})_{-1}$
where  there is a representation $({\bf 4},\overline{\bf 4})_{-2}$,
and
$({\bf 1},{\bf 4})_{-1} \otimes ({\bf 1},\overline{\bf 4})_{-3}$
where  there exists  $({\bf 1},{\bf 1})_{-4}$.
Moreover,
in the tensor product
${\bf 8} \otimes \overline{\bf 56}$, there are
six tensor products,
$({\bf 4},{\bf 1})_{1} \otimes ({\bf 4},{\bf 1})_{-3}$
from which there is $({\bf 6},{\bf 1})_{-2}$,
$({\bf 4},{\bf 1})_{1} \otimes ({\bf 6},\overline{\bf 4})_{-1}$
from which there exists $(\overline{\bf 4},\overline{\bf 4})_{0}$,
$({\bf 4},{\bf 1})_{1} \otimes (\overline{\bf 4},{\bf 6})_{1}$
from which there exists $({\bf 1},{\bf 6})_{2}$,
$({\bf 1},{\bf 4})_{-1} \otimes ({\bf 6},\overline{\bf 4})_{-1}$
where there is  $({\bf 6},{\bf 1})_{-2}$,
$({\bf 1},{\bf 4})_{-1} \otimes (\overline{\bf 4},{\bf 6})_{1}$
where  there is  $(\overline{\bf 4},\overline{\bf 4})_{0}$,
and
$({\bf 1},{\bf 4})_{-1} \otimes ({\bf 1},{\bf 4})_{3}$
where  there is  $({\bf 1},{\bf 6})_{2}$.
For the remaining two tensor products
having $\pm 4$ $U(1)$ charges,
there are trivial anticommutators
$\acomm{(\Phi^{(h_1),a}_{+\frac{3}{2}})_m}{(\Phi^{(h_2)}_{rst,\,-\frac{1}{2}})_n}_4=0=
\acomm{(\Phi^{(h_1),r}_{+\frac{3}{2}})_m}{(\Phi^{(h_2)}_{abc,\,-\frac{1}{2}})_n}_{-4}$
from the eq. $14$ of (\ref{Eqs}).}.

$\bullet$ The tenth
$\kappa \, e\, \bar{\psi}^A_{\mu} \, (\pa_{\nu} \, A^{ABCD}) \, \ga^{\nu}\,
\ga^{\mu} \, \chi^{BCD}$ and eleventh
$\kappa \, e\, \bar{\psi}^A_{\mu} \, (\pa_{\nu} \, \ga_5\, B^{ABCD}) \, \ga^{\nu}\,
\ga^{\mu} \, \chi^{BCD}$
terms (or $SU(8)$ invariant $\kappa \, e\, \bar{\psi}_{\mu}^{A} \,
\epsilon_{ABCDEFGH}(\pa_{\nu} \, \phi^{\ast EFGH}) \, \ga^{\nu}\,
\ga^{\mu} \, \chi^{BCD} + \mbox{h.c.}$ term)

There is a single derivative.
The helicities of first eight celestial anticommutators
of (\ref{eq:gravitino_photino}) are
given by
$(+\frac{3}{2},+\frac{1}{2},0)$
between the gravitino, the graviphotino and the scalar.
The helicities of the remaining celestial anticommutators
of (\ref{eq:gravitino_photino}) are denoted by
$(+\frac{3}{2},-\frac{1}{2}, +1)$
for the gravitino, the graviphotino and the graviphoton
associated with the previous Lagrangian term (ninth term)
in the subsection \ref{3halfone}.

\subsection{The gravitinos-the scalars}

The eq. $13$ of (\ref{Eqs})
provides the following celestial commutators
\bea
\comm{(\Phi^{(h_1),a}_{+\frac{3}{2}})_m}{(\Phi^{(h_2),bcdr}_{0})_n}&=&
\kappa_{+\frac{3}{2},0,+\frac{1}{2}}\,\Big((h_2-1)m-(h_1+\tfrac{1}{2})n\Big)\,\frac{1}{3!}\,\epsilon^{abcd}\epsilon^{rstu}(\Phi^{(h_1+h_2)}_{stu,\,-\frac{1}{2}})_{m+n} \, ,
\nonu\\
\comm{(\Phi^{(h_1),a}_{+\frac{3}{2}})_m}{(\Phi^{(h_2),bcrs}_{0})_n}&=&
\kappa_{+\frac{3}{2},0,+\frac{1}{2}}\,\Big((h_2-1)m-(h_1+\tfrac{1}{2})n\Big)\,\frac{1}{2!}\,\epsilon^{abcd}\epsilon^{rstu}(\Phi^{(h_1+h_2)}_{dtu,\,-\frac{1}{2}})_{m+n} \, ,
\nonu\\
\comm{(\Phi^{(h_1),a}_{+\frac{3}{2}})_m}{(\Phi^{(h_2),brst}_{0})_n}&=&
\kappa_{+\frac{3}{2},0,+\frac{1}{2}}\,\Big((h_2-1)m-(h_1+\tfrac{1}{2})n\Big)\,\frac{1}{2!}\,\epsilon^{abcd}\epsilon^{rstu}(\Phi^{(h_1+h_2)}_{cdu,\,-\frac{1}{2}})_{m+n} \, ,
\nonu\\
\comm{(\Phi^{(h_1),a}_{+\frac{3}{2}})_m}{(\Phi^{(h_2),rstu}_{0})_n}&=&
\kappa_{+\frac{3}{2},0,+\frac{1}{2}}\,\Big((h_2-1)m-(h_1+\tfrac{1}{2})n\Big)\,\frac{1}{3!}\,\epsilon^{abcd}\epsilon^{rstu}(\Phi^{(h_1+h_2)}_{bcd,\,-\frac{1}{2}})_{m+n} \, ,
\nonu\\
\comm{(\Phi^{(h_1),r}_{+\frac{3}{2}})_m}{(\Phi^{(h_2),abcd}_{0})_n}&=&
\kappa_{+\frac{3}{2},0,+\frac{1}{2}}\,\Big((h_2-1)m-(h_1+\tfrac{1}{2})n\Big)\,\frac{1}{3!}\,\epsilon^{abcd}\epsilon^{rstu}(\Phi^{(h_1+h_2)}_{stu,\,-\frac{1}{2}})_{m+n} \, ,
\nonu\\
\comm{(\Phi^{(h_1),r}_{+\frac{3}{2}})_m}{(\Phi^{(h_2),abcs}_{0})_n}&=&
-\kappa_{+\frac{3}{2},0,+\frac{1}{2}}\,\Big((h_2-1)m-(h_1+\tfrac{1}{2})n\Big)\,\frac{1}{2!}\,\epsilon^{abcd}\epsilon^{rstu}(\Phi^{(h_1+h_2)}_{dtu,\,-\frac{1}{2}})_{m+n} \, ,
\nonu\\
\comm{(\Phi^{(h_1),r}_{+\frac{3}{2}})_m}{(\Phi^{(h_2),abst}_{0})_n}&=&
\kappa_{+\frac{3}{2},0,+\frac{1}{2}}\,\Big((h_2-1)m-(h_1+\tfrac{1}{2})n\Big)\,\frac{1}{2!}\,\epsilon^{abcd}\epsilon^{rstu}(\Phi^{(h_1+h_2)}_{cdu,\,-\frac{1}{2}})_{m+n} \, ,
\nonu\\
\comm{(\Phi^{(h_1),r}_{+\frac{3}{2}})_m}{(\Phi^{(h_2),astu}_{0})_n}&=&
-\kappa_{+\frac{3}{2},0,+\frac{1}{2}}\,\Big((h_2-1)m-(h_1+\tfrac{1}{2})n\Big)\,\frac{1}{3!}\,\epsilon^{abcd}\epsilon^{rstu}(\Phi^{(h_1+h_2)}_{bcd,\,-\frac{1}{2}})_{m+n} \, .
\nonu \\
\label{eq:gravitino_scalar}
\eea
Note that the minus sign
appearing in the third from the below
and the last 
can be explained from the epsilon  in the eq. $13$
of (\ref{Eqs}). For the former,
the index $r$ passes through the $abc$ indices and this
produces the minus sign and for the latter
the index $r$ passes through the index $a$
which leads to the extra minus sign.
There are no separate $SU(4)$ global symmetries
because the indices are all mixed.
However,
the fourth and the fifth are related to
the first and the second $SU(4)$ global symmetries
respectively because the scalar having
the indices $rstu$ plays the role of a scalar in the
first $SU(4)$ global symmetry and
the scalar with the
indices $abcd$ does the role of a scalar in the
second $SU(4)$ global symmetry. Moreover, the corresponding
Lagrangian term is given by the tenth and the eleventh terms
appeared in the subsection \ref{3half2half}. In particular,
the $SU(8)$ invariant term contains the Levi Civita epsilon
which reflects the fact that the right hand sides
of (\ref{eq:gravitino_scalar}) should have this epsilon.
See also the footnote \ref{Epsilonidentity}.
The helicities of the celestial commutators
of (\ref{eq:gravitino_scalar}) are
given by
$(+\frac{3}{2},0, +\frac{1}{2})$
between the gravitino, the scalar and
the graviphotino
\footnote{
\label{20}
In
the tensor product
${\bf 8} \otimes {\bf 70}$, there are
eight tensor products,
$({\bf 4},{\bf 1})_{1} \otimes (\overline{\bf 4},
{\bf 4})_{2}$
from which there exists $({\bf 1},{\bf 4})_{3}$,
$({\bf 4},{\bf 1})_{1} \otimes ({\bf 6},{\bf 6})_{0}$
from which there is $(\overline{\bf 4},{\bf 6})_{1}$,
$({\bf 4},{\bf 1})_{1} \otimes ({\bf 4},\overline{\bf 4})_{-2}$
from which there appears $({\bf 6},\overline{\bf 4})_{-1}$,
$({\bf 4},{\bf 1})_{1} \otimes ({\bf 1},{\bf 1})_{-4}$
from which there is $({\bf 4},{\bf 1})_{-3}$,
$({\bf 1},{\bf 4})_{-1} \otimes ({\bf 1},{\bf 1})_{4}$
where there is  $({\bf 1},{\bf 4})_{3}$,
$({\bf 1},{\bf 4})_{-1} \otimes (\overline{\bf 4},{\bf 4})_{2}$
where there occurs  $(\overline{\bf 4},{\bf 6})_{1}$,
$({\bf 1},{\bf 4})_{-1} \otimes ({\bf 6},{\bf 6})_{0}$
where  there exists  $({\bf 6},\overline{\bf 4})_{-1}$,
and $({\bf 4},{\bf 1})_{1} \otimes ({\bf 1},{\bf 1})_{-4}$
where  there appears  $({\bf 4},{\bf 1})_{-3}$.
For the remaining two tensor products
having $\pm 5$ $U(1)$ charges,
there are
$\comm{(\Phi^{(h_1),a}_{+\frac{3}{2}})_m}{(\Phi^{(h_2),bcde}_{0})_n}_5=0=
\comm{(\Phi^{(h_1),r}_{+\frac{3}{2}})_m}{(\Phi^{(h_2),stuv}_{0})_n}_{-5}$
from the eq. $13$ of (\ref{Eqs}) with vanishing Levi Civita
$\epsilon$.}.

\subsection{The graviphotons-the graviphotons
\label{oneone}}

The eq. $17$ and eq. $21$ of (\ref{Eqs})
give the following celestial commutators
\bea
\comm{(\Phi^{(h_1),ab}_{+1})_m}{(\Phi^{(h_2),cd}_{+1})_n}&=&
\kappa_{+1,+1,0}\,\Big(h_2\,m-h_1\,n\Big)\,(\Phi^{(h_1+h_2),abcd}_{0})_{m+n}\, ,
\nonu\\
\comm{(\Phi^{(h_1),ab}_{+1})_m}{(\Phi^{(h_2),cr}_{+1})_n}&=&
\kappa_{+1,+1,0}\,\Big(h_2\,m-h_1\,n\Big)\,(\Phi^{(h_1+h_2),abcr}_{0})_{m+n}\, ,
\nonu\\
\comm{(\Phi^{(h_1),ab}_{+1})_m}{(\Phi^{(h_2),rs}_{+1})_n}&=&
\kappa_{+1,+1,0}\,\Big(h_2\,m-h_1\,n\Big)\,(\Phi^{(h_1+h_2),abrs}_{0})_{m+n}
\, ,
\nonu\\
\comm{(\Phi^{(h_1),ar}_{+1})_m}{(\Phi^{(h_2),bs}_{+1})_n}&=&
\kappa_{+1,+1,0}\,\Big(h_2\,m-h_1\,n\Big)\,(\Phi^{(h_1+h_2),arbs}_{0})_{m+n}
\, ,
\nonu\\
\comm{(\Phi^{(h_1),ar}_{+1})_m}{(\Phi^{(h_2),st}_{+1})_n}&=&
\kappa_{+1,+1,0}\,\Big(h_2\,m-h_1\,n\Big)\,(\Phi^{(h_1+h_2),arst}_{0})_{m+n}
\, , \nonu\\
\comm{(\Phi^{(h_1),rs}_{+1})_m}{(\Phi^{(h_2),tu}_{+1})_n}&=&
\kappa_{+1,+1,0}\,\Big(h_2\,m-h_1\,n\Big)\,(\Phi^{(h_1+h_2),rstu}_{0})_{m+n}
\, , \nonu\\
\comm{(\Phi^{(h_1),ab}_{+1})_m}{(\Phi^{(h_2)}_{cd,\,-1})_n}&=&
\kappa_{+1,-1,+2}\,\Big((h_2-2)m-h_1\,n\Big)\,\delta^{ab}_{cd}\,(\Phi^{(h_1+h_2)}_{-2})_{m+n} \, ,
\nonu\\
\comm{(\Phi^{(h_1),ar}_{+1})_m}{(\Phi^{(h_2)}_{bs,\,-1})_n}&=&
\kappa_{+1,-1,+2}\,\Big((h_2-2)m-h_1\,n\Big)\,\delta^{a}_{b}\delta^{r}_{s}\,(\Phi^{(h_1+h_2)}_{-2})_{m+n} \, ,
\nonu\\
\comm{(\Phi^{(h_1),rs}_{+1})_m}{(\Phi^{(h_2)}_{tu,\,-1})_n}&=&
\kappa_{+1,-1,+2}\,\Big((h_2-2)m-h_1\,n\Big)\,\delta^{rs}_{tu}\,(\Phi^{(h_1+h_2)}_{-2})_{m+n} \, .
\label{eq:photon_photon}
\eea
The first and
the seventh are related to the first $SU(4)$ global symmetry
while 
the sixth and
the last are related to the second $SU(4)$
global symmetry \footnote{
\label{21}
In
the tensor product ${\bf 28} \otimes {\bf 28}$, there are
six tensor products,
$({\bf 6}, {\bf 1})_2 \otimes ({\bf 6},{\bf 1})_2$
from which the representation
$({\bf 1},{\bf 1})_4$ arises,
$({\bf 6}, {\bf 1})_2 \otimes ({\bf 4},{\bf 4})_0$
where the representation $(\overline{\bf 4},{\bf 4})_2$
appears,
$({\bf 6}, {\bf 1})_2 \otimes ({\bf 1},{\bf 6})_{-2}$
from which we have
the representation $({\bf 6},{\bf 6})_{0}$,
$({\bf 4}, {\bf 4})_{0} \otimes ({\bf 4},{\bf 4})_0$
from which there exists $({\bf 6},{\bf 6})_0$,
$({\bf 4}, {\bf 4})_{0} \otimes ({\bf 1},{\bf 6})_{-2}$
where there is $({\bf 4},\overline{\bf 4})_{-2}$,
and
$({\bf 1}, {\bf 6})_{-2} \otimes ({\bf 1},{\bf 6})_{-2}$
where there appears $({\bf 1},{\bf 1})_{-4}$.
Similarly, in the tensor product
${\bf 28} \otimes \overline{\bf 28}$, there are
three tensor products,
$({\bf 6}, {\bf 1})_2 \otimes ({\bf 6},{\bf 1})_{-2}$
from which the representation
$({\bf 1},{\bf 1})_0$ arises,
$({\bf 4}, {\bf 4})_0 \otimes (\overline{\bf 4},
\overline{\bf 4})_{0}$
from which the representation
$({\bf 1},{\bf 1})_0$ occurs,
and $({\bf 1}, {\bf 6})_{-2} \otimes ({\bf 1},{\bf 6})_{2}$
from which the representation
$({\bf 1},{\bf 1})_0$ appears.
For the remaining six tensor products
having $\pm 2, \pm 4$ $U(1)$ charges,
there exist
\bea
&& \comm{(\Phi^{(h_1),ab}_{+1})_m}{(\Phi^{(h_2)}_{cr,\,-1})_n}_2 = 
0=
\comm{(\Phi^{(h_1),ab}_{+1})_m}{(\Phi^{(h_2)}_{rs,\,-1})_n}_4=
\comm{(\Phi^{(h_1),ar}_{+1})_m}{(\Phi^{(h_2)}_{bc,\,-1})_n}_{-2}
\nonu \\
&& =
\comm{(\Phi^{(h_1),ar}_{+1})_m}{(\Phi^{(h_2)}_{st,\,-1})_n}_2
=
\comm{(\Phi^{(h_1),rs}_{+1})_m}{(\Phi^{(h_2)}_{ab,\,-1})_n}_{-4}=
\comm{(\Phi^{(h_1),rs}_{+1})_m}{(\Phi^{(h_2)}_{at,\,-1})_n}_{-2}
\nonu
\eea
from eq. $21$ of (\ref{Eqs}).}.

$\bullet$ The twelfth
$\kappa \, e \, g^{\mu \rho} \, g^{\nu \si}\,
 A^{ABCD}\,
F_{\mu \nu}^{AB} \, F_{\rho \si}^{ CD}$
and thirteenth
$\kappa \, e \,
\epsilon^{\mu \nu \rho \si} \, B^{ABCD}\,
F_{\mu \nu}^{AB} \, F_{\rho \si}^{ CD}$
terms (or $SU(8)$ invariant
$\kappa \, e \, g^{\mu \rho} \, g^{\nu \si}\,
\epsilon_{ABCDEFGH} \phi^{\ast ABCD}\,
F_{\mu \nu}^{EF} \, F_{\rho \si}^{GH} +\mbox{h.c.}$ term)

In this case also, due to the presence of
the overall factor $\kappa$,
we consider the interactions between two graviphotons and scalar.
The helicities of the first six celestial commutators
of (\ref{eq:photon_photon}) are
$(+1,+1,0)$
between the graviphoton, the graviphoton and the scalar and 
the helicities $(+1,-1,+2)$ for two graviphotons
and the graviton arise in the last three celestial commutators 
of (\ref{eq:photon_photon}) where the corresponding
term of Lagrangian is given by the subsection \ref{2one}.


\subsection{The graviphotons-the graviphotinos
\label{oneonehalf}}

From the eq. $18$ and eq. $20$ of (\ref{Eqs}),
we determine the following celestial commutators
\bea
\comm{(\Phi^{(h_1),ab}_{+1})_m}{(\Phi^{(h_2),cdr}_{+\frac{1}{2}})_n}&=&
\kappa_{+1,+\frac{1}{2},+\frac{1}{2}}\,\Big((h_2-\tfrac{1}{2})m-h_1\,n\Big)\,\frac{1}{3!}\,\epsilon^{abcd}\epsilon^{rstu}\,(\Phi^{(h_1+h_2)}_{stu,\,-\frac{1}{2}})_{m+n} \, ,
\nonu\\
\comm{(\Phi^{(h_1),ab}_{+1})_m}{(\Phi^{(h_2),crs}_{+\frac{1}{2}})_n}&=&
\kappa_{+1,+\frac{1}{2},+\frac{1}{2}}\,\Big((h_2-\tfrac{1}{2})m-h_1\,n\Big)\,\frac{1}{2!}\,\epsilon^{abcd}\epsilon^{rstu}\,(\Phi^{(h_1+h_2)}_{dtu,\,-\frac{1}{2}})_{m+n}\, ,
\nonu\\
\comm{(\Phi^{(h_1),ab}_{+1})_m}{(\Phi^{(h_2),rst}_{+\frac{1}{2}})_n}&=&
\kappa_{+1,+\frac{1}{2},+\frac{1}{2}}\,\Big((h_2-\tfrac{1}{2})m-h_1\,n\Big)\,\frac{1}{2!}\,\epsilon^{abcd}\epsilon^{rstu}\,(\Phi^{(h_1+h_2)}_{cdu,\,-\frac{1}{2}})_{m+n}\, ,
\nonu\\
\comm{(\Phi^{(h_1),ar}_{+1})_m}{(\Phi^{(h_2),bcd}_{+\frac{1}{2}})_n}&=&
-\kappa_{+1,+\frac{1}{2},+\frac{1}{2}}\,\Big((h_2-\tfrac{1}{2})m-h_1\,n\Big)\,\frac{1}{3!}\,\epsilon^{abcd}\epsilon^{rstu}\,(\Phi^{(h_1+h_2)}_{stu,\,-\frac{1}{2}})_{m+n}\, ,
\nonu\\
\comm{(\Phi^{(h_1),ar}_{+1})_m}{(\Phi^{(h_2),bcs}_{+\frac{1}{2}})_n}&=&
\kappa_{+1,+\frac{1}{2},+\frac{1}{2}}\,\Big((h_2-\tfrac{1}{2})m-h_1\,n\Big)\,\frac{1}{2!}\,\epsilon^{abcd}\epsilon^{rstu}\,(\Phi^{(h_1+h_2)}_{dtu,\,-\frac{1}{2}})_{m+n} \, ,
\nonu\\
\comm{(\Phi^{(h_1),ar}_{+1})_m}{(\Phi^{(h_2),bst}_{+\frac{1}{2}})_n}&=&
-\kappa_{+1,+\frac{1}{2},+\frac{1}{2}}\,\Big((h_2-\tfrac{1}{2})m-h_1\,n\Big)\,\frac{1}{2!}\,\epsilon^{abcd}\epsilon^{rstu}\,(\Phi^{(h_1+h_2)}_{cdu,\,-\frac{1}{2}})_{m+n}\, ,
\nonu\\
\comm{(\Phi^{(h_1),ar}_{+1})_m}{(\Phi^{(h_2),stu}_{+\frac{1}{2}})_n}&=&
\kappa_{+1,+\frac{1}{2},+\frac{1}{2}}\,\Big((h_2-\tfrac{1}{2})m-h_1\,n\Big)\,\frac{1}{3!}\,\epsilon^{abcd}\epsilon^{rstu}\,(\Phi^{(h_1+h_2)}_{bcd,\,-\frac{1}{2}})_{m+n}\, ,
\nonu\\
\comm{(\Phi^{(h_1),rs}_{+1})_m}{(\Phi^{(h_2),abc}_{+\frac{1}{2}})_n}&=&
\kappa_{+1,+\frac{1}{2},+\frac{1}{2}}\,\Big((h_2-\tfrac{1}{2})m-h_1\,n\Big)\,\frac{1}{2!}\,\epsilon^{abcd}\epsilon^{rstu}\,(\Phi^{(h_1+h_2)}_{dtu,\,-\frac{1}{2}})_{m+n}\, ,
\nonu\\
\comm{(\Phi^{(h_1),rs}_{+1})_m}{(\Phi^{(h_2),abt}_{+\frac{1}{2}})_n}&=&
\kappa_{+1,+\frac{1}{2},+\frac{1}{2}}\,\Big((h_2-\tfrac{1}{2})m-h_1\,n\Big)\,\frac{1}{2!}\,\epsilon^{abcd}\epsilon^{rstu}\,(\Phi^{(h_1+h_2)}_{cdu,\,-\frac{1}{2}})_{m+n}\, ,
\nonu\\
\comm{(\Phi^{(h_1),rs}_{+1})_m}{(\Phi^{(h_2),atu}_{+\frac{1}{2}})_n}&=&
\kappa_{+1,+\frac{1}{2},+\frac{1}{2}}\,\Big((h_2-\tfrac{1}{2})m-h_1\,n\Big)\,\frac{1}{3!}\,\epsilon^{abcd}\epsilon^{rstu}\,(\Phi^{(h_1+h_2)}_{bcd,\,-\frac{1}{2}})_{m+n}\, ,
\nonu\\
\comm{(\Phi^{(h_1),ab}_{+1})_m}{(\Phi^{(h_2)}_{cde,\,-\frac{1}{2}})_n}&=&
\kappa_{+1,-\frac{1}{2},+\frac{3}{2}}\,\Big((h_2-\tfrac{3}{2})m-h_1\,n\Big)\,
3!\delta^{a}_{[c}\,(\Phi^{(h_1+h_2)}_{d,\,-\frac{3}{2}})_{m+n}\delta^b_{e]}\, ,
\nonu\\
\comm{(\Phi^{(h_1),ab}_{+1})_m}{(\Phi^{(h_2)}_{cdr,\,-\frac{1}{2}})_n}&=&
-\kappa_{+1,-\frac{1}{2},+\frac{3}{2}}\,\Big((h_2-\tfrac{3}{2})m-h_1\,n\Big)\,
\delta^{ab}_{cd}\,(\Phi^{(h_1+h_2)}_{r,\,-\frac{3}{2}})_{m+n}
\, , \nonu\\
\comm{(\Phi^{(h_1),ar}_{+1})_m}{(\Phi^{(h_2)}_{bcs,\,-\frac{1}{2}})_n}&=&
\kappa_{+1,-\frac{1}{2},+\frac{3}{2}}\,\Big((h_2-\tfrac{3}{2})m-h_1\,n\Big)\,
2!\, \delta^{r}_{s}\delta^{a}_{[b}\,(\Phi^{(h_1+h_2)}_{c],\,-\frac{3}{2}})_{m+n} \, ,
\nonu\\
\comm{(\Phi^{(h_1),ar}_{+1})_m}{(\Phi^{(h_2)}_{bst,\,-\frac{1}{2}})_n}&=&
-\kappa_{+1,-\frac{1}{2},+\frac{3}{2}}\,\Big((h_2-\tfrac{3}{2})m-h_1\,n\Big)\,
2! \, \delta^{a}_b\delta^{r}_{[s}\,(\Phi^{(h_1+h_2)}_{t],\,-\frac{3}{2}})_{m+n} \, ,
\nonu\\
\comm{(\Phi^{(h_1),rs}_{+1})_m}{(\Phi^{(h_2)}_{atu,\,-\frac{1}{2}})_n}&=&
-\kappa_{+1,-\frac{1}{2},+\frac{3}{2}}\,\Big((h_2-\tfrac{3}{2})m-h_1\,n\Big)\,
\delta^{rs}_{tu}\,(\Phi^{(h_1+h_2)}_{a,\,-\frac{3}{2}})_{m+n}
\, , \nonu\\
\comm{(\Phi^{(h_1),rs}_{+1})_m}{(\Phi^{(h_2)}_{tuv,\,-\frac{1}{2}})_n}&=&
\kappa_{+1,-\frac{1}{2},+\frac{3}{2}}\,\Big((h_2-\tfrac{3}{2})m-h_1\,n\Big)\,
3! \delta^{r}_{[t}\,(\Phi^{(h_1+h_2)}_{u,\,-\frac{3}{2}})_{m+n}\delta^s_{v]} \, .
\label{eq:photon_photino}
\eea
The eleventh is related to the first $SU(4)$ global symmetry
while 
the last is  related to the second $SU(4)$
global symmetry.

$\bullet$ The fourteenth
$\kappa \, e \, 
\epsilon^{ABCDEFGH} \, \bar{\chi}^{ABC}\,
\sigma^{\mu \nu} \, \chi^{DEF} \, F_{\mu \nu}^{ GH}$
term (or in other words, $SU(8)$ invariant $\kappa \, e \, 
\epsilon^{ABCDEFGH} \, \bar{\chi}_{ABC}\,
\sigma^{\mu \nu} \, \chi_{DEF} \, F_{\mu \nu GH}$ term)

This term of the Lagrangian is so called
Pauli moment coupling. 
Compared to the ${\cal N}=4$ supergravity,
the above first $10$ commutators of (\ref{eq:photon_photino})
appear newly in the ${\cal N}=8$ supergravity.
The eight $SU(8)$ indices appearing in
the graviphotino, the graviphotino and the graviphoton
are contracted with the Levi Civita epsilon and this feature
is present in (\ref{eq:photon_photino}). 
Note that the minus signs appearing on the right
hand sides of (\ref{eq:photon_photino}) can be
analyzed by the corresponding equations,
eq. $18$ having the epsilon and eq. $20$
with antisymmetrized Kronecker delta
of (\ref{Eqs}) \footnote{
\label{22}
In
the tensor product ${\bf 28} \otimes {\bf 56}$, there are
ten tensor products,
$({\bf 6},{\bf 1})_2 \otimes ({\bf 6},{\bf 4})_1$
from which there is $({\bf 1},{\bf 4})_3$,
$({\bf 6},{\bf 1})_2 \otimes ({\bf 4},{\bf 6})_{-1}$
from which there appears $(\overline{\bf 4},{\bf 6})_1$,
$({\bf 6},{\bf 1})_2 \otimes ({\bf 1},\overline{\bf 4})_{-3}$
from which there exists $({\bf 6},\overline{\bf 4})_{-1}$,
$({\bf 4},{\bf 4})_0 \otimes (\overline{\bf 4},{\bf 1})_3$
from which we have $({\bf 1},{\bf 4})_3$,
$({\bf 4},{\bf 4})_0 \otimes ({\bf 6},{\bf 4})_1$
from which there is a representation
$(\overline{\bf 4},{\bf 6})_1$,
$({\bf 4},{\bf 4})_0 \otimes ({\bf 4},{\bf 6})_{-1}$
from which there exists $({\bf 6},\overline{\bf 4})_{-1}$,
$({\bf 4},{\bf 4})_0 \otimes ({\bf 1},\overline{\bf 4})_{-3}$
from which there is $({\bf 4},{\bf 1})_{-3}$,
$({\bf 1},{\bf 6})_{-2} \otimes (\overline{\bf 4},{\bf 1})_3$
from which there is a representation
$(\overline{\bf 4},{\bf 6})_1$,
$({\bf 1},{\bf 6})_{-2} \otimes ({\bf 6},{\bf 4})_{1}$
from which there occurs $({\bf 6},\overline{\bf 4})_{-1}$,
and $({\bf 1},{\bf 6})_{-2} \otimes ({\bf 4},{\bf 6})_{-1}$
from which there exists $({\bf 4},{\bf 1})_{-3}$.
Moreover, in the tensor product
${\bf 28} \otimes \overline{\bf 56}$, there are
six tensor products,
$({\bf 6},{\bf 1})_2 \otimes ({\bf 4},{\bf 1})_{-3}$
where there is $(\overline{\bf 4},{\bf 1})_{-1}$,
$({\bf 6},{\bf 1})_2 \otimes ({\bf 6},\overline{\bf 4})_{-1}$
where there exists
$({\bf 1},\overline{\bf 4})_{1}$,
$({\bf 4},{\bf 4})_0 \otimes ({\bf 6},\overline{\bf 4})_{-1}$
where there is
$(\overline{\bf 4},{\bf 1})_{-1}$,
$({\bf 4},{\bf 4})_0 \otimes (\overline{\bf 4},{\bf 6})_{1}$
where there exists
$({\bf 1},\overline{\bf 4})_{1}$,
$({\bf 1},{\bf 6})_{-2} \otimes (\overline{\bf 4},{\bf 6})_{1}$
where there is $(\overline{\bf 4},{\bf 1})_{-1}$,
and
$({\bf 1},{\bf 6})_{-2} \otimes ({\bf 1},{\bf 4})_{3}$
where there exists
$({\bf 1},\overline{\bf 4})_{1}$.
For the remaining eight tensor products
having $\pm 3, \pm 5$ $U(1)$ charges,
there are
\bea
&&
\comm{(\Phi^{(h_1),ab}_{+1})_m}{(\Phi^{(h_2),cde}_{+\frac{1}{2}})_n}_5 =  0=
\comm{(\Phi^{(h_1),rs}_{+1})_m}{(\Phi^{(h_2),tuv}_{+\frac{1}{2}})_n}_{-5}=
\comm{(\Phi^{(h_1),ab}_{+1})_m}{(\Phi^{(h_2)}_{crs,-\frac{1}{2}})_n}_3
\nonu \\
&& =
\comm{(\Phi^{(h_1),ab}_{+1})_m}{(\Phi^{(h_2)}_{rst,-\frac{1}{2}})_n}_5
=
\comm{(\Phi^{(h_1),ar}_{+1})_m}{(\Phi^{(h_2)}_{bcd,-\frac{1}{2}})_n}_{-3}
=
\comm{(\Phi^{(h_1),ar}_{+1})_m}{(\Phi^{(h_2)}_{stu,-\frac{1}{2}})_n}_3
\nonu \\
&& =
\comm{(\Phi^{(h_1),rs}_{+1})_m}{(\Phi^{(h_2)}_{abc,-\frac{1}{2}})_n}_{-5}
=
\comm{(\Phi^{(h_1),rs}_{+1})_m}{(\Phi^{(h_2)}_{abt,-\frac{1}{2}})_n}_{-3},
\nonu
\eea
from the eq. $18$ and the eq. $20$ of (\ref{Eqs}).}.
Moreover, the previous Lagrangian term
(the ninth term corresponding to (\ref{eq:gravitino_photon})
in the subsection \ref{3halfone})
also appears
in the last six relations of (\ref{eq:photon_photino}).
The helicities of first ten celestial commutators
of (\ref{eq:photon_photino}) are
given by
$(+1, +\frac{1}{2},+\frac{1}{2})$
between the graviphoton, the graviphotino and the graviphotino
while
those for the remaining ones
are $(+1, -\frac{1}{2},+\frac{3}{2})$
between the graviphoton, the graviphotino and the gravitino
\footnote{The identity
\label{Epsilonidentity}
$\epsilon^{ABCDEFGH}=\frac{1}{4!}\frac{1}{4!}\delta^{ABCDEFGH}_{a\,\,b\,\,c\,\,d\,\,r\,\,s\,\,t\,\,u} \epsilon^{abcd}\epsilon^{rstu}$
is used.}.

\subsection{The graviphotons-the scalars}

The eq. $19$ of (\ref{Eqs})
leads to the following celestial commutators
\bea
\comm{(\Phi^{(h_1),ab}_{+1})_m}{(\Phi^{(h_2),cdrs}_{0})_n}&=&
\kappa_{+1,0,+1}\,\Big((h_2-1)m-h_1\,n\Big)\,\epsilon^{abcd}\epsilon^{rstu}\,\frac{1}{2!}\,(\Phi^{(h_1+h_2)}_{tu,\,-1})_{m+n} \, ,
\nonu\\
\comm{(\Phi^{(h_1),ab}_{+1})_m}{(\Phi^{(h_2),crst}_{0})_n}&=&
-\kappa_{+1,0,+1}\,\Big((h_2-1)m-h_1\,n\Big)\,\epsilon^{abcd}\epsilon^{rstu}\,(\Phi^{(h_1+h_2)}_{du,\,-1})_{m+n} \, ,
\nonu\\
\comm{(\Phi^{(h_1),ab}_{+1})_m}{(\Phi^{(h_2),rstu}_{0})_n}&=&
\kappa_{+1,0,+1}\,\Big((h_2-1)m-h_1\,n\Big)\,\epsilon^{abcd}\epsilon^{rstu}\,\frac{1}{2!}\,(\Phi^{(h_1+h_2)}_{cd,\,-1})_{m+n} \, ,
\nonu\\
\comm{(\Phi^{(h_1),ar}_{+1})_m}{(\Phi^{(h_2),bcds}_{0})_n}&=&
-\kappa_{+1,0,+1}\,\Big((h_2-1)m-h_1\,n\Big)\,\epsilon^{abcd}\epsilon^{rstu}\,\frac{1}{2!}\,(\Phi^{(h_1+h_2)}_{tu,\,-1})_{m+n} \, ,
\nonu\\
\comm{(\Phi^{(h_1),ar}_{+1})_m}{(\Phi^{(h_2),bcst}_{0})_n}&=&
-\kappa_{+1,0,+1}\,\Big((h_2-1)m-h_1\,n\Big)\,\epsilon^{abcd}\epsilon^{rstu}\,(\Phi^{(h_1+h_2)}_{du,\,-1})_{m+n} \, ,
\nonu\\
\comm{(\Phi^{(h_1),ar}_{+1})_m}{(\Phi^{(h_2),bstu}_{0})_n}&=&
-\kappa_{+1,0,+1}\,\Big((h_2-1)m-h_1\,n\Big)\,\epsilon^{abcd}\epsilon^{rstu}\,\frac{1}{2!}\,(\Phi^{(h_1+h_2)}_{cd,\,-1})_{m+n} \, ,
\nonu\\
\comm{(\Phi^{(h_1),rs}_{+1})_m}{(\Phi^{(h_2),abcd}_{0})_n}&=&
\kappa_{+1,0,+1}\,\Big((h_2-1)m-h_1\,n\Big)\,\epsilon^{abcd}\epsilon^{rstu}\,\frac{1}{2!}\,(\Phi^{(h_1+h_2)}_{tu,\,-1})_{m+n} \, ,
\nonu\\
\comm{(\Phi^{(h_1),rs}_{+1})_m}{(\Phi^{(h_2),abct}_{0})_n}&=&
-\kappa_{+1,0,+1}\,\Big((h_2-1)m-h_1\,n\Big)\,\epsilon^{abcd}\epsilon^{rstu}\,(\Phi^{(h_1+h_2)}_{du,\,-1})_{m+n} \, ,
\nonu\\
\comm{(\Phi^{(h_1),rs}_{+1})_m}{(\Phi^{(h_2),abtu}_{0})_n}&=&
\kappa_{+1,0,+1}\,\Big((h_2-1)m-h_1\,n\Big)\,\epsilon^{abcd}\epsilon^{rstu}\,\frac{1}{2!}\,(\Phi^{(h_1+h_2)}_{cd,\,-1})_{m+n} \, .
\label{eq:photon_scalar}
\eea
The third is related to the first $SU(4)$ global symmetry
while 
the seventh is related to the second $SU(4)$
global symmetry.
In this case, the corresponding
Lagrangian is given by the twelfth and the thirteen terms
(or $SU(8)$ invariant term) associated with (\ref{eq:photon_photon})
in the subsection \ref{oneone}.
The helicities are
given by
$(+1, 0,+1)$
between the graviphoton, the scalar and the graviphoton.
Note that the Levi Civita epsilon plays an important role
on the right hand sides of (\ref{eq:photon_scalar}).
The minus signs appearing on the right
hand sides of (\ref{eq:photon_scalar}) can be
realized by the corresponding equation
eq. $19$ having the epsilon of (\ref{Eqs})
\footnote{
\label{24}
In
the tensor product
${\bf 28} \otimes {\bf 70}$, there are
nine tensor products,
$({\bf 6},{\bf 1}_2) \otimes ({\bf 6},{\bf 6})_0 $
where
there occurs
$({\bf 1},{\bf 6})_2$,
$({\bf 6},{\bf 1}_2) \otimes ({\bf 4},\overline{\bf 4})_{-2} $
where there exists
$(\overline{\bf 4},\overline{\bf 4})_0$,
$({\bf 6},{\bf 1}_2) \otimes ({\bf 1},{\bf 1})_{-4} $
where there appears
$({\bf 6},{\bf 1})_{-2}$,
$({\bf 4},{\bf 4})_0 \otimes (\overline{\bf 4},{\bf 4})_{2}$
where there is
$({\bf 1},{\bf 6})_{2}$,
$({\bf 4},{\bf 4})_0 \otimes ({\bf 6},{\bf 6})_{0}$
where we have
$(\overline{\bf 4},\overline{\bf 4})_{0}$,
$({\bf 4},{\bf 4})_0 \otimes ({\bf 4},\overline{\bf 4})_{-2}$
where there exists a representation
$({\bf 6},{\bf 1})_{-2}$,
$({\bf 1},{\bf 6})_{-2} \otimes ({\bf 1},{\bf 1})_4 $
where
there appears
$({\bf 1},{\bf 6})_{2}$,
$({\bf 1},{\bf 6})_{-2} \otimes (\overline{\bf 4},{\bf 4})_2 $
where there exists
$(\overline{\bf 4},\overline{\bf 4})_{0}$,
and
$({\bf 1},{\bf 6})_{-2} \otimes ({\bf 6},{\bf 6})_0 $
where there is
$({\bf 6},{\bf 1})_{-2}$.
For the remaining six tensor products
having $\pm 4, \pm 6$ $U(1)$ charges,
there are
\bea
&& \comm{(\Phi^{(h_1),ab}_{+1})_m}{(\Phi^{(h_2),cdef}_{0})_n}_6 =  0=
\comm{(\Phi^{(h_1),ab}_{+1})_m}{(\Phi^{(h_2),cder}_{0})_n}_4=
\comm{(\Phi^{(h_1),ar}_{+1})_m}{(\Phi^{(h_2),bcde}_{0})_n}_4
\nonu \\
&& =
\comm{(\Phi^{(h_1),ar}_{+1})_m}{(\Phi^{(h_2),stuv}_{0})_n}_{-4}=
\comm{(\Phi^{(h_1),rs}_{+1})_m}{(\Phi^{(h_2),atuv}_{0})_n}_{-4}
=
\comm{(\Phi^{(h_1),rs}_{+1})_m}{(\Phi^{(h_2),tuvw}_{0})_n}_{-6},
\nonu
\eea
from the eq. $19$ of (\ref{Eqs}).}.


\subsection{The graviphotinos-the graviphotinos}

From the eq. $22$ and eq. $24$ of (\ref{Eqs}),
the following celestial anticommutators
can be determined 
\bea
\acomm{(\Phi^{(h_1),abc}_{+\frac{1}{2}})_m}{(\Phi^{(h_2),drs}_{
+\frac{1}{2}})_n}&=&\kappa_{+\frac{1}{2},+\frac{1}{2},+1}\,\Big((h_2-\tfrac{1}{2})m-(h_1-\tfrac{1}{2})n\Big)\,\epsilon^{abcd}\epsilon^{rstu}\,\frac{1}{2!}\,(\Phi^{(h_1+h_2)}_{tu,\,-1})_{m+n} \, ,
\nonu\\
\acomm{(\Phi^{(h_1),abc}_{+\frac{1}{2}})_m}{(\Phi^{(h_2),rst}_{+\frac{1}{2}})_n}&=&-\kappa_{+\frac{1}{2},+\frac{1}{2},+1}\,\Big((h_2-\tfrac{1}{2})m-(h_1-\tfrac{1}{2})n\Big)\,\epsilon^{abcd}\epsilon^{rstu}\,(\Phi^{(h_1+h_2)}_{du,\,-1})_{m+n} \, ,
\nonu\\
\acomm{(\Phi^{(h_1),abr}_{+\frac{1}{2}})_m}{(\Phi^{(h_2),cds}_{+\frac{1}{2}})_n}&=&\kappa_{+\frac{1}{2},+\frac{1}{2},+1}\,\Big((h_2-\tfrac{1}{2})m-(h_1-\tfrac{1}{2})n\Big)\,\epsilon^{abcd}\epsilon^{rstu}\,\frac{1}{2!}\,(\Phi^{(h_1+h_2)}_{tu,\,-1})_{m+n}\, ,
\nonu\\
\acomm{(\Phi^{(h_1),abr}_{+\frac{1}{2}})_m}{(\Phi^{(h_2),cst}_{+\frac{1}{2}})_n}&=&\kappa_{+\frac{1}{2},+\frac{1}{2},+1}\,\Big((h_2-\tfrac{1}{2})m-(h_1-\tfrac{1}{2})n\Big)\,\epsilon^{abcd}\epsilon^{rstu}\,(\Phi^{(h_1+h_2)}_{du,\,-1})_{m+n} \, ,
\nonu\\
\acomm{(\Phi^{(h_1),abr}_{+\frac{1}{2}})_m}{(\Phi^{(h_2),stu}_{+\frac{1}{2}})_n}&=&\kappa_{+\frac{1}{2},+\frac{1}{2},+1}\,\Big((h_2-\tfrac{1}{2})m-(h_1-\tfrac{1}{2})n\Big)\,\epsilon^{abcd}\epsilon^{rstu}\,\frac{1}{2!}\,(\Phi^{(h_1+h_2)}_{cd,\,-1})_{m+n} \, ,
\nonu\\
\acomm{(\Phi^{(h_1),ars}_{+\frac{1}{2}})_m}{(\Phi^{(h_2),btu}_{+\frac{1}{2}})_n}&=&\kappa_{+\frac{1}{2},+\frac{1}{2},+1}\,\Big((h_2-\tfrac{1}{2})m-(h_1-\tfrac{1}{2})n\Big)\,\epsilon^{abcd}\epsilon^{rstu}\,\frac{1}{2!}\,(\Phi^{(h_1+h_2)}_{cd,\,-1})_{m+n} \, ,
\nonu\\
\acomm{(\Phi^{(h_1),abc}_{+\frac{1}{2}})_m}{(\Phi^{(h_2)}_{def,\,-\frac{1}{2}})_n}&=&\kappa_{+\frac{1}{2},-\frac{1}{2},+2}\,\Big((h_2-\tfrac{3}{2})m-(h_1-\tfrac{1}{2})n\Big)\,\delta^{abc}_{def}\,(\Phi^{(h_1+h_2)}_{-2})_{m+n} \, ,
\nonu\\
\acomm{(\Phi^{(h_1),abr}_{+\frac{1}{2}})_m}{(\Phi^{(h_2)}_{cds,\,-\frac{1}{2}})_n}&=&\kappa_{+\frac{1}{2},-\frac{1}{2},+2}\,\Big((h_2-\tfrac{3}{2})m-(h_1-\tfrac{1}{2})n\Big)\,\delta^{abr}_{cds}\,(\Phi^{(h_1+h_2)}_{-2})_{m+n}\, ,
\nonu\\
\acomm{(\Phi^{(h_1),ars}_{+\frac{1}{2}})_m}{(\Phi^{(h_2)}_{btu,\,-\frac{1}{2}})_n}&=&\kappa_{+\frac{1}{2},-\frac{1}{2},+2}\,\Big((h_2-\tfrac{3}{2})m-(h_1-\tfrac{1}{2})n\Big)\,\delta^{ars}_{btu}\,(\Phi^{(h_1+h_2)}_{-2})_{m+n}\, ,
\nonu\\
\acomm{(\Phi^{(h_1),rst}_{+\frac{1}{2}})_m}{(\Phi^{(h_2)}_{uvw,\,-\frac{1}{2}})_n}&=&\kappa_{+\frac{1}{2},-\frac{1}{2},+2}\,\Big((h_2-\tfrac{3}{2})m-(h_1-\tfrac{1}{2})n\Big)\,\delta^{rst}_{uvw}\,(\Phi^{(h_1+h_2)}_{-2})_{m+n}
\, .
\label{eq:photino_photino}
\eea
The seventh is related to the first $SU(4)$ global symmetry
while 
the last is related to the second $SU(4)$
global symmetry \footnote{
\label{commexample}
Let us check the first equation of
(\ref{eq:photino_photino}). According to (\ref{abcdrstu}),
the first
split factor for the first ${\cal N}=4$ super Yang-Mills theory
contains the particle one (gluino) having the lower index
$e$ with negative helicity
$-\frac{1}{2}$ and 
the particle two (gluino) having the upper index $d$
with positive helicity
$+\frac{1}{2}$ together with $\epsilon^{abce}$ while
the split factor for the second ${\cal N}=4$ super Yang-Mills theory
contains the particle two (scalar) having the upper indices
$rs$ with zero helicity
and the particle one (gluon) with positive helicity
$+1$. From the index structure in the ${\cal N}=4$ super
Yang-Mills theory in Table $1$ of \cite{2212-1},
the  first split factor
has the Kronecker delta with epsilon for the gluon with
$-1$ helicity and the second split factor can be written as
the $\epsilon^{rstu}$ with the scalar with lower indices
$tu$ having zero helicity.
Therefore, the product between
the above gluon and the above scalar
produces the graviphotons with the helicity $-1$
together with two epsilons 
from Table $3$ of \cite{2212-1}.
This leads to the right hand side of the first equation of
(\ref{eq:photino_photino}).
In this analysis we do not consider any numerical factors.

Similarly, for the third equation of
(\ref{eq:photino_photino}),
the first
split factor
contains the particle one (scalar) having the upper index
$ab$ with zero helicity
and the particle two (scalar) having the upper index $cd$
with zero helicity
together with $\epsilon^{abcd}$ while
the second split factor 
contains the particle one (gluino) having the upper index
$r$ with $+\frac{1}{2}$ helicity and
the particle two (gluino) with 
$+\frac{1}{2}$ helicity with upper index $s$.
Then the  first split factor
has the epsilon for the gluon with
$-1$ helicity from Table 1 of \cite{2212-1}
and the second split factor
can be written as
the scalar having zero helicity.
The product between
the above gluon and the above scalar
leads to the graviphotons (from Table $3$ of
\cite{2212-1}) with the helicity $-1$
where the upper indices $rs$ can be lowered by using
$\epsilon^{rstu}$.}.
The above first $6$ anticommutators of (\ref{eq:photino_photino})
appear newly in the ${\cal N}=8$ supergravity,
compared to the ${\cal N}=4$ supergravity.
The minus sign appearing on the right hand side of
the second relation of (\ref{eq:photino_photino})
can be checked by previous analysis done before.



We already have described this coupling before
in the subsection \ref{oneonehalf}.
The helicities of first six celestial anticommutators
of (\ref{eq:photino_photino}) are
given by
$(+\frac{1}{2},+\frac{1}{2},+1)$
between the graviphotino, the graviphotino and the graviphoton.
The epsilon plays an important role on the right hand sides
of (\ref{eq:photino_photino})
\footnote{
\label{26}
In the tensor product ${\bf 56}\otimes {\bf 56}$,
there are six tensor products,  
$(\overline{\bf 4},{\bf 1})_3 \otimes ({\bf 4},{\bf 6})_{-1}$
where there is
$({\bf 1},{\bf 6})_{2}$,
$(\overline{\bf 4},{\bf 1})_3 \otimes ({\bf 1},\overline{\bf 4})_{-3}$
where there exists
$(\overline{\bf 4},\overline{\bf 4})_{0}$,
$({\bf 6},{\bf 4})_1 \otimes ({\bf 6},{\bf 4})_{1}$
where there occurs
$({\bf 1},{\bf 6})_2$,
$({\bf 6},{\bf 4})_1 \otimes ({\bf 4},{\bf 6})_{-1}$
where there appears
$(\overline{\bf 4},\overline{\bf 4})_0$,
$({\bf 6},{\bf 4})_1 \otimes ({\bf 1},\overline{\bf 4})_{-3}$
where there exists
$({\bf 6},{\bf 1})_{-2}$,
and
$({\bf 4},{\bf 6})_{-1} \otimes ({\bf 4},{\bf 6})_{-1}$,
where we have
$({\bf 6},{\bf 1})_{-2}$.
Similarly,
in the tensor
product ${\bf 56}\otimes \overline{\bf 56}$,
there are four tensor products,  
$(\overline{\bf 4},{\bf 1})_3 \otimes ({\bf 4},{\bf 1})_{-3}$
where there exists
$({\bf 1},{\bf 1})_{0}$,
$({\bf 6},{\bf 4})_1 \otimes ({\bf 6},\overline{\bf 4})_{-1}$
where there is
$({\bf 1},{\bf 1})_0$,
$({\bf 4},{\bf 6})_{-1} \otimes (\overline{\bf 4},{\bf 6})_{1}$
where there exists
$({\bf 1},{\bf 1})_0$,
and
$({\bf 1},\overline{\bf 4})_{-3} \otimes ({\bf 1},{\bf 4})_{3}$
where there is
$({\bf 1},{\bf 1})_0$.
For the remaining sixteen tensor products,
there are, with $\pm 2, \pm 4, \pm 6$ $U(1)$ charges, 
\bea
&& \acomm{(\Phi^{(h_1),abc}_{+\frac{1}{2}})_m}{(\Phi^{(h_2),def}_{
+\frac{1}{2}})_n}_6 =  0=
\acomm{(\Phi^{(h_1),abc}_{+\frac{1}{2}})_m}{(\Phi^{(h_2),der}_{
+\frac{1}{2}})_n}_4
=
\acomm{(\Phi^{(h_1),ars}_{+\frac{1}{2}})_m}{(\Phi^{(h_2),tuv}_{
+\frac{1}{2}})_n}_{-4}
\nonu \\
&&=
\acomm{(\Phi^{(h_1),rst}_{+\frac{1}{2}})_m}{(\Phi^{(h_2),uvw}_{
+\frac{1}{2}})_n}_{-6}
=
\acomm{(\Phi^{(h_1),abc}_{+\frac{1}{2}})_m}{(\Phi^{(h_2)}_{der,
-\frac{1}{2}})_n}_{2}=
\acomm{(\Phi^{(h_1),abc}_{+\frac{1}{2}})_m}{(\Phi^{(h_2)}_{drs,
-\frac{1}{2}})_n}_{4}
\nonu \\
&& =
\acomm{(\Phi^{(h_1),abc}_{+\frac{1}{2}})_m}{(\Phi^{(h_2)}_{rst,
-\frac{1}{2}})_n}_{6}
=\acomm{(\Phi^{(h_1),abr}_{+\frac{1}{2}})_m}{(\Phi^{(h_2)}_{cde,
-\frac{1}{2}})_n}_{-2}
=
\acomm{(\Phi^{(h_1),abr}_{+\frac{1}{2}})_m}{(\Phi^{(h_2)}_{cst,
-\frac{1}{2}})_n}_{4}
\nonu \\
&& =\acomm{(\Phi^{(h_1),abr}_{+\frac{1}{2}})_m}{(\Phi^{(h_2)}_{stu,
-\frac{1}{2}})_n}_{4}
=
\acomm{(\Phi^{(h_1),ars}_{+\frac{1}{2}})_m}{(\Phi^{(h_2)}_{bcd,
-\frac{1}{2}})_n}_{-2}=
\acomm{(\Phi^{(h_1),ars}_{+\frac{1}{2}})_m}{(\Phi^{(h_2)}_{bct,
-\frac{1}{2}})_n}_{-2} \nonu \\
&& =
\acomm{(\Phi^{(h_1),ars}_{+\frac{1}{2}})_m}{(\Phi^{(h_2)}_{tuv,
-\frac{1}{2}})_n}_{4}
=
\acomm{(\Phi^{(h_1),rst}_{+\frac{1}{2}})_m}{(\Phi^{(h_2)}_{abc,
-\frac{1}{2}})_n}_{-6}=
\acomm{(\Phi^{(h_1),rst}_{+\frac{1}{2}})_m}{(\Phi^{(h_2)}_{abu,
-\frac{1}{2}})_n}_{-4}
\nonu \\
&&=
\acomm{(\Phi^{(h_1),rst}_{+\frac{1}{2}})_m}{(\Phi^{(h_2)}_{auv,
-\frac{1}{2}})_n}_{-2},
\nonu
\eea
from the eq. $22$ and the eq. $24$ of (\ref{Eqs}).}.
Moreover, 
the helicities $( +\frac{1}{2}, - \frac{1}{2}, +2)$
for the graviphotino, the graviphotino and the graviton
should appear in the coupling of this three point amplitude and 
the corresponding celestial anticommutators
are given by the last four relations in
(\ref{eq:photino_photino}) where the corresponding Lagrangian
was given by previous fourth term in the subsection
\ref{22half} \footnote{
\label{missing}
%
For the fifth equation of
(\ref{eq:photino_photino}),
the first
split factor
contains the particle one (scalar) having the upper index
$ab$ with zero helicity
and the particle two (gluon)
with $+1$ helicity
and
the second split factor 
contains the particle one (gluino) having the upper index
$r$ with $+\frac{1}{2}$ helicity
and the particle two (gluino) with 
$-\frac{1}{2}$ having the lower index
$v$ together with $\epsilon^{stuv}$.
Then the  first split factor
has the scalar with
zero helicity (Table $1$ in \cite{2212-1})
where the previous upper indices $ab$ are lowered by
$\epsilon^{abcd}$
and the second split factor
has gluon
having $-1$ helicity with Kronecker delta.
The product between
the above scalar and the above gluon
gives us to the graviphotons
with the helicity $-1$ from Table $3$ in \cite{2212-1}.}.


\subsection{The graviphotinos-the scalars}

The eq. $23$ of (\ref{Eqs}),
provides the following celestial commutators
\bea
\comm{(\Phi^{(h_1),abc}_{+\frac{1}{2}})_m}{(\Phi^{(h_2),drst}_{0})_n}&=&
\kappa_{+\frac{1}{2},0,+\frac{3}{2}}\,\Big((h_2-1)m-(h_1-\tfrac{1}{2})n\Big)\,\epsilon^{abcd}\epsilon^{rstu}\,(\Phi^{(h_1+h_2)}_{u,\,-\frac{3}{2}})_{m+n}
\, ,
\nonu\\
\comm{(\Phi^{(h_1),abc}_{+\frac{1}{2}})_m}{(\Phi^{(h_2),rstu}_{0})_n}&=&
\kappa_{+\frac{1}{2},0,+\frac{3}{2}}\,\Big((h_2-1)m-(h_1-\tfrac{1}{2})n\Big)\,\epsilon^{abcd}\epsilon^{rstu}\,(\Phi^{(h_1+h_2)}_{d,\,-\frac{3}{2}})_{m+n}
\, ,
\nonu\\
\comm{(\Phi^{(h_1),abr}_{+\frac{1}{2}})_m}{(\Phi^{(h_2),cdst}_{0})_n}
&=&\kappa_{+\frac{1}{2},0,+\frac{3}{2}}\,\Big((h_2-1)m-(h_1-\tfrac{1}{2})n\Big)\,\epsilon^{abcd}\epsilon^{rstu}\,(\Phi^{(h_1+h_2)}_{u,\,-\frac{3}{2}})_{m+n} \, ,
\nonu\\
\comm{(\Phi^{(h_1),abr}_{+\frac{1}{2}})_m}{(\Phi^{(h_2),cstu}_{0})_n}&=&
-\kappa_{+\frac{1}{2},0,+\frac{3}{2}}\,\Big((h_2-1)m-(h_1-\tfrac{1}{2})n\Big)\,\epsilon^{abcd}\epsilon^{rstu}\,(\Phi^{(h_1+h_2)}_{d,\,-\frac{3}{2}})_{m+n}
\, ,
\nonu\\
\comm{(\Phi^{(h_1),ars}_{+\frac{1}{2}})_m}{(\Phi^{(h_2),bcdt}_{0})_n}
&=&
\kappa_{+\frac{1}{2},0,+\frac{3}{2}}\,\Big((h_2-1)m-(h_1-\tfrac{1}{2})n\Big)\,\epsilon^{abcd}\epsilon^{rstu}\,(\Phi^{(h_1+h_2)}_{u,\,-\frac{3}{2}})_{m+n}
\, , \nonu\\
\comm{(\Phi^{(h_1),ars}_{+\frac{1}{2}})_m}{(\Phi^{(h_2),bctu}_{0})_n}&=&
\kappa_{+\frac{1}{2},0,+\frac{3}{2}}\,\Big((h_2-1)m-(h_1-\tfrac{1}{2})n\Big)\,\epsilon^{abcd}\epsilon^{rstu}\,(\Phi^{(h_1+h_2)}_{d,\,-\frac{3}{2}})_{m+n}
\, , \nonu\\
\comm{(\Phi^{(h_1),rst}_{+\frac{1}{2}})_m}{(\Phi^{(h_2),abcd}_{0})_n}&=&
\kappa_{+\frac{1}{2},0,+\frac{3}{2}}\,\Big((h_2-1)m-(h_1-\tfrac{1}{2})n\Big)\,\epsilon^{abcd}\epsilon^{rstu}\,(\Phi^{(h_1+h_2)}_{u,\,-\frac{3}{2}})_{m+n}
\, , \nonu\\
\comm{(\Phi^{(h_1),rst}_{+\frac{1}{2}})_m}{(\Phi^{(h_2),abcu}_{0})_n}&=&
-\kappa_{+\frac{1}{2},0,+\frac{3}{2}}\,\Big((h_2-1)m-(h_1-\tfrac{1}{2})n\Big)\,\epsilon^{abcd}\epsilon^{rstu}\,(
\Phi^{(h_1+h_2)}_{d,\,-\frac{3}{2}})_{m+n} \, .
\nonu \\
\label{eq:photino_scalar}
\eea
The second is related to the first $SU(4)$ global symmetry
while 
the seventh
is related to the second $SU(4)$
global symmetry
\footnote{
\label{othermissing}
Let us consider the first equation of
(\ref{eq:photino_scalar}). According to (\ref{abcdrstu}),
the first
split factor for the first ${\cal N}=4$ super Yang-Mills theory
has the particle one (gluino) having the lower index
$e$ with negative helicity
$-\frac{1}{2}$ and 
the particle two (gluino) having the upper index $d$
with positive helicity
$+\frac{1}{2}$ together with $\epsilon^{abce}$ while
the split factor for the second ${\cal N}=4$ super Yang-Mills theory
has the particle two (gluino) having the lower index
$u$ with negative helicity $-\frac{1}{2}$ and 
the particle one (gluon) with positive helicity
$+1$ in the presence of $\epsilon^{rstu}$.
From the index structure (Table $1$ in \cite{2212-1})
in the ${\cal N}=4$ super
Yang-Mills theory, the  first split factor
has the Kronecker delta with epsilon for the gluon with
$-1$ helicity and the second split factor can be written as
the $\epsilon^{rstu}$ with the gluino having negative helicity
$-\frac{1}{2}$ with lower index $u$.
Therefore, the product between
the above gluon and the above gluino
produces the gravitinos with the helicity $-\frac{3}{2}$
from Table $3$ in \cite{2212-1}.
Any numerical factors are ignored.

Similarly,
the second equation of
(\ref{eq:photino_scalar}) provides, according to (\ref{abcdrstu}),
the first
split factor for the first ${\cal N}=4$ super Yang-Mills theory
has the particle one (gluino) having the lower index
$d$ with negative helicity
$-\frac{1}{2}$ and
the particle two (gluon)
with positive helicity
$+1$ together with $\epsilon^{abcd}$ while
the split factor for the second
${\cal N}=4$ super Yang-Mills theory
has the particle two (gluon)
with negative helicity $-1$ and 
the particle one (gluon) with positive helicity
$+1$ in the presence of $\epsilon^{rstu}$.
From the index structure in the ${\cal N}=4$ super
Yang-Mills theory, the  first split factor
has the epsilon for the gluino with
$-\frac{1}{2}$ helicity having lower index $d$ and
the second split factor can be written as
the $\epsilon^{rstu}$ with the gluon having negative helicity
$-1$ from Table $1$ in \cite{2212-1}.
Therefore, the product between
the above gluino and the above gluon
implies the gravitinos with the helicity $-\frac{3}{2}$
from Table $3$ in \cite{2212-1}.}.
The relevant Lagrangian was given in previous
tenth and eleventh terms (or corresponding $SU(8)$ invariant term)
related to (\ref{eq:gravitino_photino}) in the subsection
\ref{3half2half}.
The helicities of first ten celestial commutators
of (\ref{eq:photon_photino}) are
given by
$(+\frac{1}{2},0,+\frac{3}{2})$
between the graviphotino, the scalar and the gravitino.
The epsilon dependence appears on the right hand sides of
(\ref{eq:photino_scalar}).
Note that there exists a minus sign
on the right hand side of the fourth relation
in (\ref{eq:photino_scalar}) \footnote{
\label{29}
In the tensor product ${\bf 56} \otimes {\bf 70}$,
there are eight tensor products,
$(\overline{\bf 4},{1})_3 \otimes ({\bf 4},\overline{\bf 4})_{-2}$
where there is
$({\bf 1},\overline{\bf 4})_1$,
$(\overline{\bf 4},{\bf 1})_3 \otimes ({\bf 1},{\bf 1})_{-4}$
where there is
$(\overline{\bf 4},{\bf 1})_{-1}$,
$({\bf 6},{\bf 4})_1 \otimes ({\bf 6},{\bf 6})_{0}$
where there exists
$({\bf 1},\overline{\bf 4})_{1}$,
$({\bf 6},{\bf 4})_1 \otimes ({\bf 4},\overline{\bf 4})_{-2}$
where there is
$(\overline{\bf 4},{\bf 1})_{-1}$,
$({\bf 4},{\bf 6})_{-1} \otimes (\overline{\bf 4},{\bf 4})_{2}$
where there exists
$({\bf 1},\overline{\bf 4})_{1}$,
$({\bf 4},{\bf 6})_{-1} \otimes ({\bf 6},{\bf 6})_{0}$
where there appears
$(\overline{\bf 4},{\bf 1})_{-1}$,
$({\bf 1},\overline{\bf 4})_{-3} \otimes ({\bf 1},{\bf 1})_{4}$
where there occurs
$({\bf 1},\overline{\bf 4})_{1}$,
and $({\bf 1},\overline{\bf 4})_{-3} \otimes
(\overline{\bf 4},{\bf 4})_{2}$
where there is
$(\overline{\bf 4},{\bf 1})_{-1}$.
For the remaining  twelve tensor products, there are,
with $\pm 3, \pm 5, \pm 7$ $U(1)$ charges,
\bea
&&
\comm{(\Phi^{(h_1),abc}_{+\frac{1}{2}})_m}{(\Phi^{(h_2),defg}_{0})_n}_7 =  0=
\comm{(\Phi^{(h_1),abc}_{+\frac{1}{2}})_m}{(\Phi^{(h_2),defr}_{0})_n}_5=
\comm{(\Phi^{(h_1),abc}_{+\frac{1}{2}})_m}{(\Phi^{(h_2),ders}_{0})_n}_3
\nonu \\
&&=
\comm{(\Phi^{(h_1),abr}_{+\frac{1}{2}})_m}{(\Phi^{(h_2),cdef}_{0})_n}_5=
\comm{(\Phi^{(h_1),abr}_{+\frac{1}{2}})_m}{(\Phi^{(h_2),cdes}_{0})_n}_3
=
\comm{(\Phi^{(h_1),abr}_{+\frac{1}{2}})_m}{(\Phi^{(h_2),stuv}_{0})_n}_{-3}
\nonu \\
&& =
\comm{(\Phi^{(h_1),ars}_{+\frac{1}{2}})_m}{(\Phi^{(h_2),bcde}_{0})_n}_{5}
=
\comm{(\Phi^{(h_1),ars}_{+\frac{1}{2}})_m}{(\Phi^{(h_2),btuv}_{0})_n}_{-3}=
\comm{(\Phi^{(h_1),ars}_{+\frac{1}{2}})_m}{(\Phi^{(h_2),tuvw}_{0})_n}_{-5}
\nonu \\
&& =
\comm{(\Phi^{(h_1),rst}_{+\frac{1}{2}})_m}{(\Phi^{(h_2),abuv}_{0})_n}_{-3}=
\comm{(\Phi^{(h_1),rst}_{+\frac{1}{2}})_m}{(\Phi^{(h_2),auvw}_{0})_n}_{-5}
=
\comm{(\Phi^{(h_1),rst}_{+\frac{1}{2}})_m}{(\Phi^{(h_2),uvwx}_{0})_n}_{-7},
\nonu
\eea
from the eq. $23$ of (\ref{Eqs}).}.

\subsection{The scalars-the scalars}

From the eq. $25$ of (\ref{Eqs}),
the following celestial commutators
can be summarized by
\bea
\comm{(\Phi^{(h_1),abcd}_{0})_m}{(\Phi^{(h_2),rstu}_{0})_n}&=&
\kappa_{0,0,+2}\,\Big((h_2-1)m-(h_1-1)n\Big)\,\epsilon^{abcd}\epsilon^{rstu}\,(\Phi^{(h_1+h_2)}_{-2})_{m+n}
\, ,
\nonu\\
\comm{(\Phi^{(h_1),abcr}_{0})_m}{(\Phi^{(h_2),dstu}_{0})_n}&=&
-\kappa_{0,0,+2}\,\Big((h_2-1)m-(h_1-1)n\Big)\,\epsilon^{abcd}\epsilon^{rstu}\,(\Phi^{(h_1+h_2)}_{-2})_{m+n}
\, ,
\nonu\\
\comm{(\Phi^{(h_1),abrs}_{0})_m}{(\Phi^{(h_2),cdtu}_{0})_n}&=&
\kappa_{0,0,+2}\,\Big((h_2-1)m-(h_1-1)n\Big)\,\epsilon^{abcd}\epsilon^{rstu}\,(\Phi^{(h_1+h_2)}_{-2})_{m+n} \, .
\nonu \\
\label{eq:scalar_scalar}
\eea
The first is related to the first and the
second $SU(4)$ global symmetries.
When the index $r$ passes through the index $d$
on the right hand side of eq. $25$ in (\ref{Eqs}),
there exists an extra minus sign in the second equation
of (\ref{eq:scalar_scalar}).
The relevant Lagrangian was given by
the previous fifth and seventh terms (or relevant
$SU(8)$ invariant term) associated with (\ref{eq:graviton_scalar})
in the subsection \ref{twozero} and 
we observe that the first celestial commutator
shows the helicities $(0, 0, +2)$
for the two scalars and the graviton. The epsilon dependence
appears on the right hand sides of (\ref{eq:scalar_scalar})
\footnote{
\label{30}
In the tensor product ${\bf 70} \otimes {\bf 70}$,
there are three tensor products,
$({\bf 1},{\bf 1})_4 \otimes ({\bf 1},{\bf 1})_{-4}$
where there exists
$({\bf 1},{\bf 1})_0$,
$(\overline{\bf 4},{\bf 4})_2 \otimes ({\bf 4},
\overline{\bf 4})_{-2}$
where there is
$({\bf 1},{\bf 1})_0$,
and $({\bf 6},{\bf 6})_0 \otimes ({\bf 6},
{\bf 6})_{0}$
where there exists
$({\bf 1},{\bf 1})_0$.
For the remaining twelve tensor products,
there are, with $\pm 2, \pm 4, \pm 6, \pm 8$ $U(1)$ charges,
\bea
&& \comm{(\Phi^{(h_1),abcd}_{0})_m}{(\Phi^{(h_2),efgh}_{0})_n}_8 =  0=
\comm{(\Phi^{(h_1),abcd}_{0})_m}{(\Phi^{(h_2),efgr}_{0})_n}_6=
\comm{(\Phi^{(h_1),abcd}_{0})_m}{(\Phi^{(h_2),efrs}_{0})_n}_4
\nonu \\
&&=
\comm{(\Phi^{(h_1),abcd}_{0})_m}{(\Phi^{(h_2),erst}_{0})_n}_2
=
\comm{(\Phi^{(h_1),abcr}_{0})_m}{(\Phi^{(h_2),defs}_{0})_n}_3=
\comm{(\Phi^{(h_1),abcr}_{0})_m}{(\Phi^{(h_2),dest}_{0})_n}_2
\nonu \\
&&=
\comm{(\Phi^{(h_1),abcr}_{0})_m}{(\Phi^{(h_2),stuv}_{0})_n}_{-2}=
\comm{(\Phi^{(h_1),abrs}_{0})_m}{(\Phi^{(h_2),ctuv}_{0})_n}_{-2}
=
\comm{(\Phi^{(h_1),abrs}_{0})_m}{(\Phi^{(h_2),tuvw}_{0})_n}_{-4}
\nonu \\
&&=
\comm{(\Phi^{(h_1),arst}_{0})_m}{(\Phi^{(h_2),buvw}_{0})_n}_{-4}
=
\comm{(\Phi^{(h_1),arst}_{0})_m}{(\Phi^{(h_2),uvwx}_{0})_n}_{-6}=
\comm{(\Phi^{(h_1),rstu}_{0})_m}{(\Phi^{(h_2),vwxy}_{0})_n}_{-8},
\nonu
\eea
from the eq. $25$ of (\ref{Eqs}).}.

In summary, the fourteen terms of
the Lagrangian in the ${\cal N}=8$ $SO(8)$ supergravity \cite{dF,deWit}
have their celestial soft current algebra
characterized by (\ref{eq:graviton_graviton}),
(\ref{eq:graviton_gravitino}),
(\ref{eq:graviton_photon}),
(\ref{eq:graviton_photino}),
(\ref{eq:graviton_scalar}),
(\ref{eq:gravitino_gravitino}),
(\ref{eq:gravitino_photon}),
(\ref{eq:gravitino_photino}),
(\ref{eq:gravitino_scalar}),
(\ref{eq:photon_photon}),
(\ref{eq:photon_photino}),
(\ref{eq:photon_scalar}),
(\ref{eq:photino_photino}),
(\ref{eq:photino_scalar}), and 
(\ref{eq:scalar_scalar})
and their OPEs having the Euler beta functions are given by
Appendix $C$ explicitly. In terms of $SU(8)$ indices,
they come from the twenty five relations in (\ref{Eqs}).
Moreover, they satisfy the $U(1)$ charges introduced in the
(\ref{su4branching}). In other words, both sides of
above (anti)commutators have the same $U(1)$ charge.
Although we have described the analysis of two
${\cal N}=4$ super Yang-Mills theory amplitudes in the footnotes
\ref{commexample}, \ref{missing} and \ref{othermissing},
we can check all the other descriptions done in \cite{2212-1}
explicitly.
Moreover, several truncations from the maximal ${\cal N}=8$
supergravity are given by Appendices
$F,G,H$, and $I$
\footnote{
\label{missingsplitfactor}
The first, the third, the fifth,
the
eighth, the tenth of (\ref{eq:photino_photino})
and the first and the second of (\ref{eq:photino_scalar})
are missing in \cite{2212-1} in the sense that
there are no split factors for these cases in Appendix $B$ of \cite{2212-1}.


The eighth equation of
(\ref{eq:photino_photino}) has 
the first
split factor 
containing the particle one (scalar) having the upper index
$ab$ with zero helicity and 
the particle two (scalar) having the lower $cd$
with zero helicity
together with $\epsilon^{abef} \epsilon_{cdef}$
where the first epsilon comes from Table $1$
and the second epsilon is used for raising the lower indices
$cd$ in the scalars.
The second split factor
contains the particle one (gluino) with $+\frac{1}{2}$ helicity
and
the particle two (gluino) with $-\frac{1}{2}$ helicity
with the Kronecker delta.
Then the  first split factor
has the gluon with
$-1$ helicity and the second split factor has
the gluon with $-1$ helicity.
The product between
the gluon and the gluon
produces the graviton with the helicity $-2$.

Finally,
the final equation of
(\ref{eq:photino_photino})
has the first
split factor which  
contains the particle one (gluon) with helicity
$+1$
and the particle two (gluon) with helicity
$-1$.
The second split factor
contains the particle one (gluino) with $-\frac{1}{2}$ helicity
with lower index $t'$ and
the particle two (gluino) with helicity
$+\frac{1}{2}$ with upper index $w'$  together with
$\epsilon^{rstt'} \epsilon_{uvww'}$.
Then the  first split factor
has the gluon with
$-1$ helicity and the second split factor has
the the gluon with $-1$ helicity with Kronecker delta
from Table $1$ in \cite{2212-1}.
There is also the previous
$\epsilon^{rstt'} \epsilon_{uvww'}$.
The product between
the gluon and the gluon
produces the graviton with the helicity $+2$
from Table $3$ in \cite{2212-1}.
}.


In Tables \ref{prevtable} and \ref{latertable},
we compare what we have found in this paper
with the results of \cite{2212-1,2212-2}.
For given their results specified in the first columns
of these Tables, the second columns
provide the corresponding  celestial (anti)commutators in this
paper and the last columns describe
the missing parts of \cite{2212-2}.
For convenience, we list all the
OPEs in Appendix $C$ where the twenty five OPEs with the $SU(8)_R$
$R$ symmetry indices are given and its $117$ OPEs
\footnote{There are $68$ vanishing (anti)commutators
given by the footnotes
\ref{17}, \ref{18}, \ref{19},
\ref{20}, \ref{21},
\ref{22}, \ref{24}, \ref{26}, \ref{29}, and \ref{30}.}
in the
$SU(4) \times SU(4)$ symmetry indices are presented
explicitly.
We also point out that the thirty six OPEs with the boldface
fonts (corresponding to the number of cases of the third columns
of Tables \ref{prevtable} and \ref{latertable})
should appear in the complete description of this list,
even though some amplitudes
related to some of these OPEs
appear in the various tables of \cite{2212-1}. 

\begin{table}[tbp]
\centering
\renewcommand{\arraystretch}{1.7}
\begin{tabular}{|c|c||c| }
\hline
The equations in \cite{2212-2}  & The equations in this paper
& The missing equations in \cite{2212-2}
\\
\hline
$(2.21)$ & The eq. $1$ of (\ref{Eqs}) &
\\
\hline
$(2.25)$ & The eq. $9$  of (\ref{Eqs}) &
\\
\hline
The 1st of $(A.1)$ & The eq. $10$  of (\ref{Eqs}) &
\\
\hline
The 2nd of $(A.1)$ & The eq. $16$  of (\ref{Eqs}) &
\\
\hline
The 1st of $(A.2)$ & The eq. $17$  of (\ref{Eqs}) &
\\
\hline
The 2nd of $(A.2)$ & The eq. $21$  of (\ref{Eqs}) &
\\
\hline
The 1st of $(A.3)$ & The 6th of
(\ref{eq:photino_photino}) & 
\\
\cline{1-2}
The 2nd of 
$(A.3)$ & The 4th of
(\ref{eq:photino_photino}) &
The 1st, 3rd and 5th of (\ref{eq:photino_photino})
\\
\cline{1-2}
The 3rd of
$(A.3)$ 
& The 2nd of (\ref{eq:photino_photino}) &
\\
\hline
The 4th of
$(A.3)$ & The 9th of
(\ref{eq:photino_photino}) &
The 7th, 8th and 10th of (\ref{eq:photino_photino})
\\
\hline
$(A.4)$ & The 3rd of (\ref{eq:scalar_scalar}) &
The 1st and 2nd of (\ref{eq:scalar_scalar})
\\
\hline
The 1st of 
$(A.5)$ & The eq. $2$  of (\ref{Eqs}) &
\\
\hline
The 2nd of
$(A.5)$
& The eq. $8$   of (\ref{Eqs}) &
\\
\hline
The 1st of
$(A.6)$ & The eq. $3$  of (\ref{Eqs}) &
\\
\hline
The 2nd of 
$(A.6)$ & The eq. $7$   of (\ref{Eqs}) &
\\
\hline
\end{tabular}
\caption{
The first column denotes the various OPEs
between the graviton, the gravitinos, the graviphotons,
the graviphotinos, and scalars \cite{2212-2}
(See also Appendix $C$),
the second column
stands for the corresponding (anti)commutators between these
soft particles, and
the third column stands for the missing equations in \cite{2212-2}
and these will appear in Appendix $C$ later.
Recall that the relations in (\ref{eq:photino_photino})
come from the eq. $22$ and the eq. $24$ of (\ref{Eqs})
while the relations (\ref{eq:scalar_scalar})
come from the eq. $25$ of (\ref{Eqs}).}
\label{prevtable}
\end{table}

\begin{table}[tbp]
\centering
\renewcommand{\arraystretch}{1.7}
\begin{tabular}{|c|c||c| }
\hline
The equations in \cite{2212-2}  & The equations in this paper
& The missing equations in \cite{2212-2}
\\
\hline
The 1st of
$(A.7)$ & The 2nd of (\ref{eq:graviton_photino}) &
\\
\cline{1-2}
The 2nd of
$(A.7)$ & The 1st of
(\ref{eq:graviton_photino}) &
The 3rd, 4th, 6th-8th
of (\ref{eq:graviton_photino})
\\
\cline{1-2}
The 3rd of
$(A.7)$
& The 5th of (\ref{eq:graviton_photino}) &
\\
\hline
$(A.8) $ & The eq. $5$  of (\ref{Eqs}) &
\\
\hline
The 1st of
$(A.9)$ & The eq. $11$  of (\ref{Eqs}) &
\\
\hline
The 2nd of
$(A.9)$ &  The eq. $15$  of (\ref{Eqs}) &
\\
\hline
The 1st of 
$(A.10)$ & The eq. $12$  of (\ref{Eqs}) &
\\
\hline
The 2nd of
$(A.10)$ & The eq. $14$  of (\ref{Eqs}) &
\\
\hline
$(A.11)$ & The eq. $13$  of (\ref{Eqs}) &
\\
\hline
The 1st of
$(A.12)$ & The 1st of
(\ref{eq:photon_photino}) &
The 2nd-10th of (\ref{eq:photon_photino})
\\
\hline
The 2nd of
$(A.12)$ & The eq. $20$  of (\ref{Eqs}) &
\\
\hline
The 1st of
$(A.13)$ &  The 1st of (\ref{eq:photon_scalar}) &
The 2nd-8th of  (\ref{eq:photon_scalar})
\\
\hline
The 2nd of
$(A.13)$ & &
\\
\hline
The 1st of
$(A.14)$ & The 3rd of (\ref{eq:photino_scalar}) &
The 1st, 2nd, 5th-8th of (\ref{eq:photino_scalar})
\\
\cline{1-2}
The 2nd of
$(A.14)$ & The 4th of (\ref{eq:photino_scalar}) &
\\
\hline
\end{tabular}
\caption{This is the continuation of the previous Table
\ref{prevtable}.  The first column denotes the various OPEs
between the graviton, the gravitinos, the graviphotons,
the graviphotinos, and scalars, the second column
stands for the corresponding (anti)commutators between the soft
particles, and
the third column stands for the missing equations in \cite{2212-2}
(See also the boldface thirty six OPEs in Appendix $C$).
Recall that the relations in
(\ref{eq:graviton_photino})
come from the eq. $4$ and the eq. $6$ of (\ref{Eqs}),
the relations (\ref{eq:photon_photino}) do from
the eq. $18$ and the eq. $20$ of (\ref{Eqs}),
the relations (\ref{eq:photon_scalar}) originate from
the eq. $19$ of (\ref{Eqs})
and the relations
(\ref{eq:photino_scalar}) do from the eq. $23$ of (\ref{Eqs}).}
\label{latertable}
\end{table}

\subsection{The Jacobi identity}

Le us consider the Jacobi identity
from the celestial operators
corresponding to the graviton, the gravitinos,
the graviphotons, the graviphotinos and the
scalars.

It turns out that the couplings satisfy the
following relations \footnote{In principle, we can do this
by hands but the possible number of Jacobi identities
between the nine operators in (\ref{su4branching})
we should consider is given
by $\frac{(9+2)!}{3!(9+2-3)!}=165$ from the formula
of `combination with repetition' which is a huge number
we should check, although some of the expressions are trivially
zero.}
\bea
\kappa_ {+\frac{1}{2}, -\frac{1}{2},+2} & = & 
-\frac {\kappa_ {0, 0, +2}\, \kappa_ {+1, +\frac{1}{2}, +\frac{1}{2}} \,
\kappa_ {+\frac{3}{2}, +\frac{3}{2}, -1}} { (\kappa_{+\frac{3}{2}, 0, +\frac{1}{2}})^2}\, ,
\quad
\kappa_ {+\frac{1}{2}, 0, +\frac{3}{2}} = 
-\frac {\kappa_ {0, 0, +2}\, \kappa_ {+1, +\frac{1}{2}, +\frac{1}{2}} \,
\kappa_ {+\frac{3}{2}, +\frac{3}{2}, -1}} {\kappa_ {+\frac{3}{2}, -\frac{3}{2}, +2}\,
\kappa_ {+\frac{3}{2}, 0, +\frac{1}{2}}}\, ,
\nonu \\
\kappa_ {+\frac{1}{2}, +\frac{1}{2}, +1} & = &
-\frac { \kappa_ {0, 0, +2} 
\,(\kappa_ {+1, +\frac{1}{2}, +\frac{1}{2}})^2 \,(\kappa_ {+\frac{3}{2}, +\frac{3}{2}, -1})^2} {\kappa_ {+\frac{3}{2}, -\frac{3}{2}, +2} \,\kappa_ {+\frac{3}{2}, -1, +\frac{3}{2}}\,
(\kappa_ {+\frac{3}{2}, 0, +\frac{1}{2}})^2}\, ,
\quad 
\kappa_ {+1, -1,+2} = -\frac{\kappa_ {+\frac{3}{2}, -\frac{3}{2}, +2}\, \kappa_ {+\frac{3}{2}, -1, +\frac{3}{2}}}{\kappa_ {+\frac{3}{2}, +\frac{3}{2}, -1}}\, ,
\nonu \\
\kappa_{+1, -\frac{1}{2}, +\frac{3}{2}} & = &
\frac {\kappa_ {0, 0, +2}\, \kappa_ {+1, +\frac{1}{2}, +\frac{1}{2}} \,
\kappa_ {+\frac{3}{2}, +1, -\frac{1}{2}}\, \kappa_ {+\frac{3}{2}, +\frac{3}{2}, -1}} {
\kappa_ {+\frac{3}{2}, -\frac{3}{2},+ 2}\, (\kappa_ {+\frac{3}{2}, 0, +\frac{1}{2}})^2}\, ,
\nonu \\
\kappa_{+1, 0,+1} & = &
-\frac { \kappa_ {0, 0, +2}\, \kappa_ {+1, +\frac{1}{2}, +\frac{1}{2}} \,
\kappa_{+\frac{3}{2}, +1, -\frac{1}{2}}\, \kappa_ {+\frac{3}{2},+\frac{3}{2}, -1}} {
\kappa_{+\frac{3}{2}, -\frac{3}{2}, +2}\, \kappa_ {+\frac{3}{2}, -1, +\frac{3}{2}} \,
\kappa_ {+\frac{3}{2}, 0, +\frac{1}{2}}}\, ,
\quad
\kappa_ {+1, +1, 0} = \frac{\kappa_ {+1, +\frac{1}{2}, +\frac{1}{2}}\, \kappa_{+\frac{3}{2}, 
+1, -\frac{1}{2}}}{\kappa_ {+\frac{3}{2}, 0, +\frac{1}{2}}}\, ,
\nonu \\
\kappa_ {+\frac{3}{2}, -\frac{1}{2}, +1} & = &
\frac { \kappa_ {0, 0, +2}\, \kappa_ {+1, +\frac{1}{2}, +\frac{1}{2}}\, \kappa_ {\frac{3}{2}, +1, -\frac{1}{2}}\,
(\kappa_ {+\frac{3}{2}, +\frac{3}{2}, -1})^2} {\kappa_ {+\frac{3}{2}, -\frac{3}{2}, +2}
\,\kappa_ {+\frac{3}{2}, -1, +\frac{3}{2}} \,(\kappa_ {+\frac{3}{2}, 0, +\frac{1}{2}})^2}\, ,
\quad
\kappa_{+\frac{3}{2}, +\frac{1}{2},0} =\frac{\kappa_ {+1, +\frac{1}{2}, +\frac{1}{2}} \,
\kappa_ {+\frac{3}{2}, +\frac{3}{2}, -1}}{\kappa_ {+\frac{3}{2}, 0, +\frac{1}{2}}}\, ,
\nonu \\
\kappa_ {+2, -2, +2} & = & \kappa_ {+2, +2, -2}\, ,
\qquad
\kappa_ {+2, -\frac{3}{2}, +\frac{3}{2}} = \kappa_ {+2, +2, -2}\, ,
\qquad
\kappa_ {+2, -1, +1} = \kappa_ {+2, +2, -2}\, ,
\nonu \\
\kappa_ {+2, -\frac{1}{2}, +\frac{1}{2}} & = & \kappa_ {+2, +2, -2}\, ,
\qquad
\kappa_ {+2, 0, 0} = \kappa_{+2, +2, -2}\, ,
\qquad
\kappa_ {+2, +\frac{1}{2}, -\frac{1}{2}} = \kappa_{+2, +2, -2}\, ,
\nonu \\
\kappa_ {+2, +1, -1} & = & \kappa_ {+2, +2, -2}\, ,
\qquad 
\kappa_ {+2, +\frac{3}{2}, -\frac{3}{2}} = \kappa_{+2, +2, -2}\, .
\label{kappasol}
\eea
The twenty five (coming from
the previous relations in (\ref{Eqs}))
couplings can be written in terms of
eight unknown ones,
\bea
&& \kappa_{0,0,+2}, \qquad
\kappa_{+1,+\frac{1}{2},+\frac{1}{2}},
\qquad
\kappa_{+\frac{3}{2},+\frac{3}{2},-1},
\qquad
\kappa_{+\frac{3}{2},0,+\frac{1}{2}},
\nonu \\
&& \kappa_{+\frac{3}{2},-\frac{3}{2},+2},
\qquad
\kappa_{+\frac{3}{2},-1,+\frac{3}{2}},
\qquad
\kappa_{+\frac{3}{2},+1,-\frac{1}{2}},
\qquad
\kappa_{+2,+2,-2},
\label{Eightkappa}
\eea
which appear on the right hand sides of (\ref{kappasol})
via above seventeen relations \footnote{Let us emphasize that
the condition $s_1+s_2-s_3=2$ provides forty five cases
for given the helicities $(0,\pm \frac{1}{2}, \pm 1, \pm \frac{3}{2},
\pm 2)$ and  the twenty five couplings appear in (\ref{Eqs})
where the seventeen couplings have the relations in (\ref{kappasol})
in terms of the eight couplings (\ref{Eightkappa}).
The remaining twenty couplings can be written in terms of
the above twenty five couplings. Of course,
if we relax the condition for the helicities, then the independent
number of couplings will be increased. }.
The mathematica \cite{mathematica} program with the
Thielemans package \cite{Thielemans}
is given by Appendix $D$.
In Appendix $E$, we describe the Jacobi identity
by considering the (anti)commutators living in
the ${\cal N}=4$ supergravity first and solve
the Jacobi identity with nontrivial relations.
After that, by applying other nontrivial Jacobi identity
to the remaining (anti)commutators, the final relations are
determined and these are the same as above
\footnote{One of the solution satisfying
(\ref{kappasol}) is given by
$\kappa_{+\frac{3}{2},-\frac{3}{2},+2}=\kappa_{+1,+\frac{1}{2},+\frac{1}{2}}=
\kappa_{+\frac{1}{2},0,+\frac{3}{2}}=\kappa_{+1,-\frac{1}{2},+\frac{3}{2}}=
\kappa_{+\frac{3}{2},-\frac{1}{2},+1}=\kappa_{+\frac{3}{2},+\frac{1}{2},0}=
\kappa_{+\frac{3}{2},+1,-\frac{1}{2}}=-1$ and the remaining couplings
with $+1$.}.


After introducing the product of
Grassmann coordinates
\bea
\eta_{A_1 A_2\cdots A_n} \equiv
\frac{1}{n!}\eta_{A_1}\eta_{A_2}\cdots \eta_{A_n}\,,
\label{etaredef}
\eea
we denote the ${\cal N}=8$ superspace description
for the Laurent mode
as follows \cite{Tropper1}:
\bea
({\bf \Phi}_{{\bf s}}^{{\bf (h)}})_m(\eta)&=&
(\Phi^{(h)}_{+2})_m
+\eta_A\,(\Phi^{(h),A}_{+\frac{3}{2}})_m
+\eta_{BA}\,(\Phi^{(h),AB}_{+1})_m
+\eta_{CBA}\,(\Phi^{(h),ABC}_{+\frac{1}{2}})_m
\nonu \\
&+&\eta_{DCBA}\,(\Phi^{(h),ABCD}_{0})_m
+\frac{1}{3!}\eta_{EDCBA}\,\epsilon^{ABCDEFGH}(\Phi^{(h)}_{FGH,\, -\frac{1}{2}})_m
\nonu \\
&+ & \frac{1}{2!}\eta_{FEDCBA}\,\epsilon^{ABCDEFGH}(\Phi^{(h)}_{GH, \,-1})_m
+\eta_{GFEDCBA}\,\epsilon^{ABCDEFGH}(\Phi^{(h)}_{H, \, -\frac{3}{2}})_m
\nonu \\
&+& \eta_{HGFEDCBA}\,\epsilon^{ABCDEFGH}(\Phi^{(h)}_{-2})_m \, .
\label{etaexpansion}
\eea
The boldfac ${\bf h}$ is given by
${\bf h}=(h+2, h+\frac{3}{2}, h+1, h+\frac{1}{2}, h,
h-\frac{1}{2},h-1,h-\frac{3}{2},h-2)$ while
the boldface ${\bf s}$ is given by
${\bf s}=(+2,+\frac{3}{2},+1,+\frac{1}{2},0,-\frac{1}{2},-1,
-\frac{3}{2},-2)$.
By noticing that
\bea
({\bf \Phi_s^{(h)}})_m\bigg|_{\eta_A=0} & = & (\Phi^{(h)}_{+2})_m\,,
\nonu \\
\frac{\partial}{\partial \eta_A} ({\bf \Phi_s^{(h)}})_m\bigg|_{\eta_B=0}&=&
(\Phi^{(h),A}_{+\frac{3}{2}})_m\,,
\nonu \\
\frac{\partial}{\partial \eta_A}\frac{\partial}{\partial \eta_B}
({\bf \Phi_s^{(h)}})_m\bigg|_{\eta_C=0}&=&(\Phi^{(h),AB}_{+1})_m\,,
\nonu \\
\frac{\partial}{\partial \eta_A}\frac{\partial}{\partial \eta_B}\frac{\partial}{\partial \eta_C}
({\bf \Phi_s^{(h)}})_m\bigg|_{\eta_D=0} &= &
(\Phi^{(h),ABC}_{+\frac{1}{2}})_m\,,
\nonu \\
\frac{\partial}{\partial \eta_A}
\frac{\partial}{\partial \eta_B}
\frac{\partial}{\partial \eta_C}
\frac{\partial}{\partial \eta_D}
({\bf \Phi_s^{(h)}})_m\bigg|_{\eta_E=0}&=&(\Phi^{(h),ABCD}_{0})_m\,,
\nonu \\
\frac{\partial}{\partial \eta_A}
\frac{\partial}{\partial \eta_B}
\frac{\partial}{\partial \eta_C}
\frac{\partial}{\partial \eta_D}
\frac{\partial}{\partial \eta_E}
({\bf \Phi_s^{(h)}})_m\bigg|_{\eta_I=0}&=&
\frac{1}{3!}
\epsilon^{ABCDEFGH}(\Phi^{(h)}_{FGH, \, -\frac{1}{2}})_m\,,
\nonu \\
\frac{\partial}{\partial \eta_A}
\frac{\partial}{\partial \eta_B}
\frac{\partial}{\partial \eta_C}
\frac{\partial}{\partial \eta_D}
\frac{\partial}{\partial \eta_E}
\frac{\partial}{\partial \eta_F}
({\bf \Phi_s^{(h)}})_m\bigg|_{\eta_I=0}&=&
\frac{1}{2!}\, \epsilon^{ABCDEFGH}(\Phi^{(h)}_{GH, \, -1})_m\,,
\nonu \\
\frac{\partial}{\partial \eta_A}
\frac{\partial}{\partial \eta_B}
\frac{\partial}{\partial \eta_C}
\frac{\partial}{\partial \eta_D}
\frac{\partial}{\partial \eta_E}
\frac{\partial}{\partial \eta_F}
\frac{\partial}{\partial \eta_G}
({\bf \Phi_s^{(h)}})_m\bigg|_{\eta_I=0}&=&
\epsilon^{ABCDEFGH}(\Phi^{(h)}_{H, \, -\frac{3}{2}})_m\,,
\nonu \\
\frac{\partial}{\partial \eta_A}
\frac{\partial}{\partial \eta_B}
\frac{\partial}{\partial \eta_C}
\frac{\partial}{\partial \eta_D}
\frac{\partial}{\partial \eta_E}
\frac{\partial}{\partial \eta_F}
\frac{\partial}{\partial \eta_G}
\frac{\partial}{\partial \eta_H}
({\bf \Phi_s^{(h)}})_m\bigg|_{\eta_I=0} &= &
\epsilon^{ABCDEFGH}(\Phi^{(h)}_{-2})_m\,,
\label{compexp}
\eea
the previous (anti)commutators in (\ref{Eqs})
can be written as
\footnote{For example,
by acting
$\frac{1}{5!} \, \epsilon_{EFGHIBCD} \,
\frac{1}{\pa \eta^1_A} \frac{1}{\pa \eta^2_E}
\frac{1}{\pa \eta^2_F}
\frac{1}{\pa \eta^2_G}
\frac{1}{\pa \eta^2_H}
\frac{1}{\pa \eta^2_I}
$ on both sides of (\ref{finalsuperexp}) with (\ref{compexp}),
the left hand side becomes the left hand side of the eq. $14$ of
(\ref{Eqs}) where we use $\epsilon^{JKLMNOPQ}\,
\epsilon_{EFGHIOPQ}=3 !\,
\de_{EFGHI}^{JKLMN}$ and the right hand side becomes
the right hand side of the eq. $14$ of (\ref{Eqs}) except the minus sign
where the identity
$\epsilon^{JKLMNOPQ}\,
\epsilon_{AEFGHIPQ}=2 !\,
\de_{AEFGHI}^{JKLMNO}$
is used. Moreover, the extra minus signs
corresponding to the eqs. $16$, $20$ and $24$
of (\ref{Eqs}) appear on the right hand
sides of (\ref{finalsuperexp}). The two commutators
corresponding to the eqs. $15$ and $20$
after interchanging the two operators
on the left hand sides should have the extra minus signs.
Except of these six cases (We can absorb
these minus signs
by redefining the couplings newly),
the remaining twenty one (plus eighteen)
(anti)commutators have the plus sign on the right hand sides of
(\ref{finalsuperexp}).}
\bea
\comm{({\bf \Phi}_{{\bf s_1}}^{{\bf (h_1)}})_m(\eta_1)}{
({\bf \Phi}_{{\bf s_2}}^{{\bf (h_2)}})_n(\eta_2)}
&=& \pm 
\kappa_{{\bf s_1,s_2,-s_1-s_2+2}}\, \Big(({\bf h_2}-1)m-
({\bf h_1}-1)n\Big)\, \nonu \\
& \times & 
({\bf \Phi}_{{\bf s_1+s_2-2}}^{{\bf (h_1+h_2-s_1-s_2)}})_{m+n}(\eta_1+\eta_2) \, ,
\label{finalsuperexp}
\eea
from (\ref{etaexpansion}) and (\ref{etaredef}).
It would be interesting to observe whether
this one single relation provides the solution for
the Jacobi identity by considering the superspace description
for the Jacobi identity or not.

\section{Conclusions and outlook}

The celestial soft symmetry algebra,
which is a mathematically consistent and plausible algebra
whose the whole Jacobi identiy closes, 
from (\ref{eq:graviton_graviton}) to
(\ref{eq:scalar_scalar}) was eventually obtained.
On the one hand,
we have used  the known results \cite{2212-1,2212-2}
for the various OPEs (having the Euler beta functions) between  any two
operators among the graviton, the gravitinos,
the graviphotons, the graviphotinos and the
scalars in the ${\cal N}=8$ supergravity
to obtain the above (anti)commutators
between the soft currents by absorbing the infinities
of the Euler beta functions together with careful
analysis of $SU(8)_R$ $R$ symmetry indices appearing both sides,
along the lines of
\cite{GHPS,Strominger}. 
On the other hand, from the standard conformal field theory
in two dimensions, our starting point was based on the earlier
work by \cite{Ademolloplb} described in the introduction.
We have presented the corresponding equations in (\ref{Eqs})
where the twenty five (anti)commutators with various
structure constants exist.
In the OPE language, we don't have the double summation
terms in (\ref{Ademolloope}). It is nontrivial to
consider the eq. $17$ of (\ref{Eqs}) because this commutator
doesn't appear in the corresponding algebra for the lowest
${\cal N}=8$ multiplet ($h_1=0=h_2$).
As in the abstract or the introduction,
the first equation of (\ref{eq:graviton_graviton}),
the commutator between the soft gravitons with positive
helicities,
plays the role of wedge subalgebra of $w_{1+\infty}$ algebra
\cite{Bakas}.
In this sense, the equations (\ref{Eqs}) or
the equations from (\ref{eq:graviton_graviton}) to
(\ref{eq:scalar_scalar}) are the wedge subalgebra
of ${\cal N}=8$ supersymmetric
$w_{1+\infty}$ algebra.
From eq. $10$ to eq. $16$, the eight gravitinos
produce the remaining other multiplets explicitly.
Each mode from
the graviton, the gravitinos, the graviphoton, the graviphotinos and
the scalars satisfies some constraints although we didn't write down
them explicitly.

In this paper, we have considered the case of
$s_1+s_2-s_3=2$. The next question is 
to determine whether there exists any celestial soft symmetry
current algebra for the cases of
$s_1+s_2-s_3=3,4,5,6$ or not \cite{EJN,RSYV}.
See also \cite{AK2309,Ahn2203,Ahn2202}.
Recall that for the $s_1+s_2-s_3=2$ case,
the possible number of OPEs or (anti)commutators
between the helicities $0$, $\pm \frac{1}{2}$, $\pm 1$, $\pm \frac{3}{2}$,
and $\pm 2$ is given by forty five.
Among them, the $20$  OPEs or (anti)commutators
are not independent and we are left with twenty five
independent ones.
We should return to the original paper by \cite{PRSY}
and extract the corresponding OPEs where the right hand sides
consist of several singular terms which describe the above conditions
for the helicities (or the scaling dimension of the three point vertex
in the bulk).

It is known that the OPE coefficients for the gravitons
can be fixed \cite{PRSY} also by i) translation symmetry which leads to
the recursion relations,
ii) leading soft graviton theorem for normalization and
iii) the subsubleading
soft graviton theorem which produces other recursion relations
(no constraints from subleading soft graviton theorem).
It is natural to observe 
whether
the supersymmetric soft theorem \cite{Tropper,Liu}
or other global symmetries
can determine the OPE coefficients found in this paper or not.

\vspace{.7cm}

\centerline{\bf Acknowledgments}

CA thanks M. Pate and A. Tropper for  the discussion on
the general aspects of the celestial holography, 
T. Rahnuma and R.K. Singh for the discussion 
on their papers \cite{2212-1,2212-2}.
This work was supported by the National
Research Foundation of Korea(NRF) grant funded by the
Korea government(MSIT) 
(No. 2023R1A2C1003750).
This research was supported by Basic Science Research
Program through the National Research Foundation of Korea(NRF)
funded by the Ministry of Education(RS-2024-00446084).

\newpage

\appendix

\renewcommand{\theequation}{\Alph{section}\mbox{.}\arabic{equation}}

\section{The $SO({\cal N})$ superconformal algebra
with ${\cal N}=0$,
$1$, $2$, $\cdots$, $7$
in two dimensional conformal field theory}

The bosonic conformal algebra \cite{Virasoro,FV1}
with ${\cal N}=0$ is given by
\bea
\comm{L_m}{L_n}=(m-n)L_{m+n}\,,
\label{zerosca}
\eea
where the generator is rescaled by
\bea
O_m \equiv 2\,L_m \,.
\label{redef0}
\eea

The ${\cal N}=1$ superconformal algebra \cite{NS,Ramond,Schwarz}
with ${\cal N}=1$ is summarized
by
\bea
\comm{L_m}{L_n} & = & (m-n)L_{m+n}\,,\qquad
\comm{L_m}{G_r}=(\tfrac{1}{2}\,m-r)G_{m+r}\,,
\nonu \\
\acomm{G_r}{G_s} & = & 2 \,L_{m+n}\,,
\label{onesca}
\eea
where the rescaled generators are
\bea
O_m \equiv 2\,L_m \,,\qquad  O^{1}_r \equiv G_r\,.
\label{redef1}
\eea

The ${\cal N}=2$ superconformal algebra
\cite{Ademolloplb,Ademollonpb}
with $a,b=1,2$ 
is described by
\bea
\comm{L_m}{L_n} & =& (m-n)\,L_{m+n}\,,
\qquad
\comm{L_m}{G^a_r}=(\tfrac{1}{2}m-r)\,G^a_{m+r}\,,
\nonu \\
\comm{L_m}{J_n} & = & -n\,J_{m+n}\,,
\qquad
\acomm{G^a_r}{G^b_r}=2\delta^{ab}\,L_{r+s}+
\mathrm{i}\epsilon^{ab} (r-s)\,J_{r+s}\,,
\nonu \\
\comm{J_m}{G^a_r} & = &
\mathrm{i}\epsilon^{ab} G^b_{m+r}\,,
\label{twosca}
\eea
where 
\bea
&O_m \equiv 2\,L_m \,,\qquad
O^{a}_r \equiv G^a_r\,,\qquad
O^{ab}_m \equiv -\epsilon^{ab}\,J_m\,.
\label{redef2}
\eea

The ${\cal N}=3$ superconformal algebra \cite{Ademolloplb,RS}
with $a,b=1,2,3$ is
\bea
\comm{L_m}{L_n} & = & (m-n)\,L_{m+n}\,,
\qquad
\comm{L_m}{G^a_r}=(\tfrac{1}{2}m-r)\,G^a_{m+r}\,,
\nonu \\
\comm{L_m}{J^a_n} & = & -n\,J^a_{m+n}\,,
\qquad
\comm{L_m}{\Psi_r}=(-\tfrac{1}{2}m-r)\,\Psi_{m+r}\,,
\nonu \\
\acomm{G^a_r}{G^b_s} & = &
\delta^{ab}\,2\,L_{r+s}+\mathrm{i}\,(r-s)\,\epsilon^{abc}\,J^c_{r+s}\,,
\qquad
\comm{G^a_r}{J^b_m}=-\delta^{ab}\,m\,\Psi_{r+m}+\mathrm{i}\,\epsilon^{abc}\,G^c_{r+m}\,,
\nonu     \\
\acomm{G^a_r}{\Psi_s} & = & J^a_{r+s}\,,
\qquad
\comm{J^a_m}{J^b_n}=\mathrm{i}\,\epsilon^{abc}\,J^c_{m+n}\,,
\label{threesca}
\eea
where the generators are rescaled by
\bea
O_m \equiv 2\,L_m \,,\qquad
O^{a}_r \equiv G^a_r\,,\qquad
O^{ab}_m \equiv -\epsilon^{abc}\,J^c_m\,,
\qquad  O^{abc}_r \equiv \epsilon^{abc}\,\Psi_r\,.
\label{redef3}
\eea

The ${\cal N}=4$ superconformal algebra \cite{Schoutens,STV}
is obtained by
\bea
\comm{L_m}{L_n} & = & (m-n)\,L_{m+n}\,,
\qquad
\comm{L_m}{G^i_r}=(\tfrac{1}{2}m-r)\,G^i_{m+r}\,,
\nonu    \\
\comm{L_m}{T^{ij}_n} & = & -n\,T^{ij}_{m+n}\,,
\qquad
\comm{L_m}{\Gamma^i_r}=(-\tfrac{1}{2}m-r)\,\Gamma^i_{m+r}\,,
\nonu    \\
\comm{L_m}{\Delta_n} & = &
(-m-n)\,\Delta_{m+n}\,,
\qquad
\acomm{G^i_r}{G^j_s}=
\delta^{ij}\,2\,L_{r+s}-\mathrm{i}\,(r-s)\,T^{ij}_{r+s}\,,
\nonu     \\
\comm{G^i_r}{T^{jk}_m} & = &
\delta^{ij}\,\mathrm{i}\,G^k_{r+m}-\delta^{ik}\,\mathrm{i}\,G^j_{r+m}+\epsilon^{ijkl}\,m\,\Gamma^{l}_{r+m}\,,
\nonu \\
\acomm{G^i_r}{\Gamma^j_s} & = &
\delta^{ij}\,\mathrm{i}\,(r+s)\,\Delta_{r+s}-\frac{1}{2}\,
\epsilon^{ijkl}\, {T}^{kl}_{r+s}\,,
\qquad
\comm{G^i_r}{\Delta_m}=\mathrm{i}\,\Gamma^i_{r+m}\,,
\nonu \\
\comm{T^{ij}_m}{T^{kl}_n} & = &
-\delta^{ik}\,\mathrm{i}\,T^{jl}_{m+n}+\delta^{il}\,\mathrm{i}\,T^{jk}_{m+n}+\delta^{jk}\,\mathrm{i}\,T^{il}_{m+n}-\delta^{jl}\,\mathrm{i}\,T^{ik}_{m+n}\,,
\nonu     \\
\comm{T^{ij}_m}{\Gamma^{k}_r}& = &
-\delta^{ik}\,\mathrm{i}\,\Gamma^{j}_{m+r}+\delta^{jk}\,\mathrm{i}\,\Gamma^{i}_{m+r}\,,
\label{foursca}
\eea
where the generators are rescaled as 
\bea
O_m & \equiv & 2\,L_m \,,\qquad  O^{i}_r \equiv G^i_r\,,
\quad O^{ij}_m \equiv T^{ij}_m\,,
\quad
O^{ijk}_r  \equiv  \epsilon^{ijkl}\Gamma^{l}_r\,,
\quad O^{ijkl}_m \equiv \epsilon^{ijkl}\Delta_m\,.
\label{redef4}
\eea
This algebra can be seen from the work of \cite{AK2501}
with vanishing central charges.
%


The ${\cal N}=5,6,7$ superconformal algebra
is not known so far, compared to the previous cases,
(\ref{zerosca}), (\ref{onesca}), (\ref{twosca}),
(\ref{threesca}) and (\ref{foursca}).
By using the definition of
(\ref{Ademollo}), we present them as follows
without introducing any redefinitions done in
(\ref{redef0}), (\ref{redef1}), (\ref{redef2}), (\ref{redef3})
and (\ref{redef4}).
The ${\cal N}=5$ superconformal algebra
where $A,B,C,D, \cdots =1,2, \cdots, 5$ can be summarized by
\bea
\comm{O_m}{O_n}&=&2(m-n)\,O_{m+n} \,,
\nonu\\
\comm{O_m}{O^{A}_r}&=&(m-2r)\,O^{A}_{m+r}
\,,
\nonu\\
\comm{O_m}{O^{AB}_n}&=&-2n\,O^{AB}_{m+n}
\,,
\nonu\\
\comm{O_m}{O^{ABC}_r}&=&(-m-2r)\,O^{ABC}_{m+r}
\,,
\nonu\\
\comm{O_m}{O^{ABCD}_n}&=&2(-m-n)\,O^{ABCD}_{m+n}
\,,
\nonu\\
\comm{O_m}{O^{ABCDE}_r}&=&(-3m-2r)\,O^{ABCDE}_{m+r}
\,,
\nonu\\
\comm{O^{A}_r}{O^{B}_s}&=&\delta^{AB}\,O_{r+s}-\mathrm{i}(r-s)\,O^{AB}_{r+s}
\,,
\nonu\\
\comm{O^{A}_r}{O^{BC}_m}&=&\mathrm{i}\,2\,\delta^{A[B}O^{C]}_{r+m}+m\,O^{ABC}_{r+m}
\,,
\nonu\\
\comm{O^{A}_r}{O^{BCD}_s}&=&-3\,\delta^{A[B}O_{r+s}^{CD]}-\mathrm{i}(r+s)\,O^{ABCD}_{r+s}
\,,
\nonu\\
\comm{O^{A}_r}{O^{BCDE}_m}&=&-4\,\mathrm{i}\,\delta^{A[B}O^{CDE]}_{r+m}-(2r+m)\,O^{ABCDE}_{r+m}
\,,
\nonu\\
\comm{O^{A}_r}{O^{BCDEF}_s}&=&5\,\delta^{A[B}O_{r+s}^{CDEF]}
\,,
\nonu\\
\comm{O^{AB}_m}{O^{CD}_n}&=&-\mathrm{i}\,\delta^{AC}\,O^{BD}_{m+n}+\mathrm{i}\,\delta^{AD}\,O^{BC}_{m+n}+\mathrm{i}
\,\delta^{BC}\,O^{AD}_{m+n}-\mathrm{i}\,\delta^{BD}\,O^{AC}_{m+n}
\,,
\nonu\\
\comm{O^{AB}_m}{O^{CDE}_r}&=&-3\,\mathrm{i}\,\Big(\delta^{A[C}O^{DE]B}_{m+r}-\delta^{B[C}O^{DE]A}_{m+r}\Big)
+m\,O^{ABCDE}_{m+r}
\,,
\nonu\\
\comm{O^{AB}_m}{O^{CDEF}_n}&=&
4\,\mathrm{i}\,\Big(\delta^{A[C}O^{DEF]B}_{m+n}
-\delta^{B[C}O^{DEF]A}_{m+n}\Big)
\,,
\nonu\\
\comm{O^{AB}_m}{O^{CDEFG}_r}&=&-5\,\mathrm{i}\,\Big(\delta^{A[C}O^{DEFG]B}_{m+r}-\delta^{B[C}O^{DEFG]A}_{m+r}\Big)
\,,
\nonu\\
\comm{O^{ABC}_r}{O^{DEF}_s}&=&
3\,\Big(\delta^{A[D}O_{r+s}^{EF]BC}-\delta^{B[D}O_{r+s}^{EF]AC}+\delta^{C[D}O_{r+s}^{EF]AB}\Big)
\,,
\nonu\\
\comm{O^{ABC}_r}{O^{DEFG}_m}&=&
-4\,\mathrm{i}\,\Big(\delta^{A[D}O_{r+m}^{EFG]BC}-\delta^{B[D}O_{r+m}^{EFG]AC}
+\delta^{C[D}O_{r+m}^{EFG]AB}\Big)
\,.
\label{n5sca}
\eea

The ${\cal N}=6$ superconformal algebra where
$A,B,C,D, \cdots =1,2, \cdots, 6$ can be obtained by
\bea
\comm{O_m}{O_n}&=&2(m-n)\,O_{m+n}
\,,
\nonu\\
\comm{O_m}{O^{A}_r}&=&(m-2r)\,O^{A}_{m+r}
\,,
\nonu\\
\comm{O_m}{O^{AB}_n}&=&-2n\,O^{AB}_{m+n}
\,,
\nonu\\
\comm{O_m}{O^{ABC}_r}&=&(-m-2r)\,O^{ABC}_{m+r}
\,,
\nonu\\
\comm{O_m}{O^{ABCD}_n}&=&2(-m-n)\,O^{ABCD}_{m+n}
\,,
\nonu\\
\comm{O_m}{O^{ABCDE}_r}&=&(-3m-2r)\,O^{ABCDE}_{m+r}
\,,
\nonu\\
\comm{O_m}{O^{ABCDEF}_n}&=&2(-2m-n)\,O^{ABCDEF}_{m+n}
\,,
\nonu\\
\comm{O^{A}_r}{O^{B}_s}&=&\delta^{AB}\,O_{r+s}-\mathrm{i}(r-s)\,O^{AB}_{r+s}
\,,
\nonu\\
\comm{O^{A}_r}{O^{BC}_m}&=&\mathrm{i}\,2\,\delta^{A[B}O^{C]}_{r+m}+m\,O^{ABC}_{r+m}
\,,
\nonu\\
\comm{O^{A}_r}{O^{BCD}_s}&=&-3\,\delta^{A[B}O_{r+s}^{CD]}-\mathrm{i}(r+s)\,O^{ABCD}_{r+s}
\,,
\nonu\\
\comm{O^{A}_r}{O^{BCDE}_m}&=&-4\,\mathrm{i}\,\delta^{A[B}O^{CDE]}_{r+m}-(2r+m)\,O^{ABCDE}_{r+m}
\,,
\nonu\\
\comm{O^{A}_r}{O^{BCDEF}_s}&=&5\,\delta^{A[B}O_{r+s}^{CDEF]}+\mathrm{i}(3r+s)\,O^{ABCDEF}_{r+s}
\,,
\nonu\\
\comm{O^{A}_r}{O^{BCDEFG}_m}&=&6\,\mathrm{i}\,\delta^{A[B}O^{CDEFG]}_{r+m}
\,,
\nonu\\
\comm{O^{AB}_m}{O^{CD}_n}&=&-\mathrm{i}\,\delta^{AC}\,O^{BD}_{m+n}+\mathrm{i}\,\delta^{AD}\,O^{BC}_{m+n}+\mathrm{i}
\,\delta^{BC}\,O^{AD}_{m+n}-\mathrm{i}\,\delta^{BD}\,O^{AC}_{m+n}
\,,
\nonu\\
\comm{O^{AB}_m}{O^{CDE}_r}&=&-3\,\mathrm{i}\,\Big(\delta^{A[C}O^{DE]B}_{m+r}-\delta^{B[C}O^{DE]A}_{m+r}\Big)
+m\,O^{ABCDE}_{m+r}
\,,
\nonu\\
\comm{O^{AB}_m}{O^{CDEF}_n}&=&
4\,\mathrm{i}\,\Big(\delta^{A[C}O^{DEF]B}_{m+n}
-\delta^{B[C}O^{DEF]A}_{m+n}\Big)
-2m\,O^{ABCDEF}_{m+n}
\,,
\nonu\\
\comm{O^{AB}_m}{O^{CDEFG}_r}&=&-5\,\mathrm{i}\,\Big(\delta^{A[C}O^{DEFG]B}_{m+r}-\delta^{B[C}O^{DEFG]A}_{m+r}\Big)
\,,
\nonu\\
\comm{O^{AB}_m}{O^{CDEFGH}_n}&=&6\,\mathrm{i}\,\Big(\delta^{A[C}O^{DEFGH]B}_{m+n}-\delta^{B[C}O^{DEFGH]A}_{m+n}\Big)
\,,
\nonu\\
\comm{O^{ABC}_r}{O^{DEF}_s}&=&
3\,\Big(\delta^{A[D}O_{r+s}^{EF]BC}-\delta^{B[D}O_{r+s}^{EF]AC}+\delta^{C[D}O_{r+s}^{EF]AB}\Big)+\mathrm{i}\,(r-s)\,O_{r+s}^{ABCDEF}
\,,
\nonu\\
\comm{O^{ABC}_r}{O^{DEFG}_m}&=&
-4\,\mathrm{i}\,\Big(\delta^{A[D}O_{r+m}^{EFG]BC}-\delta^{B[D}O_{r+m}^{EFG]AC}
+\delta^{C[D}O_{r+m}^{EFG]AB}\Big)
\,,
\nonu\\
\comm{O^{ABC}_r}{O^{DEFGH}_r}&=&
-5\,\Big(\delta^{A[D}O_{r+s}^{EFGH]BC}-\delta^{B[D}O_{r+s}^{EFGH]AC}+\delta^{C[D}O_{r+s}^{EFGH]AB}\Big)
\,,
\nonu\\
\comm{O^{ABCD}_m}{O^{EFGH}_n}&=&
4\,\mathrm{i}\,\Big(\delta^{A[E}O_{m+n}^{FGH]BCD}-\delta^{B[E}O_{m+n}^{FGH]ACD}+\delta^{C[E}O_{m+n}^{FGH]ABD}\nonu \\
&- & \delta^{D[E}O_{m+n}^{FGH]ABC}\Big)
\,.
\label{n6sca}
\eea


The ${\cal N}=7$ superconformal algebra
where $A,B,C,D, \cdots =1,2, \cdots, 7$ can be determined by
\bea
\comm{O_m}{O_n}&=&2(m-n)\,O_{m+n}
\,,
\nonu\\
\comm{O_m}{O^{A}_r}&=&(m-2r)\,O^{A}_{m+r}
\,,
\nonu\\
\comm{O_m}{O^{AB}_n}&=&-2n\,O^{AB}_{m+n}
\,,
\nonu\\
\comm{O_m}{O^{ABC}_r}&=&(-m-2r)\,O^{ABC}_{m+r}
\,,
\nonu\\
\comm{O_m}{O^{ABCD}_n}&=&2(-m-n)\,O^{ABCD}_{m+n}
\,,
\nonu\\
\comm{O_m}{O^{ABCDE}_r}&=&(-3m-2r)\,O^{ABCDE}_{m+r}
\,,
\nonu\\
\comm{O_m}{O^{ABCDEF}_n}&=&2(-2m-n)\,O^{ABCDEF}_{m+n}
\,,
\nonu\\
\comm{O_m}{O^{ABCDEFG}_r}&=&(-5m-2r)\,O^{ABCDEFG}_{m+r}
\,,
\nonu\\
\comm{O^{A}_r}{O^{B}_s}&=&\delta^{AB}\,O_{r+s}-\mathrm{i}(r-s)\,O^{AB}_{r+s}
\,,
\nonu\\
\comm{O^{A}_r}{O^{BC}_m}&=&\mathrm{i}\,2\,\delta^{A[B}O^{C]}_{r+m}+m\,O^{ABC}_{r+m}
\,,
\nonu\\
\comm{O^{A}_r}{O^{BCD}_s}&=&-3\,\delta^{A[B}O_{r+s}^{CD]}-\mathrm{i}(r+s)\,O^{ABCD}_{r+s}
\,,
\nonu\\
\comm{O^{A}_r}{O^{BCDE}_m}&=&-4\,\mathrm{i}\,\delta^{A[B}O^{CDE]}_{r+m}-(2r+m)\,O^{ABCDE}_{r+m}
\,,
\nonu\\
\comm{O^{A}_r}{O^{BCDEF}_s}&=&5\,\delta^{A[B}O_{r+s}^{CDEF]}+\mathrm{i}(3r+s)\,O^{ABCDEF}_{r+s}
\,,
\nonu\\
\comm{O^{A}_r}{O^{BCDEFG}_m}&=&6\,\mathrm{i}\,\delta^{A[B}O^{CDEFG]}_{r+m}+(4r+m)\,O^{ABCDEFG}_{r+m}
\,,
\nonu\\
\comm{O^{A}_r}{O^{BCDEFGH}_s}&=&-7\,\delta^{A[B}O_{r+s}^{CDEFGH]}
\,,
\nonu\\
\comm{O^{AB}_m}{O^{CD}_n}&=&-\mathrm{i}\,\delta^{AC}\,O^{BD}_{m+n}+\mathrm{i}\,\delta^{AD}\,O^{BC}_{m+n}+\mathrm{i}
\,\delta^{BC}\,O^{AD}_{m+n}-\mathrm{i}\,\delta^{BD}\,O^{AC}_{m+n}
\,,
\nonu\\
\comm{O^{AB}_m}{O^{CDE}_r}&=&-3\,\mathrm{i}\,\Big(\delta^{A[C}O^{DE]B}_{m+r}-\delta^{B[C}O^{DE]A}_{m+r}\Big)
+m\,O^{ABCDE}_{m+r}
\,,
\nonu\\
\comm{O^{AB}_m}{O^{CDEF}_n}&=&
4\,\mathrm{i}\,\Big(\delta^{A[C}O^{DEF]B}_{m+n}
-\delta^{B[C}O^{DEF]A}_{m+n}\Big)
-2m\,O^{ABCDEF}_{m+n}
\,,
\nonu\\
\comm{O^{AB}_m}{O^{CDEFG}_r}&=&-5\,\mathrm{i}\,\Big(\delta^{A[C}O^{DEFG]B}_{m+r}-\delta^{B[C}O^{DEFG]A}_{m+r}\Big)
+3m\,O^{ABCDEFG}_{m+r}
\,,
\nonu\\
\comm{O^{AB}_m}{O^{CDEFGH}_n}&=&6\,\mathrm{i}\,\Big(\delta^{A[C}O^{DEFGH]B}_{m+n}-\delta^{B[C}O^{DEFGH]A}_{m+n}\Big)
\,,
\nonu\\
\comm{O^{AB}_m}{O^{CDEFGHI}_r}&=&-7\,\mathrm{i}\,\Big(\delta^{A[C}O^{DEFGHI]B}_{m+r}-\delta^{B[C}O^{DEFGHI]A}_{m+r}\Big)
\,,
\nonu\\
\comm{O^{ABC}_r}{O^{DEF}_s}&=&
3\,\Big(\delta^{A[D}O_{r+s}^{EF]BC}-\delta^{B[D}O_{r+s}^{EF]AC}+\delta^{C[D}O_{r+s}^{EF]AB}\Big)+\mathrm{i}\,(r-s)\,O_{r+s}^{ABCDEF}
\,,
\nonu\\
\comm{O^{ABC}_r}{O^{DEFG}_m}&=&
-4\,\mathrm{i}\,\Big(\delta^{A[D}O_{r+m}^{EFG]BC}-\delta^{B[D}O_{r+m}^{EFG]AC}
+\delta^{C[D}O_{r+m}^{EFG]AB}\Big)
\nonu\\
&&-\mathrm{i}\,(2r-m)\,O_{r+m}^{ABCDEFG}
\,,
\nonu\\
\comm{O^{ABC}_r}{O^{DEFGH}_r}&=&
-5\,\Big(\delta^{A[D}O_{r+s}^{EFGH]BC}-\delta^{B[D}O_{r+s}^{EFGH]AC}+\delta^{C[D}O_{r+s}^{EFGH]AB}\Big)
\,,
\nonu\\
\comm{O^{ABC}_r}{O^{DEFGHI}_m}&=&
6\,\mathrm{i}\,\Big(\delta^{A[D}O_{r+m}^{EFGHI]BC}-\delta^{B[D}O_{r+m}^{EFGHI]AC}+\delta^{C[D}O_{r+m}^{EFGHI]AB}\Big)
\,,
\nonu\\
\comm{O^{ABCD}_m}{O^{EFGH}_n}&=&
4\,\mathrm{i}\,\Big(\delta^{A[E}O_{m+n}^{FGH]BCD}-\delta^{B[E}O_{m+n}^{FGH]ACD}+\delta^{C[E}O_{m+n}^{FGH]ABD}\nonu \\
&- & \delta^{D[E}O_{m+n}^{FGH]ABC}\Big)
\,,
\nonu\\
\comm{O^{ABCD}_m}{O^{EFGHI}_r}&=&
5\,\mathrm{i}\,\Big(\delta^{A[E}O_{m+n}^{FGHI]BCD}-\delta^{B[E}O_{m+n}^{FGHI]ACD}+\delta^{C[E}O_{m+n}^{FGHI]ABD}
\nonu \\
&- & \delta^{D[E}O_{m+n}^{FGHI]ABC}\Big)
\,.
\label{n7sca}
\eea
In (\ref{n5sca}), (\ref{n6sca}), and (\ref{n7sca}),
there are no commutators between the operator with
four $SO({\cal N})$ (with ${\cal N}=5,6,7$) indices and the
operator with four $SO({\cal N})$ (with ${\cal N}=5,6,7$)
indices on the left hand sides 
having the mode dependent terms.
This is because the mode dependent term in (\ref{Ademollo})
depends on the number of $SO({\cal N})$ (with ${\cal N}=5,6,7$)
indices on the left hand side.
This implies that the only nonzero term appears for ${\cal N}=8$ case.
See also Appendices $F,G,H$ and $I$ which are dealing with
the truncations from the $SO(8)$ (or $SU(8)$) supergravity.

We present the algebra where
the generators with
negative conformal spins (denoted by
$\Phi^{(0)}_{ABC,-\frac{1}{2}}$, $\Phi^{(0)}_{AB,-1}$, $\Phi^{(0)}_{A,-\frac{3}{2}}$,
$\Phi^{(1)}_{A,-\frac{3}{2}}$, $\Phi^{(0)}_{-2}$ and $\Phi^{(1)}_{-2}$)
\footnote{
\label{genuineornot}
Let us describe the discussion appeared in the last paragraph of section
$1$ further.
The corresponding conformal dimension $\Delta$
is given by $\frac{5}{2}, 3, \frac{7}{2}, \frac{3}{2}, 4$ and $2$ from
Table  \ref{BigPhiassign}. It is known that the {\it genuine} operators
in the unitary celestial conformal field theory are characterized by
conformal primary operators with the conformal dimension restricted to
the principle continuous series $\Delta = 1 + \mathrm{i} \mathbb{R}$
where $\mathbb{R}$ is a real number \cite{PS}.
By analytic continuation of  principle continuous series, the
conformal soft operators arise at the specific integer (or half integer)
values
$\Delta = 1 - \mathbb{Z}_{\geq 0}=1, 0, -1, -2, \cdots $
for the bosonic operators (gravitons)
and $\Delta = \frac{1}{2} - \mathbb{Z}_{\geq 0}=\frac{1}{2}, -\frac{1}{2},
-\frac{3}{2}, \cdots $ for the fermionic operators (gravitinos)
where $\mathbb{Z}_{\geq 0}$ is a positive integer including zero \cite{DPP1,DPP2}.
For the graviphotons, graviphotinos, and scalars,
the maximum numbers of $\Delta$ are given by
$0, -\frac{1}{2}$ and $-1$ as in the introduction.
Therefore, the above six generators with negative conformal spins
cannot be obtained from the genuine operators (via
an analytic continuation) because the corresponding
conformal dimensions do not belong to these discrete series.
It is known that the {\it formal}
symmetries of the four dimensional $S$-matrix
are described by so-called Coleman-Mandula theorem \cite{CM}
which describes that the most general symmetry group of
four dimensional $S$-matrix under five constraints can be
a direct product of the Poincare group and a compact {\it finite}
dimensional internal symmetry
group. On the other hand,
the celestial holography implies the
{\it infinite}
dimensional BMS symmetries. Moreover, it is an open problem how
the Haag-Lopuszanski-Sohnius theorem \cite{HLS}
which is a supersymmetric
generalization of Coleman-Mandula theorem should be
modified in the celestial holography.
Then,
the above six generators with negative conformal spins
can be interpreted as an elements of the wedge subalgebra
of ${\cal N}=8$ supersymmetric $w_{1+\infty}$
algebra, acting on the scattering states. See also
(\ref{Negative1}) where they appear on the right hand sides of
(\ref{Negative1}).
}
appear on the left hand sides
with $h \geq 2$
\bea
\comm{(\Phi^{(h)}_{+2})_m}{(\Phi^{(0)}_{ABC,-\frac{1}{2}})_r}
    &=&\kappa_{+2,-\frac{1}{2},+\frac{1}{2}}\,\,\Big(-\tfrac{3}{2}m-(h+1)r\Big)\,(\Phi^{(h)}_{ABC,-\frac{1}{2}})_{m+r}\,\,:\text{eq.6}
\, ,
\nonu   \\
\acomm{(\Phi^{(h),A}_{+\frac{3}{2}})_r}{(\Phi^{(0)}_{BCD,-\frac{1}{2}})_s}
&
=& \kappa_{+\frac{3}{2},-\frac{1}{2},+1}\,
\Big(-\tfrac{3}{2}r-(h+\tfrac{1}{2})s\Big)\,3
\delta^{A}_{\,\,\,[B}(\Phi^{(h)}_{CD],-1})_{r+s}\,\,:\text{eq.14}\,,
\nonu 
\\
\comm{(\Phi^{(h),AB}_{+1})_m}{(\Phi^{(0)}_{CDE,-\frac{1}{2}})_r}&=&
\kappa_{+1,-\frac{1}{2},+\frac{3}{2}}\,\Big(-\tfrac{3}{2}m-h\,r\Big)\,
3! \delta^{A}_{\,\,\,[C}(\Phi^{(h)}_{D,-\frac{3}{2}})_{m+r}\delta_{E]}^{\,\,\,B}
\,\,:\text{eq.20}\, ,
\nonu \\
\acomm{(\Phi^{(h),ABC}_{+\frac{1}{2}})_r}{(\Phi^{(0)}_{DEF,-\frac{1}{2}})_s}
&=&\kappa_{+\frac{1}{2},-\frac{1}{2},+2}\,\Big(-\tfrac{3}{2}r-(h-\tfrac{1}{2})s\Big)\,
\delta^{ABC}_{DEF}\,(\Phi^{(h)}_{-2})_{r+s}
\,\,:\text{eq.24}\,,
\nonu 
\\
\comm{(\Phi^{(h)}_{+2})_m}{(\Phi^{(0)}_{AB,-1})_n}
&=&\kappa_{+2,-1,+1}\,\Big(-2m-(h+1)n\Big)\,(\Phi^{(h)}_{AB,-1})_{m+n}\,\,:\text{eq.7}
\, ,
\nonu \\
\comm{(\Phi^{(h),A}_{+\frac{3}{2}})_r}{(\Phi^{(0)}_{BC,-1})_m}
&=&
\kappa_{+\frac{3}{2},-1,+\frac{3}{2}}\,\Big(-2r-(h+\tfrac{1}{2})m\Big)\,
2!\, \delta^{A}_{\,\,\,[B}\,
(\Phi^{(h)}_{C],-\frac{3}{2}})_{r+m}\,\,:\text{eq.15}\, ,
\nonu \\
\comm{(\Phi^{(h),AB}_{+1})_m}{(\Phi^{(0)}_{CD,-1})_n}&=&
\kappa_{+1,-1,+2}\,\Big(-2m-h\,n\Big)\,
\delta^{AB}_{CD} \, (\Phi^{(h)}_{-2})_{m+n}\,\,:\text{eq.21}
\,,
\nonu    \\
\comm{(\Phi^{(h)}_{+2})_m}{(\Phi^{(0)}_{A,-\frac{3}{2}})_r}
&=&
\kappa_{+2,-\frac{3}{2},+\frac{3}{2}}\,\Big(-\tfrac{5}{2}m-(h+1)r\Big)\,(\Phi^{(h)}_{A,-\frac{3}{2}})_{m+r}\,\,:\text{eq.8}
\, ,
\nonu \\
\acomm{(\Phi^{(h),A}_{+\frac{3}{2}})_r}{(\Phi^{(0)}_{B,-\frac{3}{2}})_s}
&=&
\kappa_{+\frac{3}{2},-\frac{3}{2},+2}\,\Big(-\tfrac{5}{2}r-(h+\tfrac{1}{2})s\Big)\,\delta^{A}_{\,\,\,B}\,(\Phi^{(h)}_{-2})_{r+s}\,\,:\text{eq.16}\,,
\nonu    \\
\comm{(\Phi^{(h)}_{+2})_m}{(\Phi^{(0)}_{-2})_n}&=&
\kappa_{+2,-2,+2}\,\Big(-3m-(h+1)n\Big)\,(\Phi^{(h)}_{-2})_{m+n}\,\,:\text{eq.9}
\,  ,
\nonu    \\
\comm{(\Phi^{(h)}_{+2})_m}{(\Phi^{(1)}_{A,-\frac{3}{2}})_r}
&=&
\kappa_{+2,-\frac{3}{2},+\frac{3}{2}}\,\Big(-\tfrac{3}{2}m-(h+1)r\Big)\,(\Phi^{(h+1)}_{A,-\frac{3}{2}})_{m+r}\,\,:\text{eq.8}
\, ,
\nonu \\
\acomm{(\Phi^{(h),A}_{+\frac{3}{2}})_r}{(\Phi^{(1)}_{B,-\frac{3}{2}})_s}
&=&
\kappa_{+\frac{3}{2},-\frac{3}{2},+2}\,\Big(-\tfrac{3}{2}r-(h+\tfrac{1}{2})s\Big)\,\delta^{A}_{\,\,\,B}\,(\Phi^{(h+1)}_{-2})_{r+s}\,\,:\text{eq.16}\,,
\nonu    \\
\comm{(\Phi^{(h)}_{+2})_m}{(\Phi^{(1)}_{-2})_n}&=&
\kappa_{+2,-2,+2}\,\Big(-2m-(h+1)n\Big)\,(\Phi^{(h+1)}_{-2})_{m+n}\,\,:\text{eq.9}
\,  .
\label{Negative} 
\eea
On the right hand sides of (\ref{Negative}),
there are no negative spins due to the condition $h \geq 2$.
We have checked that there are no nonzero (anti)commutators
between the generators having
negative spins.
Moreover, we observe the generators with the negative spins (the sum of
the superscript and the subscript)
appearing on the right hand sides as follows:
\bea
\comm{(\Phi^{(0),A}_{+\frac{3}{2}})_r}{(\Phi^{(0),BCDE}_0)_m}&=&
\kappa_{+\frac{3}{2},0,+\frac{1}{2}}\,\Big(-r-\tfrac{1}{2}m\Big)\,\frac{1}{3!}\epsilon^{ABCDEFGH} (\Phi^{(0)}_{FGH,-\frac{1}{2}})_{r+m}
\, :  \text{eq.13},
\nonu \\
\comm{(\Phi^{(0),AB}_{+1})_m}{(\Phi^{(0),CDE}_{+\frac{1}{2}})_r}&=&
\kappa_{+1,+\frac{1}{2},+\frac{1}{2}}\,\Big(-\tfrac{1}{2}m\Big)\,\frac{1}{3!}\,\epsilon^{ABCDEFGH}(\Phi^{(0)}_{FGH,-\frac{1}{2}})_{m+r}
\,\, :  \text{eq.18},
\nonu \\
\comm{(\Phi^{(0),AB}_{+1})_m}{(\Phi^{(0),CDEF}_0)_n}&=&
\kappa_{+1,0,+1}\,(-m)\,\epsilon^{ABCDEFGH}\,\frac{1}{2!}(\Phi^{(0)}_{GH,-1})_{m+n}
\,\,:  \text{eq.19},
\nonu \\
\acomm{(\Phi^{(0),ABC}_{+\frac{1}{2}})_r}{(\Phi^{(0),DEF}_{+\frac{1}{2}})_s}&=&
\kappa_{+\frac{1}{2},+\frac{1}{2},+1}\,\frac{1}{2}(-r+s)\,\frac{1}{2!}\,\epsilon^{ABCDEFGH}
 (\Phi^{(0)}_{GH,-1})_{r+s}
\, \,:  \text{eq.22},
\nonu \\
\comm{(\Phi^{(0),ABC}_{+\frac{1}{2}})_r}{(\Phi^{(0),DEFG}_0)_m}&=&
\kappa_{+\frac{1}{2},0,+\frac{3}{2}}\,\Big(-r+\tfrac{1}{2}m\Big)\,\epsilon^{ABCDEFGH}
 (\Phi^{(0)}_{H,-\frac{3}{2}})_{r+m}
\,\, :  \text{eq.23},
\nonu \\
\comm{(\Phi^{(0),ABC}_{+\frac{1}{2}})_r}{(\Phi^{(1),DEFG}_0)_m}&=&
\kappa_{+\frac{1}{2},0,+\frac{3}{2}}\,\Big(+\tfrac{1}{2}m\Big)\,\epsilon^{ABCDEFGH}
(\Phi^{(1)}_{H,-\frac{3}{2}})_{r+m}
\,\, :  \text{eq.23},
\nonu \\
\comm{(\Phi^{(1),ABC}_{+\frac{1}{2}})_r}{(\Phi^{(0),DEFG}_0)_m}&=&
\kappa_{+\frac{1}{2},0,+\frac{3}{2}}\,\Big(-r-\tfrac{1}{2}m\Big)\,\epsilon^{ABCDEFGH}
 (\Phi^{(1)}_{H,-\frac{3}{2}})_{r+m}
\,\, :  \text{eq.23},
\nonu \\
\comm{(\Phi^{(0),ABCD}_0)_m}{(\Phi^{(0),EFGH}_0)_n}&=&
\kappa_{0,0,+2}\,\Big(-m+n\Big)\,\epsilon^{ABCDEFGH}\,(\Phi^{(0)}_{-2})_{m+n}
\,\,:  \text{eq.25}\,,
\nonu \\
\comm{(\Phi^{(0),ABCD}_0)_m}{(\Phi^{(1),EFGH}_0)_n}&=&
\kappa_{0,0,+2}\,(n)\,\epsilon^{ABCDEFGH}\,(\Phi^{(1)}_{-2})_{m+n}
\,\,:  \text{eq.25}\,,
\nonu \\
\comm{(\Phi^{(1),ABCD}_0)_m}{(\Phi^{(0),EFGH}_0)_n}&=&
\kappa_{0,0,+2}\,(-m)\,\epsilon^{ABCDEFGH}\,(\Phi^{(1)}_{-2})_{m+n}
\,\,:  \text{eq.25}\,.
\label{Negative1}
\eea
On the left hand sides of (\ref{Negative1}), 
the conformal spins are nonnegative and are denoted by
$\Phi_{0}^{(0),ABCD}$, $\Phi_{+\frac{1}{2}}^{(0),ABC}$, $\Phi_{+1}^{(0),AB}$,
$\Phi_{+\frac{3}{2}}^{(0),A}$, $\Phi_{0}^{(1),ABCD}$ and $
\Phi_{+\frac{1}{2}}^{(1),ABC}$.
On the right hand sides of (\ref{Negative1})
the previous negative spins
in (\ref{Negative}) occur.
We have checked that
the following operators \footnote{The missing operators in the
whole multiplets are given by
$
\Phi_{+\frac{3}{2}}^{(0),A}$,
$\Phi_{+1}^{(0),AB}$,
$\Phi_{+\frac{1}{2}}^{(0),ABC}$,
$\Phi_{0}^{(0),ABCD}$,
$\Phi^{(0)}_{ABC, -\frac{1}{2}}$,
$\Phi^{(0)}_{AB,-1}$,
$\Phi^{(0)}_{A,-\frac{3}{2}}$,
$\Phi^{(1)}_{A,-\frac{3}{2}}$,
$\Phi_{-2}^{(0)}$ and $\Phi_{-2}^{(1)}$. Here the last six operators
have negative spins.}
satisfy the (celestial soft symmetry) algebra
where there are no operators with negative spins
\bea
&& \Phi_{+2}^{(h)}, \quad h=0,1,2,\cdots, \quad
\Phi_{+\frac{3}{2}}^{(h),A},  \quad h=1,2,3, \cdots, \quad
\Phi_{+1}^{(h),AB},  \quad h=1,2,3, \cdots, 
\nonu \\
&& \Phi_{+\frac{1}{2}}^{(h),ABC},  \quad h=1,2,3, \cdots, \quad
\Phi_{0}^{(h),ABCD},  \quad h=1,2,3, \cdots, \quad
\Phi^{(h)}_{ABC, -\frac{1}{2}},   \quad h=1,2,3, \cdots, 
\nonu \\
&& \Phi^{(h)}_{AB,-1},  \quad h=1,2,3, \cdots, \quad
\Phi^{(h)}_{A,-\frac{3}{2}},  \quad h=2, 3, 4, \cdots, \quad
\Phi_{-2}^{(h)},  \quad h=2, 3, 4, \cdots.
\label{fieldcontents}
\eea
There exist the critical (or lowest) values for the superscripts $h$
of the operators in (\ref{fieldcontents}).
The previous six negative spin generators in (\ref{Negative})
can be seen from the last four operators in (\ref{fieldcontents})
by substituting the values of $h$ below the critical values. 
Note that
the two operators  $\Phi_{0}^{(1),ABCD}$ and $
\Phi_{+\frac{1}{2}}^{(1),ABC}$ appear in (\ref{Negative1}) and
(\ref{fieldcontents}) but the
other operators on the left hand sides
in the corresponding four (anti)commutators in (\ref{Negative1})
do not belong to
(\ref{fieldcontents}).

\section{The split factors and the amplitudes in the
${\cal N}=8$ supergravity}

In this Appendix, we review the works of \cite{2212-1,2212-2}.
See also the split factors for the gluons in \cite{MP,Dixon,BDPR}.

The split factor in the ${\cal N}=8$ supergravity theory
appears in the scattering amplitude when we consider
the collinear (the two particles' momenta are nearly parallel)
limit for the cubic vertex involving
the massless particles.
This split factor
in the ${\cal N}=8$ supergravity \cite{BDPR}
can be written in terms of the product of two split factors in the
two
${\cal N}=4$ super Yang-Mills theories where
the nontrivial split factors for the collinear gluons
with helicities $\pm 1$, gluinos with helicities $\pm \frac{1}{2}$ and
scalars with helicity $0$
are given by one of the complex coordinates in the celestial
sphere, the angle bracket and the square bracket of two collinear
particles \footnote{
\label{productofsplit}
  We present the mathematica program for the
split factors in the ${\cal N}=8$ supergravity.
\bea
&& \tt Clear[MM]\nonu \\
&& \tt MM=\{\{-1,+1,+1,\frac{1}{\sqrt{z (1-z)} \text{bk}(a,b)}\},\{+1,-1,+1,\frac{\sqrt{\frac{z^3}{1-z}}}{\text{bk}(a,b)}\},
\{+1,+1,-1,\frac{(1-z)^2}{\sqrt{z (1-z)} \text{bk}(a,b)}\},\nonu \\
&& \tt
\{0,+\frac{1}{2},+\frac{1}{2},\frac{1}{\text{bk}(a,b)}\},
\{+1,+\frac{1}{2},-\frac{1}{2},\frac{1-z}{\text{bk}(a,b)}\},
\{+1,-\frac{1}{2},+\frac{1}{2},\frac{z}{\text{bk}(a,b)}\},
\{+1,0,0,\frac{\sqrt{z (1-z)}}{\text{bk}(a,b)}\},\nonu \\
&& \tt
\{+\frac{1}{2},-\frac{1}{2},+1,\frac{z}{\sqrt{1-z} \text{bk}(a,b)}
\},\{+\frac{1}{2},+1,-\frac{1}{2},\frac{1-z}{\sqrt{z} \text{bk}(a,b)}
\},\{-\frac{1}{2},+\frac{1}{2},+1,\frac{1}{\sqrt{1-z} \text{bk}(a,b)}
\},\nonu \\
&& \tt
\{-\frac{1}{2},+1,+\frac{1}{2},\frac{1}{\sqrt{z} \text{bk}(a,b)}
\},\{0,0,+1,\frac{\sqrt{\frac{z}{1-z}}}{\text{bk}(a,b)}\},
\{0,+1,0,\frac{\sqrt{\frac{1-z}{z}}}{\text{bk}(a,b)}\},
\{\frac{1}{2},0,+\frac{1}{2},\frac{\sqrt{z}}{\text{bk}(a,b)}\},
\nonu \\
&& \tt
\{\frac{1}{2},+\frac{1}{2},0,\frac{\sqrt{1-z}}{\text{bk}(a,b)}
\}\};
\nonu \\
&& \tt Length[MM] \nonu \\
&& \tt Clear[sfa] \nonu \\
&& \tt sfa[1,2]=bk[1,2]Sqbk[2,1];\nonu \\
&& \tt bk[2,1]=-bk[1,2]; \nonu \\
&& \tt Sqbk[2,1]=-Sqbk[1,2]; \nonu \\
&& \tt Clear[Trz1,Trz2] \nonu \\
&& \tt
\text{Trz1}[\text{A$\_$}]\text{:=}A\text{/.}\, \{a\to 1,b\to 2\}
\nonu \\
&& \tt
\text{Trz2}[\text{A$\_$}]\text{:=}A\text{/.}\,
\{z\to 1-z,a\to 2,b\to 1\}
\nonu  \\
&& \tt
\text{Do}[\text{Print}[
\text{SPlit}_{-\{-\text{MM}[[\text{i1},1]]-\text{MM}[[\text{i2},1]]\}}
[z,\{\text{MM}[[\text{i1},2]],\text{MM}[[\text{i2},3]]\},
\{\text{MM}[[\text{i1},3]],\text{MM}[[\text{i2},2]]\},
\nonu \\
&& \tt
-\text{sfa}[1,2] \text{FullSimplify}[\text{PowerExpand}[\text{Trz1}
[\text{MM}[[\text{i1},4]]] \text{Trz2}[\text{MM}[[\text{i2},4]]]]]]
],\{\text{i1},15\},\{\text{i2},15\}]
\nonu 
\eea
Here $\tt \text{MM}$
are the split factors between the gluons, the gluinos and
the scalars in the ${\cal N}=4$ super Yang-Mills theory.
The notation $\tt \text{bk}(a,b)$
stands for $\langle a,b \rangle$ and the notation
$\tt Sqbk[a,b]$
stands for $ [ a, b]$. Among $15^2=225$ split factors, the
$25$ independent ones are presented in Tables \ref{Split1},
\ref{Split2}, \ref{Split3},
\ref{Split4} and \ref{Split5}.}.
This implies that  the split factor
in the ${\cal N}=8$ supergravity depends on the two types of
these three quantities.
More precisely,  the split factor
in the ${\cal N}=8$ supergravity   depends on
three helicities from the first split factor
in the first ${\cal N}=4$ super Yang-Mills theory and other three
helicities from the second
split factor
in the second ${\cal N}=4$ super Yang-Mills theory.
See also the second column of Table \ref{Split1}.

As in section $5$,
after parametrizing each momentum of the two collinear
particles in terms of its energy with its null vector, 
the sum of these momenta can be expressed in terms of
the sum of these energies multiplied by the above first (or second)
null vector.
Then one of the complex coordinates in the celestial two sphere
can be  given by the ratio of the energy of first particle
and the sum the energies of the two particles.
Moreover, the above angle and square brackets
can be written in terms of the two energies and
the two complex coordinates in the the celestial sphere
in the spinor helicity formalism.
Therefore, the split factor in the ${\cal N}=8$
supergravity theory can be written in terms of
\begin{itemize}
\item[]  
i) the two energies,

ii) the sum of these two energies and

iii) the two complex  coordinates (in the celestial sphere) for each
collinear particle appearing in the
two null vectors.
\end{itemize}
See also the third column of Table \ref{Split1}. 

As in section $5$,
the power of the sum of two energies plays important role of
the helicity of the collinear channel momentum inside of
$(n-1)$ point amplitude and encodes the opposite of the
subscript of the split factor
of the ${\cal N}=8$ supergravity.
See also the fourth column of Table \ref{Split1}.
It is known that the celestial amplitude can be obtained
from the conventional momentum space amplitude by so called
Mellin transform.
The $n$ point celestial correlator
can be written in terms of the complex coordinate
dependence appearing in the third column of
Table \ref{Split1}, the Euler beta function, and the $(n-1)$
point correlator. Then the OPE corresponding to
the two operators having the specific helicities
can be obtained and the power of the first energy
and the power of the second energy appear
in the first argument and second argument of the Euler beta
function respectively. 

We summarize the
the split factors,
the amplitude factors,
the power of collinear momenta,
and  the Euler beta functions
in Table \ref{Split1}.

The factorizations of the $SU(8)_R$ $R$ symmetry indices
and the corresponding helicities
$(-2,-\frac{3}{2},-1,-\frac{1}{2},0,+\frac{1}{2},+1,+\frac{3}{2},+2)$
from the ones in the two ${\cal N}=4$ super Yang-Mills theories
are given as
follows \cite{2212-1}:
\bea
{\bf zero}:(-2) &=& (-1) \otimes (-1) \, , 
\nonu \\
{\bf -one}:(a;-\tfrac{3}{2}) &=& (a;-\tfrac{1}{2}) \otimes (-1)\, ,
\qquad
{\bf one}:(r;-\tfrac{3}{2}) = (-1) \otimes (r;-\tfrac{1}{2}) \, , 
\nonu \\
{\bf -two}:(ab;-1) &= & (ab;0) \otimes (-1) \, ,
\qquad
{\bf zero}:(ar;-1) = (a;-\tfrac{1}{2}) \otimes (r;-\tfrac{1}{2})\, ,
\nonu \\
{\bf two}:(rs;-1) &=&  (-1) \otimes (rs;0) \, ,
\nonu
\\
{\bf -one}:(abr;-\tfrac{1}{2}) &=& (ab;0) \otimes (r;-\tfrac{1}{2})\, ,
\qquad
{\bf one}:(ars;-\tfrac{1}{2}) = (a;-\tfrac{1}{2}) \otimes (rs;0) \, ,
\nonu \\
{\bf three}:(rst;-\tfrac{1}{2}) &=&    
-\epsilon_{rstu} ( (-1) \otimes (u;+\tfrac{1}{2}))\, ,
{\bf -three}:(abc;-\tfrac{1}{2})
= -\epsilon_{abcd}((d ;+\tfrac{1}{2}) \otimes (-1)) \, ,
\nonu \\
{\bf zero}:(abrs;0) &=& (ab;0) \otimes (rs;0)\, ,
\qquad
{\bf four}:(abcd;0) = -\epsilon^{abcd}((-1) \otimes (+1)) \, ,
\nonu \\
{\bf -four}:(rstu;0)  &=&  -\epsilon^{rstu}((+1) \otimes (-1))\, ,
\nonu \\
{\bf two}:(abcr;0) & = & -\epsilon^{abcd}((d;-\tfrac{1}{2}) \otimes
(r;+\tfrac{1}{2})) \, ,
\nonu \\
{\bf -two}:(arst;0) & = &
-\epsilon^{rstu}((a;+\tfrac{1}{2}) \otimes (u;-\tfrac{1}{2}))\, ,
\nonu \\
{\bf one}:(abr;+\tfrac{1}{2}) &=&
(ab;0) \otimes (r;+\tfrac{1}{2}) \, ,
\qquad
{\bf -one}:(ars;+\tfrac{1}{2})  = 
(a;+\tfrac{1}{2}) \otimes (rs;0)\, ,
\nonu \\
{\bf -three}:(rst;+\tfrac{1}{2}) &=& -
\epsilon^{rstu}((+1) \otimes(u;-\tfrac{1}{2})) \, ,
{\bf three}:(abc;+\tfrac{1}{2}) = -\epsilon^{abcd}
((d;-\tfrac{1}{2}) \otimes (+1)) \, ,
\nonu \\
{\bf two}:(ab;+1) &=& (ab;0) \otimes (+1) \, ,
\qquad
{\bf zero}:(ar;+1) =
(a;+\tfrac{1}{2}) \otimes (r;+\tfrac{1}{2}) \, ,
\nonu \\
{\bf -two}:(rs;+1) & = & (+1) \otimes(rs;0)\, ,
\nonu \\
{\bf one}:(a;+\tfrac{3}{2}) &=& (a;+\tfrac{1}{2}) \otimes (+1) \, ,
\qquad
{\bf -one}:(r;+\tfrac{3}{2}) = (+1) \otimes(r;+\tfrac{1}{2})\, ,
\nonu \\
{\bf zero}:(+2) &=& (+1) \otimes (+1)\, .
\label{abcdrstu}
\eea
We are using the upper indices for the positive helicities
and the lower indices for the negative helicities
\footnote{In (\ref{abcdrstu}), we denote the $U(1)$ charges
(in the breaking of $SU(8) \rightarrow SU(4) \times SU(4)
\times U(1)$ and they appear in (\ref{su4branching}))
in front of the states of the ${\cal N}=8$ supergravity
in the boldface fonts.
In other words, for the positive helicities each index for
the first $SU(4)$ has $+1$ $U(1)$ charge and the
each index for the second $SU(4)$ has $-1$ $U(1)$ charge.
That is, the $U(1)$ charge for the states in the ${\cal N}=8$
supergravity having the positive helicity is
the number of first $SU(4)$ indices minus
the number of second $SU(4)$ indices.
For the states for
negative helicities, we simply multiply an extra minus sign
into the formula for the states of positive helicities. }.
The repeated indices on the right hand sides of (\ref{abcdrstu})
are summed. The indices $a,b,c,d, \cdots$ appear in the first
factor while the indices $r,s,t,u, \cdots$ appear in the second factor
on the right hand sides of (\ref{abcdrstu}).
Due to the minus signs in the self dual conditions
for the scalars in the two ${\cal N}=4$ super Yang-Mills theory,
there are additional minus signs in front of Levi Civitas on the
right hand sides of (\ref{abcdrstu}).
These Levi Civitas appear for the $SU(4)$ indices more than three
and four of them come from the graviphotinos and four of them
come from the scalars. Once we choose the two types of helicities
$(-1,-\frac{1}{2},0,+\frac{1}{2},+1)$
from the two ${\cal N}=4$ super Yang-Mills theories,
the corresponding
$SU(8)_R$ $R$ symmetry indices follow from the right hand sides of
(\ref{abcdrstu}).
The indices for the
scalars in the ${\cal N}=4$ super Yang-Mills theory
can be raised or lowered by using the Levi Civita.
The two types of $SU(4)$ indices in the
${\cal N}=4$ super Yang-Mills theories appear in terms of
the Levi Civita and the Kronecker delta for the
collinear limits between the gluons, the gluinos and the scalars.
Then the $n$ point amplitude for the ${\cal N}=8$ supergravity
contains these two types of $SU(4)$ indices in addition to
the various split factors in Tables \ref{Split1}, \ref{Split2},
\ref{Split3}, \ref{Split4} and \ref{Split5}
multiplied by $(n-1)$ point amplitude.

For the other split factors
in (\ref{eq:graviton_gravitino}),
there are relations between the two remaining split factors
(the second and the fourth of (\ref{eq:graviton_gravitino}))
\bea
\text{Split}_{-\frac{3}{2}}^{\text{SG}}(z,1^{+1+1},2^{+1+\frac{1}{2}})
& = &
\text{Split}_{-\frac{3}{2}}^{\text{SG}}(z,1^{+1+1},2^{+\frac{1}{2}+1})\,,
\nonu \\
\text{Split}_{+\frac{3}{2}}^{\text{SG}}(z,1^{+1+1},2^{-1-\frac{1}{2}})& = &
\text{Split}_{+\frac{3}{2}}^{\text{SG}}(z,1^{+1+1},2^{-\frac{1}{2}-1})\,,
\nonu 
\eea
in terms of previous ones (corresponding to the first and the third
of (\ref{eq:graviton_gravitino}) respectively)
appearing in Table \ref{Split1}.
Similarly, 
for the case of (\ref{eq:graviton_photon}), there exist 
the following relations between the remaining split factors
\bea
\text{Split}_{-1}^{\text{SG}}(z,1^{+1+1},2^{+\frac{1}{2}+\frac{1}{2}})
&=&
\text{Split}_{-1}^{\text{SG}}(z,1^{+1+1},2^{+1+0})=
\text{Split}_{-1}^{\text{SG}}(z,1^{+1+1},2^{0+1})\,,
\nonu \\
\text{Split}_{+1}^{\text{SG}}(z,1^{+1+1},2^{-\frac{1}{2}-\frac{1}{2}})&=&
\text{Split}_{+1}^{\text{SG}}(z,1^{+1+1},2^{-1+0})
=
\text{Split}_{+1}^{\text{SG}}(z,1^{+1+1},2^{0-1})\,,
\nonu
\eea
in terms of
the ones appearing in Table \ref{Split1}.
That is, the split factors
for the second and the third of (\ref{eq:graviton_photon})
are the same as the one for the first of (\ref{eq:graviton_photon})
and the split factors
for the fifth and the sixth of (\ref{eq:graviton_photon})
are the same as the one for 
the fourth of (\ref{eq:graviton_photon}).
For the (\ref{eq:graviton_photino}), the following relations
for the remaining split factors hold
\bea
\text{Split}_{-\frac{1}{2}}^{\text{SG}}(z,1^{+1+1},2^{0+\frac{1}{2}})&=&
\text{Split}_{-\frac{1}{2}}^{\text{SG}}(z,1^{+1+1},2^{+\frac{1}{2}+0})=
\text{Split}_{-\frac{1}{2}}^{\text{SG}}(z,1^{+1+1},2^{+1-\frac{1}{2}})
\nonu \\
&=&
\text{Split}_{-\frac{1}{2}}^{\text{SG}}(z,1^{+1+1},2^{-\frac{1}{2}+1})\, ,
\nonu \\
\text{Split}_{+\frac{1}{2}}^{\text{SG}}(z,1^{+1+1},2^{0-\frac{1}{2}})&=&
\text{Split}_{+\frac{1}{2}}^{\text{SG}}(z,1^{+1+1},2^{-\frac{1}{2}+0})=
\text{Split}_{+\frac{1}{2}}^{\text{SG}}(z,1^{+1+1},2^{-1+\frac{1}{2}})
\nonu \\
&= &
\text{Split}_{+\frac{1}{2}}^{\text{SG}}(z,1^{+1+1},2^{+\frac{1}{2}-1}) \, ,
\nonu
\eea
in terms of
the ones appearing in Table \ref{Split1}.
The split factors for the second, the third and the fourth
of (\ref{eq:graviton_photino})
are the same  as the one in the first
while
they for the sixth, the seventh and the eighth
of (\ref{eq:graviton_photino})
are the same  as the one in the fifth.
Similarly, 
for (\ref{eq:graviton_scalar}), the remaining four split factors
can be written in terms of previous one in Table \ref{Split1}
as follows:
\bea
\text{Split}_{0}^{\text{SG}}(z,1^{+1+1},2^{-\frac{1}{2}+\frac{1}{2}})
& = &
\text{Split}_{0}^{\text{SG}}(z,1^{+1+1},2^{0+0})=
\text{Split}_{0}^{\text{SG}}(z,1^{+1+1},2^{+\frac{1}{2}-\frac{1}{2}})
\nonu \\
&=&
\text{Split}_{0}^{\text{SG}}(z,1^{+1+1},2^{+1-1})
=  \text{Split}_{0}^{\text{SG}}(z,1^{+1+1},2^{-1+1}) \, .
\nonu
\eea
The split factors corresponding to the second-the fifth
of  (\ref{eq:graviton_scalar})
are the same as the one for the first.

\begin{table}[tbp]
\centering
\renewcommand{\arraystretch}{1.7}
\begin{tabular}{|c|c|c|c|c| }
\hline
Eqs. & Split factors & Amp.  & Power  & Beta functions 
\\
\hline
\hline
(\ref{eq:graviton_graviton})   &
$\text{Split}_{-2}^{\text{SG}}(z,1^{+1+1},2^{+1+1})
=
-\frac{1}{z(1-z)} \frac{[12]}{ \langle 12\rangle}
$
& $\frac{\omega_p^2}{\omega_1 \omega_2}\frac{\bar{z}_{12}}{z_{12}} $
& $p^{+2} $
& $ B(\Delta_1-1,\Delta_2-1) $
\\
& $\text{Split}_{+2}^{\text{SG}}(z,1^{+1+1},2^{-1-1})=
-\frac{(1-z)^3}{z} \frac{[12]}{ \langle 12\rangle}$
&  
$ \frac{\omega_2^3}{\omega_1 \omega_p^2}\frac{\bar{z}_{12}}{z_{12}}$
&
$ p^{-2}$
&
$B(\Delta_1-1,\Delta_2+3)$
\\
\hline
(\ref{eq:graviton_gravitino}) & $\text{Split}_{-\frac{3}{2}}^{\text{SG}}(z,1^{+1+1},2^{+\frac{1}{2}+1})=
-\frac{1}{z\sqrt{1-z}} \frac{[12]}{ \langle 12\rangle}$
& 
$\frac{\omega_p^\frac{3}{2}}{\omega_1 \omega_2^\frac{1}{2}}\frac{\bar{z}_{12}}{z_{12}}$
&
$ p^{+\frac{3}{2}}$ 
&
$ B(\Delta_1-1,\Delta_2-\frac{1}{2})$
\\
& $\text{Split}_{+\frac{3}{2}}^{\text{SG}}(z,1^{+1+1},2^{-\frac{1}{2}-1})=
-\frac{(1-z)^{\frac{5}{2}}}{z} \frac{[12]}{ \langle 12\rangle}$
&  
$  \frac{\omega_2^\frac{5}{2}}{\omega_1 \omega_p^\frac{3}{2}}\frac{\bar{z}_{12}}{z_{12}}$
&
$  p^{-\frac{3}{2}}$
&
$ B(\Delta_1-1,\Delta_2+\frac{5}{2})$
\\
\hline
(\ref{eq:graviton_photon}) &
$ \text{Split}_{-1}^{\text{SG}}(z,1^{+1+1},2^{0+1})=
-\frac{1}{z} \frac{[12]}{ \langle 12\rangle}$
&  
$  \frac{\omega_p}{\omega_1 }\frac{\bar{z}_{12}}{z_{12}}$
& 
$ p^{+1}$
&
$ B(\Delta_1-1,\Delta_2)$
\\
&$\text{Split}_{+1}^{\text{SG}}(z,1^{+1+1},2^{0-1})=
-\frac{(1-z)^2}{z} \frac{[12]}{ \langle 12\rangle}$
&
$ \frac{\omega_2^2}{\omega_1\omega_p }\frac{\bar{z}_{12}}{z_{12}}$
&
$ p^{-1}$
&
$  B(\Delta_1-1,\Delta_2+2)$
\\
\hline
(\ref{eq:graviton_photino}) &
$\text{Split}_{-\frac{1}{2}}^{\text{SG}}(z,1^{+1+1},2^{-\frac{1}{2}+1})=
-\frac{\sqrt{1-z}}{z} \frac{[12]}{ \langle 12\rangle}$
&
$ \frac{\omega_2^{\frac{1}{2}}\omega_p^{\frac{1}{2}}}{\omega_1 }\frac{\bar{z}_{12}}{z_{12}}$
& 
$  p^{+\frac{1}{2}}$
&
$B(\Delta_1-1,\Delta_2+\frac{1}{2})$
\\
&
$\text{Split}_{+\frac{1}{2}}^{\text{SG}}(z,1^{+1+1},2^{+\frac{1}{2}-1})=
-\frac{(1-z)^{\frac{3}{2}}}{z} \frac{[12]}{ \langle 12\rangle}$
& 
$  \frac{\omega_2^{\frac{3}{2}}}{\omega_1 \omega_p^{\frac{1}{2}}}\frac{\bar{z}_{12}}{z_{12}}$
&
$ p^{-\frac{1}{2}}$
&
$  B(\Delta_1-1,\Delta_2+\frac{3}{2})$
\\
\hline
(\ref{eq:graviton_scalar}) &
$\text{Split}_{0}^{\text{SG}}(z,1^{+1+1},2^{-1+1})=
-\frac{(1-z)}{z} \frac{[12]}{ \langle 12\rangle}$
&
$ \frac{\omega_2}{\omega_1 }\frac{\bar{z}_{12}}{z_{12}}$
&
$ p^{0}$
&
$ B(\Delta_1-1,\Delta_2+1) $
\\
\hline
\end{tabular}
\caption{We list the split factors for the ${\cal N}=8$
supergravity, the amplitude (Amp.) factors
which are the split factors  for the collinear limits,
the power of collinear momenta $p$
which is the opposite of the subscript of
the split factors and the Euler beta functions $B(\Delta_1+1-s_1,
\Delta_2+1-s_2)$ appearing in the OPEs between the operators of the
particles.}
\label{Split1}
\end{table}

Let us consider the next split factors
appearing in Table \ref{Split2}.
For other split factors,  
the split factors for the second and the third
of
(\ref{eq:gravitino_gravitino})
are the same as the one of the first and
the split factor for the fifth of
(\ref{eq:gravitino_gravitino})
is the same as the one for the fourth
\footnote{
We also have
$
\text{Split}_{-1}^{\text{SG}}(z,1^{+1+\frac{1}{2}},2^{+\frac{1}{2}+1}))=
\text{Split}_{-1}^{\text{SG}}(z,1^{+\frac{1}{2}+1},2^{+\frac{1}{2}+1})$.  
}
\bea
\text{Split}_{-1}^{\text{SG}}(z,1^{+1+\frac{1}{2}},2^{+1+\frac{1}{2}})
& = &
\text{Split}_{-1}^{\text{SG}}(z,1^{+\frac{1}{2}+1},2^{+1+\frac{1}{2}})
=
\text{Split}_{-1}^{\text{SG}}(z,1^{+\frac{1}{2}+1},2^{+\frac{1}{2}+1})\, ,
\nonu \\
\text{Split}_{+2}^{\text{SG}}(z,1^{+1+\frac{1}{2}},2^{-1-\frac{1}{2}})
& = & \text{Split}_{+2}^{\text{SG}}(z,1^{+\frac{1}{2}+1},2^{-\frac{1}{2}-1})\,.
\nonu
\eea

\begin{table}[tbp]
\centering
\renewcommand{\arraystretch}{1.7}
\begin{tabular}{|c|c|c|c|c| }
\hline
Eqs. & Split factors & Amp.  & Power  & Beta functions 
\\
\hline
\hline
(\ref{eq:gravitino_gravitino}) 
&$\text{Split}_{-1}^{\text{SG}}(z,1^{+\frac{1}{2}+1},2^{+\frac{1}{2}+1})=
-\frac{1}{\sqrt{z(1-z)}} \frac{[12]}{ \langle 12\rangle}$
&
$\frac{\omega_p}{\omega_1^{\frac{1}{2}} \omega_2^{\frac{1}{2}}}\frac{\bar{z}_{12}}{z_{12}}$
&
$  p^{+1}$
&
$  B(\Delta_1-\frac{1}{2},\Delta_2-\frac{1}{2}) $
\\
&
$\text{Split}_{+2}^{\text{SG}}(z,1^{+\frac{1}{2}+1},2^{-\frac{1}{2}-1})=
-\frac{(1-z)^3}{\sqrt{z(1-z)}} \frac{[12]}{ \langle 12\rangle}$
& 
$   \frac{\omega_2^{\frac{5}{2}}}{\omega_1^{\frac{1}{2}} \omega_p^{2}}\frac{\bar{z}_{12}}{z_{12}}
$ &
$ p^{-2}$
&
$B(\Delta_1-\frac{1}{2},\Delta_2+\frac{5}{2})$
\\
\hline
(\ref{eq:gravitino_photon})
&
$ \text{Split}_{-\frac{1}{2}}^{\text{SG}}(z,1^{+\frac{1}{2}+1},2^{0+1})=
-\frac{1}{\sqrt{z}} \frac{[12]}{ \langle 12\rangle} $
&
$  \frac{\omega_p^{\frac{1}{2}}}{\omega_1^{\frac{1}{2}}}\frac{\bar{z}_{12}}{z_{12}}$
&
$  p^{+\frac{1}{2}}$
&
$   B(\Delta_1-\frac{1}{2},\Delta_2)$
\\
& $\text{Split}_{+\frac{3}{2}}^{\text{SG}}(z,1^{+\frac{1}{2}+1},2^{0-1})=
-\frac{(1-z)^2}{\sqrt{z}} \frac{[12]}{ \langle 12\rangle}$
&
$\frac{\omega_2^2}{\omega_1^{\frac{1}{2}}\omega_p^{\frac{3}{2}}}\frac{\bar{z}_{12}}{z_{12}}$
&
$  p^{-\frac{3}{2}}$
&
$   B(\Delta_1-\frac{1}{2},\Delta_2+2)$
\\
\hline
(\ref{eq:gravitino_photino})
&
$\text{Split}_{0}^{\text{SG}}(z,1^{+\frac{1}{2}+1},2^{-\frac{1}{2}+1})=
-\frac{(1-z)}{\sqrt{z(1-z)}} \frac{[12]}{ \langle 12\rangle}$
&
$\frac{\omega_2^\frac{1}{2}}{\omega_1^{\frac{1}{2}}}\frac{\bar{z}_{12}}{z_{12}}$
&
$ p^{0}$
&
$   B(\Delta_1-\frac{1}{2},\Delta_2+\frac{1}{2})$
\\
& 
$  \text{Split}_{+1}^{\text{SG}}(z,1^{+\frac{1}{2}+1},2^{+\frac{1}{2}-1})=
-\frac{(1-z)^{\frac{3}{2}}}{\sqrt{z}} \frac{[12]}{ \langle 12\rangle}$
&
$  \frac{\omega_2^\frac{3}{2}}{\omega_1^{\frac{1}{2}}\omega_p}\frac{\bar{z}_{12}}{z_{12}}$
&
$   p^{-1}$
&
$  B(\Delta_1-\frac{1}{2},\Delta_2+\frac{3}{2})$
\\
\hline
(\ref{eq:gravitino_scalar})
&
$\text{Split}_{+\frac{1}{2}}^{\text{SG}}(z,1^{+\frac{1}{2}+1},2^{-\frac{1}{2}+\frac{1}{2}})=
-\frac{(1-z)}{\sqrt{z}} \frac{[12]}{ \langle 12\rangle}$
& 
$  \frac{\omega_2}{\omega_1^{\frac{1}{2}}\omega_p^{\frac{1}{2}}}\frac{\bar{z}_{12}}{z_{12}}$
&
$   p^{-\frac{1}{2}}$
&
$  B(\Delta_1-\frac{1}{2},\Delta_2+1)$
\\
\hline
\end{tabular}
\caption{ This is a continuation of previous Table \ref{Split1}.
The split factors for the ${\cal N}=8$
supergravity, the amplitude (Amp.) factors,
the power of collinear momenta $p$
and the Euler beta functions are given.}
\label{Split2}
\end{table}

Similarly,
the split factors for the second-the sixth
of (\ref{eq:gravitino_photon})
are the same as the first
as follows:
\bea
\text{Split}_{-\frac{1}{2}}^{\text{SG}}(z,1^{+\frac{1}{2}+1},2^{+\frac{1}{2}+\frac{1}{2}}) & = &
\text{Split}_{-\frac{1}{2}}^{\text{SG}}(z,1^{+\frac{1}{2}+1},2^{+1+0})
=
\text{Split}_{-\frac{1}{2}}^{\text{SG}}(z,1^{+1+\frac{1}{2}},2^{+0+1})
\nonu \\
&=&
\text{Split}_{-\frac{1}{2}}^{\text{SG}}(z,1^{+1+\frac{1}{2}},2^{+\frac{1}{2}+\frac{1}{2}})
=\text{Split}_{-\frac{1}{2}}^{\text{SG}}(z,1^{+1+\frac{1}{2}},2^{+1+0})
\nonu \\
&=& \text{Split}_{-\frac{1}{2}}^{\text{SG}}(z,1^{+\frac{1}{2}+1},2^{0+1})\,.
\nonu
\eea
The split factors for the eighth-the tenth
of (\ref{eq:gravitino_photon})
are the same as the one for the seventh
\bea
\text{Split}_{+\frac{3}{2}}^{\text{SG}}(z,1^{+\frac{1}{2}+1},2^{-\frac{1}{2}-\frac{1}{2}})& = &
\text{Split}_{+\frac{3}{2}}^{\text{SG}}(z,1^{+1+\frac{1}{2}},2^{-\frac{1}{2}-\frac{1}{2}})
=
\text{Split}_{+\frac{3}{2}}^{\text{SG}}(z,1^{+1+\frac{1}{2}},2^{-1+0})
\nonu \\
& = & \text{Split}_{+\frac{3}{2}}^{\text{SG}}(z,1^{+\frac{1}{2}+1},2^{0-1})\,.
\nonu
\eea
For the split factors
corresponding to the second-the eighth
in
(\ref{eq:gravitino_photino})
they are the same as the one for the first 
\bea
\text{Split}_{0}^{\text{SG}}(z,1^{+\frac{1}{2}+1},2^{+0+\frac{1}{2}})&=&
\text{Split}_{0}^{\text{SG}}(z,1^{+\frac{1}{2}+1},2^{+\frac{1}{2}+0})=
\text{Split}_{0}^{\text{SG}}(z,1^{+\frac{1}{2}+1},2^{+1-\frac{1}{2}})
\nonu \\
&=&
\text{Split}_{0}^{\text{SG}}(z,1^{+1+\frac{1}{2}},2^{-\frac{1}{2}+1})
=\text{Split}_{0}^{\text{SG}}(z,1^{+1+\frac{1}{2}},2^{+0+\frac{1}{2}})
\nonu \\
&=&\text{Split}_{0}^{\text{SG}}(z,1^{+1+\frac{1}{2}},2^{+\frac{1}{2}+0})
=\text{Split}_{0}^{\text{SG}}(z,1^{+1+\frac{1}{2}},2^{+1-\frac{1}{2}})
\nonu \\
&=&\text{Split}_{0}^{\text{SG}}(z,1^{+\frac{1}{2}+1},2^{-\frac{1}{2}+1})\, .
\nonu
\eea
For the split factors corresponding to the tenth-the fourteenth
there are relations 
\bea
\text{Split}_{+1}^{\text{SG}}(z,1^{+\frac{1}{2}+1},2^{+0-\frac{1}{2}})
&=&
\text{Split}_{+1}^{\text{SG}}(z,1^{+\frac{1}{2}+1},2^{-\frac{1}{2}+0})
=\text{Split}_{+1}^{\text{SG}}(z,1^{+1+\frac{1}{2}},2^{0-\frac{1}{2}})]
\nonu \\
&=&\text{Split}_{+1}^{\text{SG}}(z,1^{+1+\frac{1}{2}},2^{-\frac{1}{2}+0})
= \text{Split}_{+1}^{\text{SG}}(z,1^{1+\frac{1}{2}},2^{-1+\frac{1}{2}})
\nonu \\
&=&
\text{Split}_{+1}^{\text{SG}}(z,1^{+\frac{1}{2}+1},2^{+\frac{1}{2}-1})\, .
\nonu
\eea
Morover,
the split factors for the second-the eighth
in (\ref{eq:gravitino_scalar})
can be written as the one for the first as follows:
\bea
\text{Split}_{+\frac{1}{2}}^{\text{SG}}(z,1^{+\frac{1}{2}+1},2^{0+0})
&=&
\text{Split}_{+\frac{1}{2}}^{\text{SG}}(z,1^{+\frac{1}{2}+1},2^{+\frac{1}{2}-\frac{1}{2}})=
\text{Split}_{+\frac{1}{2}}^{\text{SG}}(z,1^{+\frac{1}{2}+1},2^{+1-1})
\nonu \\
& = & \text{Split}_{+\frac{1}{2}}^{\text{SG}}(z,1^{+1+\frac{1}{2}},2^{-1+1})
=\text{Split}_{+\frac{1}{2}}^{\text{SG}}(z,1^{+1+\frac{1}{2}},2^{-\frac{1}{2}+\frac{1}{2}})
\nonu \\
&=&\text{Split}_{+\frac{1}{2}}^{\text{SG}}(z,1^{+1+\frac{1}{2}},2^{0+0})
=\text{Split}_{+\frac{1}{2}}^{\text{SG}}(z,1^{+1+\frac{1}{2}},2^{+\frac{1}{2}-\frac{1}{2}})\nonu \\
&=&
\text{Split}_{+\frac{1}{2}}^{\text{SG}}(z,1^{+\frac{1}{2}+1},2^{-\frac{1}{2}+\frac{1}{2}}) \, .
\nonu 
\eea

\begin{table}[tbp]
\centering
\renewcommand{\arraystretch}{1.7}
\begin{tabular}{|c|c|c|c|c| }
\hline
Eqs. & Split factors & Amp.  & Power of $p$  & Beta functions 
\\
\hline
\hline
(\ref{eq:photon_photon})
& $\text{Split}_{0}^{\text{SG}}(z,1^{0+1},2^{0+1})=
-\frac{[12]}{ \langle 12\rangle}$
&
$ \frac{\bar{z}_{12}}{z_{12}}$
&
$  p^{0}$
&
$  B(\Delta_1,\Delta_2)$
\\
&
$ \text{Split}_{+2}^{\text{SG}}(z,1^{0+1},2^{0-1})=
-(1-z)^2\frac{[12]}{ \langle 12\rangle}$
&
$   \frac{\omega_2^2}{\omega_p^2}\frac{\bar{z}_{12}}{z_{12}}$
&
$ p^{-2}$
&
$   B(\Delta_1,\Delta_2+2)$
\\
\hline
(\ref{eq:photon_photino})
& $\text{Split}_{+\frac{1}{2}}^{\text{SG}}(z,1^{0+1},2^{0+\frac{1}{2}})=
-\sqrt{1-z}\frac{[12]}{ \langle 12\rangle}$
& 
$ \frac{\omega_2^\frac{1}{2}}{\omega_p^\frac{1}{2}}\frac{\bar{z}_{12}}{z_{12}}$
&
$ p^{-\frac{1}{2}}$
&
$ B(\Delta_1,\Delta_2+\frac{1}{2})$
\\
& $\text{Split}_{+\frac{3}{2}}^{\text{SG}}(z,1^{0+1},2^{+\frac{1}{2}-1})=
-(1-z)^{\frac{3}{2}}\frac{[12]}{ \langle 12\rangle}$
& 
$ \frac{\omega_2^\frac{3}{2}}{\omega_p^\frac{3}{2}}\frac{\bar{z}_{12}}{z_{12}}$
& 
$  p^{-\frac{3}{2}}$
&
$  B(\Delta_1,\Delta_2+\frac{3}{2})$
\\
\hline
(\ref{eq:photon_scalar})
&
$ \text{Split}_{+1}^{\text{SG}}(z,1^{0+1},2^{0+0})=
-(1-z)\frac{[12]}{ \langle 12\rangle}$
&
$ \frac{\omega_2}{\omega_p}\frac{\bar{z}_{12}}{z_{12}}$
&
$p^{-1}$
&
$ B(\Delta_1,\Delta_2+1)$
\\
\hline
\end{tabular}
\caption{  This is further continuation of previous Tables
\ref{Split1} and \ref{Split2}.
We present the split factors for the ${\cal N}=8$
supergravity, the amplitude (Amp.) factors,
the power of collinear momenta $p$
and the Euler beta functions as before.}
\label{Split3}
\end{table}

We consider the split factors
appearing in Table \ref{Split3}.
The split factors for the second-the 
sixth of 
(\ref{eq:photon_photon})
can be written as the one for the first
\footnote{There are
other relations $
\text{Split}_{0}^{\text{SG}}(z,1^{+1+0},2^{+\frac{1}{2}+\frac{1}{2}})
=\text{Split}_{0}^{\text{SG}}(z,1^{1+0},2^{0+1})
=\text{Split}_{0}^{\text{SG}}(z,1^{+\frac{1}{2}+\frac{1}{2}},2^{0+1})
=\text{Split}_{0}^{\text{SG}}(z,1^{0+1},2^{0+1})$.}
\bea
\text{Split}_{0}^{\text{SG}}(z,1^{0+1},2^{+\frac{1}{2}+\frac{1}{2}})
&= &
\text{Split}_{0}^{\text{SG}}(z,1^{+0+1},2^{+1+0})
=\text{Split}_{0}^{\text{SG}}(z,1^{+\frac{1}{2}+\frac{1}{2}},
2^{+\frac{1}{2}+\frac{1}{2}})
\nonu \\
& = &\text{Split}_{0}^{\text{SG}}(z,1^{+\frac{1}{2}+\frac{1}{2}},2^{+1+0})
=\text{Split}_{0}^{\text{SG}}(z,1^{+1+0},2^{+1+0})
\nonu \\
& = & \text{Split}_{0}^{\text{SG}}(z,1^{0+1},2^{0+1})\, .
\nonu 
\eea
Furthermore,
the split factors for the eighth and the 
ninth of 
(\ref{eq:photon_photon})
can be written as the one for the seventh
\bea
\text{Split}_{+2}^{\text{SG}}(z,1^{+\frac{1}{2}+\frac{1}{2}},2^{-\frac{1}{2}-\frac{1}{2}})
&=&
\text{Split}_{+2}^{\text{SG}}(z,1^{+1+0},2^{-1+0})
=
\text{Split}_{+2}^{\text{SG}}(z,1^{0+1},2^{0-1})\,.
\nonu
\eea
The split factors of the second-the tenth in
(\ref{eq:photon_photino}) can be written in terms of the
one for the first
\bea
\text{Split}_{+\frac{1}{2}}^{\text{SG}}(z,1^{0+1},2^{+\frac{1}{2}+0})
&= &
\text{Split}_{+\frac{1}{2}}^{\text{SG}}(z,1^{0+1},2^{+1-\frac{1}{2}})
=\text{Split}_{+\frac{1}{2}}^{\text{SG}}(z,1^{+\frac{1}{2}+\frac{1}{2}},2^{-\frac{1}{2}+1})
\nonu \\
& = &
\text{Split}_{+\frac{1}{2}}^{\text{SG}}(z,1^{+\frac{1}{2}+\frac{1}{2}},2^{0+\frac{1}{2}})
=\text{Split}_{+\frac{1}{2}}^{\text{SG}}(z,1^{+\frac{1}{2}+\frac{1}{2}},2^{+\frac{1}{2}+0})\nonu \\
&=&
\text{Split}_{+\frac{1}{2}}^{\text{SG}}(z,1^{+\frac{1}{2}+\frac{1}{2}},2^{+1-\frac{1}{2}})=\text{Split}_{+\frac{1}{2}}^{\text{SG}}(z,1^{+1+0},2^{-\frac{1}{2}+1})
\nonu \\
&=&\text{Split}_{+\frac{1}{2}}^{\text{SG}}(z,1^{+1+0},2^{0+\frac{1}{2}})
=\text{Split}_{+\frac{1}{2}}^{\text{SG}}(z,1^{+1+0},2^{+\frac{1}{2}+0})
\nonu \\
& = & \text{Split}_{+\frac{1}{2}}^{\text{SG}}(z,1^{0+1},2^{0+\frac{1}{2}})\, .
\nonu
\eea
Moreover, the split factors for the
twelfth-the sixteenth
can be written as the one for the eleventh
\bea
\text{Split}_{+\frac{3}{2}}^{\text{SG}}(z,1^{0+1},2^{0-\frac{1}{2}})
& = &
\text{Split}_{+\frac{3}{2}}^{\text{SG}}(z,1^{+\frac{1}{2}+\frac{1}{2}},2^{0-\frac{1}{2}})
=\text{Split}_{+\frac{3}{2}}^{\text{SG}}(z,1^{+\frac{1}{2}+\frac{1}{2}},
2^{-\frac{1}{2}+0})
\nonu \\
& = & \text{Split}_{+\frac{3}{2}}^{\text{SG}}(z,1^{+1+0},2^{-\frac{1}{2}+0})
= \text{Split}_{+\frac{3}{2}}^{\text{SG}}(z,1^{+1+0},2^{-1+\frac{1}{2}})
\nonu \\
& = & \text{Split}_{+\frac{3}{2}}^{\text{SG}}(z,1^{0+1},2^{+\frac{1}{2}-1})\, .
\nonu
\eea
For the split factors corresponding to
the second-the ninth of
(\ref{eq:photon_scalar}) they can be written in terms of
the one for the first
\bea
\text{Split}_{+1}^{\text{SG}}(z,1^{0+1},2^{+\frac{1}{2}-\frac{1}{2}}) &=&
\text{Split}_{+1}^{\text{SG}}(z,1^{0+1},2^{+1-1})
=\text{Split}_{+1}^{\text{SG}}(z,1^{+\frac{1}{2}+\frac{1}{2}},2^{-\frac{1}{2}+\frac{1}{2}})\nonu \\
&=&
\text{Split}_{+1}^{\text{SG}}(z,1^{+\frac{1}{2}+\frac{1}{2}},2^{0+0})=
\text{Split}_{+1}^{\text{SG}}(z,1^{+\frac{1}{2}+\frac{1}{2}},2^{+\frac{1}{2}-\frac{1}{2}})
\nonu \\
&=&\text{Split}_{+1}^{\text{SG}}(z,1^{1+0},2^{-1+1})=
\text{Split}_{+1}^{\text{SG}}(z,1^{+1+0},2^{-\frac{1}{2}+\frac{1}{2}})
\nonu \\
&=&
\text{Split}_{+1}^{\text{SG}}(z,1^{+1+0},2^{0+0})=
\text{Split}_{+1}^{\text{SG}}(z,1^{0+1},2^{0+0})\, .
\nonu
\eea

\begin{table}[tbp]
\centering
\renewcommand{\arraystretch}{1.7}
\begin{tabular}{|c|c|c|c|c| }
\hline
Eqs. & Split factors & Amp. & Power  & Beta functions 
\\
\hline
\hline
(\ref{eq:photino_photino})
&
$ \text{Split}_{+1}^{\text{SG}}(z,1^{-\frac{1}{2}+1},2^{+\frac{1}{2}+0})=
-\sqrt{z(1-z)}\frac{[12]}{ \langle 12\rangle}$
&  
$  \frac{\omega_1^\frac{1}{2}\omega_2^\frac{1}{2}}{\omega_p}\frac{\bar{z}_{12}}{z_{12}}$
&
$ p^{-1}$
&
$  {\scriptstyle B(\Delta_1+\frac{1}{2},\Delta_2+\frac{1}{2})}$
\\
& $ \text{Split}_{+2}^{\text{SG}}(z,1^{-\frac{1}{2}+1},2^{+\frac{1}{2}-1})=
-\sqrt{z}(1-z)^\frac{3}{2}\frac{[12]}{ \langle 12\rangle}$
&
$\frac{\omega_1^\frac{1}{2}\omega_2^\frac{3}{2}}{\omega_p^2}\frac{\bar{z}_{12}}{z_{12}}$
&
$  p^{-2}$
&
$ {\scriptstyle B(\Delta_1+\frac{1}{2},\Delta_2+\frac{3}{2})}$
\\
\hline
(\ref{eq:photino_scalar})
&
$\text{Split}_{+\frac{3}{2}}^{\text{SG}}(z,1^{-\frac{1}{2}+1},2^{+\frac{1}{2}-\frac{1}{2}})=
-\sqrt{z}(1-z)\frac{[12]}{ \langle 12\rangle}$
& $ \frac{\omega_1^\frac{1}{2}\omega_2}{\omega_p^\frac{3}{2}}\frac{\bar{z}_{12}}{z_{12}}$
& 
$ p^{-\frac{3}{2}}$
&
$  {\scriptstyle B(\Delta_1+\frac{1}{2},\Delta_2+1)}$
\\
\hline
\end{tabular}
\caption{  This is the 
continuation of previous Tables \ref{Split1}, \ref{Split2}
and \ref{Split3}.
The split factors for the ${\cal N}=8$
supergravity, the amplitude (Amp.) factors,
the power of collinear momenta $p$
and the Euler beta functions are written explicitly as before. }
\label{Split4}
\end{table}

Let us consider the Table \ref{Split4}.
The split factors corresponding to
the second-the sixth of (\ref{eq:photino_photino})
are the same as the one of the first as follows
\footnote{ We also have the following relations
$
\text{Split}_{+1}^{\text{SG}}(z,1^{+\frac{1}{2}+0},2^{-\frac{1}{2}+1})
=  \text{Split}_{+1}^{\text{SG}}(z,1^{+1-\frac{1}{2}},2^{-\frac{1}{2}+1})
=\text{Split}_{+1}^{\text{SG}}(z,1^{+1-\frac{1}{2}},2^{0+\frac{1}{2}})
=\text{Split}_{+1}^{\text{SG}}(z,1^{+\frac{1}{2}+0},2^{0+\frac{1}{2}})=
\text{Split}_{+1}^{\text{SG}}(z,1^{-\frac{1}{2}+1},2^{+\frac{1}{2}+0}) 
$.
}:
\bea
\text{Split}_{+1}^{\text{SG}}(z,1^{-\frac{1}{2}+1},2^{+1-\frac{1}{2}})
& = &
\text{Split}_{+1}^{\text{SG}}(z,1^{0+\frac{1}{2}},2^{0+\frac{1}{2}})=
\text{Split}_{+1}^{\text{SG}}(z,1^{0+\frac{1}{2}},2^{+\frac{1}{2}+0})
\nonu \\
& = & \text{Split}_{+1}^{\text{SG}}(z,1^{0+\frac{1}{2}},2^{+1-\frac{1}{2}})
=\text{Split}_{+1}^{\text{SG}}(z,1^{+\frac{1}{2}+0},2^{+\frac{1}{2}+0}) \,
\nonu \\
& = & \text{Split}_{+1}^{\text{SG}}(z,1^{-\frac{1}{2}+1},2^{+\frac{1}{2}+0}) \, .
\nonu
\eea
Moreover,
the split factors for the eighth-the tenth
can be written in terms of the one for the seventh
\bea
\text{Split}_{+2}^{\text{SG}}(z,1^{0+\frac{1}{2}},2^{0-\frac{1}{2}})
&=&
\text{Split}_{+2}^{\text{SG}}(z,1^{+\frac{1}{2}+0},2^{-\frac{1}{2}+0})
=\text{Split}_{+2}^{\text{SG}}(z,1^{+1-\frac{1}{2}},2^{-1+\frac{1}{2}})
\nonu \\
&=&
\text{Split}_{+2}^{\text{SG}}(z,1^{-\frac{1}{2}+1},2^{+\frac{1}{2}-1})\, .
\nonu
\eea
The split factors
for the second-the eighth
of (\ref{eq:photino_scalar})
can be written as the one for the first
\bea
\text{Split}_{+\frac{3}{2}}^{\text{SG}}(z,1^{-\frac{1}{2}+1},2^{+1-1})
&=&
\text{Split}_{+\frac{3}{2}}^{\text{SG}}(z,1^{0+\frac{1}{2}},2^{0+0})
=\text{Split}_{+\frac{3}{2}}^{\text{SG}}(z,1^{0+\frac{1}{2}},2^{+\frac{1}{2}-\frac{1}{2}})\nonu \\
&=& \text{Split}_{+\frac{3}{2}}^{\text{SG}}(z,1^{+\frac{1}{2}+0},2^{-\frac{1}{2}+\frac{1}{2}})=\text{Split}_{+\frac{3}{2}}^{\text{SG}}(z,1^{+\frac{1}{2}+0},2^{0+0})
\nonu \\
&=& \text{Split}_{+\frac{3}{2}}^{\text{SG}}(z,1^{+1-\frac{1}{2}},2^{-1+1})
=\text{Split}_{+\frac{3}{2}}^{\text{SG}}(z,1^{+1-\frac{1}{2}},2^{-\frac{1}{2}+\frac{1}{2}})\nonu \\
&=& 
\text{Split}_{+\frac{3}{2}}^{\text{SG}}(z,1^{-\frac{1}{2}+1},2^{+\frac{1}{2}-\frac{1}{2}}) \, .
\nonu
\eea

\begin{table}[tbp]
\centering
\renewcommand{\arraystretch}{1.7}
\begin{tabular}{|c|c|c|c|c| }
\hline
Eqs. & Split factors & Amp. & Power  & Beta functions
\\
\hline
\hline
(\ref{eq:scalar_scalar})
& $\text{Split}_{+2}^{\text{SG}}(z,1^{-1+1},2^{+1-1})=-z(1-z)\frac{[12]}{
\langle 12\rangle}$
&  $\frac{\omega_1 \omega_2}{\omega_p^2}\frac{\bar{z}_{12}}{z_{12}}$
& $p^{-2}$
&
$ B(\Delta_1+1,\Delta_2+1)$
\\
\hline
\end{tabular}
\caption{  This is the final
continuation of previous Tables \ref{Split1}, \ref{Split2}
\ref{Split3} and \ref{Split4}.
The split factor for the ${\cal N}=8$
supergravity, the amplitude (Amp.) factor,
the power of collinear momentum $p$
and the Euler beta function are written explicitly. }
\label{Split5}
\end{table}

The split factors for the second and the third
of (\ref{eq:scalar_scalar}) can be written as
the one for the first
\footnote{
  There are  $\text{Split}_{+2}^{\text{SG}}(z,1^{+1-1},2^{-1+1})
=\text{Split}_{+2}^{\text{SG}}(z,1^{+\frac{1}{2}-\frac{1}{2}},2^{-\frac{1}{2}+\frac{1}{2}})
= \text{Split}_{+2}^{\text{SG}}(z,1^{-1+1},2^{+1-1})
$.}
\bea
\text{Split}_{+2}^{\text{SG}}(z,1^{-\frac{1}{2}+\frac{1}{2}},2^{+\frac{1}{2}-\frac{1}{2}})& =&
\text{Split}_{+2}^{\text{SG}}(z,1^{0+0},2^{0+0})=
\text{Split}_{+2}^{\text{SG}}(z,1^{-1+1},2^{+1-1})\, .
\nonu
\eea

\section{The celestial OPEs after Mellin transform}

In this Appendix, the works of \cite{2212-1,2212-2} are reviewed.
We present the OPEs between the graviton,
the gravitinos, the graviphotons, the graviphotinos,
and the scalars.

$\bullet$ The graviton-the graviton

There are two relations
\begin{eqnarray}
\O_{\Delta_1,+2}(z_1,\bar{z}_1)\,\O_{\Delta_2,+2}(z_2,\bar{z}_2)&=&
\kappa_{+2,+2,-2}\,
\frac{\bar{z}_{12}}{z_{12}}\,B\bigg(\Delta_1-1,\Delta_2-1\bigg)\,\O_{\Delta_1+\Delta_2,+2}(z_2,\bar{z}_2) + \cdots,
\nonu
\\
\O_{\Delta_1,+2}(z_1,\bar{z}_1)\,\O_{\Delta_2,-2}(z_2,\bar{z}_2)&=&
\kappa_{+2,-2,+2}
\, \frac{\bar{z}_{12}}{z_{12}}\,B\bigg(\Delta_1-1 ,\Delta_2+3\bigg)\,\O_{\Delta_1+\Delta_2,-2}(z_2,\bar{z}_2)  + \cdots \, .
\nonu
\end{eqnarray}

$\bullet$ The graviton-the gravitinos

There exist the following relations
\begin{eqnarray}
\O_{\Delta_1,+2}(z_1,\bar{z}_1)\,\O^A_{\Delta_2,+\frac{3}{2}}(z_2,\bar{z}_2)&=&
\kappa_{+2,+\frac{3}{2},-\frac{3}{2}}\,
\frac{\bar{z}_{12}}{z_{12}}\,B\bigg(\Delta_1-1 ,\Delta_2-\frac{1}{2}\bigg)\,\O^A_{\Delta_1+\Delta_2,+\frac{3}{2}}(z_2,\bar{z}_2)  + \cdots,
\nonu\\
\O_{\Delta_1,+2}(z_1,\bar{z}_1)\,\O_{A,\Delta_2,-\frac{3}{2}}(z_2,\bar{z}_2)&=&
\kappa_{+2,-\frac{3}{2},+\frac{3}{2}}\, \frac{\bar{z}_{12}}{z_{12}}\,B\bigg(\Delta_1-1 ,\Delta_2+\frac{5}{2}\bigg)\,\O_{A,\Delta_1+\Delta_2,-\frac{3}{2}}(z_2,\bar{z}_2)  \nonu \\
&+& \cdots \, .
\nonu
\end{eqnarray}
For convenience,
we also present  the cases where the
$SU(8)$ indices are expanded in terms of two
$SU(4)$ indices.
They are given by
\begin{eqnarray}
\O_{\Delta_1,+2}(z_1,\bar{z}_1)\,\O^a_{\Delta_2,+\frac{3}{2}}(z_2,\bar{z}_2)&=&
\kappa_{+2,+\frac{3}{2},-\frac{3}{2}}\,
\frac{\bar{z}_{12}}{z_{12}}\,B\bigg(\Delta_1-1 ,\Delta_2-\frac{1}{2}\bigg)\,\O^a_{\Delta_1+\Delta_2,+\frac{3}{2}}(z_2,\bar{z}_2) + \cdots,
\nonu
\\
\O_{\Delta_1,+2}(z_1,\bar{z}_1)\,\O^r_{\Delta_2,+\frac{3}{2}}(z_2,\bar{z}_2)&=&
\kappa_{+2,+\frac{3}{2},-\frac{3}{2}}\,
\frac{\bar{z}_{12}}{z_{12}}\,B\bigg(\Delta_1-1 ,\Delta_2-\frac{1}{2}\bigg)\,\O^{r}_{\Delta_1+\Delta_2,+\frac{3}{2}}(z_2,\bar{z}_2) + \cdots,
\nonu
\\
\O_{\Delta_1,+2}(z_1,\bar{z}_1)\,\O_{a,\Delta_2,-\frac{3}{2}}(z_2,\bar{z}_2)&=&
\kappa_{+2,-\frac{3}{2},+\frac{3}{2}}\,\frac{\bar{z}_{12}}{z_{12}}\,B\bigg(\Delta_1-1 ,\Delta_2+\frac{5}{2}\bigg)\,\O_{a,\Delta_1+\Delta_2,-\frac{3}{2}}(z_2,\bar{z}_2) + \cdots,
\nonu
\\
\O_{\Delta_1,+2}(z_1,\bar{z}_1)\,\O_{r,\Delta_2,-\frac{3}{2}}(z_2,\bar{z}_2)&=&
\kappa_{+2,-\frac{3}{2},+\frac{3}{2}}\,\frac{\bar{z}_{12}}{z_{12}}\,B\bigg(\Delta_1-1 ,\Delta_2+\frac{5}{2}\bigg)\,\O_{r,\Delta_1+\Delta_2,-\frac{3}{2}}(z_2,\bar{z}_2) + \cdots \, .
\nonu
\end{eqnarray}

$\bullet$ The graviton-the graviphotons

In this case, we have
\begin{eqnarray}
\O_{\Delta_1,+2}(z_1,\bar{z}_1)\,\O^{AB}_{\Delta_2,+1}(z_2,\bar{z}_2)&=&
\kappa_{+2,+1,-1}\,
\frac{\bar{z}_{12}}{z_{12}}\,B\bigg(\Delta_1-1 ,\Delta_2\bigg)\,\O^{AB}_{\Delta_1+\Delta_2,+1}(z_2,\bar{z}_2) + \cdots,
\nonu
\\
\O_{\Delta_1,+2}(z_1,\bar{z}_1)\,\O_{AB,\Delta_2,-1}(z_2,\bar{z}_2)&=&
\kappa_{+2,-1,+1}\, \frac{\bar{z}_{12}}{z_{12}}\,B\bigg(\Delta_1-1 ,\Delta_2+2\bigg)\,\O_{AB,\Delta_1+\Delta_2,-1}(z_2,\bar{z}_2) \nonu\\
&+& \cdots \, .
\nonu
\end{eqnarray}
The equivalent relations are given by
the following relations
\begin{eqnarray}
\O_{\Delta_1,+2}(z_1,\bar{z}_1)\,\O^{ab}_{\Delta_2,+1}(z_2,\bar{z}_2)&=&
\kappa_{+2,+1,-1}\, \frac{\bar{z}_{12}}{z_{12}}\,B\bigg(\Delta_1-1 ,\Delta_2\bigg)\,\O^{ab}_{\Delta_1+\Delta_2,+1}(z_2,\bar{z}_2) + \cdots,
\nonu
\\
\O_{\Delta_1,+2}(z_1,\bar{z}_1)\,\O^{ar}_{\Delta_2,+1}(z_2,\bar{z}_2)&=&
\kappa_{+2,+1,-1}\,\frac{\bar{z}_{12}}{z_{12}}\,B\bigg(\Delta_1-1 ,\Delta_2\bigg)\,\O^{ar}_{\Delta_1+\Delta_2,+1}(z_2,\bar{z}_2) + \cdots,
\nonu
\\
\O_{\Delta_1,+2}(z_1,\bar{z}_1)\,\O^{rs}_{\Delta_2,+1}(z_2,\bar{z}_2)&=&
\kappa_{+2,+1,-1}\,\frac{\bar{z}_{12}}{z_{12}}\,B\bigg(\Delta_1-1 ,\Delta_2\bigg)\,\O^{rs}_{\Delta_1+\Delta_2,+1}(z_2,\bar{z}_2) + \cdots,
\nonu
\\
\O_{\Delta_1,+2}(z_1,\bar{z}_1)\,\O_{ab,\Delta_2,-1}(z_2,\bar{z}_2)&=&
\kappa_{+2,-1,+1}\,\frac{\bar{z}_{12}}{z_{12}}\,B\bigg(\Delta_1-1 ,\Delta_2+2\bigg)\,\O_{ab,\Delta_1+\Delta_2,-1}(z_2,\bar{z}_2) + \cdots,
\nonu
\\
\O_{\Delta_1,+2}(z_1,\bar{z}_1)\,\O_{ar,\Delta_2,-1}(z_2,\bar{z}_2)&=&
\kappa_{+2,-1,+1}\,\frac{\bar{z}_{12}}{z_{12}}\,B\bigg(\Delta_1-1 ,\Delta_2+2\bigg)\,\O_{ar,\Delta_1+\Delta_2,-1}(z_2,\bar{z}_2) + \cdots,
\nonu
\\
\O_{\Delta_1,+2}(z_1,\bar{z}_1)\,\O_{rs,\Delta_2,-1}(z_2,\bar{z}_2)&=&
\kappa_{+2,-1,+1}\,\frac{\bar{z}_{12}}{z_{12}}\,B\bigg(\Delta_1-1 ,\Delta_2+2\bigg)\,\O_{rs,\Delta_1+\Delta_2,-1}(z_2,\bar{z}_2) + \cdots \, .
\nonu
\end{eqnarray}

$\bullet$ The graviton-the graviphotinos

There are
\begin{eqnarray}
\O_{\Delta_1,+2}(z_1,\bar{z}_1)\,\O^{ABC}_{\Delta_2,+\frac{1}{2}}(z_2,\bar{z}_2)&=&
\kappa_{+2,+\frac{1}{2},-\frac{1}{2}}\,
\frac{\bar{z}_{12}}{z_{12}}\,B\bigg(\Delta_1-1 ,\Delta_2+\frac{1}{2}\bigg)\,\O^{ABC}_{\Delta_1+\Delta_2,+\frac{1}{2}}(z_2,\bar{z}_2) \nonu \\
&+& \cdots,
\nonu
\\
\O_{\Delta_1,+2}(z_1,\bar{z}_1)\,\O_{ABC,\Delta_2,-\frac{1}{2}}(z_2,\bar{z}_2)&=&
\kappa_{+2,-\frac{1}{2},+\frac{1}{2}}\,\frac{\bar{z}_{12}}{z_{12}}\,B\bigg(\Delta_1-1 ,\Delta_2+\frac{3}{2}\bigg)\,\O_{ABC,\Delta_1+\Delta_2,-\frac{1}{2}}(z_2,\bar{z}_2) \nonu \\
&+& \cdots \, .
\nonu
\end{eqnarray}
In the components, we have
\footnote{The missing equations in \cite{2212-2}
are denoted by the boldface in this Appendix.
These are located at the third columns in Tables \ref{prevtable}
and \ref{latertable}.}
\begin{eqnarray}
\O_{\Delta_1,+2}(z_1,\bar{z}_1)\,\O^{abc}_{\Delta_2,+\frac{1}{2}}(z_2,\bar{z}_2)&=&
\kappa_{+2,+\frac{1}{2},-\frac{1}{2}}\,\frac{\bar{z}_{12}}{z_{12}}\,B\bigg(\Delta_1-1 ,\Delta_2+\frac{1}{2}\bigg)\,\O^{abc}_{\Delta_1+\Delta_2,+\frac{1}{2}}(z_2,\bar{z}_2) \nonu \\
&+& \cdots,
\nonu
\\
\O_{\Delta_1,+2}(z_1,\bar{z}_1)\,\O^{abr}_{\Delta_2,+\frac{1}{2}}(z_2,\bar{z}_2)&=&
\kappa_{+2,+\frac{1}{2},-\frac{1}{2}}\,\frac{\bar{z}_{12}}{z_{12}}\,B\bigg(\Delta_1-1 ,\Delta_2+\frac{1}{2}\bigg)\,\O^{abr}_{\Delta_1+\Delta_2,+\frac{1}{2}}(z_2,\bar{z}_2) \nonu \\
&+& \cdots,
\nonu
\\
\bf
\O_{\Delta_1,+2}(z_1,\bar{z}_1)\,\O^{ars}_{\Delta_2,+\frac{1}{2}}(z_2,\bar{z}_2)&=&
\bf \kappa_{+2,+\frac{1}{2},-\frac{1}{2}}\,\frac{\bar{z}_{12}}{z_{12}}\,B\bigg(\Delta_1-1 ,\Delta_2+\frac{1}{2}\bigg)\,\O^{ars}_{\Delta_1+\Delta_2,+\frac{1}{2}}(z_2,\bar{z}_2) \nonu \\
&+& \cdots,
\nonu
\\
\bf
\O_{\Delta_1,+2}(z_1,\bar{z}_1)\,\O^{rst}_{\Delta_2,+\frac{1}{2}}(z_2,\bar{z}_2)&=&
\bf \kappa_{+2,+\frac{1}{2},-\frac{1}{2}}\,\frac{\bar{z}_{12}}{z_{12}}\,B\bigg(\Delta_1-1 ,\Delta_2+\frac{1}{2}\bigg)\,\O^{rst}_{\Delta_1+\Delta_2,+\frac{1}{2}}(z_2,\bar{z}_2) \nonu \\
&+& \cdots,
\nonu
\\
\O_{\Delta_1,+2}(z_1,\bar{z}_1)\,\O_{abc,\Delta_2,-\frac{1}{2}}(z_2,\bar{z}_2)&=&
\kappa_{+2,-\frac{1}{2},+\frac{1}{2}}\,\frac{\bar{z}_{12}}{z_{12}}\,B\bigg(\Delta_1-1 ,\Delta_2+\frac{3}{2}\bigg)\,\O_{abc,\Delta_1+\Delta_2,-\frac{1}{2}}(z_2,\bar{z}_2) \nonu \\
&+& \cdots,
\nonu
\\
\bf
\O_{\Delta_1,+2}(z_1,\bar{z}_1)\,\O_{abr,\Delta_2,-\frac{1}{2}}(z_2,\bar{z}_2)&=&
\bf \kappa_{+2,-\frac{1}{2},+\frac{1}{2}}\,
\frac{\bar{z}_{12}}{z_{12}}\,B\bigg(\Delta_1-1 ,\Delta_2+\frac{3}{2}\bigg)\,
\nonu \\
& \times& \bf \O_{abr,\Delta_1+\Delta_2,-\frac{1}{2}}(z_2,\bar{z}_2)
+ \cdots,
\nonu
\\
\bf
\O_{\Delta_1,+2}(z_1,\bar{z}_1)\,\O_{ars,\Delta_2,-\frac{1}{2}}(z_2,\bar{z}_2)&=&
\bf \kappa_{+2,-\frac{1}{2},+\frac{1}{2}}\,\frac{\bar{z}_{12}}{z_{12}}\,B\bigg(\Delta_1-1 ,\Delta_2+\frac{3}{2}\bigg)\,\nonu \\
& \times & \bf \O_{ars,\Delta_1+\Delta_2,-\frac{1}{2}}(z_2,\bar{z}_2) 
+ \cdots,
\nonu
\\
\bf
\O_{\Delta_1,+2}(z_1,\bar{z}_1)\,\O_{rst,\Delta_2,-\frac{1}{2}}(z_2,\bar{z}_2)&=&
\bf \kappa_{+2,-\frac{1}{2},+\frac{1}{2}}\,
\frac{\bar{z}_{12}}{z_{12}}\,B\bigg(\Delta_1-1 ,\Delta_2+\frac{3}{2}\bigg)\,
\nonu \\
& \times &
\bf \O_{rst,\Delta_1+\Delta_2,-\frac{1}{2}}(z_2,\bar{z}_2)
+ \cdots \, .
\nonu
\end{eqnarray}
The five additional OPEs appear.

$\bullet$ The graviton-the scalars

There exists
\begin{eqnarray}
\O_{\Delta_1,+2}(z_1,\bar{z}_1)\,\O^{ABCD}_{\Delta_2,0}(z_2,\bar{z}_2)&=&
\kappa_{+2,0,0} \,
\frac{\bar{z}_{12}}{z_{12}}\,B\bigg(\Delta_1-1 ,\Delta_2+1\bigg)\,\O^{ABCD}_{\Delta_1+\Delta_2,0}(z_2,\bar{z}_2) + \cdots \, .
\nonu
\end{eqnarray}
If we expand the components,
the following relations satisfy
\begin{eqnarray}
\O_{\Delta_1,+2}(z_1,\bar{z}_1)\,\O^{abcd}_{\Delta_2,0}(z_2,\bar{z}_2)&=&
\kappa_{+2,0,0} \,
\frac{\bar{z}_{12}}{z_{12}}\,B\bigg(\Delta_1-1 ,\Delta_2+1\bigg)\,\O^{abcd}_{\Delta_1+\Delta_2,0}(z_2,\bar{z}_2) + \cdots,
\nonu
\\
\O_{\Delta_1,+2}(z_1,\bar{z}_1)\,\O^{abcr}_{\Delta_2,0}(z_2,\bar{z}_2)&=&
\kappa_{+2,0,0} \,\frac{\bar{z}_{12}}{z_{12}}\,B\bigg(\Delta_1-1 ,\Delta_2+1\bigg)\,\O^{abcr}_{\Delta_1+\Delta_2,0}(z_2,\bar{z}_2) + \cdots,
\nonu\\
\O_{\Delta_1,+2}(z_1,\bar{z}_1)\,\O^{abrs}_{\Delta_2,0}(z_2,\bar{z}_2)&=&
\kappa_{+2,0,0} \,\frac{\bar{z}_{12}}{z_{12}}\,B\bigg(\Delta_1-1 ,\Delta_2+1\bigg)\,\O^{abrs}_{\Delta_1+\Delta_2,0}(z_2,\bar{z}_2) + \cdots,
\nonu\\
\O_{\Delta_1,+2}(z_1,\bar{z}_1)\,\O^{arst}_{\Delta_2,0}(z_2,\bar{z}_2)&=&
\kappa_{+2,0,0} \,\frac{\bar{z}_{12}}{z_{12}}\,B\bigg(\Delta_1-1 ,\Delta_2+1\bigg)\,\O^{arst}_{\Delta_1+\Delta_2,0}(z_2,\bar{z}_2) + \cdots,
\nonu\\
\O_{\Delta_1,+2}(z_1,\bar{z}_1)\,\O^{rstu}_{\Delta_2,0}(z_2,\bar{z}_2)&=&
\kappa_{+2,0,0} \,\frac{\bar{z}_{12}}{z_{12}}\,B\bigg(\Delta_1-1 ,\Delta_2+1\bigg)\,\O^{rstu}_{\Delta_1+\Delta_2,0}(z_2,\bar{z}_2) + \cdots \, .
\nonu
\end{eqnarray}

$\bullet$ The gravitinos-the gravitinos

There are two relations
\begin{eqnarray}
\O^A_{\Delta_1,+\frac{3}{2}}(z_1,\bar{z}_1)\,\O^B_{\Delta_2,+\frac{3}{2}}(z_2,\bar{z}_2)&=&\kappa_{+\frac{3}{2},+\frac{3}{2},-1}\, \frac{\bar{z}_{12}}{z_{12}}\,B\bigg(\Delta_1-\frac{1}{2} ,\Delta_2-\frac{1}{2}\bigg)\,\O^{AB}_{\Delta_1+\Delta_2,+1}(z_2,\bar{z}_2) + \cdots,
\nonu\\
\O^A_{\Delta_1,+\frac{3}{2}}(z_1,\bar{z}_1)\,\O_{B,\Delta_2,-\frac{3}{2}}(z_2,\bar{z}_2)&=&\kappa_{+\frac{3}{2},-\frac{3}{2},+2}\,\frac{\bar{z}_{12}}{z_{12}}\,B\bigg(\Delta_1-\frac{1}{2} ,\Delta_2+\frac{5}{2}\bigg)\,\delta^{A}_{B}\,\O_{\Delta_1+\Delta_2,-2}(z_2,\bar{z}_2) \nonu \\
&+&
\cdots \, .
\nonu
\end{eqnarray}
Equivalently, we have
\begin{eqnarray}
\O^a_{\Delta_1,+\frac{3}{2}}(z_1,\bar{z}_1)\,\O^b_{\Delta_2,+\frac{3}{2}}(z_2,\bar{z}_2)&=&\kappa_{+\frac{3}{2},+\frac{3}{2},-1}\,\frac{\bar{z}_{12}}{z_{12}}\,B\bigg(\Delta_1-\frac{1}{2} ,\Delta_2-\frac{1}{2}\bigg)\,\O^{ab}_{\Delta_1+\Delta_2,+1}(z_2,\bar{z}_2) + \cdots,
\nonu\\
\O^r_{\Delta_1,+\frac{3}{2}}(z_1,\bar{z}_1)\,\O^s_{\Delta_2,+\frac{3}{2}}(z_2,\bar{z}_2)&=&\kappa_{+\frac{3}{2},+\frac{3}{2},-1}\,\frac{\bar{z}_{12}}{z_{12}}\,B\bigg(\Delta_1-\frac{1}{2} ,\Delta_2-\frac{1}{2}\bigg)\,\O^{rs}_{\Delta_1+\Delta_2,+1}(z_2,\bar{z}_2) + \cdots,
\nonu\\
\O^a_{\Delta_1,+\frac{3}{2}}(z_1,\bar{z}_1)\,\O^r_{\Delta_2,+\frac{3}{2}}(z_2,\bar{z}_2)&=&\kappa_{+\frac{3}{2},+\frac{3}{2},-1}\,\frac{\bar{z}_{12}}{z_{12}}\,B\bigg(\Delta_1-\frac{1}{2} ,\Delta_2-\frac{1}{2}\bigg)\,\O^{ar}_{\Delta_1+\Delta_2,+1}(z_2,\bar{z}_2) + \cdots,
\nonu\\
\O^a_{\Delta_1,+\frac{3}{2}}(z_1,\bar{z}_1)\,\O_{b,\Delta_2,-\frac{3}{2}}(z_2,\bar{z}_2)&=&\kappa_{+\frac{3}{2},-\frac{3}{2},+2}\,\frac{\bar{z}_{12}}{z_{12}}\,B\bigg(\Delta_1-\frac{1}{2} ,\Delta_2+\frac{5}{2}\bigg)\,\delta^{a}_{b}\,\O_{\Delta_1+\Delta_2,-2}(z_2,\bar{z}_2) \nonu \\
&+& \cdots,
\nonu\\
\O^r_{\Delta_1,+\frac{3}{2}}(z_1,\bar{z}_1)\,\O_{s,\Delta_2,-\frac{3}{2}}(z_2,\bar{z}_2)&=&\kappa_{+\frac{3}{2},-\frac{3}{2},+2}\,\frac{\bar{z}_{12}}{z_{12}}\,B\bigg(\Delta_1-\frac{1}{2} ,\Delta_2+\frac{5}{2}\bigg)\,\delta^{r}_{s}\,\O_{\Delta_1+\Delta_2,-2}(z_2,\bar{z}_2) \nonu \\
&+&
\cdots \, .
\nonu
\end{eqnarray}

$\bullet$ The gravitinos-the graviphotons

We have the following relations
\begin{eqnarray}
\O^A_{\Delta_1,+\frac{3}{2}}(z_1,\bar{z}_1)\,\O^{BC}_{\Delta_2,+1}(z_2,\bar{z}_2)&=&
\kappa_{+\frac{3}{2},+1,-\frac{1}{2}}\,  \frac{\bar{z}_{12}}{z_{12}}\,B\bigg(\Delta_1-\frac{1}{2} ,\Delta_2\bigg)\,\O^{ABC}_{\Delta_1+\Delta_2,+\frac{1}{2}}(z_2,\bar{z}_2) + \cdots,
\nonu\\
\O^A_{\Delta_1,+\frac{3}{2}}(z_1,\bar{z}_1)\,\O_{BC,\Delta_2,-1}(z_2,\bar{z}_2)&=&
\kappa_{+\frac{3}{2},-1,+\frac{3}{2}}\, \frac{\bar{z}_{12}}{z_{12}}\,B\bigg(\Delta_1-\frac{1}{2} ,\Delta_2+2\bigg)\,\nonu \\
&\times & 2!\,\delta^A_{[B}
\O_{C],\Delta_1+\Delta_2,-\frac{3}{2}}(z_2,\bar{z}_2) + \cdots \, .
\nonu
\end{eqnarray}
Or the following relations hold
\begin{eqnarray}
\O^a_{\Delta_1,+\frac{3}{2}}(z_1,\bar{z}_1)\,\O^{bc}_{\Delta_2,+1}(z_2,\bar{z}_2)&=&
\kappa_{+\frac{3}{2},+1,-\frac{1}{2}}\,   \frac{\bar{z}_{12}}{z_{12}}\,B\bigg(\Delta_1-\frac{1}{2} ,\Delta_2\bigg)\,\O^{abc}_{\Delta_1+\Delta_2,+\frac{1}{2}}(z_2,\bar{z}_2) + \cdots,
\nonu\\
\O^a_{\Delta_1,+\frac{3}{2}}(z_1,\bar{z}_1)\,\O^{br}_{\Delta_2,+1}(z_2,\bar{z}_2)&=&
\kappa_{+\frac{3}{2},+1,-\frac{1}{2}}\, \frac{\bar{z}_{12}}{z_{12}}\,B\bigg(\Delta_1-\frac{1}{2} ,\Delta_2\bigg)\,\O^{abr}_{\Delta_1+\Delta_2,+\frac{1}{2}}(z_2,\bar{z}_2) + \cdots,
\nonu\\
\O^a_{\Delta_1,+\frac{3}{2}}(z_1,\bar{z}_1)\,\O^{rs}_{\Delta_2,+1}(z_2,\bar{z}_2)&=&
\kappa_{+\frac{3}{2},+1,-\frac{1}{2}}\, \frac{\bar{z}_{12}}{z_{12}}\,B\bigg(\Delta_1-\frac{1}{2} ,\Delta_2\bigg)\,\O^{ars}_{\Delta_1+\Delta_2,+\frac{1}{2}}(z_2,\bar{z}_2) + \cdots,
\nonu\\
\O^r_{\Delta_1,+\frac{3}{2}}(z_1,\bar{z}_1)\,\O^{ab}_{\Delta_2,+1}(z_2,\bar{z}_2)&=&
\kappa_{+\frac{3}{2},+1,-\frac{1}{2}}\, \frac{\bar{z}_{12}}{z_{12}}\,B\bigg(\Delta_1-\frac{1}{2} ,\Delta_2\bigg)\,\O^{rab}_{\Delta_1+\Delta_2,+\frac{1}{2}}(z_2,\bar{z}_2) + \cdots,
\nonu\\
\O^r_{\Delta_1,+\frac{3}{2}}(z_1,\bar{z}_1)\,\O^{as}_{\Delta_2,+1}(z_2,\bar{z}_2)&=&
\kappa_{+\frac{3}{2},+1,-\frac{1}{2}}\, \frac{\bar{z}_{12}}{z_{12}}\,B\bigg(\Delta_1-\frac{1}{2} ,\Delta_2\bigg)\,\O^{ras}_{\Delta_1+\Delta_2,+\frac{1}{2}}(z_2,\bar{z}_2) + \cdots,
\nonu\\
\O^r_{\Delta_1,+\frac{3}{2}}(z_1,\bar{z}_1)\,\O^{st}_{\Delta_2,+1}(z_2,\bar{z}_2)&=&
\kappa_{+\frac{3}{2},+1,-\frac{1}{2}}\, \frac{\bar{z}_{12}}{z_{12}}\,B\bigg(\Delta_1-\frac{1}{2} ,\Delta_2\bigg)\,\O^{rst}_{\Delta_1+\Delta_2,+\frac{1}{2}}(z_2,\bar{z}_2) + \cdots,
\nonu\\
\O^a_{\Delta_1,+\frac{3}{2}}(z_1,\bar{z}_1)\,\O_{bc,\Delta_2,-1}(z_2,\bar{z}_2)&=&
\kappa_{+\frac{3}{2},-1,+\frac{3}{2}}\, \frac{\bar{z}_{12}}{z_{12}}\,B\bigg(\Delta_1-\frac{1}{2} ,\Delta_2+2\bigg)\,2!\,\delta^a_{[b}
  \O_{c],\Delta_1+\Delta_2,-\frac{3}{2}}(z_2,\bar{z}_2) \nonu \\
&+& \cdots,
\nonu\\
\O^a_{\Delta_1,+\frac{3}{2}}(z_1,\bar{z}_1)\,\O_{br,\Delta_2,-1}(z_2,\bar{z}_2)&=&
\kappa_{+\frac{3}{2},-1,+\frac{3}{2}}\, \frac{\bar{z}_{12}}{z_{12}}\,B\bigg(\Delta_1-\frac{1}{2} ,\Delta_2+2\bigg)\,\delta^a_b\O_{r,\Delta_1+\Delta_2,-\frac{3}{2}}(z_2,\bar{z}_2) \nonu \\
&+& \cdots,
\nonu\\
\O^r_{\Delta_1,+\frac{3}{2}}(z_1,\bar{z}_1)\,\O_{as,\Delta_2,-1}(z_2,\bar{z}_2)&=&
-\kappa_{+\frac{3}{2},-1,+\frac{3}{2}}\, \frac{\bar{z}_{12}}{z_{12}}\,B\bigg(\Delta_1-\frac{1}{2} ,\Delta_2+2\bigg)\,\delta^r_s\O_{a,\Delta_1+\Delta_2,-\frac{3}{2}}(z_2,\bar{z}_2) \nonu \\
&+& \cdots,
\nonu\\
\O^r_{\Delta_1,+\frac{3}{2}}(z_1,\bar{z}_1)\,\O_{st,\Delta_2,-1}(z_2,\bar{z}_2)&=&
\kappa_{+\frac{3}{2},-1,+\frac{3}{2}}\, \frac{\bar{z}_{12}}{z_{12}}\,B\bigg(\Delta_1-\frac{1}{2} ,\Delta_2+2\bigg)\,2!\,\delta^r_{[s}
  \O_{t],\Delta_1+\Delta_2,-\frac{3}{2}}(z_2,\bar{z}_2) \nonu \\
&+& \cdots \, .
\nonu
\end{eqnarray}

$\bullet$ The gravitinos-the graviphotinos

The following relations satisfy
\begin{eqnarray}
\O^A_{\Delta_1,+\frac{3}{2}}(z_1,\bar{z}_1)\,\O^{BCD}_{\Delta_2,+\frac{1}{2}}(z_2,\bar{z}_2)&=&\kappa_{+\frac{3}{2},+\frac{1}{2},0}\,
\frac{\bar{z}_{12}}{z_{12}}\,B\bigg(\Delta_1-\frac{1}{2} ,\Delta_2+\frac{1}{2}\bigg)\,\O^{ABCD}_{\Delta_1+\Delta_2,0}(z_2,\bar{z}_2) + \cdots,
\nonu\\
\O^A_{\Delta_1,+\frac{3}{2}}(z_1,\bar{z}_1)\,\O_{BCD,\Delta_2,-\frac{1}{2}}(z_2,\bar{z}_2)&=&\kappa_{+\frac{3}{2},+\frac{1}{2},0}\,\frac{\bar{z}_{12}}{z_{12}}\,B\bigg(\Delta_1-\frac{1}{2} ,\Delta_2+\frac{3}{2}\bigg)\,\nonu \\
&\times & 3\,\delta^{A}_{[B}\O_{CD],\Delta_1+\Delta_2,-1}(z_2,\bar{z}_2)
+ \cdots \, .
\nonu
\end{eqnarray}
In other words, they can be written as 
\begin{eqnarray}
\O^a_{\Delta_1,+\frac{3}{2}}(z_1,\bar{z}_1)\,\O^{bcd}_{\Delta_2,+\frac{1}{2}}(z_2,\bar{z}_2)&=&\kappa_{+\frac{3}{2},+\frac{1}{2},0}\, \frac{\bar{z}_{12}}{z_{12}}\,B\bigg(\Delta_1-\frac{1}{2} ,\Delta_2+\frac{1}{2}\bigg)\,\O^{abcd}_{\Delta_1+\Delta_2,0}(z_2,\bar{z}_2) + \cdots,
\nonu\\
\O^a_{\Delta_1,+\frac{3}{2}}(z_1,\bar{z}_1)\,\O^{bcr}_{\Delta_2,+\frac{1}{2}}(z_2,\bar{z}_2)&=&\kappa_{+\frac{3}{2},+\frac{1}{2},0}\,\frac{\bar{z}_{12}}{z_{12}}\,B\bigg(\Delta_1-\frac{1}{2} ,\Delta_2+\frac{1}{2}\bigg)\,\O^{abcr}_{\Delta_1+\Delta_2,0}(z_2,\bar{z}_2) + \cdots,
\nonu\\
\O^a_{\Delta_1,+\frac{3}{2}}(z_1,\bar{z}_1)\,\O^{brs}_{\Delta_2,+\frac{1}{2}}(z_2,\bar{z}_2)&=&\kappa_{+\frac{3}{2},+\frac{1}{2},0}\,\frac{\bar{z}_{12}}{z_{12}}\,B\bigg(\Delta_1-\frac{1}{2} ,\Delta_2+\frac{1}{2}\bigg)\,\O^{abrs}_{\Delta_1+\Delta_2,0}(z_2,\bar{z}_2) + \cdots,
\nonu\\
\O^a_{\Delta_1,+\frac{3}{2}}(z_1,\bar{z}_1)\,\O^{rst}_{\Delta_2,+\frac{1}{2}}(z_2,\bar{z}_2)&=&\kappa_{+\frac{3}{2},+\frac{1}{2},0}\,\frac{\bar{z}_{12}}{z_{12}}\,B\bigg(\Delta_1-\frac{1}{2} ,\Delta_2+\frac{1}{2}\bigg)\,\O^{arst}_{\Delta_1+\Delta_2,0}(z_2,\bar{z}_2) + \cdots,
\nonu\\
\O^r_{\Delta_1,+\frac{3}{2}}(z_1,\bar{z}_1)\,\O^{abc}_{\Delta_2,+\frac{1}{2}}(z_2,\bar{z}_2)&=&\kappa_{+\frac{3}{2},+\frac{1}{2},0}\,\frac{\bar{z}_{12}}{z_{12}}\,B\bigg(\Delta_1-\frac{1}{2} ,\Delta_2+\frac{1}{2}\bigg)\,\O^{rabc}_{\Delta_1+\Delta_2,0}(z_2,\bar{z}_2) + \cdots,
\nonu\\
\O^r_{\Delta_1,+\frac{3}{2}}(z_1,\bar{z}_1)\,\O^{abs}_{\Delta_2,+\frac{1}{2}}(z_2,\bar{z}_2)&=&\kappa_{+\frac{3}{2},+\frac{1}{2},0}\,\frac{\bar{z}_{12}}{z_{12}}\,B\bigg(\Delta_1-\frac{1}{2} ,\Delta_2+\frac{1}{2}\bigg)\,\O^{rabs}_{\Delta_1+\Delta_2,0}(z_2,\bar{z}_2) + \cdots,
\nonu\\
\O^r_{\Delta_1,+\frac{3}{2}}(z_1,\bar{z}_1)\,\O^{ast}_{\Delta_2,+\frac{1}{2}}(z_2,\bar{z}_2)&=&\kappa_{+\frac{3}{2},+\frac{1}{2},0}\,\frac{\bar{z}_{12}}{z_{12}}\,B\bigg(\Delta_1-\frac{1}{2} ,\Delta_2+\frac{1}{2}\bigg)\,\O^{rast}_{\Delta_1+\Delta_2,0}(z_2,\bar{z}_2) + \cdots,
\nonu\\
\O^r_{\Delta_1,+\frac{3}{2}}(z_1,\bar{z}_1)\,\O^{stu}_{\Delta_2,+\frac{1}{2}}(z_2,\bar{z}_2)&=&\kappa_{+\frac{3}{2},+\frac{1}{2},0}\,\frac{\bar{z}_{12}}{z_{12}}\,B\bigg(\Delta_1-\frac{1}{2} ,\Delta_2+\frac{1}{2}\bigg)\,\O^{rstu}_{\Delta_1+\Delta_2,0}(z_2,\bar{z}_2) + \cdots,
\nonu\\
\O^a_{\Delta_1,+\frac{3}{2}}(z_1,\bar{z}_1)\,\O_{bcd,\Delta_2,-\frac{1}{2}}(z_2,\bar{z}_2)&=&\kappa_{+\frac{3}{2},-\frac{1}{2},+1}\,\frac{\bar{z}_{12}}{z_{12}}\,B\bigg(\Delta_1-\frac{1}{2} ,\Delta_2+\frac{3}{2}\bigg)\,3\,\delta^{a}_{[b}\O_{cd],\Delta_1+\Delta_2,-1}(z_2,\bar{z}_2) \nonu \\
&+& \cdots,
\nonu\\
\O^a_{\Delta_1,+\frac{3}{2}}(z_1,\bar{z}_1)\,\O_{bcr,\Delta_2,-\frac{1}{2}}(z_2,\bar{z}_2)&=&\kappa_{+\frac{3}{2},-\frac{1}{2},+1}\,\frac{\bar{z}_{12}}{z_{12}}\,B\bigg(\Delta_1-\frac{1}{2} ,\Delta_2+\frac{3}{2}\bigg)\,2!\,\delta^{a}_{[b}\O_{c]r,\Delta_1+\Delta_2,-1}(z_2,\bar{z}_2) \nonu \\
&+& \cdots,
\nonu\\
\O^a_{\Delta_1,+\frac{3}{2}}(z_1,\bar{z}_1)\,\O_{brs,\Delta_2,-\frac{1}{2}}(z_2,\bar{z}_2)&=&\kappa_{+\frac{3}{2},-\frac{1}{2},+1}\,\frac{\bar{z}_{12}}{z_{12}}\,B\bigg(\Delta_1-\frac{1}{2} ,\Delta_2+\frac{3}{2}\bigg)\,\delta^{a}_{b}\O_{rs,\Delta_1+\Delta_2,-1}(z_2,\bar{z}_2) \nonu \\
&+& \cdots,
\nonu\\
\O^r_{\Delta_1,+\frac{3}{2}}(z_1,\bar{z}_1)\,\O_{abs,\Delta_2,-\frac{1}{2}}(z_2,\bar{z}_2)&=&\kappa_{+\frac{3}{2},-\frac{1}{2},+1}\,\frac{\bar{z}_{12}}{z_{12}}\,B\bigg(\Delta_1-\frac{1}{2} ,\Delta_2+\frac{3}{2}\bigg)\,\delta^{r}_{s}\O_{ab,\Delta_1+\Delta_2,-1}(z_2,\bar{z}_2) \nonu \\
&+& \cdots,
\nonu\\
\O^r_{\Delta_1,+\frac{3}{2}}(z_1,\bar{z}_1)\,\O_{ast,\Delta_2,-\frac{1}{2}}(z_2,\bar{z}_2)&=&\kappa_{+\frac{3}{2},-\frac{1}{2},+1}\,\frac{\bar{z}_{12}}{z_{12}}\,B\bigg(\Delta_1-\frac{1}{2} ,\Delta_2+\frac{3}{2}\bigg)\,2!\,
\nonu \\
& \times & \delta^{r}_{[s}\O_{t]a,\Delta_1+\Delta_2,-1}(z_2,\bar{z}_2) \nonu \\
&+& \cdots,
\nonu\\
\O^r_{\Delta_1,+\frac{3}{2}}(z_1,\bar{z}_1)\,\O_{stu,\Delta_2,-\frac{1}{2}}(z_2,\bar{z}_2)&=&\kappa_{+\frac{3}{2},-\frac{1}{2},+1}\,\frac{\bar{z}_{12}}{z_{12}}\,B\bigg(\Delta_1-\frac{1}{2} ,\Delta_2+\frac{3}{2}\bigg)\,3\,\delta^{r}_{[s}\O_{tu],\Delta_1+\Delta_2,-1}(z_2,\bar{z}_2) \nonu \\
&+& \cdots \, .
\nonu
\end{eqnarray}

$\bullet$ The gravitinos-the scalars

We have
\begin{eqnarray}
\O^A_{\Delta_1,+\frac{3}{2}}(z_1,\bar{z}_1)\,\O^{BCDE}_{\Delta_2,0}(z_2,\bar{z}_2)
&=&\kappa_{+\frac{3}{2},0,+\frac{1}{2}}\, \frac{\bar{z}_{12}}{z_{12}}\,B\bigg(\Delta_1-\frac{1}{2} ,\Delta_2+1\bigg)\,\frac{1}{3!}\,\epsilon^{ABCDEFGH}\,
\nonu \\
&\times & \O_{FGH,\Delta_1+\Delta_2,-\frac{1}{2}}(z_2,\bar{z}_2) 
+ \cdots \, .
\nonu
\end{eqnarray}
Or the following relations can be obtained
\begin{eqnarray}
\O^a_{\Delta_1,+\frac{3}{2}}(z_1,\bar{z}_1)\,\O^{bcdr}_{\Delta_2,0}(z_2,\bar{z}_2)&=&\kappa_{+\frac{3}{2},0,+\frac{1}{2}}\,\frac{\bar{z}_{12}}{z_{12}}\,B\bigg(\Delta_1-\frac{1}{2} ,\Delta_2+1\bigg)\,\frac{1}{3!}\,\epsilon^{abcd}\epsilon^{rstu}
\nonu \\
&\times & \O_{stu,\Delta_1+\Delta_2,-\frac{1}{2}}(z_2,\bar{z}_2) 
+ \cdots,
\nonu\\
\O^a_{\Delta_1,+\frac{3}{2}}(z_1,\bar{z}_1)\,\O^{bcrs}_{\Delta_2,0}(z_2,\bar{z}_2)&=&\kappa_{+\frac{3}{2},0,+\frac{1}{2}}\,\frac{\bar{z}_{12}}{z_{12}}\,B\bigg(\Delta_1-\frac{1}{2} ,\Delta_2+1\bigg)\,\frac{1}{2!}\,\epsilon^{abcd}\epsilon^{rstu}
\nonu \\
& \times & \O_{dtu,\Delta_1+\Delta_2,-\frac{1}{2}}(z_2,\bar{z}_2)+ \cdots,
\nonu\\
\O^a_{\Delta_1,+\frac{3}{2}}(z_1,\bar{z}_1)\,\O^{brst}_{\Delta_2,0}(z_2,\bar{z}_2)&=&\kappa_{+\frac{3}{2},0,+\frac{1}{2}}\,\frac{\bar{z}_{12}}{z_{12}}\,B\bigg(\Delta_1-\frac{1}{2} ,\Delta_2+1\bigg)\,\frac{1}{2!}\,\epsilon^{abcd}\epsilon^{rstu}
\nonu \\
&\times & \O_{cdu,\Delta_1+\Delta_2,-\frac{1}{2}}(z_2,\bar{z}_2) 
+ \cdots,
\nonu\\
\O^a_{\Delta_1,+\frac{3}{2}}(z_1,\bar{z}_1)\,\O^{rstu}_{\Delta_2,0}(z_2,\bar{z}_2)&=&\kappa_{+\frac{3}{2},0,+\frac{1}{2}}\,\frac{\bar{z}_{12}}{z_{12}}\,B\bigg(\Delta_1-\frac{1}{2} ,\Delta_2+1\bigg)\,\frac{1}{3!}\,\epsilon^{abcd}\epsilon^{rstu}
\nonu \\
& \times & \O_{bcd,\Delta_1+\Delta_2,-\frac{1}{2}}(z_2,\bar{z}_2) 
+ \cdots,
\nonu\\
\O^r_{\Delta_1,+\frac{3}{2}}(z_1,\bar{z}_1)\,\O^{abcd}_{\Delta_2,0}(z_2,\bar{z}_2)&=&\kappa_{+\frac{3}{2},0,+\frac{1}{2}}\,\frac{\bar{z}_{12}}{z_{12}}\,B\bigg(\Delta_1-\frac{1}{2} ,\Delta_2+1\bigg)\,\frac{1}{3!}\,\epsilon^{abcd}\epsilon^{rstu}
\nonu \\
& \times & \O_{stu,\Delta_1+\Delta_2,-\frac{1}{2}}(z_2,\bar{z}_2) 
+\cdots,
\nonu\\
\O^r_{\Delta_1,+\frac{3}{2}}(z_1,\bar{z}_1)\,\O^{abcs}_{\Delta_2,0}(z_2,\bar{z}_2)&=&-\kappa_{+\frac{3}{2},0,+\frac{1}{2}}\,\frac{\bar{z}_{12}}{z_{12}}\,B\bigg(\Delta_1-\frac{1}{2} ,\Delta_2+1\bigg)\,\frac{1}{2!}\,\epsilon^{abcd}\epsilon^{rstu}
\nonu \\
&\times & \O_{dtu,\Delta_1+\Delta_2,-\frac{1}{2}}(z_2,\bar{z}_2) 
+ \cdots,
\nonu\\
\O^r_{\Delta_1,+\frac{3}{2}}(z_1,\bar{z}_1)\,\O^{abst}_{\Delta_2,0}(z_2,\bar{z}_2)&=&\kappa_{+\frac{3}{2},0,+\frac{1}{2}}\,\frac{\bar{z}_{12}}{z_{12}}\,B\bigg(\Delta_1-\frac{1}{2} ,\Delta_2+1\bigg)\,\frac{1}{2!}\,\epsilon^{abcd}\epsilon^{rstu}
\nonu \\
&\times & \O_{cdu,\Delta_1+\Delta_2,-\frac{1}{2}}(z_2,\bar{z}_2) 
+ \cdots,
\nonu\\
\O^r_{\Delta_1,+\frac{3}{2}}(z_1,\bar{z}_1)\,\O^{astu}_{\Delta_2,0}(z_2,\bar{z}_2)&=&-\kappa_{+\frac{3}{2},0,+\frac{1}{2}}\,\frac{\bar{z}_{12}}{z_{12}}\,B\bigg(\Delta_1-\frac{1}{2} ,\Delta_2+1\bigg)\,\frac{1}{3!}\,\epsilon^{abcd}\epsilon^{rstu}
\nonu \\
& \times & \O_{bcd,\Delta_1+\Delta_2,-\frac{1}{2}}(z_2,\bar{z}_2) 
+ \cdots \, .
\nonu
\end{eqnarray}

$\bullet$ The graviphotons-the graviphotons

Similarly, the relations are satisfied
\begin{eqnarray}
\O^{AB}_{\Delta_1,+1}(z_1,\bar{z}_1)\,\O^{CD}_{\Delta_2,+1}(z_2,\bar{z}_2)&=&
\kappa_{+1,+1,0}\,
\frac{\bar{z}_{12}}{z_{12}}\,B\bigg(\Delta_1 ,\Delta_2\bigg)\,\O^{ABCD}_{\Delta_1+\Delta_2,0}(z_2,\bar{z}_2) + \cdots,
\nonu\\
\O^{AB}_{\Delta_1,+1}(z_1,\bar{z}_1)\,\O_{CD,\Delta_2,-1}(z_2,\bar{z}_2)&=&
\kappa_{+1,-1,+2}\,\frac{\bar{z}_{12}}{z_{12}}\,B\bigg(\Delta_1 ,\Delta_2+2\bigg)\,
\delta^{AB}_{CD}
\,\O_{\Delta_1+\Delta_2,-2}(z_2,\bar{z}_2) + \cdots \, .
\nonu
\end{eqnarray}
Equivalently, the following realtions hold 
\begin{eqnarray}
  \O^{ab}_{\Delta_1,+1}(z_1,\bar{z}_1)\,\O^{cd}_{\Delta_2,+1}(z_2,\bar{z}_2)&=&
  \kappa_{+1,+1,0}\,\frac{\bar{z}_{12}}{z_{12}}\,B\bigg(\Delta_1 ,\Delta_2\bigg)\,\O^{abcd}_{\Delta_1+\Delta_2,0}(z_2,\bar{z}_2) + \cdots,
\nonu\\
\O^{ab}_{\Delta_1,+1}(z_1,\bar{z}_1)\,\O^{cr}_{\Delta_2,+1}(z_2,\bar{z}_2)&=&
\kappa_{+1,+1,0}\,\frac{\bar{z}_{12}}{z_{12}}\,B\bigg(\Delta_1 ,\Delta_2\bigg)\,\O^{abcr}_{\Delta_1+\Delta_2,0}(z_2,\bar{z}_2) + \cdots,
\nonu\\
\O^{ab}_{\Delta_1,+1}(z_1,\bar{z}_1)\,\O^{rs}_{\Delta_2,+1}(z_2,\bar{z}_2)&=&
\kappa_{+1,+1,0}\,\frac{\bar{z}_{12}}{z_{12}}\,B\bigg(\Delta_1 ,\Delta_2\bigg)\,\O^{abrs}_{\Delta_1+\Delta_2,0}(z_2,\bar{z}_2) + \cdots,
\nonu\\
\O^{ar}_{\Delta_1,+1}(z_1,\bar{z}_1)\,\O^{bs}_{\Delta_2,+1}(z_2,\bar{z}_2)&=&
\kappa_{+1,+1,0}\,\frac{\bar{z}_{12}}{z_{12}}\,B\bigg(\Delta_1 ,\Delta_2\bigg)\,\O^{arbs}_{\Delta_1+\Delta_2,0}(z_2,\bar{z}_2) + \cdots,
\nonu\\
\O^{ar}_{\Delta_1,+1}(z_1,\bar{z}_1)\,\O^{st}_{\Delta_2,+1}(z_2,\bar{z}_2)&=&
\kappa_{+1,+1,0}\,\frac{\bar{z}_{12}}{z_{12}}\,B\bigg(\Delta_1 ,\Delta_2\bigg)\,\O^{arst}_{\Delta_1+\Delta_2,0}(z_2,\bar{z}_2) + \cdots,
\nonu\\
\O^{rs}_{\Delta_1,+1}(z_1,\bar{z}_1)\,\O^{tu}_{\Delta_2,+1}(z_2,\bar{z}_2)&=&
\kappa_{+1,+1,0}\,\frac{\bar{z}_{12}}{z_{12}}\,B\bigg(\Delta_1 ,\Delta_2\bigg)\,\O^{rstu}_{\Delta_1+\Delta_2,0}(z_2,\bar{z}_2) + \cdots,
\nonu\\
\O^{ab}_{\Delta_1,+1}(z_1,\bar{z}_1)\,\O_{cd,\Delta_2,-1}(z_2,\bar{z}_2)&=&
\kappa_{+1,-1,+2}\,\frac{\bar{z}_{12}}{z_{12}}\,B\bigg(\Delta_1 ,\Delta_2+2\bigg)\,\delta^{ab}_{cd}
\,\O_{\Delta_1+\Delta_2,-2}(z_2,\bar{z}_2) + \cdots,
\nonu\\
\O^{ar}_{\Delta_1,+1}(z_1,\bar{z}_1)\,\O_{bs,\Delta_2,-1}(z_2,\bar{z}_2)&=&
\kappa_{+1,-1,+2}\,\frac{\bar{z}_{12}}{z_{12}}\,B\bigg(\Delta_1 ,\Delta_2+2\bigg)\,\delta^a_b \delta^r_s\,\O_{\Delta_1+\Delta_2,-2}(z_2,\bar{z}_2) + \cdots,
\nonu\\
\O^{rs}_{\Delta_1,+1}(z_1,\bar{z}_1)\,\O_{tu,\Delta_2,-1}(z_2,\bar{z}_2)&=&
\kappa_{+1,-1,+2}\,\frac{\bar{z}_{12}}{z_{12}}\,B\bigg(\Delta_1 ,\Delta_2+2\bigg)\,
\delta^{rs}_{tu}\,
\O_{\Delta_1+\Delta_2,-2}(z_2,\bar{z}_2) + \cdots \, .
\nonu
\end{eqnarray}

$\bullet$ The graviphotons-the graviphotinos

There exist
\begin{eqnarray}
\O^{AB}_{\Delta_1,+1}(z_1,\bar{z}_1)\,\O^{CDE}_{\Delta_2,+\frac{1}{2}}(z_2,\bar{z}_2)&=&\kappa_{+1,+\frac{1}{2},+\frac{1}{2}}\,\frac{\bar{z}_{12}}{z_{12}}\,B\bigg(\Delta_1 ,\Delta_2+\frac{1}{2}\bigg)\,\frac{1}{3!}\,\epsilon^{ABCDEFGH}\,
\nonu \\
& \times & \O_{FGH,\Delta_1+\Delta_2,-\frac{1}{2}}(z_2,\bar{z}_2)
+\cdots,
\nonu\\
\O^{AB}_{\Delta_1,+1}(z_1,\bar{z}_1)\,\O_{CDE,\Delta_2,-\frac{1}{2}}(z_2,\bar{z}_2)&=&\kappa_{+1,-\frac{1}{2},+\frac{3}{2}}\,\frac{\bar{z}_{12}}{z_{12}}\,B\bigg(\Delta_1 ,\Delta_2+\frac{3}{2}\bigg)\,3!\,
\nonu \\
& \times & \delta^{A}_{[C}\,\O_{D,\Delta_1+\Delta_2,-\frac{3}{2}}(z_2,\bar{z}_2)\delta_{E]}^{B} +
\cdots \, .
\nonu
\end{eqnarray}
Moreover, the following relations satisfy
\begin{eqnarray}
  \O^{ab}_{\Delta_1,+1}(z_1,\bar{z}_1)\,\O^{cdr}_{\Delta_2,+\frac{1}{2}}(z_2,\bar{z}_2)&=&\kappa_{+1,+\frac{1}{2},+\frac{1}{2}}\,\frac{\bar{z}_{12}}{z_{12}}\,B\bigg(\Delta_1 ,\Delta_2+\frac{1}{2}\bigg)\,\frac{1}{3!}\,\epsilon^{abcd}\epsilon^{rstu}\,
\nonu \\
& \times & \O_{stu,\Delta_1+\Delta_2,-\frac{1}{2}}(z_2,\bar{z}_2)
+ \cdots,
\nonu\\
\bf
\O^{ab}_{\Delta_1,+1}(z_1,\bar{z}_1)\,\O^{crs}_{\Delta_2,+\frac{1}{2}}(z_2,\bar{z}_2)&=&
\bf \kappa_{+1,+\frac{1}{2},+\frac{1}{2}}\,\frac{\bar{z}_{12}}{z_{12}}\,B\bigg(\Delta_1 ,\Delta_2+\frac{1}{2}\bigg)\,\frac{1}{2!}\,\epsilon^{abcd}\epsilon^{rstu}\,
\nonu \\
& \times & \bf \O_{dtu,\Delta_1+\Delta_2,-\frac{1}{2}}(z_2,\bar{z}_2)
+ \cdots,
\nonu\\
\bf
\O^{ab}_{\Delta_1,+1}(z_1,\bar{z}_1)\,\O^{rst}_{\Delta_2,+\frac{1}{2}}(z_2,\bar{z}_2)&=&
\bf \kappa_{+1,+\frac{1}{2},+\frac{1}{2}}\,
\frac{\bar{z}_{12}}{z_{12}}\,B\bigg(\Delta_1 ,\Delta_2+\frac{1}{2}\bigg)\,\frac{1}{2!}\,\epsilon^{abcd}\epsilon^{rstu}\,
\nonu \\
& \times & \bf \O_{cdu,\Delta_1+\Delta_2,-\frac{1}{2}}(z_2,\bar{z}_2)
+   \cdots,
\nonu\\
\bf
\O^{ar}_{\Delta_1,+1}(z_1,\bar{z}_1)\,\O^{bcd}_{\Delta_2,+\frac{1}{2}}(z_2,\bar{z}_2)&=&
\bf 
-\kappa_{+1,+\frac{1}{2},+\frac{1}{2}}\,\frac{\bar{z}_{12}}{z_{12}}\,B\bigg(\Delta_1 ,\Delta_2+\frac{1}{2}\bigg)\,\frac{1}{3!}\,\epsilon^{abcd}\epsilon^{rstu}\,
\nonu \\
& \times & \bf \O_{stu,\Delta_1+\Delta_2,-\frac{1}{2}}(z_2,\bar{z}_2)
+   \cdots,
\nonu \\
\bf
\O^{ar}_{\Delta_1,+1}(z_1,\bar{z}_1)\,\O^{bcs}_{\Delta_2,+\frac{1}{2}}(z_2,\bar{z}_2)&=&
\bf \kappa_{+1,+\frac{1}{2},+\frac{1}{2}}\,
\frac{\bar{z}_{12}}{z_{12}}\,B\bigg(\Delta_1 ,\Delta_2+\frac{1}{2}\bigg)\,\frac{1}{2!}\,\epsilon^{abcd}\epsilon^{rstu}\,
\nonu \\
& \times & \bf \O_{dtu,\Delta_1+\Delta_2,-\frac{1}{2}}(z_2,\bar{z}_2)
+  \cdots,
\nonu \\
\bf
\O^{ar}_{\Delta_1,+1}(z_1,\bar{z}_1)\,\O^{bst}_{\Delta_2,+\frac{1}{2}}(z_2,\bar{z}_2)&=&
\bf
-\kappa_{+1,+\frac{1}{2},+\frac{1}{2}}\,\frac{\bar{z}_{12}}{z_{12}}\,B\bigg(\Delta_1 ,\Delta_2+\frac{1}{2}\bigg)\,\frac{1}{2!}\,\epsilon^{abcd}\epsilon^{rstu}\,
\nonu \\
& \times & \bf \O_{cdu,\Delta_1+\Delta_2,-\frac{1}{2}}(z_2,\bar{z}_2)
+  \cdots,
\nonu \\
\bf
\O^{ar}_{\Delta_1,+1}(z_1,\bar{z}_1)\,\O^{stu}_{\Delta_2,+\frac{1}{2}}(z_2,\bar{z}_2)&=&
\bf \kappa_{+1,+\frac{1}{2},+\frac{1}{2}}\,
\frac{\bar{z}_{12}}{z_{12}}\,B\bigg(\Delta_1 ,\Delta_2+\frac{1}{2}\bigg)\,\frac{1}{3!}\,\epsilon^{abcd}\epsilon^{rstu}\,
\nonu \\
& \times & \bf \O_{bcd,\Delta_1+\Delta_2,-\frac{1}{2}}(z_2,\bar{z}_2)
+ \cdots,
\nonu \\
\bf
\O^{rs}_{\Delta_1,+1}(z_1,\bar{z}_1)\,\O^{abc}_{\Delta_2,+\frac{1}{2}}(z_2,\bar{z}_2)&=&
\bf \kappa_{+1,+\frac{1}{2},+\frac{1}{2}}\,
\frac{\bar{z}_{12}}{z_{12}}\,B\bigg(\Delta_1 ,\Delta_2+\frac{1}{2}\bigg)\,\frac{1}{2!}\,\epsilon^{abcd}\epsilon^{rstu}\,
\nonu \\
& \times & \bf \O_{dtu,\Delta_1+\Delta_2,-\frac{1}{2}}(z_2,\bar{z}_2)
+ \cdots,
\nonu \\
\bf
\O^{rs}_{\Delta_1,+1}(z_1,\bar{z}_1)\,\O^{abt}_{\Delta_2,+\frac{1}{2}}(z_2,\bar{z}_2)&=&
\bf \kappa_{+1,+\frac{1}{2},+\frac{1}{2}}\,
\frac{\bar{z}_{12}}{z_{12}}\,B\bigg(\Delta_1 ,\Delta_2+\frac{1}{2}\bigg)\,\frac{1}{2!}\,\epsilon^{abcd}\epsilon^{rstu}\,
\nonu \\
& \times & \bf \O_{cdu,\Delta_1+\Delta_2,-\frac{1}{2}}(z_2,\bar{z}_2)
+  \cdots,
\nonu \\
\bf \O^{rs}_{\Delta_1,+1}(z_1,\bar{z}_1)\,\O^{atu}_{\Delta_2,+\frac{1}{2}}(z_2,\bar{z}_2)&=&
\bf \kappa_{+1,+\frac{1}{2},+\frac{1}{2}}\,
\frac{\bar{z}_{12}}{z_{12}}\,B\bigg(\Delta_1 ,\Delta_2+\frac{1}{2}\bigg)\,\frac{1}{3!}\,\epsilon^{abcd}\epsilon^{rstu}\,
\nonu \\
& \times & \bf \O_{bcd,\Delta_1+\Delta_2,-\frac{1}{2}}(z_2,\bar{z}_2)
+  \cdots,
\nonu \\
\O^{ab}_{\Delta_1,+1}(z_1,\bar{z}_1)\,\O_{cde,\Delta_2,-\frac{1}{2}}(z_2,\bar{z}_2)&=&
\kappa_{+1,-\frac{1}{2},+\frac{3}{2}}\,\frac{\bar{z}_{12}}{z_{12}}\,B\bigg(\Delta_1 ,\Delta_2+\frac{3}{2}\bigg)\,3!\,\delta^{a}_{[c}\,\O_{d,\Delta_1+\Delta_2,-\frac{3}{2}}(z_2,\bar{z}_2)\delta^b_{e]} \nonu \\
&+&  \cdots,
\nonu \\
\O^{ab}_{\Delta_1,+1}(z_1,\bar{z}_1)\,\O_{cdr,\Delta_2,-\frac{1}{2}}(z_2,\bar{z}_2)&=&
-\kappa_{+1,-\frac{1}{2},+\frac{3}{2}}\,\frac{\bar{z}_{12}}{z_{12}}\,B\bigg(\Delta_1 ,\Delta_2+\frac{3}{2}\bigg)\,\delta^{ab}_{cd}\,\O_{r,\Delta_1+\Delta_2,-\frac{3}{2}}(z_2,\bar{z}_2)
\nonu \\
&+& \cdots,
\nonu \\
\O^{ar}_{\Delta_1,+1}(z_1,\bar{z}_1)\,\O_{bcs,\Delta_2,-\frac{1}{2}}(z_2,\bar{z}_2)&=&
\kappa_{+1,-\frac{1}{2},+\frac{3}{2}}\,\frac{\bar{z}_{12}}{z_{12}}\,B\bigg(\Delta_1 ,\Delta_2+\frac{3}{2}\bigg)\,2!\,\delta^{r}_{s}\delta^{a}_{[b}\,\O_{c],\Delta_1+\Delta_2,-\frac{3}{2}}(z_2,\bar{z}_2) \nonu \\
&+& \cdots,
\nonu \\
\O^{ar}_{\Delta_1,+1}(z_1,\bar{z}_1)\,\O_{bst,\Delta_2,-\frac{1}{2}}(z_2,\bar{z}_2)&=&
-\kappa_{+1,-\frac{1}{2},+\frac{3}{2}}\,\frac{\bar{z}_{12}}{z_{12}}\,B\bigg(\Delta_1 ,\Delta_2+\frac{3}{2}\bigg)\,2!\,\delta^{a}_b\delta^{r}_{[s}\,\O_{t],\Delta_1+\Delta_2,-\frac{3}{2}}(z_2,\bar{z}_2)
\nonu \\
&+&  \cdots,
\nonu \\
\O^{rs}_{\Delta_1,+1}(z_1,\bar{z}_1)\,\O_{atu,\Delta_2,-\frac{1}{2}}(z_2,\bar{z}_2)&=&
-\kappa_{+1,-\frac{1}{2},+\frac{3}{2}}\,\frac{\bar{z}_{12}}{z_{12}}\,B\bigg(\Delta_1 ,\Delta_2+\frac{3}{2}\bigg)\,\delta^{rs}_{tu}\,\O_{a,\Delta_1+\Delta_2,-\frac{3}{2}}(z_2,\bar{z}_2) \nonu \\
&+& 
\cdots,
\nonu \\
\O^{rs}_{\Delta_1,+1}(z_1,\bar{z}_1)\,\O_{tuv,\Delta_2,-\frac{1}{2}}(z_2,\bar{z}_2)&=&
\kappa_{+1,-\frac{1}{2},+\frac{3}{2}}\,\frac{\bar{z}_{12}}{z_{12}}\,B\bigg(\Delta_1 ,\Delta_2+\frac{3}{2}\bigg)\,3!\,\delta^{r}_{[t}\,\O_{u,\Delta_1+\Delta_2,-\frac{3}{2}}(z_2,\bar{z}_2)\delta^s_{v]}
\nonu \\
&+& \cdots \, .
\nonu
\end{eqnarray}
There are  nine additional OPEs.

$\bullet$ The graviphotons-the scalars

We have 
\begin{eqnarray}
\O^{AB}_{\Delta_1,+1}(z_1,\bar{z}_1)\,\O^{CDEF}_{\Delta_2,0}(z_2,\bar{z}_2)&=&
\kappa_{+1,0,+1}\,\frac{\bar{z}_{12}}{z_{12}}\,B\bigg(\Delta_1 ,\Delta_2+1\bigg)\,\epsilon^{ABCDEFGH}\,\frac{1}{2!}\,
\nonu \\
& \times & \O_{GH,\Delta_1+\Delta_2,-1}(z_2,\bar{z}_2) 
+ \cdots \, .
\nonu
\end{eqnarray}
Moreover, the following relations are satisfied
\begin{eqnarray}
  \O^{ab}_{\Delta_1,+1}(z_1,\bar{z}_1)\,\O^{cdrs}_{\Delta_2,0}(z_2,\bar{z}_2)&=&
  \kappa_{+1,0,+1}\,\frac{\bar{z}_{12}}{z_{12}}\,B\bigg(\Delta_1 ,\Delta_2+1\bigg)\,\epsilon^{abcd}\epsilon^{rstu}\,\frac{1}{2!}\,\O_{tu,\Delta_1+\Delta_2,-1}(z_2,\bar{z}_2) \nonu \\
&+ & \cdots,
\nonu\\
\bf
\O^{ab}_{\Delta_1,+1}(z_1,\bar{z}_1)\,\O^{crst}_{\Delta_2,0}(z_2,\bar{z}_2)&=&
\bf
-\kappa_{+1,0,+1}\,\frac{\bar{z}_{12}}{z_{12}}\,B\bigg(\Delta_1 ,\Delta_2+1\bigg)\,\epsilon^{abcd}\epsilon^{rstu}\,
\nonu \\
& \times & \bf \O_{du,\Delta_1+\Delta_2,-1}(z_2,\bar{z}_2) + \cdots,
\nonu\\
\bf
\O^{ab}_{\Delta_1,+1}(z_1,\bar{z}_1)\,\O^{rstu}_{\Delta_2,0}(z_2,\bar{z}_2)&=&
\bf
\kappa_{+1,0,+1}\,\frac{\bar{z}_{12}}{z_{12}}\,B\bigg(\Delta_1 ,\Delta_2+1\bigg)\,\epsilon^{abcd}\epsilon^{rstu}\,
\nonu \\
&\times &
\bf \frac{1}{2!}\,\O_{cd,\Delta_1+\Delta_2,-1}(z_2,\bar{z}_2) + \cdots,
\nonu\\
\bf
\O^{ar}_{\Delta_1,+1}(z_1,\bar{z}_1)\,\O^{bcds}_{\Delta_2,0}(z_2,\bar{z}_2)&=&
\bf
-\kappa_{+1,0,+1}\,\frac{\bar{z}_{12}}{z_{12}}\,B\bigg(\Delta_1 ,\Delta_2+1\bigg)\,\epsilon^{abcd}\epsilon^{rstu}\,\nonu \\
& \times & \bf
\frac{1}{2!}\,\O_{tu,\Delta_1+\Delta_2,-1}(z_2,\bar{z}_2) + \cdots,
\nonu\\
\bf
\O^{ar}_{\Delta_1,+1}(z_1,\bar{z}_1)\,\O^{bcst}_{\Delta_2,0}(z_2,\bar{z}_2)&=&
\bf
-\kappa_{+1,0,+1}\,\frac{\bar{z}_{12}}{z_{12}}\,B\bigg(\Delta_1 ,\Delta_2+1\bigg)\,\epsilon^{abcd}\epsilon^{rstu}\,
\nonu \\
& \times & \bf \O_{du,\Delta_1+\Delta_2,-1}(z_2,\bar{z}_2) + \cdots,
\nonu\\
\bf
\O^{ar}_{\Delta_1,+1}(z_1,\bar{z}_1)\,\O^{bstu}_{\Delta_2,0}(z_2,\bar{z}_2)&=&
\bf - \kappa_{+1,0,+1}\,\frac{\bar{z}_{12}}{z_{12}}\,B\bigg(\Delta_1 ,\Delta_2+1\bigg)\,\epsilon^{abcd}\epsilon^{rstu}\,\nonu \\
&\times&
\bf\frac{1}{2!}\,\O_{cd,\Delta_1+\Delta_2,-1}(z_2,\bar{z}_2) + \cdots,
\nonu\\
\bf
\O^{rs}_{\Delta_1,+1}(z_1,\bar{z}_1)\,\O^{abcd}_{\Delta_2,0}(z_2,\bar{z}_2)&=&
\bf
\kappa_{+1,0,+1}\,\frac{\bar{z}_{12}}{z_{12}}\,B\bigg(\Delta_1 ,\Delta_2+1\bigg)\,\epsilon^{abcd}\epsilon^{rstu}\,
\nonu \\
& \times &
\bf \frac{1}{2!}\,\O_{tu,\Delta_1+\Delta_2,-1}(z_2,\bar{z}_2) + \cdots,
\nonu\\
\bf
\O^{rs}_{\Delta_1,+1}(z_1,\bar{z}_1)\,\O^{abct}_{\Delta_2,0}(z_2,\bar{z}_2)&=&
\bf
-\kappa_{+1,0,+1}\,\frac{\bar{z}_{12}}{z_{12}}\,B\bigg(\Delta_1 ,\Delta_2+1\bigg)\,\epsilon^{abcd}\epsilon^{rstu}\,
\nonu \\
&\times & \bf \O_{du,\Delta_1+\Delta_2,-1}(z_2,\bar{z}_2) + \cdots,
\nonu\\
\bf
\O^{rs}_{\Delta_1,+1}(z_1,\bar{z}_1)\,\O^{abtu}_{\Delta_2,0}(z_2,\bar{z}_2)&=&
\bf
\kappa_{+1,0,+1}\,\frac{\bar{z}_{12}}{z_{12}}\,B\bigg(\Delta_1 ,\Delta_2+1\bigg)\,\epsilon^{abcd}\epsilon^{rstu}\nonu \\
& \times & \bf
\frac{1}{2!}\,\O_{cd,\Delta_1+\Delta_2,-1}(z_2,\bar{z}_2) \nonu \\
&+& \bf \cdots \, .
\nonu
\end{eqnarray}
The eight additional OPEs exist.

$\bullet$ The graviphotinos-the graviphotinos

Similarly, we have
\begin{eqnarray}
  \O^{ABC}_{\Delta_1,+\frac{1}{2}}(z_1,\bar{z}_1)\,\O^{DEF}_{\Delta_2,+\frac{1}{2}}(z_2,\bar{z}_2)&=&\kappa_{+\frac{1}{2},+\frac{1}{2},+1}\,
\frac{\bar{z}_{12}}{z_{12}}\,B\bigg(\Delta_1+ \frac{1}{2} ,\Delta_2+\frac{1}{2}\bigg)\,\frac{1}{2!}\,\epsilon^{ABCDEFGH}\,
\nonu \\
&\times&  \O_{GH,\Delta_1+\Delta_2,-1}(z_2,\bar{z}_2)+ \cdots,
\nonu\\
\O^{ABC}_{\Delta_1,+\frac{1}{2}}(z_1,\bar{z}_1)\,\O_{DEF,\Delta_2,-\frac{1}{2}}(z_2,\bar{z}_2)&=&\kappa_{+\frac{1}{2},-\frac{1}{2},+2}\,\frac{\bar{z}_{12}}{z_{12}}\,B\bigg(\Delta_1+ \frac{1}{2} ,\Delta_2+\frac{3}{2}\bigg)\,\delta^{ABC}_{DEF}\,
\nonu \\
& \times & \O_{\Delta_1+\Delta_2,-2}(z_2,\bar{z}_2) 
+ \cdots \, .
\nonu\
\end{eqnarray}
In components, the relations are 
\begin{eqnarray}
\bf
\O^{abc}_{\Delta_1,+\frac{1}{2}}(z_1,\bar{z}_1)\,\O^{drs}_{\Delta_2,+\frac{1}{2}}(z_2,\bar{z}_2)&=&
\bf \kappa_{+\frac{1}{2},+\frac{1}{2},+1}\,
\frac{\bar{z}_{12}}{z_{12}}\,B\bigg(\Delta_1+ \frac{1}{2} ,\Delta_2+\frac{1}{2}\bigg)\,\epsilon^{abcd}\epsilon^{rstu}\,
\nonu \\
&\times & \bf \frac{1}{2!}\,\O_{tu,\Delta_1+\Delta_2,-1}(z_2,\bar{z}_2)
+  \cdots,
\nonu\\
\O^{abc}_{\Delta_1,+\frac{1}{2}}(z_1,\bar{z}_1)\,\O^{rst}_{\Delta_2,+\frac{1}{2}}(z_2,\bar{z}_2)&=&-\kappa_{+\frac{1}{2},+\frac{1}{2},+1}\,\frac{\bar{z}_{12}}{z_{12}}\,B\bigg(\Delta_1+ \frac{1}{2} ,\Delta_2+\frac{1}{2}\bigg)\,\epsilon^{abcd}\epsilon^{rstu}\,\nonu \\
& \times & \O_{du,\Delta_1+\Delta_2,-1}(z_2,\bar{z}_2)
+ \cdots,
\nonu\\
\bf
\O^{abr}_{\Delta_1,+\frac{1}{2}}(z_1,\bar{z}_1)\,\O^{cds}_{\Delta_2,+\frac{1}{2}}(z_2,\bar{z}_2)&=&
\bf \kappa_{+\frac{1}{2},+\frac{1}{2},+1}\,
\frac{\bar{z}_{12}}{z_{12}}\,B\bigg(\Delta_1+ \frac{1}{2} ,\Delta_2+\frac{1}{2}\bigg)\,\epsilon^{abcd}\epsilon^{rstu}\,
\nonu \\
& \times &
\bf \frac{1}{2!}\,\O_{tu,\Delta_1+\Delta_2,-1}(z_2,\bar{z}_2)
+  \cdots,
\nonu\\
\O^{abr}_{\Delta_1,+\frac{1}{2}}(z_1,\bar{z}_1)\,\O^{cst}_{\Delta_2,+\frac{1}{2}}(z_2,\bar{z}_2)&=&\kappa_{+\frac{1}{2},+\frac{1}{2},+1}\,\frac{\bar{z}_{12}}{z_{12}}\,B\bigg(\Delta_1+ \frac{1}{2} ,\Delta_2+\frac{1}{2}\bigg)\,\epsilon^{abcd}\epsilon^{rstu}\,\nonu \\
& \times & \O_{du,\Delta_1+\Delta_2,-1}(z_2,\bar{z}_2) 
+ \cdots,
\nonu\\
\bf
\O^{abr}_{\Delta_1,+\frac{1}{2}}(z_1,\bar{z}_1)\,\O^{stu}_{\Delta_2,+\frac{1}{2}}(z_2,\bar{z}_2)&=&
\bf \kappa_{+\frac{1}{2},+\frac{1}{2},+1}\,
\frac{\bar{z}_{12}}{z_{12}}\,B\bigg(\Delta_1+ \frac{1}{2} ,\Delta_2+\frac{1}{2}\bigg)\,\epsilon^{abcd}\epsilon^{rstu}\,
\nonu \\
& \times &
\bf \frac{1}{2!}\,\O_{cd,\Delta_1+\Delta_2,-1}(z_2,\bar{z}_2) 
+  \cdots,
\nonu\\
\O^{ars}_{\Delta_1,+\frac{1}{2}}(z_1,\bar{z}_1)\,\O^{btu}_{\Delta_2,+\frac{1}{2}}(z_2,\bar{z}_2)&=&\kappa_{+\frac{1}{2},+\frac{1}{2},+1}\,\frac{\bar{z}_{12}}{z_{12}}\,B\bigg(\Delta_1+ \frac{1}{2} ,\Delta_2+\frac{1}{2}\bigg)\,\epsilon^{abcd}\epsilon^{rstu}\,\frac{1}{2!}\,\nonu \\
& \times & \O_{cd,\Delta_1+\Delta_2,-1}(z_2,\bar{z}_2) 
+ \cdots,
\nonu\\
\bf
\O^{abc}_{\Delta_1,+\frac{1}{2}}(z_1,\bar{z}_1)\,\O_{def,\Delta_2,-\frac{1}{2}}(z_2,\bar{z}_2)&=&
\bf \kappa_{+\frac{1}{2},-\frac{1}{2},+2}\,
\frac{\bar{z}_{12}}{z_{12}}\,B\bigg(\Delta_1+ \frac{1}{2} ,\Delta_2+\frac{3}{2}\bigg)\,\delta^{abc}_{def}\,\nonu \\
&\times & \bf
\O_{\Delta_1+\Delta_2,-2}(z_2,\bar{z}_2) 
+  \cdots,
\nonu\\
\bf
\O^{abr}_{\Delta_1,+\frac{1}{2}}(z_1,\bar{z}_1)\,\O_{cds,\Delta_2,-\frac{1}{2}}(z_2,\bar{z}_2)&=&
\bf  \kappa_{+\frac{1}{2},-\frac{1}{2},+2}\,
\frac{\bar{z}_{12}}{z_{12}}\,B\bigg(\Delta_1+ \frac{1}{2} ,\Delta_2+\frac{3}{2}\bigg)\,\delta^{abr}_{cds}\,
\nonu \\
& \times & \bf \O_{\Delta_1+\Delta_2,-2}(z_2,\bar{z}_2) 
+  \cdots,
\nonu\\
\O^{ars}_{\Delta_1,+\frac{1}{2}}(z_1,\bar{z}_1)\,\O_{btu,\Delta_2,-\frac{1}{2}}(z_2,\bar{z}_2)&=& \kappa_{+\frac{1}{2},-\frac{1}{2},+2}\,\frac{\bar{z}_{12}}{z_{12}}\,B\bigg(\Delta_1+ \frac{1}{2} ,\Delta_2+\frac{3}{2}\bigg)\,\delta^{ars}_{btu}\,
\nonu \\
& \times & \O_{\Delta_1+\Delta_2,-2}(z_2,\bar{z}_2) 
+ \cdots,
\nonu\\
\bf
\O^{rst}_{\Delta_1,+\frac{1}{2}}(z_1,\bar{z}_1)\,\O_{uvw,\Delta_2,-\frac{1}{2}}(z_2,\bar{z}_2)&=&
\bf  \kappa_{+\frac{1}{2},-\frac{1}{2},+2}\,
\frac{\bar{z}_{12}}{z_{12}}\,B\bigg(\Delta_1+ \frac{1}{2} ,\Delta_2+\frac{3}{2}\bigg)\,\delta^{rst}_{uvw}\,
\nonu \\
& \times & \bf \O_{\Delta_1+\Delta_2,-2}(z_2,\bar{z}_2)
+  \cdots \, .
\nonu
\end{eqnarray}
The six additional OPEs appear. See also the footnote
\ref{missingsplitfactor}.
The corresponding split factors should appear in Appendix $B$ of
\cite{2212-1}.  

$\bullet$ The graviphotinos-the scalars

The relations are given by
\begin{eqnarray}
\O^{ABC}_{\Delta_1,+\frac{1}{2}}(z_1,\bar{z}_1)\,\O^{DEFG}_{\Delta_2,0}(z_2,\bar{z}_2)&=&\frac{\bar{z}_{12}}{z_{12}}\,B\bigg(\Delta_1+ \frac{1}{2} ,\Delta_2+1\bigg)\,\epsilon^{ABCDEFGH}\,\O_{H,\Delta_1+\Delta_2,-\frac{3}{2}}(z_2,\bar{z}_2) \nonu \\
&+& \cdots \, .
\nonu
\end{eqnarray}
Or we have the following relations
\begin{eqnarray}
\bf
\O^{abc}_{\Delta_1,+\frac{1}{2}}(z_1,\bar{z}_1)\,\O^{drst}_{\Delta_2,0}(z_2,\bar{z}_2)&=&
\bf
\frac{\bar{z}_{12}}{z_{12}}\,B\bigg(\Delta_1+ \frac{1}{2} ,\Delta_2+1\bigg)\,\epsilon^{abcd}\epsilon^{rstu}\,\O_{u,\Delta_1+\Delta_2,-\frac{3}{2}}(z_2,\bar{z}_2) \nonu \\
&+ & \cdots,
\nonu\\
\bf
\O^{abc}_{\Delta_1,+\frac{1}{2}}(z_1,\bar{z}_1)\,\O^{rstu}_{\Delta_2,0}(z_2,\bar{z}_2)&=&
\bf
\frac{\bar{z}_{12}}{z_{12}}\,B\bigg(\Delta_1+ \frac{1}{2} ,\Delta_2+1\bigg)\,\epsilon^{abcd}\epsilon^{rstu}\,\O_{d,\Delta_1+\Delta_2,-\frac{3}{2}}(z_2,\bar{z}_2)\nonu \\
&+& \cdots,
\nonu\\
\O^{abr}_{\Delta_1,+\frac{1}{2}}(z_1,\bar{z}_1)\,\O^{cdst}_{\Delta_2,0}(z_2,\bar{z}_2)&=&\frac{\bar{z}_{12}}{z_{12}}\,B\bigg(\Delta_1+ \frac{1}{2} ,\Delta_2+1\bigg)\,\epsilon^{abcd}\epsilon^{rstu}\,\O_{u,\Delta_1+\Delta_2,-\frac{3}{2}}(z_2,\bar{z}_2) + \cdots,
\nonu\\
\O^{abr}_{\Delta_1,+\frac{1}{2}}(z_1,\bar{z}_1)\,\O^{cstu}_{\Delta_2,0}(z_2,\bar{z}_2)&=&-\frac{\bar{z}_{12}}{z_{12}}\,B\bigg(\Delta_1+ \frac{1}{2} ,\Delta_2+1\bigg)\,\epsilon^{abcd}\epsilon^{rstu}\,\O_{d,\Delta_1+\Delta_2,-\frac{3}{2}}(z_2,\bar{z}_2) + \cdots,
\nonu\\
\bf
\O^{ars}_{\Delta_1,+\frac{1}{2}}(z_1,\bar{z}_1)\,\O^{bcdt}_{\Delta_2,0}(z_2,\bar{z}_2)&=&
\bf
\frac{\bar{z}_{12}}{z_{12}}\,B\bigg(\Delta_1+ \frac{1}{2} ,\Delta_2+1\bigg)\,\epsilon^{abcd}\epsilon^{rstu}\,\O_{u,\Delta_1+\Delta_2,-\frac{3}{2}}(z_2,\bar{z}_2) \nonu \\
&+& \cdots,
\nonu\\
\bf
\O^{ars}_{\Delta_1,+\frac{1}{2}}(z_1,\bar{z}_1)\,\O^{bctu}_{\Delta_2,0}(z_2,\bar{z}_2)&=&
\bf \frac{\bar{z}_{12}}{z_{12}}\,B\bigg(\Delta_1+ \frac{1}{2} ,\Delta_2+1\bigg)\,\epsilon^{abcd}\epsilon^{rstu}\,\O_{d,\Delta_1+\Delta_2,-\frac{3}{2}}(z_2,\bar{z}_2) \nonu \\
&+& \cdots,
\nonu\\
\bf
\O^{rst}_{\Delta_1,+\frac{1}{2}}(z_1,\bar{z}_1)\,\O^{abcd}_{\Delta_2,0}(z_2,\bar{z}_2)&=&
\bf
\frac{\bar{z}_{12}}{z_{12}}\,B\bigg(\Delta_1+ \frac{1}{2} ,\Delta_2+1\bigg)\,\epsilon^{abcd}\epsilon^{rstu}\,\O_{u,\Delta_1+\Delta_2,-\frac{3}{2}}(z_2,\bar{z}_2)\nonu \\
&+& \cdots,
\nonu\\
\bf
\O^{rst}_{\Delta_1,+\frac{1}{2}}(z_1,\bar{z}_1)\,\O^{abcu}_{\Delta_2,0}(z_2,\bar{z}_2)&=&
\bf
-\frac{\bar{z}_{12}}{z_{12}}\,B\bigg(\Delta_1+ \frac{1}{2} ,\Delta_2+1\bigg)\,\epsilon^{abcd}\epsilon^{rstu}\,\O_{d,\Delta_1+\Delta_2,-\frac{3}{2}}(z_2,\bar{z}_2) \nonu \\
&+& \cdots \, .
\nonu
\end{eqnarray}
The six additional OPEs occur.
See also the footnote \ref{missingsplitfactor} and 
the corresponding split factors
for the first two OPEs should appear in Appendix $B$ of
\cite{2212-1}.  

$\bullet$ The scalars-the scalars

The following relation satisfies
\begin{eqnarray}   
\O^{ABCD}_{\Delta_1,0}(z_1,\bar{z}_1)\,\O^{EFGH}_{\Delta_2,0}(z_2,\bar{z}_2)&=&\frac{\bar{z}_{12}}{z_{12}}\,B\bigg(\Delta_1+ 1 ,\Delta_2+1\bigg)\,\epsilon^{ABCDEFGH}\,\O_{\Delta_1+\Delta_2,-2}(z_2,\bar{z}_2) + \cdots\, .
\nonu
\end{eqnarray}
Or we have 
\begin{eqnarray}   
\bf
\O^{abcd}_{\Delta_1,0}(z_1,\bar{z}_1)\,\O^{rstu}_{\Delta_2,0}(z_2,\bar{z}_2)&=&
\bf
\frac{\bar{z}_{12}}{z_{12}}\,B\bigg(\Delta_1+ 1 ,\Delta_2+1\bigg)\,\epsilon^{abcd}\epsilon^{rstu}\,\O_{\Delta_1+\Delta_2,-2}(z_2,\bar{z}_2) + \cdots,
\nonu\\
\bf
\O^{abcr}_{\Delta_1,0}(z_1,\bar{z}_1)\,\O^{dstu}_{\Delta_2,0}(z_2,\bar{z}_2)&=&
\bf
-\frac{\bar{z}_{12}}{z_{12}}\,B\bigg(\Delta_1+ 1 ,\Delta_2+1\bigg)\,\epsilon^{abcd}\epsilon^{rstu}\,\O_{\Delta_1+\Delta_2,-2}(z_2,\bar{z}_2) + \cdots,
\nonu\\
\O^{abrs}_{\Delta_1,0}(z_1,\bar{z}_1)\,\O^{cdtu}_{\Delta_2,0}(z_2,\bar{z}_2)&=&\frac{\bar{z}_{12}}{z_{12}}\,B\bigg(\Delta_1+ 1 ,\Delta_2+1\bigg)\,\epsilon^{abcd}\epsilon^{rstu}\,\O_{\Delta_1+\Delta_2,-2}(z_2,\bar{z}_2) + \cdots \, .
\nonu
\end{eqnarray}
The two additional OPEs should appear.

\section{The mathematica program on the Jacobi identity}

We present the program by the mathematica as follows:
\bea
&&
\tt <<\text{\tt OPEdefs.m}
\nonu\\ 
&&
\tt \text{\tt $\$$RecursionLimit}=\text{\tt Infinity}
\nonu\\ 
&& \text{(* This is necessary for using the Thielemans package. *)}
\nonu \\
&&
\tt \text{\tt Bosonic}[\text{\tt Phi}[\text{\tt h$\_$},+2]]
\nonu\\ 
&&
\tt \text{\tt Fermionic}\left[\text{\tt Phi}\left[\text{\tt h$\_$},\{\text{\tt A$\_$}\},+\frac{3}{2}\right]\right]
\nonu\\ 
&&
\tt \text{\tt Bosonic}[\text{\tt Phi}[\text{\tt h$\_$},\{\text{\tt A$\_$},\text{\tt B$\_$}\},+1]]
\nonu\\ 
&&
\tt \text{\tt Fermionic}\left[\text{\tt Phi}\left[\text{\tt h$\_$},\{\text{\tt A$\_$},\text{\tt B$\_$},\text{\tt CC$\_$}\},+\frac{1}{2}\right]\right]
\nonu\\ 
&&
\tt \text{\tt Bosonic}[\text{\tt Phi}[\text{\tt h$\_$},\{\text{\tt A$\_$},\text{\tt B$\_$},\text{\tt CC$\_$},\text{\tt DD$\_$}\},0]]
\nonu\\ 
&&
\tt \text{\tt Fermionic}\left[\text{\tt Phi}\left[\text{\tt h$\_$},\{\text{\tt A$\_$},\text{\tt B$\_$},\text{\tt CC$\_$}\},-\frac{1}{2}\right]\right]
\nonu\\ 
&&
\tt \text{\tt Bosonic}[\text{\tt Phi}[\text{\tt h$\_$},\{\text{\tt A$\_$},\text{\tt B$\_$}\},-1]]
\nonu\\ 
&&
\tt \text{\tt Fermionic}\left[\text{\tt Phi}\left[\text{\tt h$\_$},\{\text{\tt A$\_$}\},-\frac{3}{2}\right]\right]
\nonu\\ 
&&
\tt \text{\tt Bosonic}[\text{\tt Phi}[\text{\tt h$\_$},-2]]
\nonu\\ 
&&
\text{(*
If any two indices \
are the same, return to zero *)}
\nonu \\
&&
\tt \text{\tt Phi}[\text{\tt h$\_$},\{\text{\tt indices$\_\_$}\},\text{\tt i$\_$}]\text{\tt /;}\text{\tt Length}[\text{\tt DeleteDuplicates}[\{\text{\tt indices}\}]]<
\nonu \\
&& \text{\tt Length}[\{\text{\tt indices}\}]\text{\tt :=}0;
\nonu\\ 
&& \text{(*Define the
antisymmetry property for an arbitrary number of indices*)}
\nonu \\
&&
\tt \text{\tt Phi}[\text{\tt h$\_$},\{\text{\tt indices$\_\_$}\},\text{\tt i$\_$}]\text{\tt :=}
\text{\tt Signature}[\{\text{\tt indices}\}] \text{\tt Phi}[h,\text{\tt Sort}[\{\text{\tt indices}\}],i]\text{\tt /;}
\nonu \\
&& !\text{\tt OrderedQ}[\{\text{\tt indices}\}]
\nonu\\ 
&& \text{(* The OPEs are included. *)}
\nonu \\
&&
\tt \text{\tt OPE}[\text{\tt Phi}[\text{\tt h1$\_$},+2],\text{\tt Phi}[\text{\tt h2$\_$},+2]]
\nonu\\ 
&&=
\tt \text{\tt MakeOPE}[\{
\text{\tt kappa}[+2,+2,-2](\text{\tt h1}+\text{\tt h2}+2)\text{\tt Phi}[\text{\tt h1}+\text{\tt h2},+2],
\nonu \\  
&&
\tt \text{\tt kappa}[+2,+2,-2](\text{\tt h1}+1)\text{\tt Phi}[\text{\tt h1}+\text{\tt h2},+2]'\}];
\nonu\\ 
&&
\tt \text{\tt OPE}\left[\text{\tt Phi}[\text{\tt h1$\_$},+2],\text{\tt Phi}\left[\text{\tt h2$\_$},\{\text{\tt A$\_$}\},+\frac{3}{2}\right]\right]=
\text{\tt MakeOPE}\bigg[\bigg\{
\nonu\\ 
&&
\tt \text{\tt kappa}\left[+2,+\frac{3}{2},-\frac{3}{2}\right]\left(\text{\tt h1}+\text{\tt h2}+\frac{3}{2}\right)\text{\tt Phi}\left[\text{\tt h1}+\text{\tt h2},\{A\},+\frac{3}{2}\right],
\nonu\\ 
&&
\tt \text{\tt kappa}\left[+2,+\frac{3}{2},-\frac{3}{2}\right](\text{\tt h1}+1)\text{\tt Phi}\left[\text{\tt h1}+\text{\tt h2},\{A\},+\frac{3}{2}\right]'\bigg\}\bigg];
\nonu\\ 
&&
\tt\text{\tt OPE}[\text{\tt Phi}[\text{\tt h1$\_$},+2],\text{\tt Phi}[\text{\tt h2$\_$},\{\text{\tt A$\_$},\text{\tt B$\_$}\},+1]]=
\text{\tt MakeOPE}[\{
\nonu\\ 
&&
\tt \text{\tt kappa}[+2,+1,-1](\text{\tt h1}+\text{\tt h2}+1)\text{\tt Phi}[\text{\tt h1}+\text{\tt h2},\{A,B\},+1],
\nonu\\ 
&&
\tt\text{\tt kappa}[+2,+1,-1](\text{\tt h1}+1)\text{\tt Phi}[\text{\tt h1}+\text{\tt h2},\{A,B\},+1]'\}];
\nonu\\ 
&&
\tt \text{\tt OPE}\left[\text{\tt Phi}[\text{\tt h1$\_$},+2],\text{\tt Phi}\left[\text{\tt h2$\_$},\{\text{\tt A$\_$},\text{\tt B$\_$},\text{\tt CC$\_$}\},+\frac{1}{2}\right]\right]=
\text{\tt MakeOPE}\bigg[\bigg\{
\nonu\\
&&
\tt\text{\tt kappa}\left[+2,+\frac{1}{2},-\frac{1}{2}\right]\left(\text{\tt h1}+\text{\tt h2}+\frac{1}{2}\right)\text{\tt Phi}\left[\text{\tt h1}+\text{\tt h2},\{A,B,\text{\tt CC}\},+\frac{1}{2}\right],
\nonu\\
&&
\tt \text{\tt kappa}\left[+2,+\frac{1}{2},-\frac{1}{2}\right](\text{\tt h1}+1)\text{\tt Phi}\left[\text{\tt h1}+\text{\tt h2},\{A,B,\text{\tt CC}\},+\frac{1}{2}\right]'\bigg\}\bigg];
\nonu\\ 
&& \tt
\text{\tt OPE}[\text{\tt Phi}[\text{\tt h1$\_$},+2],
\text{\tt Phi}[\text{\tt h2$\_$},\{\text{\tt A$\_$},\text{\tt B$\_$},\text{\tt CC$\_$},\text{\tt DD$\_$}\},0]]=
\text{\tt MakeOPE}[\{
\nonu\\ 
&& \tt
\text{\tt kappa}[+2,0,0](\text{\tt h1}+\text{\tt h2})\text{\tt Phi}[\text{\tt h1}+\text{\tt h2},\{A,B,\text{\tt CC},\text{\tt DD}\},0],
\nonu\\ 
&& \tt
\text{\tt kappa}[+2,0,0](\text{\tt h1}+1)\text{\tt Phi}[\text{\tt h1}+\text{\tt h2},\{A,B,\text{\tt CC},\text{\tt DD}\},0]'\}];
\nonu\\ 
&& \tt
\text{\tt OPE}\left[\text{\tt Phi}[\text{\tt h1$\_$},+2],\text{\tt Phi}\left[\text{\tt h2$\_$},\{\text{\tt A$\_$},\text{\tt B$\_$},\text{\tt CC$\_$}\},-\frac{1}{2}\right]\right]=
\text{\tt MakeOPE}\bigg[\bigg\{
\nonu\\ 
&& \tt
\text{\tt kappa}\left[+2,-\frac{1}{2},+\frac{1}{2}\right]\left(\text{\tt h1}+\text{\tt h2}-\frac{1}{2}\right)\text{\tt Phi}\left[\text{\tt h1}+\text{\tt h2},\{A,B,\text{\tt CC}\},-\frac{1}{2}\right],
\nonu\\ 
&& \tt
\text{\tt kappa}\left[+2,-\frac{1}{2},+\frac{1}{2}\right](\text{\tt h1}+1)\text{\tt Phi}\left[\text{\tt h1}+\text{\tt h2},\{A,B,\text{\tt CC}\},-\frac{1}{2}\right]'\bigg\}\bigg];
\nonu\\ 
&& \tt
\text{\tt OPE}[\text{\tt Phi}[\text{\tt h1$\_$},+2],\text{\tt Phi}[\text{\tt h2$\_$},\{\text{\tt A$\_$},\text{\tt B$\_$}\},-1]]=
\text{\tt MakeOPE}[\{
\nonu\\ 
&& \tt
\text{\tt kappa}[+2,-1,+1](\text{\tt h1}+\text{\tt h2}-1)\text{\tt Phi}[\text{\tt h1}+\text{\tt h2},\{A,B\},-1],
\nonu\\ 
&& \tt
\text{\tt kappa}[+2,-1,+1](\text{\tt h1}+1)\text{\tt Phi}[\text{\tt h1}+\text{\tt h2},\{A,B\},-1]'\}];
\nonu\\ 
&& \tt
\text{\tt OPE}\left[\text{\tt Phi}[\text{\tt h1$\_$},+2],\text{\tt Phi}\left[\text{\tt h2$\_$},\{\text{\tt A$\_$}\},-\frac{3}{2}\right]\right]=
\text{\tt MakeOPE}\bigg[\bigg\{
\nonu\\ 
&& \tt
\text{\tt kappa}\left[+2,-\frac{3}{2},+\frac{3}{2}\right]\left(\text{\tt h1}+\text{\tt h2}-\frac{3}{2}\right)\text{\tt Phi}\left[\text{\tt h1}+\text{\tt h2},\{A\},-\frac{3}{2}\right],
\nonu\\ 
&& \tt
\text{\tt kappa}\left[+2,-\frac{3}{2},+\frac{3}{2}\right](\text{\tt h1}+1)\text{\tt Phi}\left[\text{\tt h1}+\text{\tt h2},\{A\},-\frac{3}{2}\right]'\bigg\}\bigg];
\nonu\\ 
&& \tt
\text{\tt OPE}[\text{\tt Phi}[\text{\tt h1$\_$},+2],\text{\tt Phi}[\text{\tt h2$\_$},-2]]=
\text{\tt MakeOPE}[\{
\nonu\\ 
&& \tt
\text{\tt kappa}[+2,-2,+2](\text{\tt h1}+\text{\tt h2}-2)\text{\tt Phi}[\text{\tt h1}+\text{\tt h2},-2],
\nonu \\
&& \tt
\text{\tt kappa}[+2,-2,+2](\text{\tt h1}+1)\text{\tt Phi}[\text{\tt h1}+\text{\tt h2},-2]'\}];
\nonu\\ 
&& \tt
\text{\tt OPE}\left[\text{\tt Phi}\left[\text{\tt h1$\_$},\{\text{\tt A$\_$}\},+\frac{3}{2}\right],\text{\tt Phi}\left[\text{\tt h2$\_$},\{\text{\tt B$\_$}\},+\frac{3}{2}\right]\right]=
\text{\tt MakeOPE}\bigg[\bigg\{
\nonu\\ 
&& \tt
\text{\tt kappa}\left[+\frac{3}{2},+\frac{3}{2},-1\right](\text{\tt h1}+\text{\tt h2}+1)\text{\tt Phi}[\text{\tt h1}+\text{\tt h2},\{A,B\},+1],
\nonu\\ 
&& \tt
\text{\tt kappa}\left[+\frac{3}{2},+\frac{3}{2},-1\right]\left(\text{\tt h1}+\frac{1}{2}\right)\text{\tt Phi}[\text{\tt h1}+\text{\tt h2},\{A,B\},+1]'\bigg\}\bigg];
\nonu\\ 
&& \tt
\text{\tt OPE}\left[\text{\tt Phi}\left[\text{\tt h1$\_$},\{\text{\tt A$\_$}\},+\frac{3}{2}\right],\text{\tt Phi}[\text{\tt h2$\_$},\{\text{\tt B$\_$},\text{\tt CC$\_$}\},+1]\right]=
\text{\tt MakeOPE}\bigg[\bigg\{
\nonu\\ 
&& \tt
\text{\tt kappa}\left[+\frac{3}{2},+1,-\frac{1}{2}\right]\left(\text{\tt h1}+\text{\tt h2}+\frac{1}{2}\right)\text{\tt Phi}\left[\text{\tt h1}+\text{\tt h2},\{A,B,\text{\tt CC}\},+\frac{1}{2}\right],
\nonu\\ 
&& \tt
\text{\tt kappa}\left[+\frac{3}{2},+1,-\frac{1}{2}\right]\left(\text{\tt h1}+\frac{1}{2}\right)\text{\tt Phi}\left[\text{\tt h1}+\text{\tt h2},\{A,B,\text{\tt CC}\},+\frac{1}{2}\right]'\bigg\}\bigg];
\nonu\\ 
&& \tt
\text{\tt OPE}\left[\text{\tt Phi}\left[\text{\tt h1$\_$},\{\text{\tt A$\_$}\},+\frac{3}{2}\right],\text{\tt Phi}\left[\text{\tt h2$\_$},\{\text{\tt B$\_$},\text{\tt CC$\_$},\text{\tt DD$\_$}\},+\frac{1}{2}\right]\right]=
\text{\tt MakeOPE}\bigg[\bigg\{
\nonu\\ 
&& \tt
\text{\tt kappa}\left[+\frac{3}{2},+\frac{1}{2},0\right](\text{\tt h1}+\text{\tt h2})\text{\tt Phi}[\text{\tt h1}+\text{\tt h2},\{A,B,\text{\tt CC},\text{\tt DD}\},0],
\nonu\\ 
&& \tt
\text{\tt kappa}\left[+\frac{3}{2},+\frac{1}{2},0\right]\left(\text{\tt h1}+\frac{1}{2}\right)\text{\tt Phi}[\text{\tt h1}+\text{\tt h2},\{A,B,\text{\tt CC},\text{\tt DD}\},0]'\bigg]\}\bigg]];
\nonu\\ 
&& \tt
\text{\tt OPE}\left[\text{\tt Phi}\left[\text{\tt h1$\_$},\{\text{\tt A$\_$}\},+\frac{3}{2}\right],\text{\tt Phi}[\text{\tt h2$\_$},\{\text{\tt B$\_$},\text{\tt CC$\_$},\text{\tt DD$\_$},\text{\tt EE$\_$}\},0]\right]=
\text{\tt MakeOPE}\bigg[\bigg\{
\nonu\\ 
&& \tt
\text{\tt kappa}\left[+\frac{3}{2},0,+\frac{1}{2}\right]\left(\text{\tt h1}+\text{\tt h2}-\frac{1}{2}\right)\frac{1}{3!}
\sum _{F=1}^8 \bigg(\sum _{G=1}^8 
\bigg(\sum _{H=1}^8 \text{\tt Epsilon}[A,B,\text{\tt CC},\text{\tt DD},\text{\tt EE},F,G,H]
\nonu\\ 
&& \tt
\text{\tt Phi}\left[\text{\tt h1}+\text{\tt h2},\{F,G,H\},-\frac{1}{2}\right]\bigg)\bigg),
\text{\tt kappa}\left[+\frac{3}{2},0,+\frac{1}{2}\right]\left(\text{\tt h1}+\frac{1}{2}\right)
\frac{1}{3!}
\nonu\\ 
&& \tt
\sum_{F=1}^8 \bigg(\sum_{G=1}^8 \bigg(\sum_{H=1}^8 \text{\tt Epsilon}[A,B,\text{\tt CC},\text{\tt DD},\text{\tt EE},F,G,H]
\text{\tt Phi}\left[\text{\tt h1}+\text{\tt h2},\{F,G,H\},-\frac{1}{2}\right]'\bigg)\bigg)\bigg\}\bigg];
\nonu\\ 
&& \tt
\text{\tt OPE}\left[\text{\tt Phi}\left[\text{\tt h1$\_$},\{\text{\tt A$\_$}\},+\frac{3}{2}\right],\text{\tt Phi}\left[\text{\tt h2$\_$},\{\text{\tt B$\_$},\text{\tt CC$\_$},\text{\tt DD$\_$}\},-\frac{1}{2}\right]\right]=
\text{\tt MakeOPE}\bigg[\bigg\{
\nonu\\ 
&& \tt
\text{\tt kappa}\left[+\frac{3}{2},-\frac{1}{2},+1\right](\text{\tt h1}+\text{\tt h2}-1)3
\bigg(\frac{1}{6} ( \text{\tt Delta}[A,B]\text{\tt Phi}[\text{\tt h1}+\text{\tt h2},\{\text{\tt CC},\text{\tt DD}\},-1]
\nonu\\ 
&& \tt-
\text{\tt Delta}[A,B]\text{\tt Phi}[\text{\tt h1}+\text{\tt h2},\{\text{\tt DD},\text{\tt CC}\},-1]
- \text{\tt Delta}[A,\text{\tt CC}]\text{\tt Phi}[\text{\tt h1}+\text{\tt h2},\{B,\text{\tt DD}\},-1]
\nonu\\ 
&& \tt
+\text{\tt Delta}[A,\text{\tt CC}]\text{\tt Phi}[\text{\tt h1}+\text{\tt h2},\{\text{\tt DD},B\},-1]
+\text{\tt Delta}[A,\text{\tt DD}]\text{\tt Phi}[\text{\tt h1}+\text{\tt h2},\{B,\text{\tt CC}\},-1]
\nonu\\ 
&& \tt
-\text{\tt Delta}[A,\text{\tt DD}]\text{\tt Phi}[\text{\tt h1}+\text{\tt h2},\{\text{\tt CC},B\},-1])),
\nonu\\ 
&& \tt
\text{\tt kappa}\left[+\frac{3}{2},-\frac{1}{2},+1\right]\left(\text{\tt h1}+\frac{1}{2}\right)3
\bigg(\frac{1}{6} ( \text{\tt Delta}[A,B]\text{\tt Phi}[\text{\tt h1}+\text{\tt h2},
\{\text{\tt CC},\text{\tt DD}\},-1]
\nonu\\ 
&& \tt
- \text{\tt Delta}[A,B]\text{\tt Phi}[\text{\tt h1}+\text{\tt h2},\{\text{\tt DD},\text{\tt CC}\},-1]
- \text{\tt Delta}[A,\text{\tt CC}]\text{\tt Phi}[\text{\tt h1}+\text{\tt h2},\{B,\text{\tt DD}\},-1]
\nonu\\ 
&& \tt
+ \text{\tt Delta}[A,\text{\tt CC}]\text{\tt Phi}[\text{\tt h1}+\text{\tt h2},\{\text{\tt DD},B\},-1]
+ \text{\tt Delta}[A,\text{\tt DD}]\text{\tt Phi}[\text{\tt h1}+\text{\tt h2},\{B,\text{\tt CC}\},-1]
\nonu\\ 
&& \tt
- \text{\tt Delta}[A,\text{\tt DD}]\text{\tt Phi}[\text{\tt h1}+\text{\tt h2},\{\text{\tt CC},B\},-1]))'\bigg\}\bigg];
\nonu\\ 
&& \tt
\text{\tt OPE}\left[\text{\tt Phi}
\left[\text{\tt h1$\_$},\{\text{\tt A$\_$}\},+\frac{3}{2}\right],
\text{\tt Phi}[\text{\tt h2$\_$},
\{\text{\tt B$\_$},\text{\tt CC$\_$}\},-1]\right]=
\text{\tt MakeOPE}\bigg[\bigg\{
\nonu\\ 
&& \tt
\text{\tt kappa}\left[+\frac{3}{2},-1,+\frac{3}{2}\right]\left(\text{\tt h1}+\text{\tt h2}-\frac{3}{2}\right)
\bigg(
\nonu\\ 
&& \tt
\text{\tt Delta}[A,B]\text{\tt Phi}\left[\text{\tt h1}+\text{\tt h2},\{\text{\tt CC}\},-\frac{3}{2}\right]
-\text{\tt Delta}[A,\text{\tt CC}]\text{\tt Phi}\left[\text{\tt h1}+\text{\tt h2},\{B\},-\frac{3}{2}\right]\bigg),
\nonu\\ 
&& \tt
\text{\tt kappa}\left[+\frac{3}{2},-1,+\frac{3}{2}\right]\left(\text{\tt h1}+\frac{1}{2}\right)
\left(\text{\tt Delta}[A,B]\text{\tt Phi}\left[\text{\tt h1}+\text{\tt h2},\{\text{\tt CC}\},-\frac{3}{2}\right]' \right.
\nonu\\ 
&& \tt
-\text{\tt Delta}[A,\text{\tt CC}]\text{\tt Phi}\left[\text{\tt h1}+\text{\tt h2},\{B\},-\frac{3}{2}\right]'\bigg)\bigg\}\bigg];
\nonu\\ 
&& \tt
\text{\tt OPE}\left[\text{\tt Phi}\left[\text{\tt h1$\_$},\{\text{\tt A$\_$}\},+\frac{3}{2}\right],\text{\tt Phi}\left[\text{\tt h2$\_$},\{\text{\tt B$\_$}\},-\frac{3}{2}\right]\right]=
\text{\tt MakeOPE}\bigg[\bigg\{
\nonu\\ 
&& \tt
\text{\tt kappa}\left[+\frac{3}{2},-\frac{3}{2},+2\right]\text{\tt Delta}[A,B](\text{\tt h1}+\text{\tt h2}-2) \text{\tt Phi}[\text{\tt h1}+\text{\tt h2},-2],
\nonu\\ 
&& \tt
\text{\tt kappa}\left[+\frac{3}{2},-\frac{3}{2},+2\right]\text{\tt Delta}[A,B]\left(\text{\tt h1}+\frac{1}{2}\right)\text{\tt Phi}[\text{\tt h1}+\text{\tt h2},-2]'\bigg\}\bigg];
\nonu\\ 
&& \tt
\text{\tt OPE}[\text{\tt Phi}[\text{\tt h1$\_$},\{\text{\tt A$\_$},\text{\tt B$\_$}\},+1],\text{\tt Phi}[\text{\tt h2$\_$},\{\text{\tt CC$\_$},\text{\tt DD$\_$}\},+1]]=
\text{\tt MakeOPE}[\{
\nonu\\ 
&& \tt
\text{\tt kappa}[+1,+1,0](\text{\tt h1}+\text{\tt h2})\text{\tt Phi}[\text{\tt h1}+\text{\tt h2},\{A,B,\text{\tt CC},\text{\tt DD}\},0],
\nonu\\ 
&& \tt
\text{\tt kappa}[+1,+1,0](\text{\tt h1})\text{\tt Phi}[\text{\tt h1}+\text{\tt h2},\{A,B,\text{\tt CC},\text{\tt DD}\},0]'\}];
\nonu\\ 
&& \tt
\text{\tt OPE}\left[\text{\tt Phi}[\text{\tt h1$\_$},\{\text{\tt A$\_$},\text{\tt B$\_$}\},+1],\text{\tt Phi}\left[\text{\tt h2$\_$},\{\text{\tt CC$\_$},\text{\tt DD$\_$},\text{\tt EE$\_$}\},+\frac{1}{2}\right]\right]=
\text{\tt MakeOPE}\bigg[\bigg\{
\nonu\\ 
&& \tt
\text{\tt kappa}\left[+1,+\frac{1}{2},+\frac{1}{2}\right]\left(\text{\tt h1}+\text{\tt h2}-\frac{1}{2}\right)\frac{1}{3!}
\sum _{F=1}^8 \bigg(\sum _{G=1}^8 \bigg(\sum _{H=1}^8 \text{\tt Epsilon}[A,B,\text{\tt CC},\text{\tt DD},\text{\tt EE},F,G,H]
\nonu\\ 
&& \tt
\text{\tt Phi}\left[\text{\tt h1}+\text{\tt h2},\{F,G,H\},-\frac{1}{2}\right]\bigg)\bigg),
\text{\tt kappa}\left[+1,+\frac{1}{2},+\frac{1}{2}\right](\text{\tt h1})\frac{1}{3!}
\sum _{F=1}^8 \bigg(\sum _{G=1}^8 \bigg(\sum _{H=1}^8 
\nonu\\ 
&& \tt
\text{\tt Epsilon}[A,B,\text{\tt CC},\text{\tt DD},\text{\tt EE},F,G,H]\text{\tt Phi}\left[\text{\tt h1}+\text{\tt h2},\{F,G,H\},-\frac{1}{2}\right]'\bigg)\bigg)\bigg\}\bigg];
\nonu\\ 
&& \tt
\text{\tt OPE}[\text{\tt Phi}[\text{\tt h1$\_$},\{\text{\tt A$\_$},\text{\tt B$\_$}\},+1],\text{\tt Phi}[\text{\tt h2$\_$},\{\text{\tt CC$\_$},\text{\tt DD$\_$},\text{\tt EE$\_$},\text{\tt F$\_$}\},0]]=
\text{\tt MakeOPE}\bigg[\bigg\{
\nonu\\ 
&& \tt
\text{\tt kappa}[+1,0,+1](\text{\tt h1}+\text{\tt h2}-1)\frac{1}{2}
\bigg(\sum _{H=1}^8 \bigg(\sum _{G=1}^8 \text{\tt Epsilon}[A,B,\text{\tt CC},\text{\tt DD},\text{\tt EE},F,G,H]
\nonu\\ 
&& \tt
\text{\tt Phi}[\text{\tt h1}+\text{\tt h2},\{G,H\},-1]\bigg)\bigg),
\text{\tt kappa}[+1,0,+1](\text{\tt h1})\nonu \\
&& \tt \frac{1}{2}
\bigg(\sum _{H=1}^8 \bigg(\sum _{G=1}^8 \text{\tt Epsilon}[A,B,\text{\tt CC},\text{\tt DD},\text{\tt EE},F,G,H]
\nonu\\ 
&& \tt
\text{\tt Phi}[\text{\tt h1}+\text{\tt h2},\{G,H\},-1]'\bigg)\bigg)\bigg\}\bigg];
\nonu\\ 
&& \tt
\text{\tt OPE}\left[\text{\tt Phi}[\text{\tt h1$\_$},
\{\text{\tt A$\_$},\text{\tt B$\_$}\},+1],
\text{\tt Phi}\left[\text{\tt h2$\_$},\{\text{\tt CC$\_$},
\text{\tt DD$\_$},\text{\tt EE$\_$}\},-\frac{1}{2}\right]\right]=
\text{\tt MakeOPE}\bigg[\bigg\{
\nonu\\ 
&& \tt
\text{\tt kappa}\left[+1,-\frac{1}{2},+\frac{3}{2}\right]\left(\text{\tt h1}+\text{\tt h2}-\frac{3}{2}\right)3!
\nonu \\
&& \tt \frac{1}{3!}
\left(\text{\tt Delta}[A,\text{\tt CC}]\text{\tt Phi}\left[\text{\tt h1}+\text{\tt h2},\{\text{\tt DD}\},-\frac{3}{2}\right]\text{\tt Delta}[\text{\tt EE},B] \right.
\nonu\\ 
&& \tt
-\text{\tt Delta}[A,\text{\tt CC}]\text{\tt Phi}\left[\text{\tt h1}+\text{\tt h2},\{\text{\tt EE}\},-\frac{3}{2}\right]\text{\tt Delta}[\text{\tt DD},B]
\nonu\\ 
&& \tt
-\text{\tt Delta}[A,\text{\tt DD}]\text{\tt Phi}\left[\text{\tt h1}+\text{\tt h2},\{\text{\tt CC}\},-\frac{3}{2}\right]\text{\tt Delta}[\text{\tt EE},B]
\nonu\\ 
&& \tt
+\text{\tt Delta}[A,\text{\tt DD}]\text{\tt Phi}\left[\text{\tt h1}+\text{\tt h2},\{\text{\tt EE}\},-\frac{3}{2}\right]\text{\tt Delta}[\text{\tt CC},B]
\nonu\\ 
&& \tt
+\text{\tt Delta}[A,\text{\tt EE}]\text{\tt Phi}\left[\text{\tt h1}+\text{\tt h2},\{\text{\tt CC}\},-\frac{3}{2}\right]\text{\tt Delta}[\text{\tt DD},B]
\nonu\\ 
&& \tt
-\text{\tt Delta}[A,\text{\tt EE}]\text{\tt Phi}\left[\text{\tt h1}+\text{\tt h2},\{\text{\tt DD}\},-\frac{3}{2}\right]\text{\tt Delta}[\text{\tt CC},B]\bigg),
\nonu\\ 
&& \tt
\text{\tt kappa}\left[+1,-\frac{1}{2},+\frac{3}{2}\right](\text{\tt h1})3 !\frac{1}{3!}
\left(\text{\tt Delta}[A,\text{\tt CC}]\text{\tt Phi}\left[\text{\tt h1}+\text{\tt h2},\{\text{\tt DD}\},-\frac{3}{2}\right]'\text{\tt Delta}[\text{\tt EE},B] \right.
\nonu\\ 
&& \tt
-\text{\tt Delta}[A,\text{\tt CC}]\text{\tt Phi}\left[\text{\tt h1}+\text{\tt h2},\{\text{\tt EE}\},-\frac{3}{2}\right]'\text{\tt Delta}[\text{\tt DD},B]
\nonu\\ 
&& \tt
-\text{\tt Delta}[A,\text{\tt DD}]\text{\tt Phi}\left[\text{\tt h1}+\text{\tt h2},\{\text{\tt CC}\},-\frac{3}{2}\right]'\text{\tt Delta}[\text{\tt EE},B]
\nonu\\ 
&& \tt
+\text{\tt Delta}[A,\text{\tt DD}]\text{\tt Phi}\left[\text{\tt h1}+\text{\tt h2},\{\text{\tt EE}\},-\frac{3}{2}\right]'\text{\tt Delta}[\text{\tt CC},B]
\nonu\\ 
&& \tt
+\text{\tt Delta}[A,\text{\tt EE}]\text{\tt Phi}\left[\text{\tt h1}+\text{\tt h2},\{\text{\tt CC}\},-\frac{3}{2}\right]'\text{\tt Delta}[\text{\tt DD},B]
\nonu\\ 
&& \tt
-
\text{\tt Delta}[A,\text{\tt EE}]\text{\tt Phi}\left[\text{\tt h1}+\text{\tt h2},\{\text{\tt DD}\},-\frac{3}{2}\right]'\text{\tt Delta}[\text{\tt CC},B]\bigg)\bigg\}\bigg];
\nonu\\ 
&& \tt
\text{\tt OPE}[\text{\tt Phi}[\text{\tt h1$\_$},\{\text{\tt A$\_$},\text{\tt B$\_$}\},+1],\text{\tt Phi}[\text{\tt h2$\_$},\{\text{\tt CC$\_$},\text{\tt DD$\_$}\},-1]]=
\text{\tt MakeOPE}[\{
\nonu\\ 
&& \tt
\text{\tt kappa}[+1,-1,+2](\text{\tt h1}+\text{\tt h2}-2)
(\text{\tt Delta}[A,\text{\tt CC}]\text{\tt Delta}[B,\text{\tt DD}]-\text{\tt Delta}[A,\text{\tt DD}]\text{\tt Delta}[B,\text{\tt CC}])
\nonu\\ 
&& \tt
\text{\tt Phi}[\text{\tt h1}+\text{\tt h2},-2],
\text{\tt kappa}[+1,-1,+2](\text{\tt h1})
(\text{\tt Delta}[A,\text{\tt CC}]\text{\tt Delta}[B,\text{\tt DD}]
\nonu \\
&&\tt -\text{\tt Delta}[A,\text{\tt DD}]\text{\tt Delta}[B,\text{\tt CC}])
] \tt \text{\tt Phi}[\text{\tt h1}+\text{\tt h2},-2]'\}];
\nonu\\ 
&& \tt
\text{\tt OPE}\left[\text{\tt Phi}\left[\text{\tt h1$\_$},\{\text{\tt A$\_$},\text{\tt B$\_$},\text{\tt CC$\_$}\},+\frac{1}{2}\right],\text{\tt Phi}\left[\text{\tt h2$\_$},\{\text{\tt DD$\_$},\text{\tt EE$\_$},\text{\tt F$\_$}\},+\frac{1}{2}\right]\right]=
\text{\tt MakeOPE}\bigg[\bigg\{
\nonu\\ 
&& \tt
\text{\tt kappa}\left[+\frac{1}{2},+\frac{1}{2},+1\right](\text{\tt h1}+\text{\tt h2}-1)\frac{1}{2}
\sum _{G=1}^8 \bigg(\sum _{H=1}^8 \text{\tt Epsilon}[A,B,\text{\tt CC},\text{\tt DD},\text{\tt EE},F,G,H]
\nonu\\ 
&& \tt
\text{\tt Phi}[\text{\tt h1}+\text{\tt h2},\{G,H\},-1]\bigg),
\text{\tt kappa}\left[+\frac{1}{2},+\frac{1}{2},+1\right]\left(\text{\tt h1} -\frac{1}{2}\right)\frac{1}{2}
\nonu\\ 
&& \tt
\sum _{G=1}^8 \left(\sum _{H=1}^8 \text{\tt Epsilon}[A,B,\text{\tt CC},\text{\tt DD},\text{\tt EE},F,G,H]\text{\tt Phi}[\text{\tt h1}+\text{\tt h2},\{G,H\},-1]'\right)\bigg\}\bigg];
\nonu\\ 
&& \tt
\text{\tt OPE}\left[\text{\tt Phi}\left[\text{\tt h1$\_$},\{\text{\tt A$\_$},\text{\tt B$\_$},\text{\tt CC$\_$}\},+\frac{1}{2}\right],\text{\tt Phi}[\text{\tt h2$\_$},\{\text{\tt DD$\_$},\text{\tt EE$\_$},\text{\tt F$\_$},\text{\tt G$\_$}\},0]\right]=
\text{\tt MakeOPE}\bigg[\bigg\{
\nonu\\ 
&& \tt
\text{\tt kappa}\left[+\frac{1}{2},0,+\frac{3}{2}\right]\left(\left(\text{\t h1}+\text{\tt h2}-\frac{3}{2}\right)\right)
\bigg(\sum _{H=1}^8 \text{\tt Epsilon}[A,B,\text{\tt CC},\text{\tt DD},\text{\tt EE},F,G,H]
\nonu \\
&& \tt
\text{\tt Phi}\left[\text{\tt h1}+\text{\tt h2},\{H\},-\frac{3}{2}\right]\bigg),
\text{\tt kappa}\left[+\frac{1}{2},0,+\frac{3}{2}\right]\left(\left(\text{\tt h1}-\frac{1}{2}\right)\right)
\nonu \\
&& \tt
\bigg(\sum _{H=1}^8 \text{\tt Epsilon}[A,B,\text{\tt CC},\text{\tt DD},\text{\tt EE},F,G,H]
\text{\tt Phi}\left[\text{\tt h1}+\text{\tt h2},\{H\},-\frac{3}{2}\right]'\bigg)\bigg\}\bigg];
\nonu\\ 
&& \tt
\text{\tt OPE}\left[\text{\tt Phi}\left[\text{\tt h1$\_$},\{\text{\tt A$\_$},\text{\tt B$\_$},\text{\tt CC$\_$}\},+\frac{1}{2}\right],\text{\tt Phi}\left[\text{\tt h2$\_$},\{\text{\tt DD$\_$},\text{\tt EE$\_$},\text{\tt F$\_$}\},-\frac{1}{2}\right]\right]=
\text{\tt MakeOPE}\bigg[\bigg\{
\nonu\\ 
&& \tt
\text{\tt kappa}\left[+\frac{1}{2},-\frac{1}{2},+2\right](\text{\tt h1}+\text{\tt h2}-2)\frac{1}{5!}
\sum _{G=1}^8 \bigg(\sum _{H=1}^8 \bigg(\sum _{I=1}^8 \bigg(\sum _{J=1}^8
\bigg(
\nonu \\
&& \tt \sum _{K=1}^8 (\text{\tt Epsilon}[A,B,\text{\tt CC},G,H,I,J,K]
\nonu\\ 
&& \tt
\text{\tt Epsilon}[\text{\tt DD},\text{\tt EE},F,G,H,I,J,K]\bigg)\bigg)\bigg)\bigg)\text{\tt Phi}[\text{\tt h1}+\text{\tt h2},-2],
\text{\tt kappa}\left[+\frac{1}{2},-\frac{1}{2},+2\right]\left(\text{\tt h1}-\frac{1}{2}\right)\frac{1}{5!}
\nonu\\ 
&& \tt
\sum _{G=1}^8 \left(\sum _{H=1}^8 \left(\sum _{I=1}^8 \left(\sum _{J=1}^8
\right. \right. \right. \nonu \\
&& \tt \left. \left. \left.
\left(\sum _{K=1}^8 (\text{\tt Epsilon}[A,B,\text{\tt CC},G,H,I,J,K]
\text{\tt Epsilon}[\text{\tt DD},\text{\tt EE},F,G,H,I,J,K]
\right)\right)\right)\right)
\nonu\\ 
&& \tt
\text{\tt Phi}[\text{\tt h1}+\text{\tt h2},-2]'\bigg\}\bigg];
\nonu\\ 
&& \tt
\text{\tt OPE}[\text{\tt Phi}[\text{\tt h1$\_$},\{\text{\tt A$\_$},\text{\tt B$\_$},\text{\tt CC$\_$},\text{\tt DD$\_$}\},0],\text{\tt Phi}[\text{\tt h2$\_$},\{\text{\tt EE$\_$},\text{\tt F$\_$},\text{\tt G$\_$},\text{\tt H$\_$}\},0]]=
\text{\tt MakeOPE}[\{
\nonu\\ 
&& \tt
\text{\tt kappa}[0,0,+2](\text{\tt h1}+\text{\tt h2}-2)\text{\tt Epsilon}[A,B,\text{\tt CC},\text{\tt DD},\text{\tt EE},F,G,H]
\text{\tt Phi}[\text{\tt h1}+\text{\tt h2},-2],
\nonu\\ 
&& \tt
\text{\tt kappa}[0,0,+2](\text{\tt h1}-1)
\text{\tt Epsilon}[A,B,\text{\tt CC},\text{\tt DD},\text{\tt EE},F,G,H]\text{\tt Phi}[\text{\tt h1}+\text{\tt h2},-2]'\}];
\nonu\\ 
&&
\tt \text{\tt Clear}[\text{\tt IndexList}] 
\nonu\\
&&
\tt
\text{\tt IndexList}[n\_]\:=\text{\tt Subsets}[\{1,2,3,4,5,6,7,8\},\{n\}];
\nonu\\
&&
\tt \text{\tt Clear}[\text{\tt CurrentSet}]
\nonu\\
&&
\tt \text{\tt CurrentSet}[\text{\tt h$\_$}]\text{\tt :=}
\nonu\\
&&
\tt \text{\tt Flatten}\left[\left\{\text{\tt Phi}[h,+2],\text{\tt Table}\left[\text{\tt Phi}\left[h,I,+\frac{3}{2}\right],\{I,\text{\tt IndexList}[1]\}\right],\right.\right.
\nonu\\
&&
\tt \text{\tt Table}[\text{\tt Phi}[h,I,+1],\{I,\text{\tt IndexList}[2]\}],
\nonu\\
&&
\tt \text{\tt Table}\left[\text{\tt Phi}\left[h,I,+\frac{1}{2}\right],\{I,\text{\tt IndexList}[3]\}\right],\nonu\\
&&
\tt \text{\tt Table}[\text{\tt Phi}[h,I,0],\{I,\text{\tt IndexList}[4]\}],
\nonu\\
&&
\tt \text{\tt Table}\left[\text{\tt Phi}\left[h,I,-\frac{1}{2}\right],\{I,\text{\tt IndexList}[5]\}\right],\nonu\\
&&
\tt \text{\tt Table}[\text{\tt Phi}[h,I,-1],\{I,\text{\tt IndexList}[6]\}],\nonu\\
&&
\tt \left.\left.\text{\tt Table}\left[\text{\tt Phi}\left[h,I,-\frac{3}{2}\right],\{I,\text{\tt IndexList}[7]\}\right],\text{\tt Phi}[h,-2]\right\}\right];
\nonu\\
&& \text{(*The table for the Jacobi when the first operator is
given by $\Phi_{+2}^{(h_1)}$} \nonu \\
&& \text{and the other two
are arbitrary operators*)}
\nonu \\
&&
\tt \text{\tt Clear}[\text{\tt JTableN1}]
\nonu\\
&& \tt \text{\tt JTableN1}=
\nonu\\
&& \tt
\text{\tt Table}[\text{\tt OPESimplify}[\text{\tt OPEJacobi}[\text{\tt CurrentSet}[\text{\tt h1}][[\text{\tt I1}]],\text{\tt CurrentSet}[\text{\tt h2}][[\text{\tt I2}]],
\nonu\\
&& \tt 
\text{\tt CurrentSet}[\text{\tt h3}][[\text{\tt I3}]]],\text{\tt Simplify}],\{\text{\tt I1},1\},\{\text{\tt I2},256\},\{\text{\tt I3},\text{\tt I2},256\}];
\nonu\\
&& \text{(*The table for the Jacobi when the first operator is
given by $\Phi_{+\frac{3}{2}}^{(h_1),A=1}$ with fixed index}
\nonu \\
&& \text{ $A=1$
and the other two
are arbitrary operators. }
\nonu \\
&& \text{ For different $A$, the relations are the
same *)}
\nonu \\
&&
\tt \text{\tt Clear}[\text{\tt JTableN2}]
\nonu\\
&&
\tt \text{\tt JTableN2}=
\nonu\\
&&
\tt \text{\tt Table}[\text{\tt OPESimplify}[\text{\tt OPEJacobi}[\text{\tt CurrentSet}[\text{\tt h1}][[\text{\tt I1}]],\text{\tt CurrentSet}[\text{\tt h2}][[\text{\tt I2}]],
\nonu\\
&&
\tt \text{\tt CurrentSet}[\text{\tt h3}][[\text{\tt I3}]]],\text{\tt Simplify}],\{\text{\tt I1},2,2\},\{\text{\tt I2},\text{\tt I1},256\},
\{\text{\tt I3},\text{\tt I2},256\}];
\nonu\\
&& \text{(*The table for the Jacobi when the first operator is
given by $\Phi_{+1}^{(h_1),AB=12}$} \nonu \\
&& \text{and the other two
are arbitrary operators*)}
\nonu \\
&&
\text{\tt Clear}[\text{\tt JTableN3}]
\nonu\\
&&
\tt\text{\tt JTableN3}=
\nonu\\
&&
\tt \text{\tt Table}[\text{\tt OPESimplify}[\text{\tt OPEJacobi}[\text{\tt CurrentSet}[\text{\tt h1}][[\text{\tt I1}]],\text{\tt CurrentSet}[\text{\tt h2}][[\text{\tt I2}]],
\nonu\\
&&
\tt
\text{\tt CurrentSet}[\text{\tt h3}][[\text{\tt I3}]]],\text{\tt Simplify}],\{\text{\tt I1},10,10\},\{\text{\tt I2},\text{\tt I1},256\},
 \{\text{\tt I3},\text{\tt I2},256\}];
\nonu\\
&& \text{(*The table for the Jacobi when the first operator is
given by $\Phi_{+\frac{1}{2}}^{(h_1),ABC=123}$} \nonu \\
&& \text{and the other two
are arbitrary operators*)}
\nonu \\
&&
\tt \text{\tt Clear}[\text{\tt JTableN4}]
\nonu\\
&&
\tt \text{\tt JTableN4}=
\nonu\\
&&
\tt \text{\tt Table}[\text{\tt OPESimplify}[\text{\tt OPEJacobi}[\text{\tt CurrentSet}[\text{\tt h1}][[\text{\tt I1}]],\text{\tt CurrentSet}[\text{\tt h2}][[\text{\tt I2}]],
\nonu\\
&&
\tt
\text{\tt CurrentSet}[\text{\tt h3}][[\text{\tt I3}]]],\text{\tt Simplify}],\{\text{\tt I1},38,38\},\{\text{\tt I2},\text{\tt I1},256\},
\{\text{\tt I3},\text{\tt I2},256\}];
\nonu\\
&& \text{(*The table for the Jacobi when the first operator is
given by $\Phi_{0}^{(h_1),ABCD=1234}$} \nonu \\
&& \text{and the other two
are arbitrary operators*)}
\nonu \\
&&
\tt \text{\tt Clear}[\text{\tt JTableN5}]
\nonu\\
&&
\tt
\text{\tt JTableN5}=
\nonu\\
&&
\tt
\text{\tt Table}[\text{\tt OPESimplify}[\text{\tt OPEJacobi}[\text{\tt CurrentSet}[\text{\tt h1}][[\text{\tt I1}]],\text{\tt CurrentSet}[\text{\tt h2}][[\text{\tt I2}]],
\nonu\\
&&
\tt
\text{\tt CurrentSet}[\text{\tt h3}][[\text{\tt I3}]]],\text{\tt Simplify}],\{\text{\tt I1},94,94\},\{\text{\tt I2},\text{\tt I1},256\},\{\text{\tt I3},\text{\tt I2},256\}];
\nonu\\
&& \text{(* We can check that for the first operators
having negative helicities there are }
\nonu \\
&& \text{trivial zeros in the
Jacobi identity. We collect the following nontrivial
Jacobi identities.  *)}
\nonu \\
&&
\tt \text{\tt TANZ}=\text{\tt Union}[\text{\tt Flatten}[\{\text{\tt JTableN1},\text{\tt JTableN2},\text{\tt JTableN3},\text{\tt JTableN4},\text{\tt JTableN5}\}]];
\nonu\\
&&
\text{(*Among nonzero $6991$ relations, we select
the first nonzero Jacobi relation denoted by}
\nonu \\
&& \text{$\tt TANZ[[2]]$.
The $\tt TANZ[[1]]$ vanishes. *)}
\nonu \\
&&
\tt \text{\tt Clear}[\text{\tt solstep1}]
\nonu\\
&&
\tt
\text{\tt solstep1}=\text{\tt Simplify}\left[\text{\tt Solve}\left[\text{\tt TANZ}[[2]]==0,\text{\tt kappa}\left[\frac{1}{2},-\frac{1}{2},2\right]\right]\right]
\nonu\\
&&
\tt
\text{\tt Clear}[\text{\tt SUB}]
\nonu\\
&&
\tt
\text{\tt SUB}=\left\{\text{\tt kappa}\left[\frac{1}{2},-\frac{1}{2},2\right]\to -\frac{\text{\tt kappa}\left[\frac{1}{2},\frac{1}{2},1\right] \text{\tt kappa}[1,-1,2]}{\text{\tt kappa}\left[1,\frac{1}{2},\frac{1}{2}\right]}\right\};
\nonu\\
&& \text{(* Now we substitute the solution
by ${\tt SUB}$ into the previous ${\tt TANZ}$. *)}
\nonu \\
&&
\tt
\text{\tt Clear}[\text{\tt TSAN}]
\nonu\\
&&
\tt
\text{\tt TSAN}=\text{\tt Union}[\text{\tt Flatten}[\text{\tt Simplify}[\text{\tt GetCoefficients}[\text{\tt TANZ}]\text{\tt /.}\text{\tt SUB}]]];
\nonu\\
&&
\text{(* We select
the first nonzero Jacobi relation denoted by
${\tt TSAN[[2]]}$. *)}
\nonu \\
&&
\tt \text{\tt Clear}[\text{\tt solstep2}]
\nonu\\
&&
\tt \text{\tt solstep2}=\text{\tt Simplify}[\text{\tt Solve}[\text{\tt TSAN}[[2]]==0,\text{\tt kappa}[0,0,2]]]
\nonu\\
&&
\tt 
\text{\tt Union}[\text{\tt Flatten}[\{\text{\tt SUB}\text{\tt /.}\text{\tt solstep2}[[1]],\text{\tt solstep2}[[1]]\}]]
\nonu\\
&&
\tt 
\text{\tt SUB}=\%;
\nonu\\
&& \text{(* Now we substitute the solution
by new ${\tt SUB}$ into the previous ${\tt TANZ}$. *)}
\nonu \\
&&
\tt 
\text{\tt Clear}[\text{\tt TSAN}]
\nonu\\
&&
\tt 
\text{\tt TSAN}=\text{\tt Union}[\text{\tt Flatten}[\text{\tt Simplify}[\text{\tt GetCoefficients}[\text{\tt TANZ}]\text{\tt /.}\text{\tt SUB}]]];
\nonu\\
&&
\text{(* We repeat the previous procedure until
the ${\tt solstep17}$. *)}
\nonu \\
&&
\tt 
\text{\tt Clear}[\text{\tt solstep3}]
\nonu\\
&&
\tt 
\text{\tt solstep3}=\text{\tt Simplify}[\text{\tt Solve}[\text{\tt TSAN}[[2]]==0,\text{\tt kappa}[1,1,0]]]
\nonu\\
&&
\tt 
\text{\tt Union}[\text{\tt Flatten}[\{\text{\tt SUB}\text{\tt /.}\text{\tt solstep3}[[1]],\text{\tt solstep3}[[1]]\}]]
\nonu\\
&&
\tt 
\text{\tt SUB}=\%;
\nonu\\
&&
\tt 
\text{\tt Clear}[\text{\tt TSAN}]
\nonu\\
&&
\tt 
\text{\tt TSAN}=\text{\tt Union}[\text{\tt Flatten}[\text{\tt Simplify}[\text{\tt GetCoefficients}[\text{\tt TANZ}]\text{\tt /.}\text{\tt SUB}]]];
\nonu\\
&&
\tt \text{\tt Clear}[\text{\tt solstep4}]
\nonu\\ && 
\tt 
\text{\tt solstep4}=\text{\tt Simplify}[\text{\tt Solve}[\text{\tt TSAN}[[2]]==0,\text{\tt kappa}[1,-1,2]]]
\nonu\\ 
&& 
\tt \text{\tt Union}[\text{\tt Flatten}[\{\text{\tt SUB}\text{\tt /.}\text{\tt solstep4}[[1]],\text{\tt solstep4}[[1]]\}]]
\nonu\\
&& \tt \text{\tt SUB}=\%;
\nonu\\
&& \tt 
\text{\tt Clear}[\text{\tt TSAN}]
\nonu\\
&& \tt 
\text{\tt TSAN}=\text{\tt Union}[\text{\tt Flatten}[\text{\tt Simplify}[\text{\tt GetCoefficients}[\text{\tt TANZ}]\text{\tt /.}\text{\tt SUB}]]];
\nonu\\
&& \tt \text{\tt Clear}[\text{\tt solstep5}]
\nonu\\ 
&& \tt 
\text{\tt solstep5}=\text{\tt Simplify}\left[\text{\tt Solve}\left[\text{\tt TSAN}[[2]]==0,\text{\tt kappa}\left[\frac{3}{2},-1,\frac{3}{2}\right]\right]\right]
\nonu\\ 
&& \tt \text{\tt Union}[\text{\tt Flatten}[\{\text{\tt SUB}\text{\tt /.}\text{\tt solstep5}[[1]],\text{\tt solstep5}[[1]]\}]]
\nonu\\ 
&& \tt \text{\tt SUB}=\%;
\nonu\\ 
&& \tt 
\text{\tt Clear}[\text{\tt TSAN}]
\nonu\\ 
&& \tt 
\text{\tt TSAN}=\text{\tt Union}[\text{\tt Flatten}[\text{\tt Simplify}[\text{\tt GetCoefficients}[\text{\tt TANZ}]\text{\tt /.}\text{\tt SUB}]]];
\nonu\\
&& \tt \text{\tt Clear}[\text{\tt solstep6}]
\nonu\\ 
&& \tt 
\text{\tt solstep6}=\text{\tt Simplify}\left[\text{\tt Solve}\left[\text{\tt TSAN}[[2]]==0,\text{\tt kappa}\left[\frac{3}{2},\frac{1}{2},0\right]\right]\right]
\nonu\\ 
&& \tt \text{\tt Union}[\text{\tt Flatten}[\{\text{\tt SUB}\text{\tt /.}\text{\tt solstep6}[[1]],\text{\tt solstep6}[[1]]\}]]
\nonu\\ 
&& \tt \text{\tt SUB}=\%;
\nonu\\ 
&& \tt 
\text{\tt Clear}[\text{\tt TSAN}]
\nonu\\ 
&& \tt 
\text{\tt TSAN}=\text{\tt Union}[\text{\tt Flatten}[\text{\tt Simplify}[\text{\tt GetCoefficients}[\text{\tt TANZ}]\text{\tt /.}\text{\tt SUB}]]];
\nonu\\
&& \tt \text{\tt Clear}[\text{\tt solstep7}]
\nonu\\ 
&& \tt 
\text{\tt solstep7}=\text{\tt Simplify}\left[\text{\tt Solve}\left[\text{\tt TSAN}[[2]]==0,\text{\tt kappa}\left[\frac{3}{2},-\frac{1}{2},1\right]\right]\right]
\nonu\\ && \tt \text{\tt Union}[\text{\tt Flatten}[\{\text{\tt SUB}\text{\tt /.}\text{\tt solstep7}[[1]],\text{\tt solstep7}[[1]]\}]]
\nonu\\ 
&& \tt \text{\tt SUB}=\%;
\nonu\\ 
&& \tt 
\text{\tt Clear}[\text{\tt TSAN}]
\nonu\\ && \tt 
\text{\tt TSAN}=\text{\tt Union}[\text{\tt Flatten}[\text{\tt Simplify}[\text{\tt GetCoefficients}[\text{\tt TANZ}]\text{\tt /.}\text{\tt SUB}]]];
\nonu\\
&& \tt \text{\tt Clear}[\text{\tt solstep8}]
\nonu\\ 
&& \tt 
\text{\tt solstep8}=\text{\tt Simplify}\left[\text{\tt Solve}\left[\text{\tt TSAN}[[2]]==0,\text{\tt kappa}\left[\frac{3}{2},1,-\frac{1}{2}\right]\right]\right]
\nonu\\
&& \tt \text{\tt Union}[\text{\tt Flatten}[\{\text{\tt SUB}\text{\tt /.}\text{\tt solstep8}[[1]],\text{\tt solstep8}[[1]]\}]]
\nonu\\
&& \tt \text{\tt SUB}=\%;
\nonu\\ 
&& \tt 
\text{\tt Clear}[\text{\tt TSAN}]
\nonu\\ 
&& 
\tt 
\text{\tt TSAN}=\text{\tt Union}[\text{\tt Flatten}[\text{\tt Simplify}[\text{\tt GetCoefficients}[\text{\tt TANZ}]\text{\tt /.}\text{\tt SUB}]]];
\nonu\\
&& \tt \text{\tt Clear}[\text{\tt solstep9}]
\nonu\\ 
&& 
\tt 
\text{\tt solstep9}=\text{\tt Simplify}\left[\text{\tt Solve}\left[\text{\tt TSAN}[[2]]==0,\text{\tt kappa}\left[\frac{3}{2},\frac{3}{2},-1\right]\right]\right]
\nonu\\ 
&& 
\tt \text{\tt Union}[\text{\tt Flatten}[\{\text{\tt SUB}\text{\tt /.}\text{\tt solstep9}[[1]],\text{\tt solstep9}[[1]]\}]]
\nonu\\ 
&& \tt \text{\tt SUB}=\%;
\nonu\\ 
&& \tt 
\text{\tt Clear}[\text{\tt TSAN}]
\nonu\\ && \tt 
\text{\tt TSAN}=\text{\tt Union}[\text{\tt Flatten}[\text{\tt Simplify}[\text{\tt GetCoefficients}[\text{\tt TANZ}]\text{\tt /.}\text{\tt SUB}]]];
\nonu\\
&& \tt \text{\tt Clear}[\text{\tt solstep10}]
\nonu\\ 
&& \tt 
\text{\tt solstep10}=\text{\tt Simplify}[\text{\tt Solve}[\text{\tt TSAN}[[2]]==0,\text{\tt kappa}[2,-2,2]]]
\nonu\\ 
&& 
\tt \text{\tt Union}[\text{\tt Flatten}[\{\text{\tt SUB}\text{\tt /.}\text{\tt solstep10}[[1]],\text{\tt solstep10}[[1]]\}]]
\nonu\\ 
&& \tt \text{\tt SUB}=\%;
\nonu\\ 
&& \tt 
\text{\tt Clear}[\text{\tt TSAN}]
\nonu\\ 
&& \tt 
\text{\tt TSAN}=\text{\tt Union}[\text{\tt Flatten}[\text{\tt Simplify}[\text{\tt GetCoefficients}[\text{\tt TANZ}]\text{\tt /.}\text{\tt SUB}]]];
\nonu\\
&& \tt 
\text{\tt Clear}[\text{\tt solstep11}]
\nonu\\ 
&& 
\tt 
\text{\tt solstep11}=\text{\tt Simplify}\left[\text{\tt Solve}\left[\text{\tt TSAN}[[2]]==0,\text{\tt kappa}\left[2,-\frac{3}{2},\frac{3}{2}\right]\right]\right]
\nonu\\ 
&& 
\tt \text{\tt Union}[\text{\tt Flatten}[\{\text{\tt SUB}\text{\tt /.}\text{\tt solstep11}[[1]],\text{\tt solstep11}[[1]]\}]]
\nonu\\ 
&& \tt \text{\tt SUB}=\%;
\nonu\\ 
&& \tt 
\text{\tt Clear}[\text{\tt TSAN}]
\nonu\\ 
&& \tt 
\text{\tt TSAN}=\text{\tt Union}[\text{\tt Flatten}[\text{\tt Simplify}[\text{\tt GetCoefficients}[\text{\tt TANZ}]\text{\tt /.}\text{\tt SUB}]]];
\nonu\\
&& \tt \text{\tt Clear}[\text{\tt solstep12}]
\nonu\\ 
&& \tt 
\text{\tt solstep12}=\text{\tt Simplify}[\text{\tt Solve}[\text{\tt TSAN}[[2]]==0,\text{\tt kappa}[2,-1,1]]]
\nonu\\
&& \tt \text{\tt Union}[\text{\tt Flatten}[\{\text{\tt SUB}\text{\tt /.}\text{\tt solstep12}[[1]],\text{\tt solstep12}[[1]]\}]]
\nonu\\ 
&& \tt \text{\tt SUB}=\%;
\nonu\\ 
&& \tt 
\text{\tt Clear}[\text{\tt TSAN}]
\nonu\\ 
&& \tt 
\text{\tt TSAN}=\text{\tt Union}[\text{\tt Flatten}[\text{\tt Simplify}[\text{\tt GetCoefficients}[\text{\tt TANZ}]\text{\tt /.}\text{\tt SUB}]]];
\nonu\\
&& \tt \text{\tt Clear}[\text{\tt solstep13}]
\nonu\\ 
&& \tt 
\text{\tt solstep13}=\text{\tt Simplify}\left[\text{\tt Solve}\left[\text{\tt TSAN}[[2]]==0,\text{\tt kappa}\left[2,-\frac{1}{2},\frac{1}{2}\right]\right]\right]
\nonu\\ 
&& 
\tt \text{\tt Union}[\text{\tt Flatten}[\{\text{\tt SUB}\text{\tt /.}\text{\tt solstep13}[[1]],\text{\tt solstep13}[[1]]\}]]
\nonu\\ 
&& \tt \text{\tt SUB}=\%;
\nonu\\ 
&& \tt 
\text{\tt Clear}[\text{\tt TSAN}]
\nonu\\ 
&& \tt 
\text{\tt TSAN}=\text{\tt Union}[\text{\tt Flatten}[\text{\tt Simplify}[\text{\tt GetCoefficients}[\text{\tt TANZ}]\text{\tt /.}\text{\tt SUB}]]];
\nonu\\
&& 
\tt \text{\tt Clear}[\text{\tt solstep14}]
\nonu\\ 
&& \tt 
\text{\tt solstep14}=\text{\tt Simplify}\left[\text{\tt Solve}\left[\text{\tt TSAN}[[2]]==0,\text{\tt kappa}\left[2,\frac{1}{2},-\frac{1}{2}\right]\right]\right]
\nonu\\ 
&& \tt \text{\tt Union}[\text{\tt Flatten}[\{\text{\tt SUB}\text{\tt /.}\text{\tt solstep14}[[1]],\text{\tt solstep14}[[1]]\}]]
\nonu\\ 
&& \tt \text{\tt SUB}=\%;
\nonu\\ 
&& \tt 
\text{\tt Clear}[\text{\tt TSAN}]
\nonu\\ 
&& \tt 
\text{\tt TSAN}=\text{\tt Union}[\text{\tt Flatten}[\text{\tt Simplify}[\text{\tt GetCoefficients}[\text{\tt TANZ}]\text{\tt /.}\text{\tt SUB}]]];
\nonu\\
&& \tt \text{\tt Clear}[\text{\tt solstep15}]
\nonu\\ 
&& \tt 
\text{\tt solstep15}=\text{\tt Simplify}[\text{\tt Solve}[\text{\tt TSAN}[[2]]==0,\text{\tt kappa}[2,1,-1]]]
\nonu\\ 
&& 
\tt \text{\tt Union}[\text{\tt Flatten}[\{\text{\tt SUB}\text{\tt /.}\text{\tt solstep15}[[1]],\text{\tt solstep15}[[1]]\}]]
\nonu\\ 
&& \tt \text{\tt SUB}=\%;
\nonu\\ 
&& \tt 
\text{\tt Clear}[\text{\tt TSAN}]
\nonu\\ 
&& \tt 
\text{\tt TSAN}=\text{\tt Union}[\text{\tt Flatten}[\text{\tt Simplify}[\text{\tt GetCoefficients}[\text{\tt TANZ}]\text{\tt /.}\text{\tt SUB}]]];
\nonu\\
&& \tt \text{\tt Clear}[\text{\tt solstep16}]
\nonu\\ 
&& \tt 
\text{\tt solstep16}=\text{\tt Simplify}\left[\text{\tt Solve}\left[\text{\tt TSAN}[[2]]==0,\text{\tt kappa}\left[2,\frac{3}{2},-\frac{3}{2}\right]\right]\right]
\nonu\\ 
&& 
\tt \text{\tt Union}[\text{\tt Flatten}[\{\text{\tt SUB}\text{\tt /.}\text{\tt solstep16}[[1]],\text{\tt solstep16}[[1]]\}]]
\nonu\\ 
&& \tt \text{\tt SUB}=\%;
\nonu\\ 
&& \tt 
\text{\tt Clear}[\text{\tt TSAN}]
\nonu\\ 
&& \tt 
\text{\tt TSAN}=\text{\tt Union}[\text{\tt Flatten}[\text{\tt Simplify}[\text{\tt GetCoefficients}[\text{\tt TANZ}]\text{\tt /.}\text{\tt SUB}]]];
\nonu\\
&& \tt \text{\tt Clear}[\text{\tt solstep17}]
\nonu\\ 
&& \tt 
\text{\tt solstep17}=\text{\tt Simplify}[\text{\tt Solve}[\text{\tt TSAN}[[2]]==0,\text{\tt kappa}[2,2,-2]]]
\nonu\\
&&
\text{(* This leads to the final solution. *)}
\nonu \\
&& \tt \text{\tt Union}[\text{\tt Flatten}[\{\text{\tt SUB}\text{\tt /.}\text{\tt solstep17}[[1]],\text{\tt solstep17}[[1]]\}]]
\nonu
\eea
At the final stage,  
we obtain the final solution for the
couplings (\ref{kappasol})
\footnote{The defining OPEs are written in terms of
the arbitrary $h_1$ and $h_2$ and the Jacobi identities
inside the Thielemans package can be checked for
the general $h_1$, $h_2$ and $h_3$ appearing in the three operators.}.

\section{More on the Jacobi identity}

We can also obtain (\ref{kappasol})
by considering the Jacobi identities
for the operators having the indices $A,B,C,D, \cdots =
1,2,3,4$ with nonzero $\Phi_0^{(h),5678}$
and it turns out that
there exist the following relations for the couplings
\bea
\kappa_{0, 0,+2}& = &
\frac{ \kappa_{+1, -\frac{1}{2}, +\frac{3}{2}} \,
\kappa_{+\frac{3}{2}, -\frac{3}{2},+2} \,\kappa_{+\frac{3}{2}, 0, +\frac{1}{2}}}{
\kappa_{+1,+1, 0}\, \kappa_{+\frac{3}{2}, +\frac{3}{2}, -1}}\,, 
\nonu\\ 
\kappa_{+\frac{1}{2}, -\frac{1}{2},+2}
&= & \frac{ \kappa_{+\frac{3}{2}, -\frac{3}{2},+2}\,\kappa_{+\frac{1}{2}, 0, +\frac{3}{2}}}{
\kappa_{+\frac{3}{2}, 0, +\frac{1}{2}}}\,, 
\qquad
\kappa_{+\frac{3}{2}, -\frac{1}{2},+1}= -\frac{\kappa_{+\frac{3}{2}, +\frac{3}{2}, -1} \,\kappa_{+1, 0,+1}}{ \kappa_{+\frac{3}{2}, 0, +\frac{1}{2}}}\,, 
\nonu\\ 
\kappa_{+\frac{3}{2}, +\frac{1}{2}, 0} &=& -\frac{
\kappa_{+1,+1, 0} \,\kappa_{+\frac{3}{2}, +\frac{3}{2}, -1}\,\kappa_{+\frac{1}{2}, 0, +\frac{3}{2}}}{ \kappa_{+1, -\frac{1}{2}, +\frac{3}{2}}\, \kappa_{+\frac{3}{2}, 0, +\frac{1}{2}}}\,, 
\qquad  \kappa_{+2, +\frac{1}{2}, -\frac{1}{2}}= \kappa_{+2, 0, 0}\,, 
\nonu\\
\kappa_{+2, -\frac{1}{2}, +\frac{1}{2}} &=& \kappa_{+2, 0, 0}\,, 
\qquad 
\kappa_{+1, -1,+2}= \frac{
\kappa_{+1, -\frac{1}{2}, +\frac{3}{2}}\, \kappa_{+\frac{3}{2}, -\frac{3}{2},+2} \,\kappa_{+\frac{3}{2}, 0, +\frac{1}{2}}}{
\kappa_{+\frac{3}{2}, +\frac{3}{2}, -1}\, \kappa_{+1, 0,+1} }\,, 
\nonu\\
\kappa_{+\frac{3}{2}, -1, +\frac{3}{2}}
&=& -\frac{\kappa_{+1, -\frac{1}{2}, +\frac{3}{2}}\, \kappa_{+\frac{3}{2}, 0, +\frac{1}{2}}}{  \kappa_{+1, 0,+1}}\,,
\qquad 
\kappa_{+\frac{3}{2},+1, -\frac{1}{2}}= -\frac{
\kappa_{+1, -\frac{1}{2}, +\frac{3}{2}}\, \kappa_{+\frac{3}{2}, 0, +\frac{1}{2}}}{
\kappa_{+\frac{1}{2}, 0, +\frac{3}{2}}}\,,
\nonu\\
\kappa_{+2,+1, -1} &=& \kappa_{+2, 0, 0}\,, \qquad 
\kappa_{+2, -1,+1}= \kappa_{+2, 0, 0}\,, \qquad 
\kappa_{+2, +\frac{3}{2}, -\frac{3}{2}}= \kappa_{+2, 0, 0}\,, 
\nonu\\
\kappa_{+2, -\frac{3}{2}, +\frac{3}{2}} &=& \kappa_{+2, 0, 0}\,, 
\qquad 
\kappa_{+2,+2, -2}= \kappa_{+2, 0, 0}\,,\qquad \kappa_{+2,-2, +2}=
\kappa_{+2, 0, 0}\,.
\label{interrelation}
\eea

Moreover, 
from the Jacobi identity between 
$(\Phi^{(h_1),A}_{+\frac{3}{2}},\Phi^{(h_2),B}_{+\frac{3}{2}},
\Phi^{(h_3),CDE}_{+\frac{1}{2}})$,
we obtain the relation
\bea
\Big(\kappa_{+1,-\frac{1}{2},+\frac{3}{2}}\,\kappa_{+1,+\frac{1}{2},+\frac{1}{2}}
+\kappa_{+\frac{1}{2},0,+\frac{3}{2}}\,\kappa_{+1,+1,0}\Big)=0\,,
\label{further1}
\eea
and
from the Jacobi identity between
$(\Phi^{(h_1),A}_{+\frac{3}{2}},\Phi^{(h_2),BC}_{+1},\Phi^{(h_3),DEF}_{+\frac{1}{2}})$,
the following relation satisfies
\bea
 \bigg(
\kappa_{+\frac{1}{2},+\frac{1}{2},+1}\,
\kappa_{+1,-\frac{1}{2},+\frac{3}{2}}\,
(\kappa_{+\frac{3}{2},0,+\frac{1}{2}})^2
+\kappa_{+\frac{1}{2},0,+\frac{3}{2}}\,\kappa_{+1,0,+1}\,
\kappa_{+1,+\frac{1}{2},+\frac{1}{2}}\,\kappa_{+\frac{3}{2},+\frac{3}{2},-1}
\bigg)=0\,.
\label{further2}
\eea
Then we obtain the final relations,
by combining (\ref{interrelation}) with (\ref{further1})
and (\ref{further2}),
between the couplings
as follows:
\bea
\kappa_{0, 0,+2} 
& = &
\frac{(\kappa_{+\frac{1}{2}, 0, +\frac{3}{2}})^2\, \kappa_{+1, 0,+1} \,\kappa_{+\frac{3}{2}, -\frac{3}{2},+2}}{ \kappa_{+\frac{1}{2}, +\frac{1}{2},+1}\, \kappa_{+1, -\frac{1}{2}, +\frac{3}{2}} \,\kappa_{+\frac{3}{2}, 0, +\frac{1}{2}}}\,, 
\nonu \\
\kappa_{+\frac{1}{2}, -\frac{1}{2},+2} 
& = &
\frac{\kappa_{+\frac{3}{2}, -\frac{3}{2},+2} \,\kappa_{+\frac{1}{2}, 0, +\frac{3}{2}} }{\kappa_{+\frac{3}{2}, 0, +\frac{1}{2}}}\,, 
\qquad 
\kappa_{+\frac{3}{2}, -\frac{1}{2},+1} 
= \frac{\kappa_{+\frac{1}{2}, +\frac{1}{2},+1}\, \kappa_{+1, -\frac{1}{2}, +\frac{3}{2}} \,\kappa_{+\frac{3}{2}, 0, +\frac{1}{2}}}{
 \kappa_{+\frac{1}{2}, 0, +\frac{3}{2}}\, \kappa_{+1, +\frac{1}{2}, +\frac{1}{2}}}\,, 
\nonu \\
\kappa_{+\frac{3}{2}, +\frac{1}{2}, 0} 
&=&
-\frac{\kappa_{+\frac{1}{2}, +\frac{1}{2},+1}\, \kappa_{+1, -\frac{1}{2}, +\frac{3}{2}} \,\kappa_{+\frac{3}{2}, 0, +\frac{1}{2}}}{
\kappa_{+\frac{1}{2}, 0, +\frac{3}{2}}\, \kappa_{+1, 0,+1}}\,, \qquad 
\kappa_{+2, +\frac{1}{2}, -\frac{1}{2}} = \kappa_{+2, 0, 0}\,, 
\nonu \\
\kappa_{+2, -\frac{1}{2}, +\frac{1}{2}} & = & \kappa_{+2, 0, 0}\,,
\qquad
\kappa_{+1, -1,+2} = -\frac{
  \kappa_{+\frac{1}{2}, 0, +\frac{3}{2}}\, \kappa_{+1, +\frac{1}{2}, +\frac{1}{2}} \,\kappa_{+\frac{3}{2}, -\frac{3}{2},+2}}{
  \kappa_{+\frac{1}{2}, +\frac{1}{2},+1} \,\kappa_{+\frac{3}{2}, 0, +\frac{1}{2}}}\,, 
\nonu \\
\kappa_{+\frac{3}{2}, -1, +\frac{3}{2}} 
&=&
-\frac{\kappa_{+1, -\frac{1}{2}, +\frac{3}{2}}\, \kappa_{+\frac{3}{2}, 0, +\frac{1}{2}}}{
  \kappa_{+1, 0,+1}}\,, 
\qquad
\kappa_{+\frac{3}{2},+1, -\frac{1}{2}} 
= -\frac{\kappa_{+1, -\frac{1}{2}, +\frac{3}{2}}\, \kappa_{+\frac{3}{2}, 0, +\frac{1}{2}}}{
  \kappa_{+\frac{1}{2}, 0, +\frac{3}{2}}}\,, 
\nonu \\
\kappa_{+1,+1, 0} & = &
-\frac{\kappa_{+1, -\frac{1}{2}, +\frac{3}{2}}\, \kappa_{+1, +\frac{1}{2}, +\frac{1}{2}}}{ \kappa_{+\frac{1}{2}, 0, +\frac{3}{2}}}\,, 
\nonu \\
\kappa_{+\frac{3}{2}, +\frac{3}{2}, -1} 
& = & -\frac{
  \kappa_{+\frac{1}{2}, +\frac{1}{2},+1} \,\kappa_{+1, -\frac{1}{2}, +\frac{3}{2}} \,(\kappa_{+\frac{3}{2}, 0, +\frac{1}{2}})^2}{\kappa_{+\frac{1}{2}, 0, +\frac{3}{2}} \,\kappa_{+1, 0,+1}\, \kappa_{+1, +\frac{1}{2}, +\frac{1}{2}}}\,, 
\nonu \\
\kappa_{+2, -1,+1} & = & \kappa_{+2, 0, 0}\,, \qquad
\kappa_{+2,+1, -1} = \kappa_{+2, 0, 0}\,, \qquad
\kappa_{+2, +\frac{3}{2}, -\frac{3}{2}} = \kappa_{+2, 0, 0}\,, \qquad
\nonu \\
\kappa_{+2,-\frac{3}{2}, +\frac{3}{2}} &=& \kappa_{+2, 0, 0}\,, \qquad
\kappa_{+2, -2,+2} = \kappa_{+2, 0, 0}\,, \qquad
\kappa_{+2,+2, -2} = \kappa_{+2, 0, 0}\,,
\label{otherrelation}
\eea
which is the same as the ones in (\ref{kappasol})
by rearranging the above relations (\ref{otherrelation}).
This implies that compared to the case of
${\cal N}=4$ supergravity, the present case of
${\cal N}=8$ supergravity
can be understood from the fact that by considering
the additional relations in the eq. $18$ and the eq. $22$,
the extended celestial soft symmetry algebra
exists.

\section{The celestial holography in the ${\cal N}=7$ supergravity }


We can decompose the $SU(8)$ indices into the $SU(7)$
indices 
and it turns out, from (\ref{Eqs}),  that
the corresponding (anti)commutators can be summarized
by
\bea
\comm{(\Phi^{(h_1)}_{+2})_m}{(\Phi^{(h_2)}_{+2})_n}&=&
\kappa_{+2,+2,-2}\,\Big((h_2+1)m-(h_1+1)n\Big)\,(\Phi^{(h_1+h_2)}_{+2})_{m+n}
\,\,\, :\text{eq.1},
\nonu \\
\comm{(\Phi^{(h_1)}_{+2})_m}{(\Phi^{(h_2),\hat{A}}_{+\frac{3}{2}})_r}&=&
\kappa_{+2,+\frac{3}{2},-\frac{3}{2}}\,\Big((h_2+\tfrac{1}{2})m-(h_1+1)r\Big)\,(\Phi^{(h_1+h_2),\hat{A}}_{+\frac{3}{2}})_{m+r}
\nonu \\
&:& \text{eq.2-1},
\nonu \\
\comm{(\Phi^{(h_1)}_{+2})_m}{(\Phi^{(h_2),8}_{+\frac{3}{2}})_r}&=&
\kappa_{+2,+\frac{3}{2},-\frac{3}{2}}\,\Big((h_2+\tfrac{1}{2})m-(h_1+1)r\Big)\,(\Phi^{(h_1+h_2),8}_{+\frac{3}{2}})_{m+r}
\nonu \\
&:& \text{eq.2-2},
\nonu \\
\comm{(\Phi^{(h_1)}_{+2})_m}{(\Phi^{(h_2),\hat{A}\hat{B}}_{+1})_n}&=&
\kappa_{+2,+1,-1}\,\Big(h_2\,m-(h_1+1)n \Big)\,(\Phi^{(h_1+h_2),\hat{A}\hat{B}}_{+1})_{m+n}
\,\,\, :\text{eq.3-1},
\nonu \\
\comm{(\Phi^{(h_1)}_{+2})_m}{(\Phi^{(h_2),\hat{A}8}_{+1})_n}&=&
\kappa_{+2,+1,-1}\,\Big(h_2\,m-(h_1+1)n \Big)\,(\Phi^{(h_1+h_2),\hat{A}8}_{+1})_{m+n}
\,\,\, :\text{eq.3-2},
\nonu \\
\comm{(\Phi^{(h_1)}_{+2})_m}{(\Phi^{(h_2),\hat{A}\hat{B}\hat{C}}_{+\frac{1}{2}})_r}
&=&\kappa_{+2,+\frac{1}{2},-\frac{1}{2}}\,\Big((h_2-\tfrac{1}{2})m-(h_1+1)r\Big)\,
(\Phi^{(h_1+h_2),\hat{A}\hat{B}\hat{C}}_{+\frac{1}{2}})_{m+r}
\nonu \\
&:& \text{eq.4-1},
\nonu \\   
\comm{(\Phi^{(h_1)}_{+2})_m}{(\Phi^{(h_2),\hat{A}\hat{B}8}_{+\frac{1}{2}})_r}
&=&\kappa_{+2,+\frac{1}{2},-\frac{1}{2}}\,\Big((h_2-\tfrac{1}{2})m-(h_1+1)r\Big)\,
(\Phi^{(h_1+h_2),\hat{A}\hat{B}8}_{+\frac{1}{2}})_{m+r}
\nonu \\
&:& \text{eq.4-2},
\nonu \\   
\comm{(\Phi^{(h_1)}_{+2})_m}{(\Phi^{(h_2),\hat{A}\hat{B}\hat{C}\hat{D}}_{0})_n}
&=&\kappa_{+2,0,0}\,\Big((h_2-1)m-(h_1+1)n\Big)\,(\Phi^{(h_1+h_2),\hat{A}\hat{B}\hat{C}\hat{D}}_{0})_{m+n}
\nonu \\
&:&\text{eq.5-1},
\nonu \\
\comm{(\Phi^{(h_1)}_{+2})_m}{(\Phi^{(h_2),\hat{A}\hat{B}\hat{C}8}_{0})_n}
&=&\kappa_{+2,0,0}\,\Big((h_2-1)m-(h_1+1)n\Big)\,(\Phi^{(h_1+h_2),
\hat{A}\hat{B}\hat{C}8}_{0})_{m+n}
\nonu \\
&: & \text{eq.5-2},
\nonu \\
\comm{(\Phi^{(h_1)}_{+2})_m}{(\Phi^{(h_2)}_{\hat{A}\hat{B}\hat{C},-\frac{1}{2}})_r}
&=&\kappa_{+2,-\frac{1}{2},+\frac{1}{2}}\,\,\Big((h_2-\tfrac{3}{2})m-(h_1+1)r\Big)\,(\Phi^{(h_1+h_2)}_{\hat{A}\hat{B}\hat{C},-\frac{1}{2}})_{m+r}
\,\,\, :\text{eq.6-1},
\nonu \\
\comm{(\Phi^{(h_1)}_{+2})_m}{(\Phi^{(h_2)}_{\hat{A}\hat{B}8,-\frac{1}{2}})_r}
&=&\kappa_{+2,-\frac{1}{2},+\frac{1}{2}}\,\,\Big((h_2-\tfrac{3}{2})m-(h_1+1)r\Big)\,(\Phi^{(h_1+h_2)}_{\hat{A}\hat{B}8,-\frac{1}{2}})_{m+r}
\,\,\, :\text{eq.6-2},
\nonu \\
\comm{(\Phi^{(h_1)}_{+2})_m}{(\Phi^{(h_2)}_{\hat{A}\hat{B},-1})_n}
&=&\kappa_{+2,-1,+1}\,\Big((h_2-2)m-(h_1+1)n\Big)\,(\Phi^{(h_1+h_2)}_{
\hat{A}\hat{B},-1})_{m+n}
\, :\text{eq.7-1},
\nonu    \\
\comm{(\Phi^{(h_1)}_{+2})_m}{(\Phi^{(h_2)}_{\hat{A}8,-1})_n}
&=&\kappa_{+2,-1,+1}\,\Big((h_2-2)m-(h_1+1)n\Big)\,(\Phi^{(h_1+h_2)}_{
\hat{A}8,-1})_{m+n}
\, :\text{eq.7-2},
\nonu    \\
\comm{(\Phi^{(h_1)}_{+2})_m}{(\Phi^{(h_2)}_{\hat{A},-\frac{3}{2}})_r}
&=&
\kappa_{+2,-\frac{3}{2},+\frac{3}{2}}\,\Big((h_2-\tfrac{5}{2})m-(h_1+1)r\Big)\,(\Phi^{(h_1+h_2)}_{\hat{A},-\frac{3}{2}})_{m+r}
\, :\text{eq.8-1},
\nonu    \\
\comm{(\Phi^{(h_1)}_{+2})_m}{(\Phi^{(h_2)}_{8,-\frac{3}{2}})_r}
&=&
\kappa_{+2,-\frac{3}{2},+\frac{3}{2}}\,\Big((h_2-\tfrac{5}{2})m-(h_1+1)r\Big)\,(\Phi^{(h_1+h_2)}_{8,-\frac{3}{2}})_{m+r}
\, :\text{eq.8-2},
\nonu    \\
\comm{(\Phi^{(h_1)}_{+2})_m}{(\Phi^{(h_2)}_{-2})_n}&=&
\kappa_{+2,-2,+2}\,\Big((h_2-3)m-(h_1+1)n\Big)\,(\Phi^{(h_1+h_2)}_{-2})_{m+n}
\, :\text{eq.9},
\nonu \\
\acomm{(\Phi^{(h_1),\hat{A}}_{+\frac{3}{2}})_r}{(\Phi^{(h_2),\hat{B}}_{+\frac{3}{2}})_s}
&=&\kappa_{+\frac{3}{2},+\frac{3}{2},-1}\,
\Big((h_2+\tfrac{1}{2})r-(h_1+\tfrac{1}{2})s\Big)\,(\Phi^{(h_1+h_2),\hat{A}\hat{B}}_{+1})_{r+s}\,,
\nonu \\
&:& \text{eq.10-1},
\nonu \\
\acomm{(\Phi^{(h_1),\hat{A}}_{+\frac{3}{2}})_r}{(\Phi^{(h_2),8}_{+\frac{3}{2}})_s}
&=&\kappa_{+\frac{3}{2},+\frac{3}{2},-1}\,
\Big((h_2+\tfrac{1}{2})r-(h_1+\tfrac{1}{2})s\Big)\,
(\Phi^{(h_1+h_2),\hat{A}8}_{+1})_{r+s}\,,
\nonu \\
&:& \text{eq.10-2},
\nonu \\
\comm{(\Phi^{(h_1),\hat{A}}_{+\frac{3}{2}})_r}{(\Phi^{(h_2),\hat{B}\hat{C}}_{+1})_m}
&=&\kappa_{+\frac{3}{2},+1,-\frac{1}{2}}\,\Big(h_2\,r-(h_1+\tfrac{1}{2})m\Big)\,(\Phi^{(h_1+h_2),\hat{A}\hat{B}\hat{C}}_{+\frac{1}{2}})_{r+m}
\,\,\, :\text{eq.11-1},
\nonu \\
\comm{(\Phi^{(h_1),\hat{A}}_{+\frac{3}{2}})_r}{(\Phi^{(h_2),\hat{B}8}_{+1})_m}
&=&\kappa_{+\frac{3}{2},+1,-\frac{1}{2}}\,\Big(h_2\,r-(h_1+\tfrac{1}{2})m\Big)\,(\Phi^{(h_1+h_2),\hat{A}\hat{B}8}_{+\frac{1}{2}})_{r+m}
\,\,\, :\text{eq.11-2},
\nonu \\
\comm{(\Phi^{(h_1),8}_{+\frac{3}{2}})_r}{(\Phi^{(h_2),\hat{B}\hat{C}}_{+1})_m}
&=&\kappa_{+\frac{3}{2},+1,-\frac{1}{2}}\,\Big(h_2\,r-(h_1+\tfrac{1}{2})m\Big)\,(\Phi^{(h_1+h_2),8\hat{B}\hat{C}}_{+\frac{1}{2}})_{r+m}
\,\,\, :\text{eq.11-3},
\nonu \\
\acomm{(\Phi^{(h_1),\hat{A}}_{+\frac{3}{2}})_r}{(\Phi^{(h_2),\hat{B}\hat{C}\hat{D}}_{+\frac{1}{2}})_s}
&=&\kappa_{+\frac{3}{2},+\frac{1}{2},0}\,\Big((h_2-\tfrac{1}{2})r-(h_1+\tfrac{1}{2})s\Big)(\Phi^{(h_1+h_2),\hat{A}\hat{B}\hat{C}\hat{D}}_0)_{r+s}
\nonu \\
&:& \text{eq.12-1},
\nonu \\
\acomm{(\Phi^{(h_1),\hat{A}}_{+\frac{3}{2}})_r}{
(\Phi^{(h_2),\hat{B}\hat{C}8}_{+\frac{1}{2}})_s}
&=&\kappa_{+\frac{3}{2},+\frac{1}{2},0}\,\Big((h_2-\tfrac{1}{2})r-(h_1+\tfrac{1}{2})s\Big)(\Phi^{(h_1+h_2),\hat{A}\hat{B}\hat{C}8}_0)_{r+s}
\nonu \\
&:& \text{eq.12-2},
\nonu \\
\acomm{(\Phi^{(h_1),8}_{+\frac{3}{2}})_r}{(\Phi^{(h_2),\hat{B}\hat{C}\hat{D}}_{+\frac{1}{2}})_s}
&=&\kappa_{+\frac{3}{2},+\frac{1}{2},0}\,\Big((h_2-\tfrac{1}{2})r-(h_1+\tfrac{1}{2})s\Big)(\Phi^{(h_1+h_2),8\hat{B}\hat{C}\hat{D}}_0)_{r+s}
\nonu \\
&:& \text{eq.12-3},
\nonu \\
\comm{(\Phi^{(h_1),\hat{A}}_{+\frac{3}{2}})_r}{
(\Phi^{(h_2),\hat{B}\hat{C}\hat{D}\hat{E}}_0)_m}&=&
\kappa_{+\frac{3}{2},0,+\frac{1}{2}}\,\Big((h_2-1)r-(h_1+\tfrac{1}{2})m\Big)\,
\frac{1}{2!}\,
\epsilon^{\hat{A}\hat{B}\hat{C}\hat{D}\hat{E}\hat{F}\hat{G}8}\nonu \\
& \times & (\Phi^{(h_1+h_2)}_{
\hat{F}\hat{G}8,-\frac{1}{2}})_{r+m}
\,\,\, :\text{eq.13-1},
\nonu \\
\comm{(\Phi^{(h_1),\hat{A}}_{+\frac{3}{2}})_r}{(\Phi^{(h_2),\hat{B}\hat{C}\hat{D}8}_0)_m}&=&
\kappa_{+\frac{3}{2},0,+\frac{1}{2}}\,\Big((h_2-1)r-(h_1+\tfrac{1}{2})m\Big)\,
\frac{1}{3!}\,
\epsilon^{\hat{A}\hat{B}\hat{C}\hat{D}8\hat{F}\hat{G}\hat{H}}
\nonu \\
& \times & (\Phi^{(h_1+h_2)}_{
\hat{F}\hat{G}\hat{H},-\frac{1}{2}})_{r+m}
\,\,\, : \text{eq.13-2},
\nonu \\
\comm{(\Phi^{(h_1),8}_{+\frac{3}{2}})_r}{
(\Phi^{(h_2),\hat{B}\hat{C}\hat{D}\hat{E}}_0)_m}&=&
\kappa_{+\frac{3}{2},0,+\frac{1}{2}}\,\Big((h_2-1)r-(h_1+\tfrac{1}{2})m\Big)\,
\frac{1}{3!}\,
\epsilon^{8\hat{B}\hat{C}\hat{D}\hat{E}\hat{F}\hat{G}\hat{H}}\nonu \\
& \times & (\Phi^{(h_1+h_2)}_{
\hat{F}\hat{G}\hat{H},-\frac{1}{2}})_{r+m}
\,\,\, :\text{eq.13-3},
\nonu \\
\acomm{(\Phi^{(h_1),\hat{A}}_{+\frac{3}{2}})_r}{(\Phi^{(h_2)}_{
\hat{B}\hat{C}\hat{D},-\frac{1}{2}})_s}
&
=& \kappa_{+\frac{3}{2},-\frac{1}{2},+1}\,
\Big((h_2-\tfrac{3}{2})r-(h_1+\tfrac{1}{2})s\Big)\,3
\delta^{\hat{A}}_{\,\,\,[\hat{B}}(\Phi^{(h_1+h_2)}_{\hat{C}\hat{D}],-1})_{r+s}\,
\nonu \\
&:& \text{eq.14-1},
\nonu \\
\acomm{(\Phi^{(h_1),\hat{A}}_{+\frac{3}{2}})_r}{(\Phi^{(h_2)}_{
\hat{B}\hat{C}8,-\frac{1}{2}})_s}
&
=& \kappa_{+\frac{3}{2},-\frac{1}{2},+1}\,
\Big((h_2-\tfrac{3}{2})r-(h_1+\tfrac{1}{2})s\Big)\,3
\delta^{\hat{A}}_{\,\,\,[\hat{B}}(\Phi^{(h_1+h_2)}_{\hat{C}8],-1})_{r+s}\,
\nonu \\
&:& \text{eq.14-2},
\nonu \\
\acomm{(\Phi^{(h_1),8}_{+\frac{3}{2}})_r}{(\Phi^{(h_2)}_{
\hat{B}\hat{C}8,-\frac{1}{2}})_s}
&
=& \kappa_{+\frac{3}{2},-\frac{1}{2},+1}\,
\Big((h_2-\tfrac{3}{2})r-(h_1+\tfrac{1}{2})s\Big)\,3
\delta^{8}_{\,\,\,[\hat{B}}(\Phi^{(h_1+h_2)}_{\hat{C}8],-1})_{r+s}\,
\nonu \\
&:& \text{eq.14-3},
\nonu \\
\comm{(\Phi^{(h_1),\hat{A}}_{+\frac{3}{2}})_r}{(\Phi^{(h_2)}_{\hat{B}\hat{C},-1})_m}
&=&
\kappa_{+\frac{3}{2},-1,+\frac{3}{2}}\,\Big((h_2-2)r-(h_1+\tfrac{1}{2})m\Big)\,
\nonu \\
& \times & 2!\, \delta^{\hat{A}}_{\,\,\,[\hat{B}}\,
(\Phi^{(h_1+h_2)}_{\hat{C}],-\frac{3}{2}})_{r+m}\, :\text{eq.15-1},
\nonu \\
\comm{(\Phi^{(h_1),8}_{+\frac{3}{2}})_r}{(\Phi^{(h_2)}_{\hat{B}8,-1})_m}
&=&
\kappa_{+\frac{3}{2},-1,+\frac{3}{2}}\,\Big((h_2-2)r-(h_1+\tfrac{1}{2})m\Big)\,
\nonu \\
& \times & 2!\, \delta^{8}_{\,\,\,[\hat{B}}\,
(\Phi^{(h_1+h_2)}_{8],-\frac{3}{2}})_{r+m}\, :\text{eq.15-2},
\nonu \\
\comm{(\Phi^{(h_1),\hat{A}}_{+\frac{3}{2}})_r}{(\Phi^{(h_2)}_{\hat{B}8,-1})_m}
&=&
\kappa_{+\frac{3}{2},-1,+\frac{3}{2}}\,\Big((h_2-2)r-(h_1+\tfrac{1}{2})m\Big)\,
\nonu \\
& \times & 2!\, \delta^{\hat{A}}_{\,\,\,[\hat{B}}\,
(\Phi^{(h_1+h_2)}_{8],-\frac{3}{2}})_{r+m}\, :\text{eq.15-3},
\nonu \\
\acomm{(\Phi^{(h_1),\hat{A}}_{+\frac{3}{2}})_r}{(\Phi^{(h_2)}_{\hat{B},-\frac{3}{2}})_s}
&=&
\kappa_{+\frac{3}{2},-\frac{3}{2},+2}\,\Big((h_2-\tfrac{5}{2})r-(h_1+\tfrac{1}{2})s\Big)\,\delta^{\hat{A}}_{\,\,\,\hat{B}}\,(\Phi^{(h_1+h_2)}_{-2})_{r+s}
\nonu \\
&:& \text{eq.16-1},
\nonu \\
\acomm{(\Phi^{(h_1),8}_{+\frac{3}{2}})_r}{(\Phi^{(h_2)}_{8,-\frac{3}{2}})_s}
&=&
\kappa_{+\frac{3}{2},-\frac{3}{2},+2}\,\Big((h_2-\tfrac{5}{2})r-(h_1+\tfrac{1}{2})s\Big)\,\delta^{8}_{\,\,\,8}\,(\Phi^{(h_1+h_2)}_{-2})_{r+s}
\nonu \\
&:& \text{eq.16-2},
\nonu \\
\comm{(\Phi^{(h_1),\hat{A}\hat{B}}_{+1})_m}{(\Phi^{(h_2),\hat{C}\hat{D}}_{+1})_n}&=&
\kappa_{+1,+1,0}\,\Big(h_2\,m-h_1\,n\Big)\,(\Phi^{(h_1+h_2),\hat{A}\hat{B}\hat{C}\hat{D}}_{0})_{m+n}
\,\,\, :\text{eq.17-1},
\nonu \\
\comm{(\Phi^{(h_1),\hat{A}\hat{B}}_{+1})_m}{(\Phi^{(h_2),\hat{C}8}_{+1})_n}&=&
\kappa_{+1,+1,0}\,\Big(h_2\,m-h_1\,n\Big)\,
(\Phi^{(h_1+h_2),\hat{A}\hat{B}\hat{C}8}_{0})_{m+n}
\,\,\, :\text{eq.17-2},
\nonu \\
\comm{(\Phi^{(h_1),\hat{A}8}_{+1})_m}{(\Phi^{(h_2),\hat{C}\hat{D}}_{+1})_n}&=&
\kappa_{+1,+1,0}\,\Big(h_2\,m-h_1\,n\Big)\,(\Phi^{(h_1+h_2),\hat{A}8\hat{C}\hat{D}}_{0})_{m+n}
\,\,\, :\text{eq.17-3},
\nonu \\
\comm{(\Phi^{(h_1),\hat{A}\hat{B}}_{+1})_m}{(\Phi^{(h_2),\hat{C}\hat{D}\hat{E}}_{+\frac{1}{2}})_r}&=&
\kappa_{+1,+\frac{1}{2},+\frac{1}{2}}\,\Big((h_2-\tfrac{1}{2})m-h_1\,r\Big)\,
\frac{1}{2!}\,
\epsilon^{\hat{A}\hat{B}\hat{C}\hat{D}\hat{E}\hat{F}\hat{G}8}\nonu \\
& \times & (\Phi^{(h_1+h_2)}_{
\hat{F}\hat{G}8,
-\frac{1}{2}})_{m+r}
:\text{eq.18-1},
\nonu \\
\comm{(\Phi^{(h_1),\hat{A}\hat{B}}_{+1})_m}{(\Phi^{(h_2),\hat{C}\hat{D}8}_{+\frac{1}{2}})_r}&=&
\kappa_{+1,+\frac{1}{2},+\frac{1}{2}}\,\Big((h_2-\tfrac{1}{2})m-h_1\,r\Big)\,
\frac{1}{3!}\,
\epsilon^{\hat{A}\hat{B}\hat{C}\hat{D}8\hat{F}\hat{G}\hat{H}}
\nonu \\
& \times & (\Phi^{(h_1+h_2)}_{\hat{F}\hat{G}\hat{H},-\frac{1}{2}})_{m+r}
 :\text{eq.18-2},
\nonu \\
\comm{(\Phi^{(h_1),\hat{A}8}_{+1})_m}{(\Phi^{(h_2),
\hat{C}\hat{D}\hat{E}}_{+\frac{1}{2}})_r}&=&
\kappa_{+1,+\frac{1}{2},+\frac{1}{2}}\,\Big((h_2-\tfrac{1}{2})m-h_1\,r\Big)\,
\frac{1}{3!}\,
\epsilon^{\hat{A}8\hat{C}\hat{D}\hat{E}\hat{F}\hat{G}\hat{H}}
\nonu \\
& \times & (\Phi^{(h_1+h_2)}_{\hat{F}\hat{G}\hat{H},-\frac{1}{2}})_{m+r}
 :\text{eq.18-3},
\nonu \\
\comm{(\Phi^{(h_1),\hat{A}\hat{B}}_{+1})_m}{
(\Phi^{(h_2),\hat{C}\hat{D}\hat{E}\hat{F}}_0)_n}&=&
\kappa_{+1,0,+1}\,\Big((h_2-1)m-h_1\,n\Big)\,
\epsilon^{\hat{A}\hat{B}\hat{C}\hat{D}\hat{E}\hat{F}8}\,\frac{1}{2!}\,
(\Phi^{(h_1+h_2)}_{\hat{F}8,-1})_{m+n}
\nonu \\
&: & \text{eq.19-1},
\nonu \\
\comm{(\Phi^{(h_1),\hat{A}\hat{B}}_{+1})_m}{(\Phi^{(h_2),\hat{C}\hat{D}\hat{E}8}_0)_n}&=&
\kappa_{+1,0,+1}\,\Big((h_2-1)m-h_1\,n\Big)\,
\epsilon^{\hat{A}\hat{B}\hat{C}\hat{D}\hat{E}8\hat{G}\hat{H}}\,\frac{1}{2!}\,
(\Phi^{(h_1+h_2)}_{\hat{G}\hat{H},-1})_{m+n}
\nonu \\
&: & \text{eq.19-2},
\nonu \\
\comm{(\Phi^{(h_1),\hat{A}8}_{+1})_m}{(\Phi^{(h_2),\hat{B}\hat{C}\hat{D}\hat{E}}_0)_n}&=&
\kappa_{+1,0,+1}\,\Big((h_2-1)m-h_1\,n\Big)\,
\epsilon^{\hat{A}8\hat{B}\hat{C}\hat{D}\hat{E}\hat{F}\hat{G}}\,\frac{1}{2!}\,
(\Phi^{(h_1+h_2)}_{\hat{F}\hat{G},-1})_{m+n}
\nonu \\
&: & \text{eq.19-3},
\nonu \\
\comm{(\Phi^{(h_1),\hat{A}\hat{B}}_{+1})_m}{(\Phi^{(h_2)}_{\hat{C}\hat{D}\hat{E},-\frac{1}{2}})_r}&=&
\kappa_{+1,-\frac{1}{2},+\frac{3}{2}}\,\Big((h_2-\tfrac{3}{2})m-h_1\,r\Big)\,
3! \delta^{\hat{A}}_{\,\,\,[\hat{C}}(\Phi^{(h_1+h_2)}_{\hat{D},-\frac{3}{2}})_{m+r}\delta_{\hat{E}]}^{\,\,\,\hat{B}}
\, \nonu \\
&: & \text{eq.20-1},
\nonu \\
\comm{(\Phi^{(h_1),\hat{A}8}_{+1})_m}{(\Phi^{(h_2)}_{\hat{C}\hat{D}8,-\frac{1}{2}})_r}&=&
\kappa_{+1,-\frac{1}{2},+\frac{3}{2}}\,\Big((h_2-\tfrac{3}{2})m-h_1\,r\Big)\,
3! \delta^{\hat{A}}_{\,\,\,[\hat{C}}
(\Phi^{(h_1+h_2)}_{\hat{D},-\frac{3}{2}})_{m+r}\delta_{8]}^{\,\,\,8}
\, \nonu \\
&: & \text{eq.20-2},
\nonu \\
\comm{(\Phi^{(h_1),\hat{A}\hat{B}}_{+1})_m}{(\Phi^{(h_2)}_{\hat{C}\hat{D}8,-\frac{1}{2}})_r}&=&
\kappa_{+1,-\frac{1}{2},+\frac{3}{2}}\,\Big((h_2-\tfrac{3}{2})m-h_1\,r\Big)\,
3! \delta^{\hat{A}}_{\,\,\,[\hat{C}}
(\Phi^{(h_1+h_2)}_{\hat{D},-\frac{3}{2}})_{m+r}\delta_{8]}^{\,\,\,\hat{B}}
\, \nonu \\
&: & \text{eq.20-3},
\nonu \\
\comm{(\Phi^{(h_1),\hat{A}\hat{B}}_{+1})_m}{(\Phi^{(h_2)}_{\hat{C}\hat{D},-1})_n}&=&
\kappa_{+1,-1,+2}\,\Big((h_2-2)m-h_1\,n\Big)\,
\delta^{\hat{A}\hat{B}}_{\hat{C}\hat{D}} \, (\Phi^{(h_1+h_2)}_{-2})_{m+n}
\, : \text{eq.21-1},
\nonu \\
\comm{(\Phi^{(h_1),\hat{A}8}_{+1})_m}{(\Phi^{(h_2)}_{\hat{A}8,-1})_n}&=&
\kappa_{+1,-1,+2}\,\Big((h_2-2)m-h_1\,n\Big)\,
\delta^{\hat{A}8}_{\hat{A}8} \, (\Phi^{(h_1+h_2)}_{-2})_{m+n}
\, : \text{eq.21-2},
\nonu \\
\acomm{(\Phi^{(h_1),\hat{A}\hat{B}\hat{C}}_{+\frac{1}{2}})_r}{(\Phi^{(h_2),\hat{D}\hat{E}
\hat{F}}_{+\frac{1}{2}})_s}&=&
\kappa_{+\frac{1}{2},+\frac{1}{2},+1}\,\Big((h_2-\tfrac{1}{2})r-(h_1-\tfrac{1}{2})s\Big)\,\frac{1}{2!}\,\epsilon^{\hat{A}\hat{B}\hat{C}\hat{D}\hat{E}
\hat{F}\hat{G}8}
\nonu \\
& \times & (\Phi^{(h_1+h_2)}_{\hat{G}8,-1})_{r+s}
\, :  \text{eq.22-1},
\nonu \\
\acomm{(\Phi^{(h_1),\hat{A}\hat{B}\hat{C}}_{+\frac{1}{2}})_r}{(\Phi^{(h_2),\hat{D}\hat{E}8}_{+\frac{1}{2}})_s}&=&
\kappa_{+\frac{1}{2},+\frac{1}{2},+1}\,\Big((h_2-\tfrac{1}{2})r-(h_1-\tfrac{1}{2})s\Big)\,\frac{1}{2!}\,\epsilon^{\hat{A}\hat{B}\hat{C}\hat{D}\hat{E}8
\hat{G}\hat{H}}
\nonu \\
& \times & (\Phi^{(h_1+h_2)}_{\hat{G}\hat{H},-1})_{r+s}
\, :  \text{eq.22-2},
\nonu \\
\acomm{(\Phi^{(h_1),\hat{A}\hat{B}8}_{+\frac{1}{2}})_r}{(\Phi^{(h_2),\hat{D}
\hat{E}\hat{F}}_{+\frac{1}{2}})_s}&=&
\kappa_{+\frac{1}{2},+\frac{1}{2},+1}\,\Big((h_2-\tfrac{1}{2})r-(h_1-\tfrac{1}{2})s\Big)\,\frac{1}{2!}\,\epsilon^{\hat{A}\hat{B}8\hat{D}\hat{E}\hat{F}
\hat{G}\hat{H}}
\nonu \\
& \times & (\Phi^{(h_1+h_2)}_{\hat{G}\hat{H},-1})_{r+s}
\, :  \text{eq.22-3},
\nonu \\
\comm{(\Phi^{(h_1),\hat{A}\hat{B}\hat{C}}_{+\frac{1}{2}})_r}{(\Phi^{(h_2),
\hat{D}\hat{E}\hat{F}\hat{G}}_0)_m}&=&
\kappa_{+\frac{1}{2},0,+\frac{3}{2}}\,\Big((h_2-1)r-(h_1-\tfrac{1}{2})m\Big)\,
\epsilon^{\hat{A}\hat{B}\hat{C}\hat{D}\hat{E}\hat{F}\hat{G}8}
\nonu \\
& \times & (\Phi^{(h_1+h_2)}_{8,-\frac{3}{2}})_{r+m}
\, :  \text{eq.23-1},
\nonu \\
\comm{(\Phi^{(h_1),\hat{A}\hat{B}\hat{C}}_{+\frac{1}{2}})_r}{(\Phi^{(h_2),
\hat{D}\hat{E}\hat{F}8}_0)_m}&=&
\kappa_{+\frac{1}{2},0,+\frac{3}{2}}\,\Big((h_2-1)r-(h_1-\tfrac{1}{2})m\Big)\,
\epsilon^{\hat{A}\hat{B}\hat{C}\hat{D}\hat{E}\hat{F}\hat{G}8\hat{H}}
\nonu \\
& \times & (\Phi^{(h_1+h_2)}_{\hat{H},-\frac{3}{2}})_{r+m}
\, :  \text{eq.23-2},
\nonu \\
\comm{(\Phi^{(h_1),\hat{A}\hat{B}8}_{+\frac{1}{2}})_r}{(\Phi^{(h_2),
\hat{D}\hat{E}\hat{F}\hat{G}}_0)_m}&=&
\kappa_{+\frac{1}{2},0,+\frac{3}{2}}\,\Big((h_2-1)r-(h_1-\tfrac{1}{2})m\Big)\,
\epsilon^{\hat{A}\hat{B}8\hat{D}\hat{E}\hat{F}\hat{G}\hat{H}}
\nonu \\
& \times & (\Phi^{(h_1+h_2)}_{\hat{H},-\frac{3}{2}})_{r+m}
\, :  \text{eq.23-3},
\nonu \\
\acomm{(\Phi^{(h_1),\hat{A}\hat{B}\hat{C}}_{+\frac{1}{2}})_r}{
(\Phi^{(h_2)}_{\hat{D}\hat{E}\hat{F},-\frac{1}{2}})_s}
&=&\kappa_{+\frac{1}{2},-\frac{1}{2},+2}\,\Big((h_2-\tfrac{3}{2})r-(h_1-\tfrac{1}{2})s\Big)\,
\delta^{\hat{A}\hat{B}\hat{C}}_{\hat{D}\hat{E}\hat{F}}\,(\Phi^{(h_1+h_2)}_{-2})_{r+s}
\nonu \\
&: & \text{eq.24-1},
\nonu \\
\acomm{(\Phi^{(h_1),\hat{A}\hat{B}8}_{+\frac{1}{2}})_r}{
(\Phi^{(h_2)}_{\hat{D}\hat{E}8,-\frac{1}{2}})_s}
&=&\kappa_{+\frac{1}{2},-\frac{1}{2},+2}\,\Big((h_2-\tfrac{3}{2})r-(h_1-\tfrac{1}{2})s\Big)\,
\delta^{\hat{A}\hat{B}8}_{\hat{D}\hat{E}8}\,(\Phi^{(h_1+h_2)}_{-2})_{r+s}
\nonu \\
&: & \text{eq.24-2},
\nonu \\
\comm{(\Phi^{(h_1),\hat{A}\hat{B}\hat{C}\hat{D}}_0)_m}{
(\Phi^{(h_2),\hat{E}\hat{F}\hat{G}8}_0)_n}&=&
\kappa_{0,0,+2}\,\Big((h_2-1)m-(h_1-1)n\Big)
\epsilon^{\hat{A}\hat{B}\hat{C}\hat{D}\hat{E}\hat{F}\hat{G}8}\,(\Phi^{(h_1+h_2)}_{-2})_{m+n}
\,\,\,
\nonu \\
& : & \text{eq.25}\,,
\label{su7anticomm}
\eea
where $\hat{A}, \hat{B}, \cdots =1,2, \cdots, 7$
\footnote{
\label{su7foot}
The tensor product of $SU(7)$ \cite{FKS} in the lower representations
is given by  
\bea
{\bf 7} \otimes {\bf 7} &=& {\bf 21} \oplus \cdots,
\qquad
{\bf 7} \otimes {\bf 21} = {\bf 35} \oplus \cdots,    
\qquad
{\bf 7} \otimes {\bf 35} = \overline{\bf 35} \oplus \cdots,
\qquad
{\bf 7} \otimes \overline{\bf 7} = {\bf 1} \oplus \cdots,
\nonu \\
{\bf 7} \otimes \overline{\bf 21} &=& \overline{\bf 7} \oplus \cdots,
\qquad
{\bf 7} \otimes \overline{\bf 35} =
\overline{\bf 21} \oplus \cdots,      
\qquad
{\bf 21} \otimes {\bf 21} = \overline{\bf 35} \oplus \cdots,
\qquad
{\bf 21} \otimes {\bf 35} = \overline{\bf 21} \oplus \cdots,
\nonu \\
{\bf 21} \otimes \overline{\bf 7} &=& {\bf 7} \oplus \cdots,
\qquad
{\bf 21} \otimes \overline{\bf 21} = {\bf 1} \oplus \cdots,
\qquad
{\bf 21} \otimes \overline{\bf 35} = \overline{\bf 7} \oplus \cdots,
\qquad
{\bf 35} \otimes {\bf 35} = \overline{\bf 7} \oplus \cdots,
\nonu \\
{\bf 35} \otimes \overline{\bf 7} &=& {\bf 21} \oplus \cdots,
\qquad
{\bf 35} \otimes \overline{\bf 21} = {\bf 21} \oplus \cdots,
\qquad
{\bf 35} \otimes \overline{\bf 35} = {\bf 1} \oplus \cdots \, .
\nonu
\eea
The branching rule of $SU(8) \rightarrow SU(7) \times U(1)$
is given by
\bea
{\bf 8} & \rightarrow & {\bf 1}_{-7} \oplus {\bf 7}_{1} \, ,
\qquad
{\bf 28}  \rightarrow  {\bf 7}_{-6} \oplus {\bf 21}_{2} \, ,  
\qquad
{\bf 56}  \rightarrow  {\bf 21}_{-5} \oplus {\bf 35}_{3} \, ,
\qquad
{\bf 70}  \rightarrow  {\bf 35}_{-4} \oplus \overline{\bf 35}_{4} \, ,
\nonu \\
\overline{\bf 56} & \rightarrow & \overline{\bf 21}_{5} \oplus
\overline {\bf 35}_{-3} \, ,
\qquad
\overline{\bf 28}  \rightarrow  \overline{\bf 7}_{6} \oplus
\overline{\bf 21}_{-2} \, ,  
\qquad
\overline{\bf 8}  \rightarrow  {\bf 1}_{7} \oplus \overline{\bf 7}_{-1} \, .
\nonu       
\eea}.
Note that according to the branching rule of
$SU(8) \rightarrow SU(7) \times U(1)$ in the footnote \ref{su7foot},
there are no vanishing operators (due to the
fact that there are no additional nonabelian groups
except the $SU(7)$),
compared to other three truncations
which will appear later.
There are
operators having the index $8$ in (\ref{su7anticomm})
\bea
\Phi^{(h),8}_{+\frac{3}{2}}, \qquad
\Phi^{(h)}_{8, -\frac{3}{2}}, \qquad
\Phi^{(h),\hat{A}8}_{+1}, \qquad
\Phi^{(h)}_{\hat{A}8,-1}, \qquad
\Phi^{(h),\hat{A}\hat{B}8}_{+\frac{1}{2}}, \qquad
\Phi^{(h)}_{\hat{A}\hat{B}8, -\frac{1}{2}}, \qquad
\Phi^{(h),\hat{A}\hat{B}\hat{C}8}_{0}.
\label{8index}
\eea
In (\ref{8index}), the first two play the role of
scalars (${\bf 1}_{-7}$ and ${\bf 1}_{7}$ ),
the next two correspond to the fundamental and antifundamental
representations
(${\bf 7}_{-6}$ and $\overline{\bf 7}_{6}$),
the next two correspond to the antisymmetric representations
(${\bf 21}_{-5}$ and $\overline{\bf 21}_{5}$),
and the last plays the role of
the antisymmetric representation (${\bf 35}_{-4}$) where the
Young tableaux is given by three boxes in one column.

\section{The celestial holography in the ${\cal N}=6$ supergravity }


We can decompose the $SU(8)$ indices into the $SU(6)$
indices 
and it turns out that
the corresponding (anti)commutators can be described
by
\footnote{
\label{su6foot}
The tensor product of $SU(6)$ in the lower representations
\cite{FKS}
is given by  
\bea
{\bf 6} \otimes {\bf 6} &=& {\bf 15} \oplus \cdots,
\qquad
{\bf 6} \otimes {\bf 15} = {\bf 20} \oplus \cdots,    
\qquad
{\bf 6} \otimes {\bf 20} = \overline{\bf 15} \oplus \cdots,
\qquad
{\bf 6} \otimes \overline{\bf 6} = {\bf 1} \oplus \cdots,
\nonu \\
{\bf 6} \otimes \overline{\bf 15} &=&
\overline{\bf 6} \oplus \cdots,
\qquad
{\bf 15} \otimes {\bf 15} = \overline{\bf 15} \oplus \cdots,    
\qquad
{\bf 15} \otimes {\bf 20} = \overline{\bf 6} \oplus \cdots,
\qquad
{\bf 15} \otimes \overline{\bf 6} = {\bf 6} \oplus \cdots,
\nonu \\
{\bf 15} \otimes \overline{\bf 15} &=& {\bf 1} \oplus \cdots,
\qquad
{\bf 20} \otimes {\bf 20} = {\bf 1} \oplus \cdots,
\qquad
{\bf 20} \otimes \overline{\bf 6} =
{\bf 15} \oplus \cdots,
\qquad
{\bf 20} \otimes \overline{\bf 15} =
{\bf 6} \oplus \cdots.
\nonu
\eea
The branching rule of $SU(8) \rightarrow SU(6) \times SU(2)
\times U(1)$
is given by
\bea
{\bf 8} & \rightarrow & ({\bf 1},{\bf 2})_{-3} \oplus
[({\bf 6},{\bf 1})_{1}],
\qquad
{\bf 28} \rightarrow [({\bf 1},{\bf 1})_{-6}]
\oplus ({\bf 6},{\bf 2})_{-2}
\oplus [({\bf 15},{\bf 1})_{2}],
\nonu \\
{\bf 56} & \rightarrow &
[({\bf 6},{\bf 1})_{-5}] \oplus ({\bf 15},{\bf 2})_{-1}
\oplus [({\bf 20},{\bf 1})_{3}],
\qquad
{\bf 70} \rightarrow [({\bf 15},{\bf 1})_{-4}] \oplus
[(\overline{\bf 15},{\bf 1})_{4}]
\oplus ({\bf 20},{\bf 2})_{0},
\nonu \\
\overline{\bf 56} & \rightarrow &
[(\overline{\bf 6},{\bf 1})_{5}] \oplus (\overline{\bf 15},{\bf 2})_{1}
\oplus [({\bf 20},{\bf 1})_{-3}],
\qquad
\overline{\bf 28} \rightarrow [({\bf 1},{\bf 1})_{6}]
\oplus (\overline{\bf 6},{\bf 2})_{2}
\oplus [(\overline{\bf 15},{\bf 1})_{-2}]\, ,
\nonu \\
\overline{\bf 8} & \rightarrow &
({\bf 1},{\bf 2})_{3} \oplus [(\overline{\bf 6},{\bf 1})_{-1}]
\, ,
\nonu
\eea
where the bracket means the singlet representations under the
$SU(2)$
which are the gravitinos, the graviphotons, the gravitinos
and the scalars. Note that $\overline{\bf 20}={\bf 20}$ in $SU(6)$.
The ${\cal N}=6$ supergravity is realized by a double copy
\cite{BCJ} (See also \cite{KLT})
of the ${\cal N}=4$ super Yang-Mills theory \cite{BSS} and
the ${\cal N}=2$ super Yang-Mills theory \cite{SS}.
This implies that the constructions given
by Tables \ref{Split1}, \ref{Split2}, \ref{Split3}, \ref{Split4},
and \ref{Split5} can be done similarly
by taking the second  ${\cal N}=2$ super Yang-Mills theory
instead
of ${\cal N}=4$ super Yang-Mills theory.
See also \cite{FKMVY}.}
\bea
\comm{(\Phi^{(h_1)}_{+2})_m}{(\Phi^{(h_2)}_{+2})_n}&=&
\kappa_{+2,+2,-2}\,\Big((h_2+1)m-(h_1+1)n\Big)\,(\Phi^{(h_1+h_2)}_{+2})_{m+n}
\,\,\, :\text{eq.1},
\nonu \\
\comm{(\Phi^{(h_1)}_{+2})_m}{(\Phi^{(h_2),\hat{A}}_{+\frac{3}{2}})_r}&=&
\kappa_{+2,+\frac{3}{2},-\frac{3}{2}}\,\Big((h_2+\tfrac{1}{2})m-(h_1+1)r\Big)\,(\Phi^{(h_1+h_2),\hat{A}}_{+\frac{3}{2}})_{m+r}
\,\,\, :\text{eq.2},
\nonu \\
\comm{(\Phi^{(h_1)}_{+2})_m}{(\Phi^{(h_2),\hat{A}\hat{B}}_{+1})_n}&=&
\kappa_{+2,+1,-1}\,\Big(h_2\,m-(h_1+1)n \Big)\,(\Phi^{(h_1+h_2),\hat{A}\hat{B}}_{+1})_{m+n}
\,\,\, :\text{eq.3-1},
\nonu \\
\comm{(\Phi^{(h_1)}_{+2})_m}{(\Phi^{(h_2),78}_{+1})_n}&=&
\kappa_{+2,+1,-1}\,\Big(h_2\,m-(h_1+1)n \Big)\,(\Phi^{(h_1+h_2),78}_{+1})_{m+n}
\,\,\, :\text{eq.3-2},
\nonu \\
\comm{(\Phi^{(h_1)}_{+2})_m}{(\Phi^{(h_2),\hat{A}\hat{B}\hat{C}}_{+\frac{1}{2}})_r}
&=&\kappa_{+2,+\frac{1}{2},-\frac{1}{2}}\,\Big((h_2-\tfrac{1}{2})m-(h_1+1)r\Big)\,
(\Phi^{(h_1+h_2),\hat{A}\hat{B}\hat{C}}_{+\frac{1}{2}})_{m+r}
\nonu \\
&:& \text{eq.4-1},
\nonu \\   
\comm{(\Phi^{(h_1)}_{+2})_m}{(\Phi^{(h_2),\hat{A}78}_{+\frac{1}{2}})_r}
&=&\kappa_{+2,+\frac{1}{2},-\frac{1}{2}}\,\Big((h_2-\tfrac{1}{2})m-(h_1+1)r\Big)\,
(\Phi^{(h_1+h_2),\hat{A}78}_{+\frac{1}{2}})_{m+r}
\nonu \\
&:& \text{eq.4-2},
\nonu \\   
\comm{(\Phi^{(h_1)}_{+2})_m}{(\Phi^{(h_2),\hat{A}\hat{B}\hat{C}\hat{D}}_{0})_n}
&=&\kappa_{+2,0,0}\,\Big((h_2-1)m-(h_1+1)n\Big)\,(\Phi^{(h_1+h_2),\hat{A}\hat{B}\hat{C}\hat{D}}_{0})_{m+n}
\nonu \\
&:& \text{eq.5-1},
\nonu \\
\comm{(\Phi^{(h_1)}_{+2})_m}{(\Phi^{(h_2),\hat{A}\hat{B}78}_{0})_n}
&=&\kappa_{+2,0,0}\,\Big((h_2-1)m-(h_1+1)n\Big)\,(\Phi^{(h_1+h_2),
\hat{A}\hat{B}78}_{0})_{m+n}
\nonu \\
&:& \text{eq.5-2},
\nonu \\
\comm{(\Phi^{(h_1)}_{+2})_m}{(\Phi^{(h_2)}_{\hat{A}\hat{B}\hat{C},-\frac{1}{2}})_r}
&=&\kappa_{+2,-\frac{1}{2},+\frac{1}{2}}\,\,\Big((h_2-\tfrac{3}{2})m-(h_1+1)r\Big)\,(\Phi^{(h_1+h_2)}_{\hat{A}\hat{B}\hat{C},-\frac{1}{2}})_{m+r}
\,\,\, :\text{eq.6-1},
\nonu \\
\comm{(\Phi^{(h_1)}_{+2})_m}{(\Phi^{(h_2)}_{\hat{A}78,-\frac{1}{2}})_r}
&=&\kappa_{+2,-\frac{1}{2},+\frac{1}{2}}\,\,\Big((h_2-\tfrac{3}{2})m-(h_1+1)r\Big)\,(\Phi^{(h_1+h_2)}_{\hat{A}78,-\frac{1}{2}})_{m+r}
\,\,\, :\text{eq.6-2},
\nonu \\
\comm{(\Phi^{(h_1)}_{+2})_m}{(\Phi^{(h_2)}_{\hat{A}\hat{B},-1})_n}
&=&\kappa_{+2,-1,+1}\,\Big((h_2-2)m-(h_1+1)n\Big)\,(\Phi^{(h_1+h_2)}_{
\hat{A}\hat{B},-1})_{m+n}
\, :\text{eq.7-1},
\nonu    \\
\comm{(\Phi^{(h_1)}_{+2})_m}{(\Phi^{(h_2)}_{78,-1})_n}
&=&\kappa_{+2,-1,+1}\,\Big((h_2-2)m-(h_1+1)n\Big)\,(\Phi^{(h_1+h_2)}_{
78,-1})_{m+n}
\, :\text{eq.7-2},
\nonu    \\
\comm{(\Phi^{(h_1)}_{+2})_m}{(\Phi^{(h_2)}_{\hat{A},-\frac{3}{2}})_r}
&=&
\kappa_{+2,-\frac{3}{2},+\frac{3}{2}}\,\Big((h_2-\tfrac{5}{2})m-(h_1+1)r\Big)\,(\Phi^{(h_1+h_2)}_{\hat{A},-\frac{3}{2}})_{m+r}
\, :\text{eq.8},
\nonu    \\
\comm{(\Phi^{(h_1)}_{+2})_m}{(\Phi^{(h_2)}_{-2})_n}&=&
\kappa_{+2,-2,+2}\,\Big((h_2-3)m-(h_1+1)n\Big)\,(\Phi^{(h_1+h_2)}_{-2})_{m+n}
\, :\text{eq.9},
\nonu \\
\acomm{(\Phi^{(h_1),\hat{A}}_{+\frac{3}{2}})_r}{(\Phi^{(h_2),\hat{B}}_{+\frac{3}{2}})_s}
&=&\kappa_{+\frac{3}{2},+\frac{3}{2},-1}\,
\Big((h_2+\tfrac{1}{2})r-(h_1+\tfrac{1}{2})s\Big)\,(\Phi^{(h_1+h_2),\hat{A}\hat{B}}_{+1})_{r+s}\,,
\nonu \\
&:& \text{eq.10},
\nonu \\
\comm{(\Phi^{(h_1),\hat{A}}_{+\frac{3}{2}})_r}{(\Phi^{(h_2),\hat{B}\hat{C}}_{+1})_m}
&=&\kappa_{+\frac{3}{2},+1,-\frac{1}{2}}\,\Big(h_2\,r-(h_1+\tfrac{1}{2})m\Big)\,(\Phi^{(h_1+h_2),\hat{A}\hat{B}\hat{C}}_{+\frac{1}{2}})_{r+m}
\,\,\, :\text{eq.11-1},
\nonu \\
\comm{(\Phi^{(h_1),\hat{A}}_{+\frac{3}{2}})_r}{(\Phi^{(h_2),78}_{+1})_m}
&=&\kappa_{+\frac{3}{2},+1,-\frac{1}{2}}\,\Big(h_2\,r-(h_1+\tfrac{1}{2})m\Big)\,(\Phi^{(h_1+h_2),\hat{A}78}_{+\frac{1}{2}})_{r+m}
\,\,\, :\text{eq.11-2},
\nonu \\
\acomm{(\Phi^{(h_1),\hat{A}}_{+\frac{3}{2}})_r}{(\Phi^{(h_2),\hat{B}\hat{C}\hat{D}}_{+\frac{1}{2}})_s}
&=&\kappa_{+\frac{3}{2},+\frac{1}{2},0}\,\Big((h_2-\tfrac{1}{2})r-(h_1+\tfrac{1}{2})s\Big)(\Phi^{(h_1+h_2),\hat{A}\hat{B}\hat{C}\hat{D}}_0)_{r+s}
\nonu \\
&:& \text{eq.12-1},
\nonu \\
\acomm{(\Phi^{(h_1),\hat{A}}_{+\frac{3}{2}})_r}{
(\Phi^{(h_2),\hat{B}78}_{+\frac{1}{2}})_s}
&=&\kappa_{+\frac{3}{2},+\frac{1}{2},0}\,\Big((h_2-\tfrac{1}{2})r-(h_1+\tfrac{1}{2})s\Big)(\Phi^{(h_1+h_2),\hat{A}\hat{B}78}_0)_{r+s}
\nonu \\
&:& \text{eq.12-2},
\nonu \\
\comm{(\Phi^{(h_1),\hat{A}}_{+\frac{3}{2}})_r}{(\Phi^{(h_2),\hat{B}\hat{C}\hat{D}\hat{E}}_0)_m}&=&
\kappa_{+\frac{3}{2},0,+\frac{1}{2}}\,\Big((h_2-1)r-(h_1+\tfrac{1}{2})m\Big)\,
\epsilon^{\hat{A}\hat{B}\hat{C}\hat{D}\hat{E}\hat{F}78}
\nonu \\
&\times & (\Phi^{(h_1+h_2)}_{\hat{F}78,-\frac{1}{2}})_{r+m}
\,\,\, :\text{eq.13-1},
\nonu \\
\comm{(\Phi^{(h_1),\hat{A}}_{+\frac{3}{2}})_r}{(\Phi^{(h_2),\hat{B}\hat{C}78}_0)_m}&=&
\kappa_{+\frac{3}{2},0,+\frac{1}{2}}\,\Big((h_2-1)r-(h_1+\tfrac{1}{2})m\Big)\,
\frac{1}{3!}\,
\epsilon^{\hat{A}\hat{B}\hat{C}78\hat{F}\hat{G}\hat{H}}\nonu \\
& \times & (\Phi^{(h_1+h_2)}_{
\hat{F}\hat{G}\hat{H},-\frac{1}{2}})_{r+m}
\,\,\, : \text{eq.13-2},
\nonu \\
\acomm{(\Phi^{(h_1),\hat{A}}_{+\frac{3}{2}})_r}{(\Phi^{(h_2)}_{
\hat{B}\hat{C}\hat{D},-\frac{1}{2}})_s}
&
=& \kappa_{+\frac{3}{2},-\frac{1}{2},+1}\,
\Big((h_2-\tfrac{3}{2})r-(h_1+\tfrac{1}{2})s\Big)\,3
\delta^{\hat{A}}_{\,\,\,[\hat{B}}(\Phi^{(h_1+h_2)}_{\hat{C}\hat{D}],-1})_{r+s}\,
\nonu \\
&:& \text{eq.14-1},
\nonu \\
\acomm{(\Phi^{(h_1),\hat{A}}_{+\frac{3}{2}})_r}{(\Phi^{(h_2)}_{
\hat{B}78,-\frac{1}{2}})_s}
&
=& \kappa_{+\frac{3}{2},-\frac{1}{2},+1}\,
\Big((h_2-\tfrac{3}{2})r-(h_1+\tfrac{1}{2})s\Big)\,3
\delta^{\hat{A}}_{\,\,\,[\hat{B}}(\Phi^{(h_1+h_2)}_{78],-1})_{r+s}\,
\nonu \\
&:& \text{eq.14-2},
\nonu \\
\comm{(\Phi^{(h_1),\hat{A}}_{+\frac{3}{2}})_r}{(\Phi^{(h_2)}_{\hat{B}\hat{C},-1})_m}
&=&
\kappa_{+\frac{3}{2},-1,+\frac{3}{2}}\,\Big((h_2-2)r-(h_1+\tfrac{1}{2})m\Big)\,
\nonu \\
& \times & 2!\, \delta^{\hat{A}}_{\,\,\,[\hat{B}}\,
(\Phi^{(h_1+h_2)}_{\hat{C}],-\frac{3}{2}})_{r+m}\, :\text{eq.15},
\nonu \\
\acomm{(\Phi^{(h_1),\hat{A}}_{+\frac{3}{2}})_r}{(\Phi^{(h_2)}_{\hat{B},-\frac{3}{2}})_s}
&=&
\kappa_{+\frac{3}{2},-\frac{3}{2},+2}\,\Big((h_2-\tfrac{5}{2})r-(h_1+\tfrac{1}{2})s\Big)\,\delta^{\hat{A}}_{\,\,\,\hat{B}}\,(\Phi^{(h_1+h_2)}_{-2})_{r+s}
\, :\text{eq.16},
\nonu \\
\comm{(\Phi^{(h_1),\hat{A}\hat{B}}_{+1})_m}{(\Phi^{(h_2),\hat{C}\hat{D}}_{+1})_n}&=&
\kappa_{+1,+1,0}\,\Big(h_2\,m-h_1\,n\Big)\,(\Phi^{(h_1+h_2),\hat{A}\hat{B}\hat{C}\hat{D}}_{0})_{m+n}
\,\,\, :\text{eq.17-1},
\nonu \\
\comm{(\Phi^{(h_1),\hat{A}\hat{B}}_{+1})_m}{(\Phi^{(h_2),78}_{+1})_n}&=&
\kappa_{+1,+1,0}\,\Big(h_2\,m-h_1\,n\Big)\,
(\Phi^{(h_1+h_2),\hat{A}\hat{B}78}_{0})_{m+n}
\,\,\, :\text{eq.17-2},
\nonu \\
\comm{(\Phi^{(h_1),\hat{A}\hat{B}}_{+1})_m}{(\Phi^{(h_2),\hat{C}\hat{D}\hat{E}}_{+\frac{1}{2}})_r}&=&
\kappa_{+1,+\frac{1}{2},+\frac{1}{2}}\,\Big((h_2-\tfrac{1}{2})m-h_1\,r\Big)\,
\epsilon^{\hat{A}\hat{B}\hat{C}\hat{D}\hat{E}\hat{F}78}(\Phi^{(h_1+h_2)}_{\hat{F}78,
  -\frac{1}{2}})_{m+r}
\nonu \\
&:&\text{eq.18-1},
\nonu \\
\comm{(\Phi^{(h_1),\hat{A}\hat{B}}_{+1})_m}{(\Phi^{(h_2),\hat{C}78}_{+\frac{1}{2}})_r}&=&
\kappa_{+1,+\frac{1}{2},+\frac{1}{2}}\,\Big((h_2-\tfrac{1}{2})m-h_1\,r\Big)\,
\frac{1}{3!}\,
\epsilon^{\hat{A}\hat{B}\hat{C}78\hat{F}\hat{G}\hat{H}}(\Phi^{(h_1+h_2)}_{\hat{F}\hat{G}\hat{H},-\frac{1}{2}})_{m+r}
\nonu \\
&:&\text{eq.18-2},
\nonu \\
\comm{(\Phi^{(h_1),78}_{+1})_m}{(\Phi^{(h_2),
\hat{C}\hat{D}\hat{E}}_{+\frac{1}{2}})_r}&=&
\kappa_{+1,+\frac{1}{2},+\frac{1}{2}}\,\Big((h_2-\tfrac{1}{2})m-h_1\,r\Big)\,
\frac{1}{3!}\,
\epsilon^{78\hat{C}\hat{D}\hat{E}\hat{F}\hat{G}\hat{H}}(\Phi^{(h_1+h_2)}_{\hat{F}\hat{G}\hat{H},-\frac{1}{2}})_{m+r}
\nonu \\
&:& \text{eq.18-3},
\nonu \\
\comm{(\Phi^{(h_1),\hat{A}\hat{B}}_{+1})_m}{(\Phi^{(h_2),\hat{C}\hat{D}78}_0)_n}&=&
\kappa_{+1,0,+1}\,\Big((h_2-1)m-h_1\,n\Big)\,
\epsilon^{\hat{A}\hat{B}\hat{C}\hat{D}78\hat{G}\hat{H}}\,\frac{1}{2!}\,
(\Phi^{(h_1+h_2)}_{\hat{G}\hat{H},-1})_{m+n}
\nonu \\
&: & \text{eq.19-1},
\nonu \\
\comm{(\Phi^{(h_1),78}_{+1})_m}{(\Phi^{(h_2),\hat{A}\hat{B}\hat{C}\hat{D}}_0)_n}&=&
\kappa_{+1,0,+1}\,\Big((h_2-1)m-h_1\,n\Big)\,
\epsilon^{78\hat{A}\hat{B}\hat{C}\hat{D}\hat{G}\hat{H}}\,\frac{1}{2!}\,
(\Phi^{(h_1+h_2)}_{\hat{G}\hat{H},-1})_{m+n}
\nonu \\
&: & \text{eq.19-2},
\nonu \\
\comm{(\Phi^{(h_1),\hat{A}\hat{B}}_{+1})_m}{
(\Phi^{(h_2),\hat{C}\hat{D}\hat{E}\hat{F}}_0)_n}&=&
\kappa_{+1,0,+1}\,\Big((h_2-1)m-h_1\,n\Big)\,
\epsilon^{\hat{A}\hat{B}\hat{C}\hat{D}\hat{E}\hat{F}78}\,
(\Phi^{(h_1+h_2)}_{78,-1})_{m+n}
\nonu \\
&: & \text{eq.19-3},
\nonu \\
\comm{(\Phi^{(h_1),\hat{A}\hat{B}}_{+1})_m}{(\Phi^{(h_2)}_{\hat{C}\hat{D}\hat{E},-\frac{1}{2}})_r}&=&
\kappa_{+1,-\frac{1}{2},+\frac{3}{2}}\,\Big((h_2-\tfrac{3}{2})m-h_1\,r\Big)\,
3! \delta^{\hat{A}}_{\,\,\,[\hat{C}}(\Phi^{(h_1+h_2)}_{\hat{D},-\frac{3}{2}})_{m+r}\delta_{\hat{E}]}^{\,\,\,\hat{B}}
\, \nonu \\
&: & \text{eq.20-1},
\nonu \\
\comm{(\Phi^{(h_1),78}_{+1})_m}{(\Phi^{(h_2)}_{\hat{C}78,-\frac{1}{2}})_r}&=&
\kappa_{+1,-\frac{1}{2},+\frac{3}{2}}\,\Big((h_2-\tfrac{3}{2})m-h_1\,r\Big)\,
3! \delta^{7}_{\,\,\,[\hat{C}}
(\Phi^{(h_1+h_2)}_{7,-\frac{3}{2}})_{m+r}\delta_{8]}^{\,\,\,8}
\, \nonu \\
&: & \text{eq.20-2},
\nonu \\
\comm{(\Phi^{(h_1),\hat{A}\hat{B}}_{+1})_m}{(\Phi^{(h_2)}_{\hat{C}\hat{D},-1})_n}&=&
\kappa_{+1,-1,+2}\,\Big((h_2-2)m-h_1\,n\Big)\,
\delta^{\hat{A}\hat{B}}_{\hat{C}\hat{D}} \, (\Phi^{(h_1+h_2)}_{-2})_{m+n}
\, : \text{eq.21-1},
\nonu \\
\comm{(\Phi^{(h_1),78}_{+1})_m}{(\Phi^{(h_2)}_{78,-1})_n}&=&
\kappa_{+1,-1,+2}\,\Big((h_2-2)m-h_1\,n\Big)\,
\delta^{78}_{78} \, (\Phi^{(h_1+h_2)}_{-2})_{m+n}
\, : \text{eq.21-2},
\nonu \\
\acomm{(\Phi^{(h_1),\hat{A}\hat{B}\hat{C}}_{+\frac{1}{2}})_r}{(\Phi^{(h_2),\hat{D}\hat{E}
\hat{F}}_{+\frac{1}{2}})_s}&=&
\kappa_{+\frac{1}{2},+\frac{1}{2},+1}\,\Big((h_2-\tfrac{1}{2})r-(h_1-\tfrac{1}{2})s\Big)\,\epsilon^{\hat{A}\hat{B}\hat{C}\hat{D}
\hat{E}\hat{F}78}
\nonu \\
& \times & (\Phi^{(h_1+h_2)}_{78,-1})_{r+s}
\, :  \text{eq.22-1},
\nonu \\
\acomm{(\Phi^{(h_1),\hat{A}\hat{B}\hat{C}}_{+\frac{1}{2}})_r}{(\Phi^{(h_2),\hat{D}78}_{+\frac{1}{2}})_s}&=&
\kappa_{+\frac{1}{2},+\frac{1}{2},+1}\,\Big((h_2-\tfrac{1}{2})r-(h_1-\tfrac{1}{2})s\Big)\,\frac{1}{2!}\,\epsilon^{\hat{A}\hat{B}\hat{C}\hat{D}78
\hat{G}\hat{H}}
\nonu \\
& \times & (\Phi^{(h_1+h_2)}_{\hat{G}\hat{H},-1})_{r+s}
\, :  \text{eq.22-2},
\nonu \\
\acomm{(\Phi^{(h_1),\hat{A}78}_{+\frac{1}{2}})_r}{(\Phi^{(h_2),\hat{D}\hat{E}\hat{F}}_{+\frac{1}{2}})_s}&=&
\kappa_{+\frac{1}{2},+\frac{1}{2},+1}\,\Big((h_2-\tfrac{1}{2})r-(h_1-\tfrac{1}{2})s\Big)\,\frac{1}{2!}\,\epsilon^{\hat{A}78\hat{D}\hat{E}\hat{F}
\hat{G}\hat{H}}
\nonu \\
& \times & (\Phi^{(h_1+h_2)}_{\hat{G}\hat{H},-1})_{r+s}
\, :  \text{eq.22-3},
\nonu \\
\comm{(\Phi^{(h_1),\hat{A}\hat{B}\hat{C}}_{+\frac{1}{2}})_r}{(\Phi^{(h_2),
\hat{D}\hat{E}78}_0)_m}&=&
\kappa_{+\frac{1}{2},0,+\frac{3}{2}}\,\Big((h_2-1)r-(h_1-\tfrac{1}{2})m\Big)\,
\epsilon^{\hat{A}\hat{B}\hat{C}\hat{D}\hat{E}78\hat{H}}
\nonu \\
& \times & (\Phi^{(h_1+h_2)}_{\hat{H},-\frac{3}{2}})_{r+m}
\, :  \text{eq.23-1},
\nonu \\
\comm{(\Phi^{(h_1),\hat{A}78}_{+\frac{1}{2}})_r}{(\Phi^{(h_2),
\hat{D}\hat{E}\hat{F}\hat{G}}_0)_m}&=&
\kappa_{+\frac{1}{2},0,+\frac{3}{2}}\,\Big((h_2-1)r-(h_1-\tfrac{1}{2})m\Big)\,
\epsilon^{\hat{A}78\hat{D}\hat{E}\hat{F}\hat{G}\hat{H}}
\nonu \\
& \times & (\Phi^{(h_1+h_2)}_{\hat{H},-\frac{3}{2}})_{r+m}
\, :  \text{eq.23-2},
\nonu \\
\acomm{(\Phi^{(h_1),\hat{A}\hat{B}\hat{C}}_{+\frac{1}{2}})_r}{
(\Phi^{(h_2)}_{\hat{D}\hat{E}\hat{F},-\frac{1}{2}})_s}
&=&\kappa_{+\frac{1}{2},-\frac{1}{2},+2}\,\Big((h_2-\tfrac{3}{2})r-(h_1-\tfrac{1}{2})s\Big)\,
\delta^{\hat{A}\hat{B}\hat{C}}_{\hat{D}\hat{E}\hat{F}}\,(\Phi^{(h_1+h_2)}_{-2})_{r+s}
\nonu \\
&: & \text{eq.24-1},
\nonu \\
\acomm{(\Phi^{(h_1),\hat{A}78}_{+\frac{1}{2}})_r}{
(\Phi^{(h_2)}_{\hat{A}78,-\frac{1}{2}})_s}
&=&\kappa_{+\frac{1}{2},-\frac{1}{2},+2}\,\Big((h_2-\tfrac{3}{2})r-(h_1-\tfrac{1}{2})s\Big)\,
\delta^{\hat{A}78}_{\hat{A}78}\,(\Phi^{(h_1+h_2)}_{-2})_{r+s}
\nonu \\
&: & \text{eq.24-2},
\nonu \\
\comm{(\Phi^{(h_1),\hat{A}\hat{B}\hat{C}\hat{D}}_0)_m}{
(\Phi^{(h_2),\hat{E}\hat{F}78}_0)_n}&=&
\kappa_{0,0,+2}\,\Big((h_2-1)m-(h_1-1)n\Big)
\epsilon^{\hat{A}\hat{B}\hat{C}\hat{D}\hat{E}\hat{F}78}\,(\Phi^{(h_1+h_2)}_{-2})_{m+n}
\nonu \\
& : & \text{eq.25}\,,
\label{su6result}
\eea
\enlargethispage{3pt}
where $\hat{A}, \hat{B}, \cdots =1,2, \cdots,6$.
Note that
there are following nonzero quantities
having the indices $78$ appearing in (\ref{su6result})
\bea
\Phi_{+ 1}^{(h),78}, \qquad
\Phi_{78,-1}^{(h)}, \qquad
\Phi_{+ \frac{1}{2}}^{(h),\hat{A}78}, \qquad
\Phi_{\hat{A}78,-\frac{1}{2}}^{(h)}, \qquad
\Phi_{0}^{(h),\hat{A}\hat{B}78}
\neq 0,
\label{nonzerosu6}
\eea
in addition to the particle contents in the
${\cal N}=6$ supergravity denoted by the brackets in the
footnote \ref{su6foot}.
In the footnote \ref{su6foot},
the representations,
$({\bf 1},{\bf 1})_{-6}$, $({\bf 1},{\bf 1})_{6}$,
$({\bf 6},{\bf 1})_{-5}$,
$(\overline{\bf 6},{\bf 1})_{5}$, and 
$({\bf 15},{\bf 1})_{-4}$
correspond to each element of (\ref{nonzerosu6}) respectively.
In other words, the
first two plays the role of the scalars,
the third corresponds to the fundamental representation,
the fourth corresponds to the antifundamental representation,
and the last plays the role of antisymmetric representation
(two Young tableaux boxes in one column)
of $SU(6)$ respectively.


\section{The  celestial holography in the
${\cal N}=5$ supergravity}


We can decompose the $SU(8)$ indices into the $SU(5)$
indices 
and it turns out that
the corresponding (anti)commutators can be obtained
by
\footnote{
The tensor product of
$SU(5)$ in the lower representations
\cite{FKS}
is 
\bea
{\bf 5} \otimes {\bf 5} &=& {\bf 10} \oplus \cdots,
\qquad
{\bf 5} \otimes {\bf 10} = \overline{\bf 10} \oplus \cdots,    
\qquad
{\bf 5} \otimes \overline{\bf 5} = {\bf 1} \oplus \cdots,
\qquad
{\bf 5} \otimes \overline{\bf 10} = \overline{\bf 5} \oplus \cdots,
\nonu \\
{\bf 10} \otimes {\bf 10} &=& \overline{\bf 5} \oplus \cdots,
\qquad
{\bf 10} \otimes \overline{\bf 5} ={\bf 5} \oplus \cdots,
\qquad
{\bf 10} \otimes \overline{\bf 10} = {\bf 1} \oplus \cdots.
\nonu
\eea}
\bea
\comm{(\Phi^{(h_1)}_{+2})_m}{(\Phi^{(h_2)}_{+2})_n}&=&
\kappa_{+2,+2,-2}\,\Big((h_2+1)m-(h_1+1)n\Big)\,(\Phi^{(h_1+h_2)}_{+2})_{m+n}
\,\,\, :\text{eq.1},
\nonu \\
\comm{(\Phi^{(h_1)}_{+2})_m}{(\Phi^{(h_2),\hat{A}}_{+\frac{3}{2}})_r}&=&
\kappa_{+2,+\frac{3}{2},-\frac{3}{2}}\,\Big((h_2+\tfrac{1}{2})m-(h_1+1)r\Big)\,(\Phi^{(h_1+h_2),\hat{A}}_{+\frac{3}{2}})_{m+r}
\,\,\, :\text{eq.2},
\nonu \\
\comm{(\Phi^{(h_1)}_{+2})_m}{(\Phi^{(h_2),\hat{A}\hat{B}}_{+1})_n}&=&
\kappa_{+2,+1,-1}\,\Big(h_2\,m-(h_1+1)n \Big)\,(\Phi^{(h_1+h_2),\hat{A}\hat{B}}_{+1})_{m+n}
\,\,\, :\text{eq.3},
\nonu \\
\comm{(\Phi^{(h_1)}_{+2})_m}{(\Phi^{(h_2),\hat{A}\hat{B}\hat{C}}_{+\frac{1}{2}})_r}
&=&\kappa_{+2,+\frac{1}{2},-\frac{1}{2}}\,\Big((h_2-\tfrac{1}{2})m-(h_1+1)r\Big)\,
(\Phi^{(h_1+h_2),\hat{A}\hat{B}\hat{C}}_{+\frac{1}{2}})_{m+r}
\nonu \\
&:& \text{eq.4-1},
\nonu \\   
\comm{(\Phi^{(h_1)}_{+2})_m}{(\Phi^{(h_2),678}_{+\frac{1}{2}})_r}
&=&\kappa_{+2,+\frac{1}{2},-\frac{1}{2}}\,\Big((h_2-\tfrac{1}{2})m-(h_1+1)r\Big)\,
(\Phi^{(h_1+h_2),678}_{+\frac{1}{2}})_{m+r}
\nonu \\
&:& \text{eq.4-2},
\nonu \\   
\comm{(\Phi^{(h_1)}_{+2})_m}{(\Phi^{(h_2),\hat{A}\hat{B}\hat{C}\hat{D}}_{0})_n}
&=&\kappa_{+2,0,0}\,\Big((h_2-1)m-(h_1+1)n\Big)\,(\Phi^{(h_1+h_2),\hat{A}\hat{B}\hat{C}\hat{D}}_{0})_{m+n}
\nonu \\
&:& \text{eq.5-1},
\nonu \\
\comm{(\Phi^{(h_1)}_{+2})_m}{(\Phi^{(h_2),\hat{A}678}_{0})_n}
&=&\kappa_{+2,0,0}\,\Big((h_2-1)m-(h_1+1)n\Big)\,(\Phi^{(h_1+h_2),
\hat{A}678}_{0})_{m+n}
\nonu \\
&:& \text{eq.5-2},
\nonu \\
\comm{(\Phi^{(h_1)}_{+2})_m}{(\Phi^{(h_2)}_{\hat{A}\hat{B}\hat{C},-\frac{1}{2}})_r}
&=&\kappa_{+2,-\frac{1}{2},+\frac{1}{2}}\,\,\Big((h_2-\tfrac{3}{2})m-(h_1+1)r\Big)\,(\Phi^{(h_1+h_2)}_{\hat{A}\hat{B}\hat{C},-\frac{1}{2}})_{m+r}
\,\,\, :\text{eq.6-1},
\nonu \\
\comm{(\Phi^{(h_1)}_{+2})_m}{(\Phi^{(h_2)}_{678,-\frac{1}{2}})_r}
&=&\kappa_{+2,-\frac{1}{2},+\frac{1}{2}}\,\,\Big((h_2-\tfrac{3}{2})m-(h_1+1)r\Big)\,(\Phi^{(h_1+h_2)}_{678,-\frac{1}{2}})_{m+r}
\,\,\, :\text{eq.6-2},
\nonu \\
\comm{(\Phi^{(h_1)}_{+2})_m}{(\Phi^{(h_2)}_{\hat{A}\hat{B},-1})_n}
&=&\kappa_{+2,-1,+1}\,\Big((h_2-2)m-(h_1+1)n\Big)\,(\Phi^{(h_1+h_2)}_{
\hat{A}\hat{B},-1})_{m+n}
\, :\text{eq.7},
\nonu    \\
\comm{(\Phi^{(h_1)}_{+2})_m}{(\Phi^{(h_2)}_{\hat{A},-\frac{3}{2}})_r}
&=&
\kappa_{+2,-\frac{3}{2},+\frac{3}{2}}\,\Big((h_2-\tfrac{5}{2})m-(h_1+1)r\Big)\,(\Phi^{(h_1+h_2)}_{\hat{A},-\frac{3}{2}})_{m+r}
\, :\text{eq.8},
\nonu    \\
\comm{(\Phi^{(h_1)}_{+2})_m}{(\Phi^{(h_2)}_{-2})_n}&=&
\kappa_{+2,-2,+2}\,\Big((h_2-3)m-(h_1+1)n\Big)\,(\Phi^{(h_1+h_2)}_{-2})_{m+n}
\, :\text{eq.9},
\nonu \\
\acomm{(\Phi^{(h_1),\hat{A}}_{+\frac{3}{2}})_r}{(\Phi^{(h_2),\hat{B}}_{+\frac{3}{2}})_s}
&=&\kappa_{+\frac{3}{2},+\frac{3}{2},-1}\,
\Big((h_2+\tfrac{1}{2})r-(h_1+\tfrac{1}{2})s\Big)\,(\Phi^{(h_1+h_2),\hat{A}\hat{B}}_{+1})_{r+s}\,,
\,\,\, :\text{eq.10},
\nonu \\
\comm{(\Phi^{(h_1),\hat{A}}_{+\frac{3}{2}})_r}{(\Phi^{(h_2),\hat{B}\hat{C}}_{+1})_m}
&=&\kappa_{+\frac{3}{2},+1,-\frac{1}{2}}\,\Big(h_2\,r-(h_1+\tfrac{1}{2})m\Big)\,(\Phi^{(h_1+h_2),\hat{A}\hat{B}\hat{C}}_{+\frac{1}{2}})_{r+m}
\,\,\, :\text{eq.11},
\nonu \\
\acomm{(\Phi^{(h_1),\hat{A}}_{+\frac{3}{2}})_r}{(\Phi^{(h_2),\hat{B}\hat{C}\hat{D}}_{+\frac{1}{2}})_s}
&=&\kappa_{+\frac{3}{2},+\frac{1}{2},0}\,\Big((h_2-\tfrac{1}{2})r-(h_1+\tfrac{1}{2})s\Big)(\Phi^{(h_1+h_2),\hat{A}\hat{B}\hat{C}\hat{D}}_0)_{r+s}
\nonu \\
&:& \text{eq.12-1},
\nonu \\
\acomm{(\Phi^{(h_1),\hat{A}}_{+\frac{3}{2}})_r}{(\Phi^{(h_2),678}_{+\frac{1}{2}})_s}
&=&\kappa_{+\frac{3}{2},+\frac{1}{2},0}\,\Big((h_2-\tfrac{1}{2})r-(h_1+\tfrac{1}{2})s\Big)(\Phi^{(h_1+h_2),\hat{A}678}_0)_{r+s}
\nonu \\
&:& \text{eq.12-2},
\nonu \\
\comm{(\Phi^{(h_1),\hat{A}}_{+\frac{3}{2}})_r}{(\Phi^{(h_2),\hat{B}\hat{C}\hat{D}\hat{E}}_0)_m}&=&
\kappa_{+\frac{3}{2},0,+\frac{1}{2}}\,\Big((h_2-1)r-(h_1+\tfrac{1}{2})m\Big)\,\epsilon^{\hat{A}\hat{B}\hat{C}\hat{D}\hat{E}678}(\Phi^{(h_1+h_2)}_{678,-\frac{1}{2}})_{r+m}
\nonu \\
&:& \text{eq.13-1},
\nonu \\
\comm{(\Phi^{(h_1),\hat{A}}_{+\frac{3}{2}})_r}{(\Phi^{(h_2),\hat{B}678}_0)_m}&=&
\kappa_{+\frac{3}{2},0,+\frac{1}{2}}\,\Big((h_2-1)r-(h_1+\tfrac{1}{2})m\Big)\,
\frac{1}{3!} \, \epsilon^{\hat{A}\hat{B}678\hat{F}\hat{G}\hat{H}}
\nonu \\
& \times & (\Phi^{(h_1+h_2)}_{
\hat{F}\hat{G}\hat{H},-\frac{1}{2}})_{r+m}
\, \, \, : \text{eq.13-2},
\nonu \\
\acomm{(\Phi^{(h_1),\hat{A}}_{+\frac{3}{2}})_r}{(\Phi^{(h_2)}_{
\hat{B}\hat{C}\hat{D},-\frac{1}{2}})_s}
&
=& \kappa_{+\frac{3}{2},-\frac{1}{2},+1}\,
\Big((h_2-\tfrac{3}{2})r-(h_1+\tfrac{1}{2})s\Big)\,3
\delta^{\hat{A}}_{\,\,\,[\hat{B}}(\Phi^{(h_1+h_2)}_{\hat{C}\hat{D}],-1})_{r+s}\,
\nonu \\
&:& \text{eq.14},
\nonu \\
\comm{(\Phi^{(h_1),\hat{A}}_{+\frac{3}{2}})_r}{(\Phi^{(h_2)}_{\hat{B}\hat{C},-1})_m}
&=&
\kappa_{+\frac{3}{2},-1,+\frac{3}{2}}\,\Big((h_2-2)r-(h_1+\tfrac{1}{2})m\Big)\,
\nonu \\
& \times & 2!\, \delta^{\hat{A}}_{\,\,\,[\hat{B}}\,
(\Phi^{(h_1+h_2)}_{\hat{C}],-\frac{3}{2}})_{r+m}\, :\text{eq.15},
\nonu \\
\acomm{(\Phi^{(h_1),\hat{A}}_{+\frac{3}{2}})_r}{(\Phi^{(h_2)}_{\hat{B},-\frac{3}{2}})_s}
&=&
\kappa_{+\frac{3}{2},-\frac{3}{2},+2}\,\Big((h_2-\tfrac{5}{2})r-(h_1+\tfrac{1}{2})s\Big)\,\delta^{\hat{A}}_{\,\,\,\hat{B}}\,(\Phi^{(h_1+h_2)}_{-2})_{r+s}
\, :\text{eq.16},
\nonu \\
\comm{(\Phi^{(h_1),\hat{A}\hat{B}}_{+1})_m}{(\Phi^{(h_2),\hat{C}\hat{D}}_{+1})_n}&=&
\kappa_{+1,+1,0}\,\Big(h_2\,m-h_1\,n\Big)\,(\Phi^{(h_1+h_2),\hat{A}\hat{B}\hat{C}\hat{D}}_{0})_{m+n}
\,\,\, :\text{eq.17},
\nonu \\
\comm{(\Phi^{(h_1),\hat{A}\hat{B}}_{+1})_m}{(\Phi^{(h_2),\hat{C}\hat{D}\hat{E}}_{+\frac{1}{2}})_r}&=&
\kappa_{+1,+\frac{1}{2},+\frac{1}{2}}\,\Big((h_2-\tfrac{1}{2})m-h_1\,r\Big)\,
\epsilon^{\hat{A}\hat{B}\hat{C}\hat{D}\hat{E}678}(\Phi^{(h_1+h_2)}_{678,
-\frac{1}{2}})_{m+r}
\nonu \\
&:& \text{eq.18-1},
\nonu \\
\comm{(\Phi^{(h_1),\hat{A}\hat{B}}_{+1})_m}{(\Phi^{(h_2),678}_{+\frac{1}{2}})_r}&=&
\kappa_{+1,+\frac{1}{2},+\frac{1}{2}}\,\Big((h_2-\tfrac{1}{2})m-h_1\,r\Big)\,
\frac{1}{3!}\,
\epsilon^{\hat{A}\hat{B}678\hat{F}\hat{G}\hat{H}}(\Phi^{(h_1+h_2)}_{\hat{F}\hat{G}\hat{H},-\frac{1}{2}})_{m+r}
\nonu \\
&:& \text{eq.18-2},
\nonu \\
\comm{(\Phi^{(h_1),\hat{A}\hat{B}}_{+1})_m}{(\Phi^{(h_2),\hat{C}678}_0)_n}&=&
\kappa_{+1,0,+1}\,\Big((h_2-1)m-h_1\,n\Big)\,
\epsilon^{\hat{A}\hat{B}\hat{C}678\hat{G}\hat{H}}\,\frac{1}{2!}\,
(\Phi^{(h_1+h_2)}_{\hat{G}\hat{H},-1})_{m+n}
\nonu \\
&: & \text{eq.19},
\nonu \\
\comm{(\Phi^{(h_1),\hat{A}\hat{B}}_{+1})_m}{(\Phi^{(h_2)}_{\hat{C}\hat{D}\hat{E},-\frac{1}{2}})_r}&=&
\kappa_{+1,-\frac{1}{2},+\frac{3}{2}}\,\Big((h_2-\tfrac{3}{2})m-h_1\,r\Big)\,
3! \delta^{\hat{A}}_{\,\,\,[\hat{C}}(\Phi^{(h_1+h_2)}_{\hat{D},-\frac{3}{2}})_{m+r}\delta_{\hat{E}]}^{\,\,\,\hat{B}}
\, \nonu \\
&: & \text{eq.20},
\nonu \\
\comm{(\Phi^{(h_1),\hat{A}\hat{B}}_{+1})_m}{(\Phi^{(h_2)}_{\hat{C}\hat{D},-1})_n}&=&
\kappa_{+1,-1,+2}\,\Big((h_2-2)m-h_1\,n\Big)\,
\delta^{\hat{A}\hat{B}}_{\hat{C}\hat{D}} \, (\Phi^{(h_1+h_2)}_{-2})_{m+n}
\, : \text{eq.21},
\nonu \\
\acomm{(\Phi^{(h_1),\hat{A}\hat{B}\hat{C}}_{+\frac{1}{2}})_r}{(\Phi^{(h_2),678}_{+\frac{1}{2}})_s}&=&
\kappa_{+\frac{1}{2},+\frac{1}{2},+1}\,\Big((h_2-\tfrac{1}{2})r-(h_1-\tfrac{1}{2})s\Big)\,\frac{1}{2!}\,\epsilon^{\hat{A}\hat{B}\hat{C}678
\hat{G}\hat{H}}
\nonu \\
& \times & (\Phi^{(h_1+h_2)}_{\hat{G}\hat{H},-1})_{r+s}
\, :  \text{eq.22},
\nonu \\
\comm{(\Phi^{(h_1),\hat{A}\hat{B}\hat{C}}_{+\frac{1}{2}})_r}{(\Phi^{(h_2),
\hat{D}678}_0)_m}&=&
\kappa_{+\frac{1}{2},0,+\frac{3}{2}}\,\Big((h_2-1)r-(h_1-\tfrac{1}{2})m\Big)\,
\epsilon^{\hat{A}\hat{B}\hat{C}\hat{D}678\hat{H}}
\nonu \\
& \times & (\Phi^{(h_1+h_2)}_{\hat{H},-\frac{3}{2}})_{r+m}
\, :  \text{eq.23-1},
\nonu \\
\comm{(\Phi^{(h_1),678}_{+\frac{1}{2}})_r}{(\Phi^{(h_2),
\hat{D}\hat{E}\hat{F}\hat{G}}_0)_m}&=&
\kappa_{+\frac{1}{2},0,+\frac{3}{2}}\,\Big((h_2-1)r-(h_1-\tfrac{1}{2})m\Big)\,
\epsilon^{678\hat{D}\hat{E}\hat{F}\hat{G}\hat{H}}
\nonu \\
& \times & (\Phi^{(h_1+h_2)}_{\hat{H},-\frac{3}{2}})_{r+m}
\, :  \text{eq.23-2},
\nonu \\
\acomm{(\Phi^{(h_1),\hat{A}\hat{B}\hat{C}}_{+\frac{1}{2}})_r}{
(\Phi^{(h_2)}_{\hat{D}\hat{E}\hat{F},-\frac{1}{2}})_s}
&=&\kappa_{+\frac{1}{2},-\frac{1}{2},+2}\,\Big((h_2-\tfrac{3}{2})r-(h_1-\tfrac{1}{2})s\Big)\,
\delta^{\hat{A}\hat{B}\hat{C}}_{\hat{D}\hat{E}\hat{F}}\,(\Phi^{(h_1+h_2)}_{-2})_{r+s}
\nonu \\
&: & \text{eq.24-1},
\nonu \\
\acomm{(\Phi^{(h_1),678}_{+\frac{1}{2}})_r}{
(\Phi^{(h_2)}_{678,-\frac{1}{2}})_s}
&=&\kappa_{+\frac{1}{2},-\frac{1}{2},+2}\,\Big((h_2-\tfrac{3}{2})r-(h_1-\tfrac{1}{2})s\Big)\,
\delta^{678}_{678}\,(\Phi^{(h_1+h_2)}_{-2})_{r+s}
\nonu \\
&: & \text{eq.24-2},
\nonu \\
\comm{(\Phi^{(h_1),\hat{A}\hat{B}\hat{C}\hat{D}}_0)_m}{
(\Phi^{(h_2),\hat{E}678}_0)_n}&=&
\kappa_{0,0,+2}\,\Big((h_2-1)m-(h_1-1)n\Big)
\epsilon^{\hat{A}\hat{B}\hat{C}\hat{D}\hat{E}678}\,(\Phi^{(h_1+h_2)}_{-2})_{m+n}
\nonu \\
& : & \text{eq.25}\,,
\label{su5result}
\eea
where $\hat{A}, \hat{B}, \cdots =1,2, \cdots, 5$.
Note that
there are following nonzero quantities
having the indices $678$ appearing in (\ref{su5result})
\bea
\Phi_{+ \frac{1}{2}}^{(h),678}, \qquad
\Phi_{678,-\frac{1}{2}}^{(h)}, \qquad
\Phi_{0}^{(h),\hat{A}678}
\neq 0, 
\label{nonzerosu5}
\eea
as well as
the particle contents in the
${\cal N}=5$ supergravity denoted by the brackets in the
footnote \ref{su5foot}.
According to the footnote
\ref{su5foot}, the first, the second and the third 
of (\ref{nonzerosu5}) correspond to
$({\bf 1},{\bf 1})_{-15}$, $({\bf 1},{\bf 1})_{15}$
and $({\bf 5},{\bf 1})_{-12}$ respectively.
In other words, from this
truncation, the first two play the role of the
scalars (no $SU(5)$ indices)
while the last plays the role of the fundamental representation of
$SU(5)$
\footnote{\label{su5foot}
The branching rule of $SU(8) \rightarrow SU(5) \times SU(3)
\times U(1)$ is given by
\bea
{\bf 8} &\rightarrow &
({\bf 1},{\bf 3})_{-5} \oplus [({\bf 5},{\bf 1})_3]   \, ,
\qquad
{\bf 28} \rightarrow 
({\bf 1},\overline{\bf 3})_{-10} \oplus ({\bf 5},{\bf 3})_{-2}
\oplus [({\bf 10},{\bf 1})_6]\, ,
\nonu \\
{\bf 56} &\rightarrow & 
[({\bf 1},{\bf 1})_{-15}]\oplus ({\bf 5},\overline{\bf 3})_{-7}
\oplus [(\overline{\bf 10},{\bf 1})_9]\oplus
({\bf 10},{\bf 3})_1 \, ,
\nonu \\
{\bf 70} &\rightarrow & 
[({\bf 5},{\bf 1})_{-12}] \oplus [(\overline{\bf 5},{\bf 1})_{12}]
\oplus ({\bf 10},\overline{\bf 3})_{-4} \oplus
(\overline{\bf 10},{\bf 3})_4 \, ,
\nonu \\
\overline{\bf 56} &\rightarrow & 
[({\bf 1},{\bf 1})_{15}] \oplus (\overline{\bf 5},{\bf 3})_{7}
\oplus [({\bf 10},{\bf 1})_{-9}] \oplus
(\overline{\bf 10},\overline{\bf 3})_{-1} \, ,
\nonu \\
\overline{\bf 28} & \rightarrow & 
({\bf 1},{\bf 3})_{10} \oplus (\overline{\bf 5},\overline{\bf 3})_{2}
\oplus [(\overline{\bf 10},{\bf 1})_{-6}] \, ,
\qquad
\overline{\bf 8} \rightarrow 
({\bf 1},\overline{\bf 3})_{5} \oplus [(\overline{\bf 5},{\bf 1})_{-3}]
\, ,
\nonu 
\eea
where the bracket denotes the singlets under the
$SU(3)$ which are the gravitinos, the graviphotons, the gravitinos
and the scalars.
The ${\cal N}=5$ supergravity is realized by a double copy
of the ${\cal N}=4$ super Yang-Mills theory \cite{BSS} and
the ${\cal N}=1$ super Yang-Mills theory \cite{WZ}.
The constructions given
by Tables \ref{Split1}, \ref{Split2}, \ref{Split3}, \ref{Split4},
and \ref{Split5} can be performed similarly
by taking the second  ${\cal N}=1$ super Yang-Mills theory
instead
of ${\cal N}=4$ super Yang-Mills theory.}.




\section{The
celestial holography in the
${\cal N}=4$ supergravity}


Let us collect the relevant (anti)commutators
for the ${\cal N}=4$ supergravity as follows
%
%
\bea
\comm{(\Phi^{(h_1)}_{+2})_m}{(\Phi^{(h_2)}_{+2})_n}&=&
\kappa_{+2,+2,-2}\,\Big((h_2+1)m-(h_1+1)n\Big)\,
(\Phi^{(h_1+h_2)}_{+2})_{m+n}\, ,
\nonu \\
\comm{(\Phi^{(h_1)}_{+2})_m}{(\Phi^{(h_2)}_{-2})_n}&=&
\kappa_{+2,-2,+2}\,\Big((h_2-3)m-(h_1+1)n\Big)\,
(\Phi^{(h_1+h_2)}_{-2})_{m+n} \, ,
\nonu \\
%
%
\comm{(\Phi^{(h_1)}_{+2})_m}{(\Phi^{(h_2),a}_{+\frac{3}{2}})_n}&=&
\kappa_{+2,+\frac{3}{2},-\frac{3}{2}}\,\Big((h_2+\tfrac{1}{2})m-(h_1+1)n\Big)\,(\Phi^{(h_1+h_2),a}_{+\frac{3}{2}})_{m+n} \, ,
\nonu
\\
\bf
\comm{(\Phi^{(h_1)}_{+2})_m}{(\Phi^{(h_2),r}_{+\frac{3}{2}})_n}&=&
\bf
\kappa_{+2,+\frac{3}{2},-\frac{3}{2}}\,\Big((h_2+\tfrac{1}{2})m-(h_1+1)n\Big)\,(\Phi^{(h_1+h_2),r}_{+\frac{3}{2}})_{m+n} \, ,
\nonu
\\
\comm{(\Phi^{(h_1)}_{+2})_m}{(\Phi^{(h_2)}_{a,\,-\frac{3}{2}})_n}&=&
\kappa_{+2,-\frac{3}{2},+\frac{3}{2}}\,\Big((h_2-\tfrac{5}{2})m-(h_1+1)n\Big)\,(\Phi^{(h_1+h_2)}_{a,\,-\frac{3}{2}})_{m+n} \, ,
\nonu
\\
\bf 
\comm{(\Phi^{(h_1)}_{+2})_m}{(\Phi^{(h_2)}_{r,\,-\frac{3}{2}})_n}&=&
\bf
\kappa_{+2,-\frac{3}{2},+\frac{3}{2}}\,\Big((h_2-\tfrac{5}{2})m-(h_1+1)n\Big)\,(\Phi^{(h_1+h_2)}_{r,\,-\frac{3}{2}})_{m+n}\, .
\nonu \\
%
%
\comm{(\Phi^{(h_1)}_{+2})_m}{(\Phi^{(h_2,ab)}_{+1})_n}&=&
\kappa_{+2,+1,-1}\,\Big(h_2\,m-(h_1+1)n\Big)\,(\Phi^{(h_1+h_2),ab}_{+1})_{m+n} \, ,
\nonu
\\
\bf 
\comm{(\Phi^{(h_1)}_{+2})_m}{(\Phi^{(h_2),rs}_{\,+1})_n}&=&
\bf 
\kappa_{+2,+1,-1}\,\Big(h_2\,m-(h_1+1)n\Big)\,(\Phi^{(h_1+h_2),rs}_{+1})_{m+n} \, ,
\nonu
\\
\comm{(\Phi^{(h_1)}_{+2})_m}{(\Phi^{(h_2)}_{ab,\,-1})_n}&=&
\kappa_{+2,-1,+1}\,\Big((h_2-2)m-(h_1+1)n\Big)\,(\Phi^{(h_1+h_2)}_{ab,\,-1})_{m+n} \, ,
\nonu
\\
\bf 
\comm{(\Phi^{(h_1)}_{+2})_m}{(\Phi^{(h_2)}_{rs,\,-1})_n}&=&
\bf
\kappa_{+2,-1,+1}\,\Big((h_2-2)m-(h_1+1)n\Big)\,(\Phi^{(h_1+h_2)}_{rs,\,-1})_{m+n} \, ,
\nonu \\
%
%
\comm{(\Phi^{(h_1)}_{+2})_m}{(\Phi^{(h_2),abc}_{+\frac{1}{2}})_n}&=&
\kappa_{+2,+\frac{1}{2},-\frac{1}{2}}\,\Big((h_2-\tfrac{1}{2})m-(h_1+1)n\Big)\,(\Phi^{(h_1+h_2),abc}_{+\frac{1}{2}})_{m+n} \, ,
\nonu
\\
\bf 
\comm{(\Phi^{(h_1)}_{+2})_m}{(\Phi^{(h_2),rst}_{+\frac{1}{2}})_n}&=&
\bf
\kappa_{+2,+\frac{1}{2},-\frac{1}{2}}\,\Big((h_2-\tfrac{1}{2})m-(h_1+1)n\Big)\,(\Phi^{(h_1+h_2),rst}_{+\frac{1}{2}})_{m+n} \, ,
\nonu
\\
\comm{(\Phi^{(h_1)}_{+2})_m}{(\Phi^{(h_2)}_{abc,\,-\frac{1}{2}})_n}&=&
\kappa_{+2,-\frac{1}{2},+\frac{1}{2}}\,\Big((h_2-\tfrac{3}{2})m-(h_1+1)n\Big)\,(\Phi^{(h_1+h_2)}_{abc,\,-\frac{1}{2}})_{m+n} \, ,
\nonu
\\
\bf
\comm{(\Phi^{(h_1)}_{+2})_m}{(\Phi^{(h_2)}_{rst,\,-\frac{1}{2}})_n}&=&
\bf
\kappa_{+2,-\frac{1}{2},+\frac{1}{2}}\,\Big((h_2-\tfrac{3}{2})m-(h_1+1)n\Big)\,(\Phi^{(h_1+h_2)}_{rst,\,-\frac{1}{2}})_{m+n} \, ,
\nonu \\
%
%
\comm{(\Phi^{(h_1)}_{+2})_m}{(\Phi^{(h_2),abcd}_{0})_n}&=&
\kappa_{+2,0,0}\,\Big((h_2-1)m-(h_1+1)n\Big)\,(\Phi^{(h_1+h_2),abcd}_{0})_{m+n} \, ,
\nonu
\\
\comm{(\Phi^{(h_1)}_{+2})_m}{(\Phi^{(h_2),rstu}_{0})_n}&=&
\kappa_{+2,0,0}\,\Big((h_2-1)m-(h_1+1)n\Big)\,(\Phi^{(h_1+h_2),rstu}_{0})_{m+n} \, ,
\nonu \\
%
%
\acomm{(\Phi^{(h_1),a}_{+\frac{3}{2}})_m}{(\Phi^{(h_2),b}_{+\frac{3}{2}})_n}
&= & \kappa_{+\frac{3}{2},+\frac{3}{2},-1}\,\Big((h_2+\tfrac{1}{2})m-(h_1+\tfrac{1}{2})n\Big)\,(\Phi^{(h_1+h_2),ab}_{+1})_{m+n} \, ,
\nonu\\
\bf
\acomm{(\Phi^{(h_1),r}_{+\frac{3}{2}})_m}{(\Phi^{(h_2),s}_{+\frac{3}{2}})_n}
&= &
\bf
\kappa_{+\frac{3}{2},+\frac{3}{2},-1}\,\Big((h_2+\tfrac{1}{2})m-(h_1+\tfrac{1}{2})n\Big)\,(\Phi^{(h_1+h_2),rs}_{+1})_{m+n} \, ,
\nonu\\
\acomm{(\Phi^{(h_1),a}_{+\frac{3}{2}})_m}{(\Phi^{(h_2)}_{b,\,-\frac{3}{2}})_n}
&=&\kappa_{+\frac{3}{2},-\frac{3}{2},+2}\,\Big((h_2-\tfrac{5}{2})m-(h_1+\tfrac{1}{2})n\Big)\delta^{a}_{b}\,(\Phi^{(h_1+h_2)}_{-2})_{m+n} \, ,
\nonu\\
\bf
\acomm{(\Phi^{(h_1),r}_{+\frac{3}{2}})_m}{(\Phi^{(h_2)}_{s,\,-\frac{3}{2}})_n}
& = &
\bf
\kappa_{+\frac{3}{2},-\frac{3}{2},+2}\,\Big((h_2-\tfrac{5}{2})m-(h_1+\tfrac{1}{2})n\Big)\delta^{r}_{s}\,(\Phi^{(h_1+h_2)}_{-2})_{m+n} \, ,
\nonu \\
%
%
\comm{(\Phi^{(h_1),a}_{+\frac{3}{2}})_m}{(\Phi^{(h_2),bc}_{+1})_n}&=&\kappa_{+\frac{3}{2},+1,-\frac{1}{2}}\,\Big(h_2\,m-(h_1+\tfrac{1}{2})n\Big)\,(\Phi^{(h_1+h_2),abc}_{+\frac{1}{2}})_{m+n} \, ,
\nonu\\
\bf
\comm{(\Phi^{(h_1),r}_{+\frac{3}{2}})_m}{(\Phi^{(h_2),st}_{+1})_n}&=&
\bf
\kappa_{+\frac{3}{2},+1,-\frac{1}{2}}\,\Big(h_2\,m-(h_1+\tfrac{1}{2})n\Big)\,(\Phi^{(h_1+h_2),rst}_{+\frac{1}{2}})_{m+n} \, ,
\nonu\\
\comm{(\Phi^{(h_1),a}_{+\frac{3}{2}})_m}{(\Phi^{(h_2)}_{bc,\,-1})_n}&=&
\kappa_{+\frac{3}{2},-1,+\frac{3}{2}}\,\Big((h_2-2)m-(h_1+\tfrac{1}{2})n\Big)\,
2!\, \delta^a_{[b}(\Phi^{(h_1+h_2)}_{c],\,-\frac{3}{2}})_{m+n} \, ,
\nonu\\
\bf 
\comm{(\Phi^{(h_1),r}_{+\frac{3}{2}})_m}{(\Phi^{(h_2)}_{st,\,-1})_n}&=&
\bf
\kappa_{+\frac{3}{2},-1,+\frac{3}{2}}\,\Big((h_2-2)m-(h_1+\tfrac{1}{2})n\Big)\,
2!\, \delta^r_{[s}(\Phi^{(h_1+h_2)}_{t],\,-\frac{3}{2}})_{m+n} \, ,
\nonu \\
%
%
\acomm{(\Phi^{(h_1),a}_{+\frac{3}{2}})_m}{(\Phi^{(h_2),bcd}_{+\frac{1}{2}})_n}
& = & \kappa_{+\frac{3}{2},+\frac{1}{2},0}\,\Big((h_2-\tfrac{1}{2})m-(h_1+\tfrac{1}{2})n\Big)\,(\Phi^{(h_1+h_2),abcd}_{0})_{m+n} \, ,
\nonu\\
\bf
\acomm{(\Phi^{(h_1),r}_{+\frac{3}{2}})_m}{(\Phi^{(h_2),stu}_{+\frac{1}{2}})_n}
& = &
\bf
\kappa_{+\frac{3}{2},+\frac{1}{2},0}\,\Big((h_2-\tfrac{1}{2})m-(h_1+\tfrac{1}{2})n\Big)\,(\Phi^{(h_1+h_2),rstu}_{0})_{m+n} \, ,
\nonu\\
\acomm{(\Phi^{(h_1),a}_{+\frac{3}{2}})_m}{(\Phi^{(h_2)}_{bcd,\,-\frac{1}{2}})_n}
&= & \kappa_{+\frac{3}{2},-\frac{1}{2},+1}\,\Big((h_2-\tfrac{3}{2})m-(h_1+\tfrac{1}{2})n\Big)\,3\delta^{a}_{[b}(\Phi^{(h_1+h_2)}_{cd],\,-1})_{m+n} \, ,
\nonu\\
\bf
\acomm{(\Phi^{(h_1),r}_{+\frac{3}{2}})_m}{(\Phi^{(h_2)}_{stu,\,-\frac{1}{2}})_n}
& = &
\bf
\kappa_{+\frac{3}{2},-\frac{1}{2},+1}\,\Big((h_2-\tfrac{3}{2})m-(h_1+\tfrac{1}{2})n\Big)\,3\delta^{r}_{[s}(\Phi^{(h_1+h_2)}_{tu],\,-1})_{m+n} \, ,
\nonu \\
%
%
%
\comm{(\Phi^{(h_1),a}_{+\frac{3}{2}})_m}{(\Phi^{(h_2),rstu}_{0})_n}&=&
\kappa_{+\frac{3}{2},0,+\frac{1}{2}}\,\Big((h_2-1)m-(h_1+\tfrac{1}{2})n\Big)\,\frac{1}{3!}\,\epsilon^{abcd}\epsilon^{rstu}(\Phi^{(h_1+h_2)}_{bcd,\,-\frac{1}{2}})_{m+n} \, ,
\nonu\\
\bf
\comm{(\Phi^{(h_1),r}_{+\frac{3}{2}})_m}{(\Phi^{(h_2),abcd}_{0})_n}&=&
\bf
\kappa_{+\frac{3}{2},0,+\frac{1}{2}}\,\Big((h_2-1)m-(h_1+\tfrac{1}{2})n\Big)\,\frac{1}{3!}\,\epsilon^{abcd}\epsilon^{rstu}\nonu \\
& \times & \bf (\Phi^{(h_1+h_2)}_{stu,\,-\frac{1}{2}})_{m+n} \, ,
\nonu\\
%
%
%
\comm{(\Phi^{(h_1),ab}_{+1})_m}{(\Phi^{(h_2),cd}_{+1})_n}&=&
\kappa_{+1,+1,0}\,\Big(h_2\,m-h_1\,n\Big)\,(\Phi^{(h_1+h_2),abcd}_{0})_{m+n}\, ,
\nonu\\
\bf 
\comm{(\Phi^{(h_1),rs}_{+1})_m}{(\Phi^{(h_2),tu}_{+1})_n}&=&
\bf
\kappa_{+1,+1,0}\,\Big(h_2\,m-h_1\,n\Big)\,(\Phi^{(h_1+h_2),rstu}_{0})_{m+n}
\, , \nonu\\
\comm{(\Phi^{(h_1),ab}_{+1})_m}{(\Phi^{(h_2)}_{cd,\,-1})_n}&=&
\kappa_{+1,-1,+2}\,\Big((h_2-2)m-h_1\,n\Big)\,\delta^{ab}_{cd}\,(\Phi^{(h_1+h_2)}_{-2})_{m+n} \, ,
\nonu\\
\bf
\comm{(\Phi^{(h_1),rs}_{+1})_m}{(\Phi^{(h_2)}_{tu,\,-1})_n}&=&
\bf
\kappa_{+1,-1,+2}\,\Big((h_2-2)m-h_1\,n\Big)\,\delta^{rs}_{tu}\,(\Phi^{(h_1+h_2)}_{-2})_{m+n} \, ,
\nonu \\
\comm{(\Phi^{(h_1),ab}_{+1})_m}{(\Phi^{(h_2)}_{cde,\,-\frac{1}{2}})_n}&=&
\kappa_{+1,-\frac{1}{2},+\frac{3}{2}}\,\Big((h_2-\tfrac{3}{2})m-h_1\,n\Big)\,
3!\delta^{a}_{[c}\,(\Phi^{(h_1+h_2)}_{d,\,-\frac{3}{2}})_{m+n}\delta^b_{e]}\, ,
\nonu\\
\bf
\comm{(\Phi^{(h_1),rs}_{+1})_m}{(\Phi^{(h_2)}_{tuv,\,-\frac{1}{2}})_n}&=&
\bf
\kappa_{+1,-\frac{1}{2},+\frac{3}{2}}\,\Big((h_2-\tfrac{3}{2})m-h_1\,n\Big)\,
3! \delta^{r}_{[t}\,(\Phi^{(h_1+h_2)}_{u,\,-\frac{3}{2}})_{m+n}\delta^s_{v]} \, ,
\nonu \\
%
%
%
\comm{(\Phi^{(h_1),ab}_{+1})_m}{(\Phi^{(h_2),rstu}_{0})_n}&=&
\kappa_{+1,0,+1}\,\Big((h_2-1)m-h_1\,n\Big)\,\epsilon^{abcd}\epsilon^{rstu}\,\frac{1}{2!}\,(\Phi^{(h_1+h_2)}_{cd,\,-1})_{m+n} \, ,
\nonu\\
\bf
\comm{(\Phi^{(h_1),rs}_{+1})_m}{(\Phi^{(h_2),abcd}_{0})_n}&=&
\bf
\kappa_{+1,0,+1}\,\Big((h_2-1)m-h_1\,n\Big)\,\epsilon^{abcd}\epsilon^{rstu}\,\frac{1}{2!}\,(\Phi^{(h_1+h_2)}_{tu,\,-1})_{m+n} \, ,
\nonu\\
\acomm{(\Phi^{(h_1),abc}_{+\frac{1}{2}})_m}{(\Phi^{(h_2)}_{def,\,-\frac{1}{2}})_n}&=&\kappa_{+\frac{1}{2},-\frac{1}{2},+2}\,\Big((h_2-\tfrac{3}{2})m-(h_1-\tfrac{1}{2})n\Big)\,\delta^{abc}_{def}\,(\Phi^{(h_1+h_2)}_{-2})_{m+n} \, ,
\nonu\\
\bf
\acomm{(\Phi^{(h_1),rst}_{+\frac{1}{2}})_m}{(\Phi^{(h_2)}_{uvw,\,-\frac{1}{2}})_n}&=&
\bf
\kappa_{+\frac{1}{2},-\frac{1}{2},+2}\,\Big((h_2-\tfrac{3}{2})m-(h_1-\tfrac{1}{2})n\Big)\,\delta^{rst}_{uvw}\,(\Phi^{(h_1+h_2)}_{-2})_{m+n}
\, ,
\nonu \\
%
%
%
%
\comm{(\Phi^{(h_1),abc}_{+\frac{1}{2}})_m}{(\Phi^{(h_2),rstu}_{0})_n}&=&
\kappa_{+\frac{1}{2},0,+\frac{3}{2}}\,\Big((h_2-1)m-(h_1-\tfrac{1}{2})n\Big)\,\epsilon^{abcd}\epsilon^{rstu}\,(\Phi^{(h_1+h_2)}_{d,\,-\frac{3}{2}})_{m+n}
\, ,
\nonu\\
\bf
\comm{(\Phi^{(h_1),rst}_{+\frac{1}{2}})_m}{(\Phi^{(h_2),abcd}_{0})_n}&=&
\bf
\kappa_{+\frac{1}{2},0,+\frac{3}{2}}\,\Big((h_2-1)m-(h_1-\tfrac{1}{2})n\Big)\,\epsilon^{abcd}\epsilon^{rstu}\,(\Phi^{(h_1+h_2)}_{u,\,-\frac{3}{2}})_{m+n}
\, , \nonu\\
%
%
%
%
\comm{(\Phi^{(h_1),abcd}_{0})_m}{(\Phi^{(h_2),rstu}_{0})_n}&=&
\kappa_{0,0,+2}\,\Big((h_2-1)m-(h_1-1)n\Big)\,\epsilon^{abcd}\epsilon^{rstu}\,(\Phi^{(h_1+h_2)}_{-2})_{m+n}
\, .
\nonu \\
\label{su4result}
\eea
There are two ${\cal N}=4$ celestial soft symmetry algebras
related to the two $SU(4)$ symmetries (without a boldface notation
and with a boldface notation) in (\ref{su4result})
\footnote{ The tensor product of $SU(4)$
\cite{FKS}
in the lower representations is given by
\bea
{\bf 4} \otimes {\bf 4} &=& {\bf 6} \oplus \cdots,
\qquad
{\bf 4} \otimes {\bf 6} = \overline{\bf 4} \oplus \cdots,    
\qquad
{\bf 4} \otimes \overline{\bf 4} = {\bf 1} \oplus \cdots,
\qquad
{\bf 6} \otimes {\bf 6} = {\bf 1} \oplus \cdots,
\nonu \\
{\bf 6} \otimes \overline{\bf 4} &=& {\bf 4} \oplus \cdots.
\nonu
\eea
As described in (\ref{su4branching}), for example,
the singlets in the second $SU(4)$ are given by
the gravitinos (${\bf 4}$ and $\overline{\bf 4}$),
the graviphotons (${\bf 6}$ and ${\bf 6}$), the graviphotinos
($\overline{\bf 4}$ and ${\bf 4}$), and the scalars (${\bf 1}$ and ${\bf 1}$)
as well as the graviton (${\bf 1}$).
The ${\cal N}=4$ supergravity is realized by a double copy
of the ${\cal N}=4$ super Yang-Mills theory and
the ${\cal N}=0$ nonsupersymmetric Yang-Mills theory.}.
Note that the first two, the fifteenth, the sixteenth
and the last commutators are common to both cases.
These commutators should appear in both symmetry algebras.
For the first case where the operators
contain the indices $a,b,c,d, \cdots =
1,2,3,4$, the following scalar is nonvanishing
\bea
\Phi_{0}^{(h),5678} \neq 0,
\label{5678}
\eea
and for the second case
where the operators
contain the indices $r,s,t,u, \cdots =
5,6,7,8$,
the scalar having the indices $1234$ is nonzero 
\bea
\Phi_{0}^{(h),1234} \neq 0\, .
\label{1234}
\eea
In \cite{AK2501}, the helicity of (\ref{5678}) is denoted by
$-0$ while the helicity of (\ref{1234}) is denoted by
$+0$. The scalar and pseudoscalar of ${\cal N}=4$ supergravity
are linear combination of these with $\pm 0$ helicities.



\begin{thebibliography}{99}

\bibitem{Ademolloplb}
M.~Ademollo, L.~Brink, A.~D'Adda, R.~D'Auria, E.~Napolitano, S.~Sciuto, E.~Del Giudice, P.~Di Vecchia, S.~Ferrara and F.~Gliozzi, \textit{et al.}
``Supersymmetric Strings and Color Confinement,''
Phys. Lett. B \textbf{62}, 105-110 (1976)
doi:10.1016/0370-2693(76)90061-7

\bibitem{PRSY}
M.~Pate, A.~M.~Raclariu, A.~Strominger and E.~Y.~Yuan,
``Celestial operator products of gluons and gravitons,''
Rev. Math. Phys. \textbf{33}, no.09, 2140003 (2021)
doi:10.1142/S0129055X21400031
[arXiv:1910.07424 [hep-th]].

\bibitem{GHPS}
A.~Guevara, E.~Himwich, M.~Pate and A.~Strominger,
``Holographic symmetry algebras for gauge theory and gravity,''
 JHEP \textbf{11}, 152 (2021)
doi:10.1007/JHEP11(2021)152
[arXiv:2103.03961 [hep-th]].

\bibitem{Strominger}
A.~Strominger,
``$w_{1+\infty}$ Algebra and the Celestial Sphere: Infinite Towers of Soft Graviton, Photon, and Gluon Symmetries,''
Phys. Rev. Lett. \textbf{127}, no.22, 221601 (2021)
doi:10.1103/PhysRevLett.127.221601
[arXiv:2105.14346 [hep-th]].
  
\bibitem{Bakas}
I.~Bakas,
``The Large n Limit of Extended Conformal Symmetries,''
Phys. Lett. B \textbf{228}, 57 (1989)
doi:10.1016/0370-2693(89)90525-X

\bibitem{FSTZ}
A.~Fotopoulos, S.~Stieberger, T.~R.~Taylor and B.~Zhu,
``Extended Super BMS Algebra of Celestial CFT,''
JHEP \textbf{09}, 198 (2020)
doi:10.1007/JHEP09(2020)198
[arXiv:2007.03785 [hep-th]].

\bibitem{Ahn2111}
C.~Ahn,
``Towards a supersymmetric w1+\ensuremath{\infty} symmetry in the celestial conformal field theory,''
Phys. Rev. D \textbf{105}, no.8, 086028 (2022)
doi:10.1103/PhysRevD.105.086028
[arXiv:2111.04268 [hep-th]].

\bibitem{PRSS}
C.~N.~Pope, L.~J.~Romans, E.~Sezgin and X.~Shen,
``W topological matter and gravity,''
Phys. Lett. B \textbf{256}, 191-198 (1991)
doi:10.1016/0370-2693(91)90672-D

\bibitem{dF}
B.~de Wit and D.~Z.~Freedman,
``On SO(8) Extended Supergravity,''
Nucl. Phys. B \textbf{130}, 105-113 (1977)
doi:10.1016/0550-3213(77)90395-9

\bibitem{CJnpb}
E.~Cremmer and B.~Julia,
``The SO(8) Supergravity,''
Nucl. Phys. B \textbf{159}, 141-212 (1979)
doi:10.1016/0550-3213(79)90331-6

\bibitem{CJplb}
E.~Cremmer and B.~Julia,
``The N=8 Supergravity Theory. 1. The Lagrangian,''
Phys. Lett. B \textbf{80}, 48 (1978)
doi:10.1016/0370-2693(78)90303-9

\bibitem{2212-2}
N.~Banerjee, T.~Rahnuma and R.~K.~Singh,
``Asymptotic symmetry algebra of N=8 supergravity,''
Phys. Rev. D \textbf{109}, no.4, 046010 (2024)
doi:10.1103/PhysRevD.109.046010
[arXiv:2212.12133 [hep-th]].

\bibitem{BSS}
L.~Brink, J.~H.~Schwarz and J.~Scherk,
``Supersymmetric Yang-Mills Theories,''
Nucl. Phys. B \textbf{121}, 77-92 (1977)
doi:10.1016/0550-3213(77)90328-5

\bibitem{2212-1}
N.~Banerjee, T.~Rahnuma and R.~K.~Singh,
``Soft and collinear limits in $ \mathcal{N} $ = 8 supergravity using double copy formalism,''
JHEP \textbf{04}, 126 (2023)
doi:10.1007/JHEP04(2023)126
[arXiv:2212.11480 [hep-th]].

\bibitem{Tropper}
A.~Tropper,
``Supersymmetric Soft Theorems,''
[arXiv:2404.03717 [hep-th]].

\bibitem{deWit}
B.~de Wit,
``Properties of SO(8) Extended Supergravity,''
Nucl. Phys. B \textbf{158}, 189-212 (1979)
doi:10.1016/0550-3213(79)90195-0

\bibitem{Prabhu}
K.~Prabhu,
``Novel supersymmetric extension of BMS symmetries at null infinity,''
Phys. Rev. D \textbf{105}, no.6, 064054 (2022)
doi:10.1103/PhysRevD.105.064054
[arXiv:2112.07186 [gr-qc]].

\bibitem{BvM}
H.~Bondi, M.~G.~J.~van der Burg and A.~W.~K.~Metzner,
``Gravitational waves in general relativity. 7. Waves from axisymmetric isolated systems,''
Proc. Roy. Soc. Lond. A \textbf{269}, 21-52 (1962)
doi:10.1098/rspa.1962.0161

\bibitem{Sachs}
R.~K.~Sachs,
``Gravitational waves in general relativity. 8. Waves in asymptotically flat space-times,''
Proc. Roy. Soc. Lond. A \textbf{270}, 103-126 (1962)
doi:10.1098/rspa.1962.0206

\bibitem{Raclariu}
A.~M.~Raclariu,
``Lectures on Celestial Holography,''
[arXiv:2107.02075 [hep-th]].

\bibitem{Pasterski}
S.~Pasterski,
``Lectures on celestial amplitudes,''
Eur. Phys. J. C \textbf{81}, no.12, 1062 (2021)
doi:10.1140/epjc/s10052-021-09846-7
[arXiv:2108.04801 [hep-th]].

\bibitem{PPR}
S.~Pasterski, M.~Pate and A.~M.~Raclariu,
``Celestial Holography,''
[arXiv:2111.11392 [hep-th]].

\bibitem{Donnay}
L.~Donnay,
``Celestial holography: An asymptotic symmetry perspective,''
Phys. Rept. \textbf{1073}, 1-41 (2024)
doi:10.1016/j.physrep.2024.04.003
[arXiv:2310.12922 [hep-th]].

\bibitem{Strominger1703}
A.~Strominger,
``Lectures on the Infrared Structure of Gravity and Gauge Theory,''
[arXiv:1703.05448 [hep-th]].

\bibitem{Thielemans}
K.~Thielemans,
``A Mathematica package for computing operator product expansions,''
Int. J. Mod. Phys. C \textbf{2}, 787-798 (1991)
doi:10.1142/S0129183191001001

\bibitem{mathematica}
  Wolfram Research, Inc., Mathematica, Version 13.0.0, Champaign, IL
(2021).

\bibitem{MPR}
T.~McLoughlin, A.~Puhm and A.~M.~Raclariu,
``The SAGEX review on scattering amplitudes chapter 11: soft theorems and celestial amplitudes,''
J. Phys. A \textbf{55}, no.44, 443012 (2022)
doi:10.1088/1751-8121/ac9a40
[arXiv:2203.13022 [hep-th]].
  
\bibitem{HPS}
E.~Himwich, M.~Pate and K.~Singh,
``Celestial operator product expansions and w$_{1+\infty}$
symmetry for all spins,''
JHEP \textbf{01}, 080 (2022)
doi:10.1007/JHEP01(2022)080
[arXiv:2108.07763 [hep-th]].

\bibitem{EF}
H.~Elvang and D.~Z.~Freedman,
``Note on graviton MHV amplitudes,''
JHEP \textbf{05}, 096 (2008)
doi:10.1088/1126-6708/2008/05/096
[arXiv:0710.1270 [hep-th]].

\bibitem{PT}
S.~J.~Parke and T.~R.~Taylor,
``An Amplitude for $n$ Gluon Scattering,''
Phys. Rev. Lett. \textbf{56}, 2459 (1986)
doi:10.1103/PhysRevLett.56.2459

\bibitem{KLT}
H.~Kawai, D.~C.~Lewellen and S.~H.~H.~Tye,
``A Relation Between Tree Amplitudes of Closed and Open Strings,''
Nucl. Phys. B \textbf{269}, 1-23 (1986)
doi:10.1016/0550-3213(86)90362-7

\bibitem{Puhm}
A.~Puhm,
``Conformally Soft Theorem in Gravity,''
JHEP \textbf{09}, 130 (2020)
doi:10.1007/JHEP09(2020)130
[arXiv:1905.09799 [hep-th]].

\bibitem{ST}
S.~Stieberger and T.~R.~Taylor,
``Strings on Celestial Sphere,''
Nucl. Phys. B \textbf{935}, 388-411 (2018)
doi:10.1016/j.nuclphysb.2018.08.019
[arXiv:1806.05688 [hep-th]].

\bibitem{Weinberg}
S.~Weinberg,
``Infrared photons and gravitons,''
Phys. Rev. \textbf{140}, B516-B524 (1965)
doi:10.1103/PhysRev.140.B516

\bibitem{BDPR}
Z.~Bern, L.~J.~Dixon, M.~Perelstein and J.~S.~Rozowsky,
``Multileg one loop gravity amplitudes from gauge theory,''
Nucl. Phys. B \textbf{546}, 423-479 (1999)
doi:10.1016/S0550-3213(99)00029-2
[arXiv:hep-th/9811140 [hep-th]].

\bibitem{ADHPZ}
S.~Abreu, L.~J.~Dixon, E.~Herrmann, B.~Page and M.~Zeng,
``The two-loop five-point amplitude in $ \mathcal{N} $ = 8 supergravity,''
JHEP \textbf{03}, 123 (2019)
doi:10.1007/JHEP03(2019)123
[arXiv:1901.08563 [hep-th]].


\bibitem{Dixon9601}
L.~J.~Dixon,
``Calculating scattering amplitudes efficiently,''
[arXiv:hep-ph/9601359 [hep-ph]].

\bibitem{KLR}
R.~Kallosh, C.~H.~Lee and T.~Rube,
``N=8 Supergravity 4-point Amplitudes,''
JHEP \textbf{02}, 050 (2009)
doi:10.1088/1126-6708/2009/02/050
[arXiv:0811.3417 [hep-th]].

\bibitem{BEF}
M.~Bianchi, H.~Elvang and D.~Z.~Freedman,
``Generating Tree Amplitudes in N=4 SYM and N = 8 SG,''
JHEP \textbf{09}, 063 (2008)
doi:10.1088/1126-6708/2008/09/063
[arXiv:0805.0757 [hep-th]].

\bibitem{Liu}
Z.~W.~Liu,
``Soft theorems in maximally supersymmetric theories,''
Eur. Phys. J. C \textbf{75}, no.3, 105 (2015)
doi:10.1140/epjc/s10052-015-3304-1
[arXiv:1410.1616 [hep-th]].

\bibitem{AHHH}
N.~Arkani-Hamed, T.~C.~Huang and Y.~t.~Huang,
``Scattering amplitudes for all masses and spins,''
JHEP \textbf{11}, 070 (2021)
doi:10.1007/JHEP11(2021)070
[arXiv:1709.04891 [hep-th]].

\bibitem{BHP}
A.~Ball, Y.~Hu and S.~Pasterski,
``Multicollinear singularities in celestial CFT,''
JHEP \textbf{02}, 219 (2024)
doi:10.1007/JHEP02(2024)219
[arXiv:2309.16602 [hep-th]].

\bibitem{GHP}
A.~Guevara, Y.~Hu and S.~Pasterski,
``Multiparticle contributions to the celestial OPE,''
JHEP \textbf{07}, 178 (2025)
doi:10.1007/JHEP07(2025)178
[arXiv:2402.18798 [hep-th]].

\bibitem{Pate}
M.~Calkins and M.~Pate,
``Multi-particle Celestial Operator Product
Expansions from the Boundary,''
[arXiv:2601.04329 [hep-th]].

\bibitem{YangLee}
C.~N.~Yang and T.~D.~Lee,
``Statistical theory of equations of state and phase transitions. 1. Theory of condensation,''
Phys. Rev. \textbf{87}, 404-409 (1952)
doi:10.1103/PhysRev.87.404

\bibitem{YangLee1}
T.~D.~Lee and C.~N.~Yang,
``Statistical theory of equations of state and phase transitions. 2. Lattice gas and Ising model,''
Phys. Rev. \textbf{87}, 410-419 (1952)
doi:10.1103/PhysRev.87.410

\bibitem{Cardy}
J.~L.~Cardy,
``Conformal Invariance and the Yang-lee Edge Singularity in Two-dimensions,''
Phys. Rev. Lett. \textbf{54}, 1354-1356 (1985)
doi:10.1103/PhysRevLett.54.1354

\bibitem{Fisher}
M.~E.~Fisher,
``Yang-Lee Edge Singularity and phi**3 Field Theory,''
Phys. Rev. Lett. \textbf{40}, 1610-1613 (1978)
doi:10.1103/PhysRevLett.40.1610

\bibitem{Jiang2105}
H.~Jiang,
``Celestial superamplitude in $ \mathcal{N} $ = 4 SYM theory,''
JHEP \textbf{08}, 031 (2021)
doi:10.1007/JHEP08(2021)031
[arXiv:2105.10269 [hep-th]].

\bibitem{Tropper1}
A.~Tropper,
``Symmetries of the Celestial Supersphere,''
[arXiv:2412.13113 [hep-th]].

\bibitem{FKS}
R.~Feger, T.~W.~Kephart and R.~J.~Saskowski,
``LieART 2.0 {\textendash} A Mathematica application for Lie Algebras and Representation Theory,''
Comput. Phys. Commun. \textbf{257}, 107490 (2020)
doi:10.1016/j.cpc.2020.107490
[arXiv:1912.10969 [hep-th]].

\bibitem{AK2501}
C.~Ahn and M.~H.~Kim,
``A supersymmetric $w_{1+\infty }$ symmetry, the extended supergravity and the celestial holography,''
Eur. Phys. J. C \textbf{85}, no.8, 834 (2025)
doi:10.1140/epjc/s10052-025-14479-1
[arXiv:2501.11471 [hep-th]].

\bibitem{Das}
A.~K.~Das,
``SO(4) Invariant Extended Supergravity,''
Phys. Rev. D \textbf{15}, 2805 (1977)
doi:10.1103/PhysRevD.15.2805

\bibitem{CS1}
E.~Cremmer and J.~Scherk,
``Modified Interaction of the Scalar Multiplet Coupled to Supergravity,''
Phys. Lett. B \textbf{69}, 97-100 (1977)
doi:10.1016/0370-2693(77)90142-3

\bibitem{CS2}
E.~Cremmer and J.~Scherk,
``Algebraic Simplifications in Supergravity Theories,''
Nucl. Phys. B \textbf{127}, 259-268 (1977)
doi:10.1016/0550-3213(77)90214-0

\bibitem{CSF}
E.~Cremmer, J.~Scherk and S.~Ferrara,
``SU(4) Invariant Supergravity Theory,''
Phys. Lett. B \textbf{74}, 61-64 (1978)
doi:10.1016/0370-2693(78)90060-6

\bibitem{AK2407}
C.~Ahn and M.~H.~Kim,
``A supersymmetric extension of w$_{1+\infty}$
algebra in the celestial holography,''
JHEP \textbf{09}, 081 (2024)
doi:10.1007/JHEP09(2024)081
[arXiv:2407.05601 [hep-th]].

\bibitem{BBJ}
Z.~Bern, C.~Boucher-Veronneau and H.~Johansson,
``N \ensuremath{>}= 4 Supergravity Amplitudes from Gauge Theory at One Loop,''
Phys. Rev. D \textbf{84}, 105035 (2011)
doi:10.1103/PhysRevD.84.105035
[arXiv:1107.1935 [hep-th]].

\bibitem{MRSV}
J.~Mago, L.~Ren, A.~Y.~Srikant and A.~Volovich,
``Deformed $w_{1+\infty}$ Algebras in the Celestial CFT,''
SIGMA \textbf{19}, 044 (2023)
doi:10.3842/SIGMA.2023.044
[arXiv:2111.11356 [hep-th]].

\bibitem{Jiang2108}
H.~Jiang,
``Holographic chiral algebra: supersymmetry, infinite Ward identities, and EFTs,''
JHEP \textbf{01}, 113 (2022)
doi:10.1007/JHEP01(2022)113
[arXiv:2108.08799 [hep-th]].

\bibitem{AMS}
T.~Adamo, L.~Mason and A.~Sharma,
``Celestial $w_{1+\infty}$ Symmetries from Twistor Space,''
SIGMA \textbf{18}, 016 (2022)
doi:10.3842/SIGMA.2022.016
[arXiv:2110.06066 [hep-th]].

\bibitem{DeWitt}
Bryce S.~DeWitt,
``Quantum Theory of Gravity. III. Applications of
the Covariant Theory,''
Phys. Rev.  \textbf{162}, no.5, 1239 (1967)
doi:10.1103/PhysRev.162.1239

\bibitem{BG}
F.~A.~Berends and R.~Gastmans,
``On the High-Energy Behavior in Quantum Gravity,''
Nucl. Phys. B \textbf{88}, 99-108 (1975)
doi:10.1016/0550-3213(75)90528-3

\bibitem{Woodard}
R.~P.~Woodard,
``The Vierbein Is Irrelevant in Perturbation Theory,''
Phys. Lett. B \textbf{148}, 440-444 (1984)
doi:10.1016/0370-2693(84)90734-2

\bibitem{CSS}
S.~Y.~Choi, J.~S.~Shim and H.~S.~Song,
``Factorization and polarization in linearized gravity,''
Phys. Rev. D \textbf{51}, 2751-2769 (1995)
doi:10.1103/PhysRevD.51.2751
[arXiv:hep-th/9411092 [hep-th]].

\bibitem{EJN}
H.~Elvang, C.~R.~T.~Jones and S.~G.~Naculich,
``Soft Photon and Graviton Theorems in Effective Field Theory,''
Phys. Rev. Lett. \textbf{118}, no.23, 231601 (2017)
doi:10.1103/PhysRevLett.118.231601
[arXiv:1611.07534 [hep-th]].

\bibitem{RSYV}
L.~Ren, M.~Spradlin, A.~Yelleshpur Srikant and A.~Volovich,
``On effective field theories with celestial duals,''
JHEP \textbf{08}, 251 (2022)
doi:10.1007/JHEP08(2022)251
[arXiv:2206.08322 [hep-th]].

\bibitem{AK2309}
C.~Ahn and M.~H.~Kim,
``The $ \mathcal{N} $ = 2, 4 supersymmetric linear
W$_{\infty}$[{\ensuremath{\lambda}}] algebras for
generic {\ensuremath{\lambda}} parameter,''
JHEP \textbf{02}, 006 (2024)
doi:10.1007/JHEP02(2024)006
[arXiv:2309.01537 [hep-th]].

\bibitem{Ahn2203}
C.~Ahn,
``The $ \mathcal{N} $ = 2 supersymmetric w$_{1+\infty}$
symmetry in the two-dimensional SYK models,''
JHEP \textbf{05}, 115 (2022)
doi:10.1007/JHEP05(2022)115
[arXiv:2203.03105 [hep-th]].

\bibitem{Ahn2202}
C.~Ahn,
``A deformed supersymmetric $w_{1+\infty }$ symmetry in the celestial conformal field theory,''
Eur. Phys. J. C \textbf{82}, no.7, 630 (2022)
doi:10.1140/epjc/s10052-022-10582-9
[arXiv:2202.02949 [hep-th]].


\bibitem{Virasoro}
M.~A.~Virasoro,
``Subsidiary conditions and ghosts in dual resonance models,''
Phys. Rev. D \textbf{1}, 2933-2936 (1970)
doi:10.1103/PhysRevD.1.2933

\bibitem{FV1}
S.~Fubini and G.~Veneziano,
``Algebraic treatment of subsidiary conditions in dual resonance models,''
Annals Phys. \textbf{63}, 12-27 (1971)
doi:10.1016/0003-4916(71)90295-8


\bibitem{NS}
A.~Neveu and J.~H.~Schwarz,
``Factorizable dual model of pions,''
Nucl. Phys. B \textbf{31}, 86-112 (1971)
doi:10.1016/0550-3213(71)90448-2

\bibitem{Ramond}
P.~Ramond,
``Dual Theory for Free Fermions,''
Phys. Rev. D \textbf{3}, 2415-2418 (1971)
doi:10.1103/PhysRevD.3.2415

\bibitem{Schwarz}
J.~H.~Schwarz,
``Dual resonance theory,''
Phys. Rept. \textbf{8}, 269-335 (1973)
doi:10.1016/0370-1573(73)90003-3
  
\bibitem{Ademollonpb}
M.~Ademollo, L.~Brink, A.~D'Adda, R.~D'Auria, E.~Napolitano, S.~Sciuto, E.~Del Giudice, P.~Di Vecchia, S.~Ferrara and F.~Gliozzi, \textit{et al.}
``Dual String with U(1) Color Symmetry,''
Nucl. Phys. B \textbf{111}, 77-110 (1976)
doi:10.1016/0550-3213(76)90483-1

\bibitem{RS}
P.~Ramond and J.~H.~Schwarz,
``Classification of Dual Model Gauge Algebras,''
Phys. Lett. B \textbf{64}, 75-77 (1976)
doi:10.1016/0370-2693(76)90361-0

\bibitem{Schoutens}
K.~Schoutens,
``O(n) Extended Superconformal Field Theory in Superspace,''
Nucl. Phys. B \textbf{295}, 634-652 (1988)
doi:10.1016/0550-3213(88)90539-1

\bibitem{STV}
A.~Sevrin, W.~Troost and A.~Van Proeyen,
``Superconformal Algebras in Two-Dimensions with N=4,''
Phys. Lett. B \textbf{208}, 447-450 (1988)
doi:10.1016/0370-2693(88)90645-4

\bibitem{PS}
S.~Pasterski and S.~H.~Shao,
``Conformal basis for flat space amplitudes,''
Phys. Rev. D \textbf{96}, no.6, 065022 (2017)
doi:10.1103/PhysRevD.96.065022
[arXiv:1705.01027 [hep-th]].

\bibitem{DPP1}
L.~Donnay, S.~Pasterski and A.~Puhm,
``Asymptotic Symmetries and Celestial CFT,''
JHEP \textbf{09}, 176 (2020)
doi:10.1007/JHEP09(2020)176
[arXiv:2005.08990 [hep-th]].

\bibitem{DPP2}
L.~Donnay, S.~Pasterski and A.~Puhm,
``Goldilocks modes and the three scattering bases,''
JHEP \textbf{06}, 124 (2022)
doi:10.1007/JHEP06(2022)124
[arXiv:2202.11127 [hep-th]].

\bibitem{CM}
S.~R.~Coleman and J.~Mandula,
``All Possible Symmetries of the S Matrix,''
Phys. Rev. \textbf{159}, 1251-1256 (1967)
doi:10.1103/PhysRev.159.1251

\bibitem{HLS}
R.~Haag, J.~T.~Lopuszanski and M.~Sohnius,
``All Possible Generators of Supersymmetries of the s Matrix,''
Nucl. Phys. B \textbf{88}, 257 (1975)
doi:10.1016/0550-3213(75)90279-5

\bibitem{MP}
M.~L.~Mangano and S.~J.~Parke,
``Multiparton amplitudes in gauge theories,''
Phys. Rept. \textbf{200}, 301-367 (1991)
doi:10.1016/0370-1573(91)90091-Y
[arXiv:hep-th/0509223 [hep-th]].

\bibitem{Dixon}
L.~J.~Dixon,
``Calculating scattering amplitudes efficiently,''
[arXiv:hep-ph/9601359 [hep-ph]].

\bibitem{BCJ}
Z.~Bern, J.~J.~M.~Carrasco and H.~Johansson,
``Perturbative Quantum Gravity as a Double Copy of Gauge Theory,''
Phys. Rev. Lett. \textbf{105}, 061602 (2010)
doi:10.1103/PhysRevLett.105.061602
[arXiv:1004.0476 [hep-th]].


\bibitem{SS}
A.~Salam and J.~A.~Strathdee,
``Supersymmetry and Nonabelian Gauges,''
Phys. Lett. B \textbf{51}, 353-355 (1974)
doi:10.1016/0370-2693(74)90226-3

\bibitem{FKMVY}
D.~Z.~Freedman, R.~Kallosh, D.~Murli, A.~Van Proeyen and Y.~Yamada,
``Absence of U(1) Anomalous Superamplitudes in $\mathcal{N}\geq 5$ Supergravities,''
JHEP \textbf{05}, 067 (2017)
doi:10.1007/JHEP05(2017)067
[arXiv:1703.03879 [hep-th]].

\bibitem{WZ}
J.~Wess and B.~Zumino,
``Supergauge Transformations in Four-Dimensions,''
Nucl. Phys. B \textbf{70}, 39-50 (1974)
doi:10.1016/0550-3213(74)90355-1

\end{thebibliography}
\end{document}